UNIVERSITÀ DEGLI STUDI ROMA TRE

Ph.D. in Physics

XXXVII$^{th}$ Cycle

*Ph.D. Thesis*

---

# Study of the response of a large volume TPC prototype for the CYGNO experiment at LNGS

---

Ph.D. candidate: Rita Antonietti

Adviser: Prof. Fabrizio Petrucci

# Contents

# Introduction

Astrophysical and cosmological observations have provided strong evidence that a significant portion of the Universe is composed of non-luminous matter, known as Dark Matter (DM). The Standard Model (SM) of particle physics successfully describes the fundamental particles and their interactions but does not account for any Dark Matter particle candidate. One of the most supported hypotheses suggests that DM is composed of particles different from those in the SM. Among the most promising candidates are Weakly Interactive Massive Particles (WIMPs), which are non-relativistic particles that interact weakly with SM particles. A stable, weakly interacting particle in thermal equilibrium in the early Universe would be capable of explaining the observed relic DM density.
The motion of the Sun and the Earth with respect to the rest frame of the Galaxy produces an apparent "wind" of WIMPs coming from the direction of the Cygnus constellation. Assuming DM can interact with SM particles, it is possible to exploit the weak interaction with regular matter measuring the recoils induced by DM interactions. This is known as direct detection. The fundamental strategy is to expose a large amount of instrumented mass and wait for DM to produce recoils in it. Direct detection experiments look for nuclear recoils of low energy, 1-100 keV, with an expected rate below 1 event/kg/year. The low expected event rate implies detectors with extremely challenging requirements on the background reduction techniques, such as operating detectors underground to suppress cosmic rays, using radiopure materials, and implementing active or passive shielding.
The CYGNO project aims to build a large O(30m$^3$) directional detector for rare event searches, including Dark Matter. The detector uses a gaseous Time Projection Chamber (TPC), which is sensitive to recoil topology and allows direction measurement. The TPC is filled with a gas mixture rich in helium and fluorine at atmospheric pressure, making it sensitive to both Spin-Independent and Spin-Dependent interactions. A triple Gas Electron Multiplier (GEM) stack provides ionization signal amplification, and the signals are optically read out using photomultiplier tubes (PMTs) and scientific CMOS cameras, enabling 3D track reconstruction. Several CYGNO prototypes have been built and tested in both overground and underground environments to assess their performance.
Currently, the CYGNO project is at the end of the R&D phase with the Long Imaging ModulE (LIME), the largest prototype built, comprising a 50 liters active volume. After commissioning overground at Laboratori Nazionali di Frascati (LNF), LIME was installed underground at Laboratori Nazionali del Gran Sasso (LNGS) in February 2022. This underground operation marks a significant milestone in the CYGNO roadmap towards constructing a large-scale TPC for directional Dark Matter searches. LIME is the first CYGNO prototype to be tested in a low-background environment.
This work has been carried out in the context of the CYGNO experiment with the goal of studying the calibration and stability of the response of the LIME prototype.

Chapter 1 provides an overview of the theoretical framework and experimental observations that have led to the hypothesis of Dark Matter. It also presents an overview of WIMP candidates and detection methods.

In Chapter 2, the WIMP-nucleus scattering is described in the context of the direct detection of DM. In addition a focus on the background sources and reduction techniques are reported.

In Chapter 3, the CYGNO project and the concept of the detector are described. The most important results obtained with several prototypes are shown and an outlook on the future prospect is reported.

In Chapter 4, the LIME prototype is described. Moreover a description of the clustering algorithm is provided.

Chapter 5 explores the effect of environmental variations on light yield, based on both overground and underground data.

In Chapter 6, a novel technique for the equalization of the data independently from the environmental condition is described. Results based on long data taking periods underground are presented.

Chapter 7 presents a preliminary study of Dark Matter detection limits with the LIME prototype.

Finally, the Conclusions summarize the results obtained from this work.

# 1 Physics - Dark Matter

Over the past century, astrophysical and cosmological observations have suggested that something is missing in the description of the universe. The Standard Model (SM) has several limitations, including its inability to incorporate gravity and its inadequate explanation for phenomena such as neutrino oscillations and the lack of strong CP violation. Addition, the SM does not account for a large amount of non visible matter, known as Dark Matter (DM), which spans all the scales of the Universe, from single galaxies to the largest structures. Since the 1930s, there are growing evidence that our cosmos is dominated by this new form of non-baryonic cold matter [1], which binds galaxies and clusters together and influences cosmic structures up to the largest observed scales.

This chapter provides a comprehensive overview of the evidence supporting the existence of Dark Matter across various scales, as outlined in Section 1.1. It also introduces potential particle candidates (Section 1.2) and discusses methods for detecting Dark Matter (Section 1.3).

## 1.1 Evidence for Dark Matter

In 1904, observing the dark regions in the sky, Lord Kelvin attempted an estimation of the total amount of matter in our galaxy by applying the theory of gasses for the astronomical bodies observed in the Milky Way. He reasoned that if the stars in our galaxy can be treated like particles in a gas, then the number of stars could be inferred from the relationship between their velocity dispersion and the size of the system. The number of stars founded from his calculations was about 10 times greater than the astronomical observations. He concluded that many of our stars might be dark bodies.

In 1933, the Swiss-American astronomer Fritz Zwicky [2] used a telescope at the Mount Wilson Observatory in California to measure the radial velocities of galaxies (then called nebulae) in the Coma cluster, a rich cluster of galaxies about 320 million light-years away from Earth. Applying the viral theorem, Zwicky discovered an unexpectedly large velocity dispersion, suggesting that the cluster's density was much higher than the one derived from luminous matter. In particular, he found a mass-to-light ratio of around 400 solar masses per solar luminosity, thus exceeding the ratio in the solar neighborhood by two orders of magnitude. These studies led to conclude that a significant amount of non-luminous matter, which he believed to be ordinary matter in a non-shining form, must exist in the cluster.
In the 1970's, Vera Rubin et al. and Albert Bosma measured the rotation curves of spiral galaxies, in particular the rotation curve of the Andromeda Galaxy [3], using optical images. They discovered further evidence of missing mass: the rotation curves remained flat at large distances from the galactic center instead of declining as expected based solely on the distribution of visible matter. By the early 1980s, evidence for this non-luminous matter was firmly established on galactic scales, rotation curves of galaxies well beyond their optical radii were measured in radio emission using the 21 cm line of neutral hydrogen gas.

### 1.1.1 Galactic rotation curves

The most compelling and direct evidence for Dark Matter at galactic scales comes from observations of rotation curves in spiral galaxies, such as the Milky Way [4; 5]. In these galaxies, the majority of stars reside in the central bulge, while the surrounding spiral arms, made up of stars and gas clouds, follow nearly circular path. The rotation curve represents the orbital velocities of stars and gas as function of their distance from the galaxy's center [1]. Considering a rotating galaxy as a closed system, an object of mass $m$ at radius $R$ from the center is subject to a centripetal force equal to the gravitational force exerted by the total mass $M(R)$ enclosed within that radius. This is expressed as:

$$\frac{mv^2}{R} = \frac{GM(R)m}{R^2} \tag{1.1}$$

where $G$ is the gravitational constant and $v$ is the circular velocity of the object with mass $m$. Therefore, the rotational velocity of the constituents of a galaxy can be expressed as a function of R and the total mass $M(R)$ as:

$$v = \sqrt{\frac{GM(R)}{R}} \tag{1.2}$$

It is reasonable to assume that the large majority of the mass of a galaxy is located within a characteristic radius $R_c$, estimated by measurement to fall between 5 and 10 kpc. So, for distances $R < R_c$ the mass density can be considered uniform, therefore $M(R)$ grows linearly with the volume and the velocity rises linearly with $R$. Instead, for $R > R_c$ the mass $M(R)$ remains constant and the velocity should decrease as $\sqrt{R}$. The rotation curve can be measured via the Doppler shift of the 21 cm emission line of neutral hydrogen. A wide array of rotation curves, extending well beyond the optical limit and reaching down to 200 kpc, has been measured, exhibiting similar results. The experimental results on the rotation curve of the galaxy NGC6503 [6] is shown in Fig. 1.1. While the orbital velocity increases linearly at small radii, it remains flat at larger radii, contrary to expectation. This flat rotation curve indicates the existence of an additional and invisible mass in the form of a dark galactic halo surrounding spiral galaxy. This dark halo, which does not interact electromagnetically, appears to be consistent at first order to a spherical distribution with a density proportional to $R^{-2}$. A modified version of Newton's law, called MOND, was proposed to explain these observations without the need to introduce a new type of matter. This theory will be illustrated in Section 1.2.1

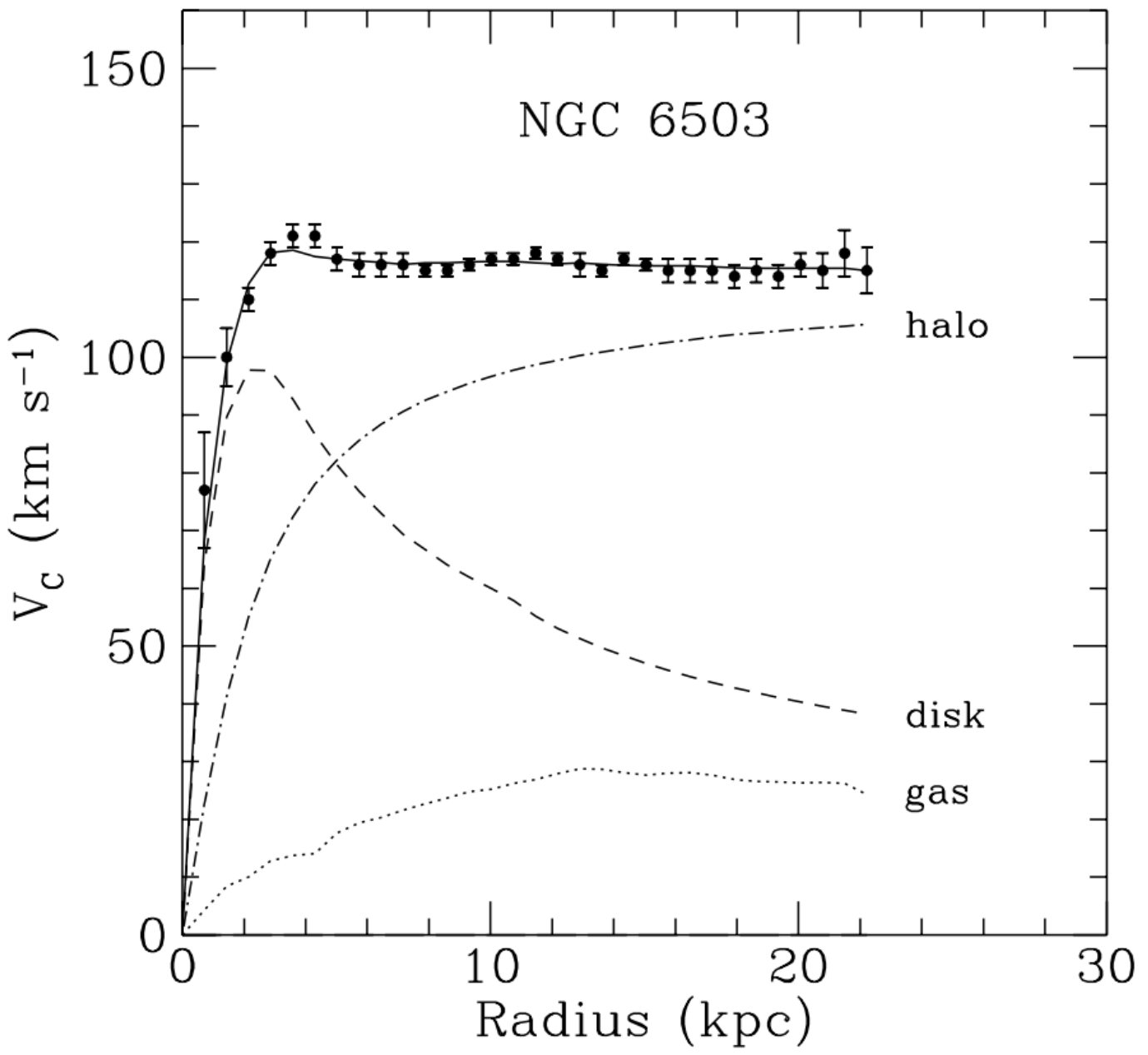

**Figure 1.1** Rotation curve of NGC 6503. The dotted, dashed and dash-dotted lines are the contributions of gas, disk and dark matter, respectively [6].

### 1.1.2 Gravitational lensing

Another compelling piece of evidence for the existence of Dark Matter comes from measurements of gravitational lensing. According to Einstein's theory of General Relativity, gravitational lensing occurs when a massive objects lies along the line of sight between an observer on the Earth and the object under study, as illustrated in Fig. 1.2. The light-rays are deflected through their path due to the gravitational field. The degree of light bending depends on the mass of the lensing object; more massive objects generate stronger gravitational fields and greater light deflection. Gravitational lensing provides an alternative method to estimate the mass of galaxy clusters, distinct from traditional techniques like the viral theorem or rotational velocity measurements.

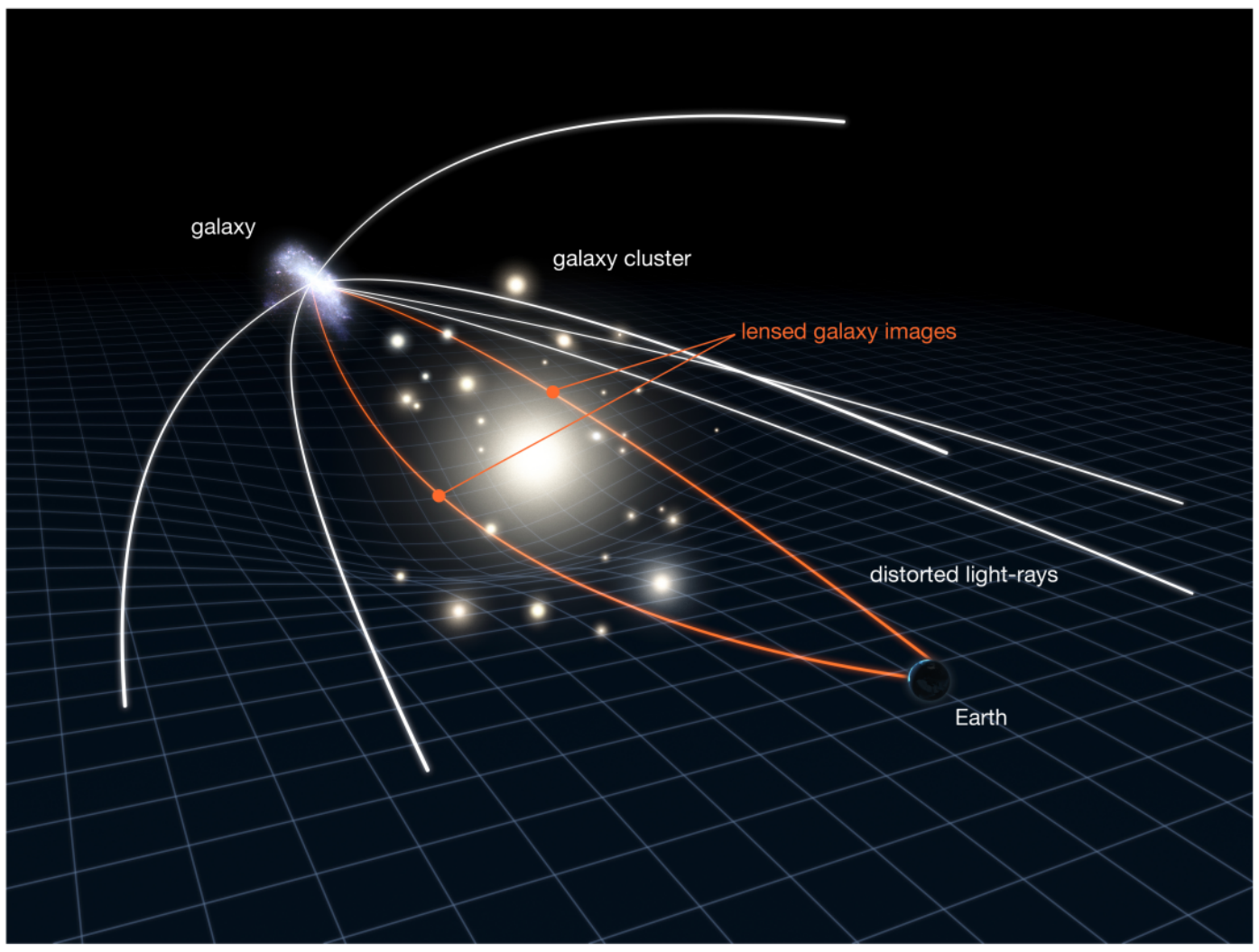

**Figure 1.2** A picture illustrating the gravitation lensing effect. Image credit: NASA, ESA & L. Calçada [7].

In some cases, gravitational lensing can produce multiple images of a single object, such as arcs or rings, around the massive lens. Assuming spherical symmetry, the deflection angle $\alpha$ for a light ray passing at a distance $b$ from the lens depends on the mass $M$ of the lensing object and is given by [8]:

$$\alpha = \frac{4GM}{bc^2} \tag{1.3}$$

where $G$ is the gravitational constant and $c$ the light speed in vacuum. This equation allows us to measure the mass causing the deflection.

Gravitational lensing effects can be categorized into strong and weak lensing. Strong lensing occurs when the mass density of the lens is large enough and its position change in the sky is small compared to the Earth's velocity resulting in significant deflection by a large angle. This effect often leads to observable phenomena like double images, arcs or rings. The effect are resolvable with telescope in single images [9]. Though strong lensing is easily detectable, it is relatively rare due to the scarcity of highly massive clusters.

On the other hand, weak lensing is more common and results from smaller deflection angles. Directly measuring weak lensing can be challenging, it reveals a mass distribution with a diffuse component centered on galaxy clusters, atop which peaks corresponding to visible bodies are superimposed. This suggests that most of the cluster's mass resides in a smooth Dark Matter halo that extends beyond the visible galaxies, accounting for the majority of the cluster's total mass.

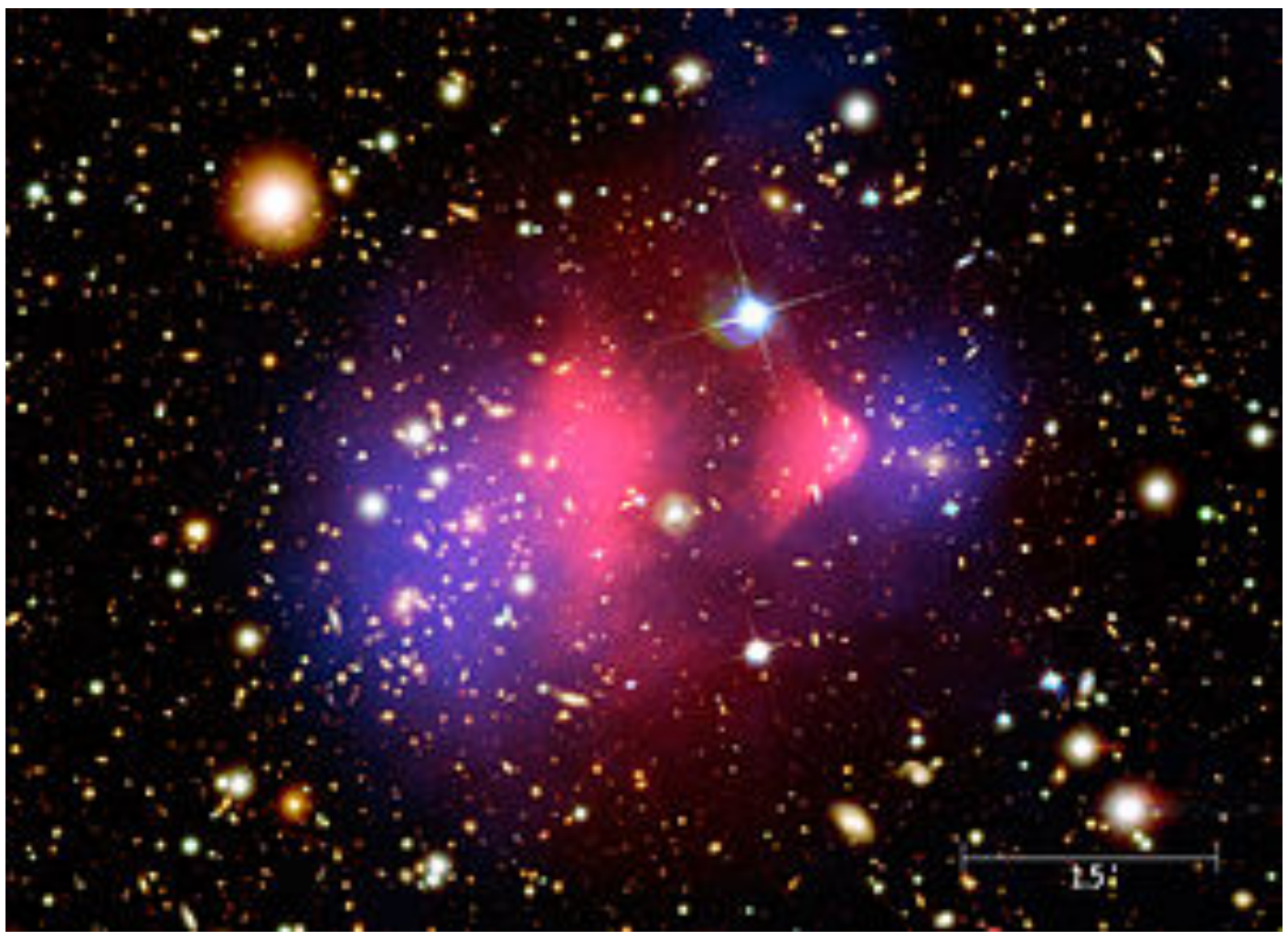

**Figure 1.3** The optical image, captured by the Magellan and Hubble Space Telescopes, displays background galaxies in shades of orange and white. In pink the hot intracluster gas, which contains most of the normal matter in the cluster. Gravitational lensing, seen as distortions in the background galaxies, reveals that Dark Matter (blue) makes up the majority of the cluster's mass [12].

Weak lensing produces slight distortions in galaxy images due to light bending from less massive objects [10], and both strong and weak lensing analyses are often combined for a more comprehensive understanding.

In addition to galaxies and clusters, smaller objects such as individual stars or single planets can bend the light coming from an object in the background, temporarily increasing the observed brightness in an effect known as microlensing [11]. Microlensing has been used to study the structure of our galaxy and was proposed to demonstrate the existence of Massive Compact Halo Objects (MACHOs), ordinary matter objects like neutron stars, black holes or brown dwarfs that could account for a fraction of Dark Matter.

A striking demonstration of Dark Matter's presence comes from studying merging galaxy clusters. By comparing the X-ray emission from hot gas to the mass distribution inferred from gravitational lensing, one can identify regions dominated by Dark Matter. An example is the Bullet Cluster, a system of two galaxy clusters that recently collided. The X-ray-emitting hot gas, representing the bulk of the visible matter, is shown in pink in Fig. 1.3, while the DM distribution, inferred from gravitational lensing, is highlighted in blue. During the collision, the hot gas particles, which make up the baryonic matter, interact through gravitational and electromagnetic forces, causing them to declare and lag behind. In contrast, DM interacts only through gravity and thus passes through unaffected, resulting in a spatial separation between the baryonic and Dark Matter components. This offset provides strong evidence that Dark Matter constitutes the majority of the mass in galaxy clusters, far exceeding the amount of baryonic matter.

### 1.1.3 Cosmological scale

The standard Model of Cosmology, also known as $\Lambda$CDM [13] is currently the most successful model for describing the composition and evolution of the universe on large scale. It is based on a few key assumption: the Universe is composed of radiation (photons and neutrinos), ordinary matter (baryons and leptons), non relativistic (cold) DM responsible for structure formation and cosmological constant $\Lambda$ [14]. This cosmological constant, associated with a Dark Energy or vacuum energy, represents a homogeneous form of energy which is responsible for the observed accelerated expansion. Dark energy or vacuum energy remains constant in density, even as the Universe expands. General Relativity describes the behavior of gravity on cosmological scales, and for the Cosmological Principle, the Universe is statistically homogeneous and isotropic in space and matter at sufficiently large scales ($\geq$ 100 Mpc).
The $\Lambda$CDM model is governed by six independent parameters:

- the baryon and cold dark matter energy densities;
- the angular diameter distance to the sound horizon at last scattering $\theta_S$, which is the scale defined by the distance sound waves could have traveled in the time before recombination;
- the amplitude and tilt of primordial scalar fluctuations, i.e. the initial magnitude and spectral index of density variations in the early Universe, related to the large scale structure formation;
- the reionization optical depth, i.e. the probability of a photon to scatter during the cosmic reionization epoch.

The cosmological principle implies a Friedmann Lemaître Robertson Walker metric, that allows to express the space-time evolution and its geometry as a function of the energy density of some constituents fluid. Therefore the velocity of the expansion of the Universe can be expressed as:

$$H^2 = \left(\frac{\dot{a}}{a}\right)^2 = \frac{8\pi G}{3}\rho(a) - \frac{k}{a^2} \quad (1.4)$$

where $H$ is the Hubble parameter, $a$ is the scale factor of the Universe, $\rho$ is the energy density of the fluid and $k$ represents the curvature of the 4-dimensional space-time. The scale factor describes the dimension of the Universe.
If $k > 0$, the geometry is elliptical and the Universe is said to be closed, if $k < 0$ the geometry is hyperbolic and the universe is said to be open [15]. Setting in Eq. 1.4 the curvature to zero ($k$=0), which corresponds to a flat universe, the so-called critical density is obtained:

$$\rho_c = \frac{3H_0^2}{8\pi G} \approx 9.20 \times 10^{-27} \mathrm{kgm}^{-3} \tag{1.5}$$

The energy density of different components can be expressed as dimensionless densities, normalized to the critical density:

$$\Omega_i = \frac{\rho_i}{\rho_c} \tag{1.6}$$

where *i* represents the different energy components.
The Friedmann equation can then be rewritten as:

$$H^2 = H_0^2 \left[ \Omega_{r,0} \left( \frac{a_0}{a} \right)^4 + \Omega_{m,0} \left( \frac{a_0}{a} \right)^3 + \Omega_{k,0} \left( \frac{a_0}{a} \right)^2 + \Omega_\Lambda \right] \tag{1.7}$$

Here, the subscript 0 denotes the present time, $\Omega_r$ is the radiation density, $\Omega_m$ is the matter density, which includes both baryons and DM contributions, $\Omega_k$ is the spatial curvature energy density and $\Omega_\Lambda$ is the dark energy contribution. If the total dimensionless energy density equals 1 the universe is flat.
Measurements able to constrain $\Omega$ parameters provides information on the structure of the Universe and help to quantify the amount of DM.
A primordial phase of cosmic inflation, a period of rapid accelerated expansion is assumed following the Big Bang singularity, a point of infinite temperature and density. During this period, Gaussian scale-invariant primordial fluctuations are produced from quantum fluctuations in the inflationary epoch.

**Cosmic Microwave Background (CMB)**

A crucial role in determining the DM abundance in the Universe comes from studies of cosmic microwave background (CMB) radiation. First discovered in 1965 [16], the CMB radiation originates from the decoupling of photons during the recombination epoch. At this stage, the Universe became transparent to electromagnetic radiation, allowing photons to travel freely. These photons persist today and the CMB is observed with a temperature of (2.726 $\pm$ 0.010) K. Small inhomogeneities in the distribution of its temperature correspond to fluctuations of the matter density in the early Universe, which led to the formation of the observed large structures.

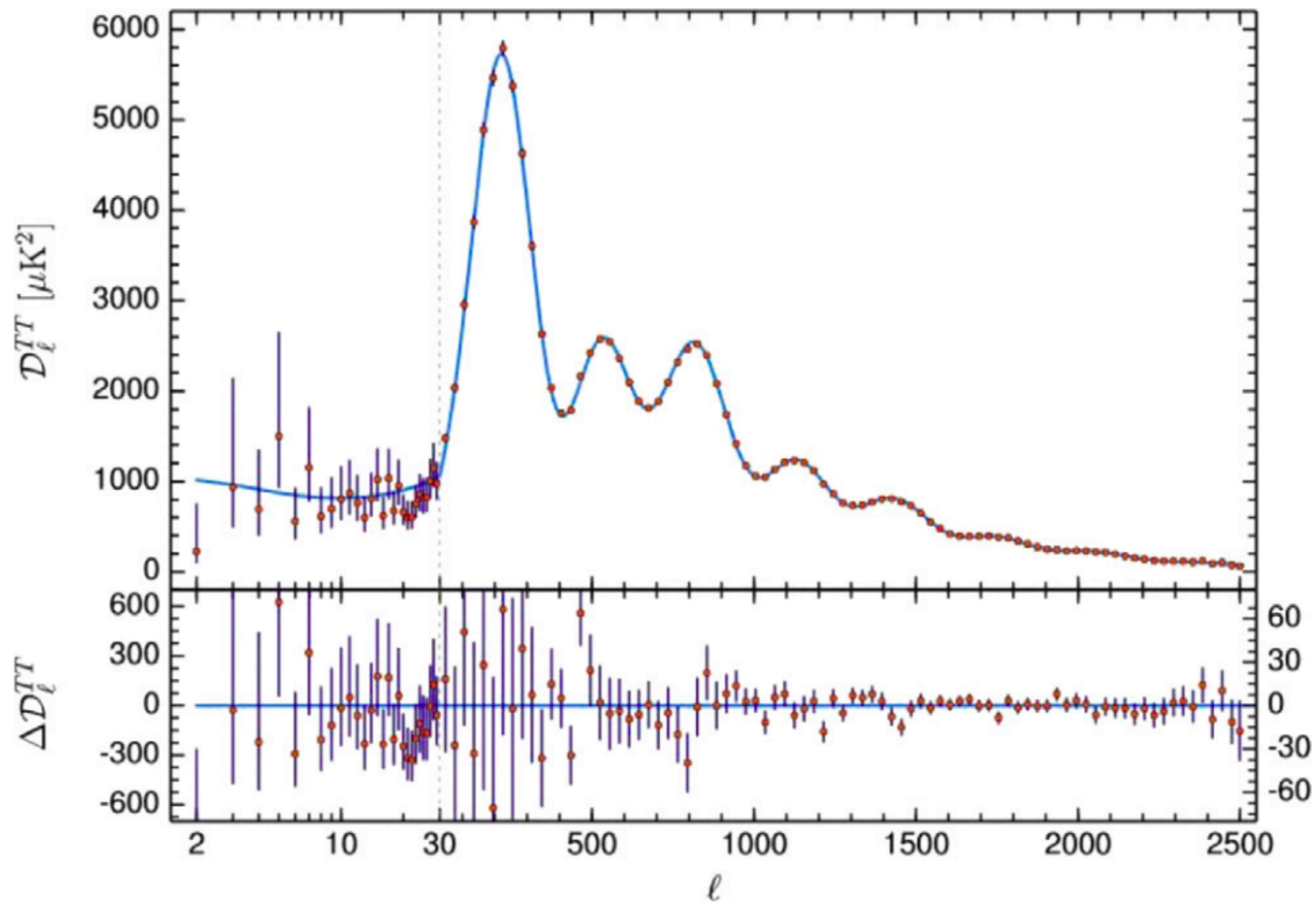

**Figure 1.4** Temperature anisotropies spectrum in spherical harmonics presented by PLANCK collaboration. The blue line shows the ΛCDM model fit to the experimental data [17].

The Planck experiment provided detailed measurements of the angular power spectrum of temperature fluctuations within the CMB, which quantifies how each region deviates from the mean temperature, expressed through spherical harmonics. These small anisotropies stem from variations in particle densities during the early Universe as they transitioned between different state. The intensity and scale of these fluctuation depend on the types of particles and components present at that time.

By analyzing the power spectrum, cosmologists have determined that our Universe is spatially flat and undergoing accelerated expansion. Moreover, the evidence indicates that Dark Matter is non-baryonic, supported by measurements of the abundance of light elements produced during primordial nucleosynthesis.

In Fig. 1.4 are shown the temperature anisotropies as presented by PLANCK collaboration [17] as a function of the spherical harmonics. The blue line represents the fit of the ΛCDM model, the simplest parametric model of the Big Bang that includes both Dark Matter and Dark Energy. The first intense peak in the spectrum corresponds to the oscillation frequency of energy and matter density, and its position is strongly related to the value of total curvature of the Universe at the recombination time. The intensity of the subsequent peaks depends on the total matter content of the Universe, while the shape of the temperature spectrum is highly sensitive to cosmological parameters. The best fit gives us strong constraints on the energy density components of the Universe. The relative heights of the acoustic peaks allow for an estimation of the baryonic matter content, which in turn enables a calculation of the total Dark Matter density in the Universe. According to the analysis done by the PLANCK collaboration [17] suggests that the Universe is flat, the relative densities are summarized in Table 1.1.

| Component | Parameter | 68% Limits |
|---|---|---|
| Radiation | $\Omega_r$ | $\sim 9 \times 10^{-5}$ |
| Baryonic matter | $\Omega_b$ | 0.0489 ± 0.0003 |
| Non-Baryonic Matter | $\Omega_{nb}$ | 0.2607± 0.0019 |
| Dark Energy | $\Omega_\Lambda$ | 0.6889 ± 0.0056 |
| Total | $\Omega_{tot}$ | 0.9993 ± 0.0019 |

**Table 1.1** Summary of the $\Omega$ parameters from the $\Lambda$CDM model obtained from Planck Collaboration [17].

## 1.2 Dark Matter candidates

Astronomical and cosmological observations based on gravitational probes are not consistent with the current picture of the Universe particles and forces content. Several theories have been proposed to explain these anomalies and are briefly discussed in the following sections. Although Dark Matter particles have never been observed, any DM models have to satisfy specific properties based on indirect astronomical and cosmological observations [18]. Firstly, at the time of decoupling from baryons, DM have been non-relativistic, meaning it consists of cold Dark Matter (CDM). Hot dark matter, which would have a mass in the range of a few tens of eV, is predicted to contribute only a small fraction of the total DM density. Even if some arguments favour warm Dark Matter (WDM) due to its ability to match observations of large-scale structures in the Universe, most simulations and models still support a CDM scenario.

Secondly, properties of the cosmic microwave background imply that DM is non-baryonic. Since DM particles do not emit photons (otherwise they would become visible) they must be electrically neutral. Strong interaction is also ruled out, since DM would lose energy and concentrate more in the galactic centres than what is observed.

Experimental observations supporting the hypothesis that DM interacts other than gravitationally do not exist. DM particles should interact with ordinary matter preferably only weakly, this weak interaction does not refer to the SM electro-weak interaction, but it could be any form of sub-weak strength type of interaction. DM particles could also interact with themselves, and this type of self-interactions is in fact rather poorly constrained. Dark Matter particles should also be either completely stable, or have a lifetime far exceeding the age of the universe since we still observe its gravitational influence today. Recent analysis suggests a lower bound for DM lifetimes of at least 160 Gyr.

None of the Standard Model particles satisfy the requirement for Dark Matter particle. While neutrinos are not baryonic, neutral and only weakly interactive particles, their predicted relic density does not match observational data. Therefore, it is clear that an extension beyond the Standard Model is required to account for Dark Matter.

### 1.2.1 Modified Newtonian Dynamics

Astrophysical and cosmological studies of the Universe are traditionally based on the application of Newtonian dynamics and General Relativity. However, in 1978, it was observed that rotation velocities of stars in galaxies remain roughly constant as their distance from the galactic center increases. [19]. This observation was in contradiction with expectation, as according to Newtonian dynamics, objects outside the visible mass distribution should have velocities $v \propto 1/\sqrt{r}$, where $r$ is the distance from the center. To explain this discrepancy, a uniformly distributed halo of Dark Matter was proposed, which could account for both the velocities in galaxy clusters and the rotation velocities of objects far from the luminous matter in galaxies [20].
An alternative explanation involves modifying the laws of gravity to align with these observations. One such theory is Modified Newtonian Dynamics (MOND) [21], introduced in 1983, and its relativistic extension, TeVeS (Tensor-Vector-Scalar gravity) [22]. MOND attempts to explain the rotation curve of galaxies by altering Newton's gravitational force law. It introduces a new fundamental acceleration constant, $a_0$, which defines the transition between the Newtonian regime and the so-called deep-MOND regime. When the system's acceleration is smaller than $a_0$, the gravitational force scales differently than in standard Newtonian gravity. For accelerations larger than $a_0$, standard dynamics is recovered. This acceleration constant, $a_0$, is estimated from galactic rotation curve measurements to be about $10^{-10}$cm/s$^2$. MOND can explain the dynamics of galaxies and galaxy clusters, leaving the Solar System unaffected.
However, MOND fails or needs unrealistic parameters to fit observations on larger scales such as the formation of cosmic structure or the CMB structure and violates fundamental laws like momentum conservation and the cosmological principles [23]. While TeVes can solve some of the conceptual problems of MOND, but it introduces other problems, such as generating an unstable Universe [24] or failing to simultaneously explain both gravitational lensing and galactic rotation curves [25].

### 1.2.2 Weakly Interactive Massive Particles (WIMPs)

One of the most studied and supported Dark Matter candidates are the Weakly Interactive Massive Particles (WIMPs) .WIMPs are neutral, massive, stable, weakly -interacting particles that were in thermal equilibrium in the early Universe and have the correct present-day density to account for Dark Matter. Furthermore, well-motivated extensions of the Standard Model of particle physics provide concrete WIMP candidates [26]. WIMPs are expected to interact with regular matter at the order of the weak interaction scale or below, allowing to relate the WIMP mass and interaction cross section to the present-day relic density.
In the early universe, DM and SM particles were in thermal equilibrium. At this stage, WIMP production and self-annihilation balanced each other:

$$\chi + \overline{\chi} \longleftrightarrow \phi + \phi^* \tag{1.8}$$

$$\phi + \chi \longleftrightarrow \phi + \chi \tag{1.9}$$

where $\phi$ represents a SM particles and $\phi^*$ its antiparticle and $\chi$ represents a generic WIMP. As the Universe expanded and cooled down, the DM production rate from the annihilation of SM particles fell below the expansion rate of the Universe $H$. The number of DM particles is therefore fixed and this thermal relic abundance survives until now. This process, known as freeze-out, is the most widely accepted mechanism for the formation of WIMPs.
The number density of WIMPs $n_\chi$, can be described by the Boltzmann equation:

$$\frac{dn_\chi}{dt} = -Hn_\chi - \langle \sigma_A v \rangle \left( n_\chi^2 - n_{\chi,eq}^2 \right) \tag{1.10}$$

with $H$ is the Hubble parameter, $n_{\chi,eq}$ is the number density in equilibrium and $\langle \sigma_A v \rangle$ is the thermal average of the annihilation cross section flux. The equation shows how the number density diminishes in time due to the spatial expansion of the Universe and it is modified by the annihilation and generation of DM particles.
In cosmology, the total entropy of the Universe remains conserved during the thermal equilibrium and its density $s$ is often used to describe the comoving volume of the Universe. The comoving number density of Dark Matter $Y_\chi$, can be expressed as:

$$\mathrm{Y}_\chi := \frac{n_\chi}{s} \tag{1.11}$$

The freeze-out moment, which is crucial in determining the present-day WIMP abundance, occurs when:

$$n_{\chi,FO} \langle \sigma_A v \rangle \sim H \tag{1.12}$$

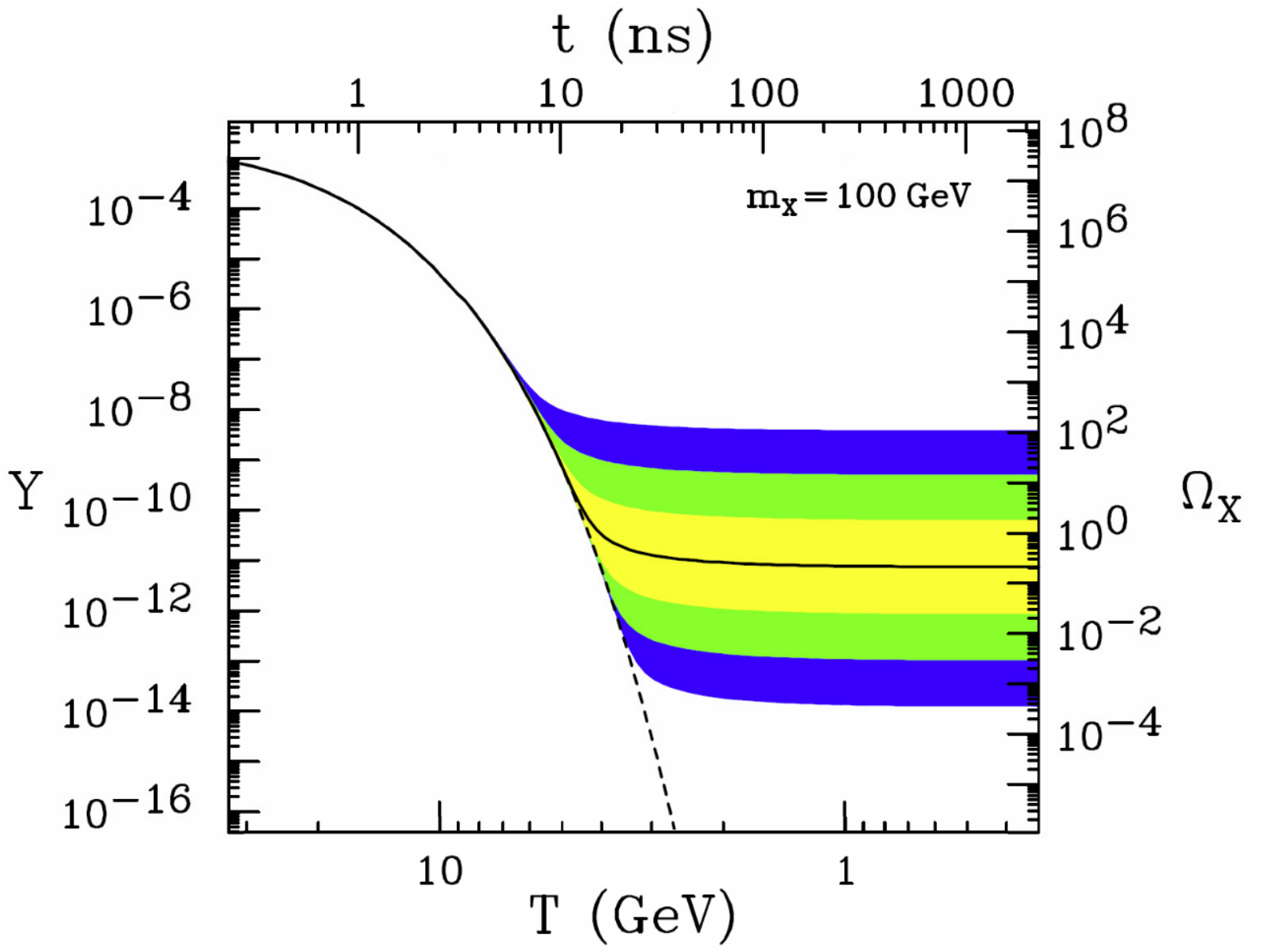

**Figure 1.5** The comoving number density Y as a function of temperature during the Universe's evolution for a WIMP with a mass of 100 GeV/c$^2$. The solid black line shows the Boltzman equation solution corresponding to the measured DM density parameter $\Omega_{nb}$. The colored bands indicate the effect of varying the annihilation cross section by factors of 10, 100 and 1000 [27].

Fig. 1.5 illustrates the comoving number density $Y_\chi$ as a function of temperature during the Universe's evolution. The solid black line represents the solution to the Boltzmann equation, which corresponds to the observed Dark Matter density parameter $\Omega_{nb}$. The colored bands represent the effect of varying the annihilation cross section by factors of 10, 100 and 1000, illustrating how changes in WIMP interaction rates affect the final DM abundance.

Dark Matter particles can be classified based on how relativistic they were at the time they decoupled from thermal equilibrium. Hot Dark Matter (HDM), with masses up to a few tens of eV, remained relativistic at decoupling. Due to its large mean free path, HDM could not form small clumps like galaxies, making it an unsuitable candidate for explaining the Universe's structure. HDM is only expected to make up a small fraction of the total DM density, constrained by CMB observations. A familiar example of HDM is neutrinos with very small mass.

Cold Dark Matter (CDM), on the other hand, decoupled from the thermal plasma early in the Universe and had low velocities at the onset of matter dominance, allowing density perturbations to grow and form the large-scale structures we observe today. Non-baryonic CDM is the most favored scenario.

Warm Dark Matter (WDM) [4], with masses around a few keV, represents an intermediate case. It was relativistic at decoupling but could still form structures on sufficiently large scales. While WDM could explain certain features of the Universe's structure, the most widely supported models continue to favor CDM.
WIMPs are a subset of CDM candidates, characterized by their low velocity at freeze-out, ensuring they do not impact galaxy formation of the large-scale structure.

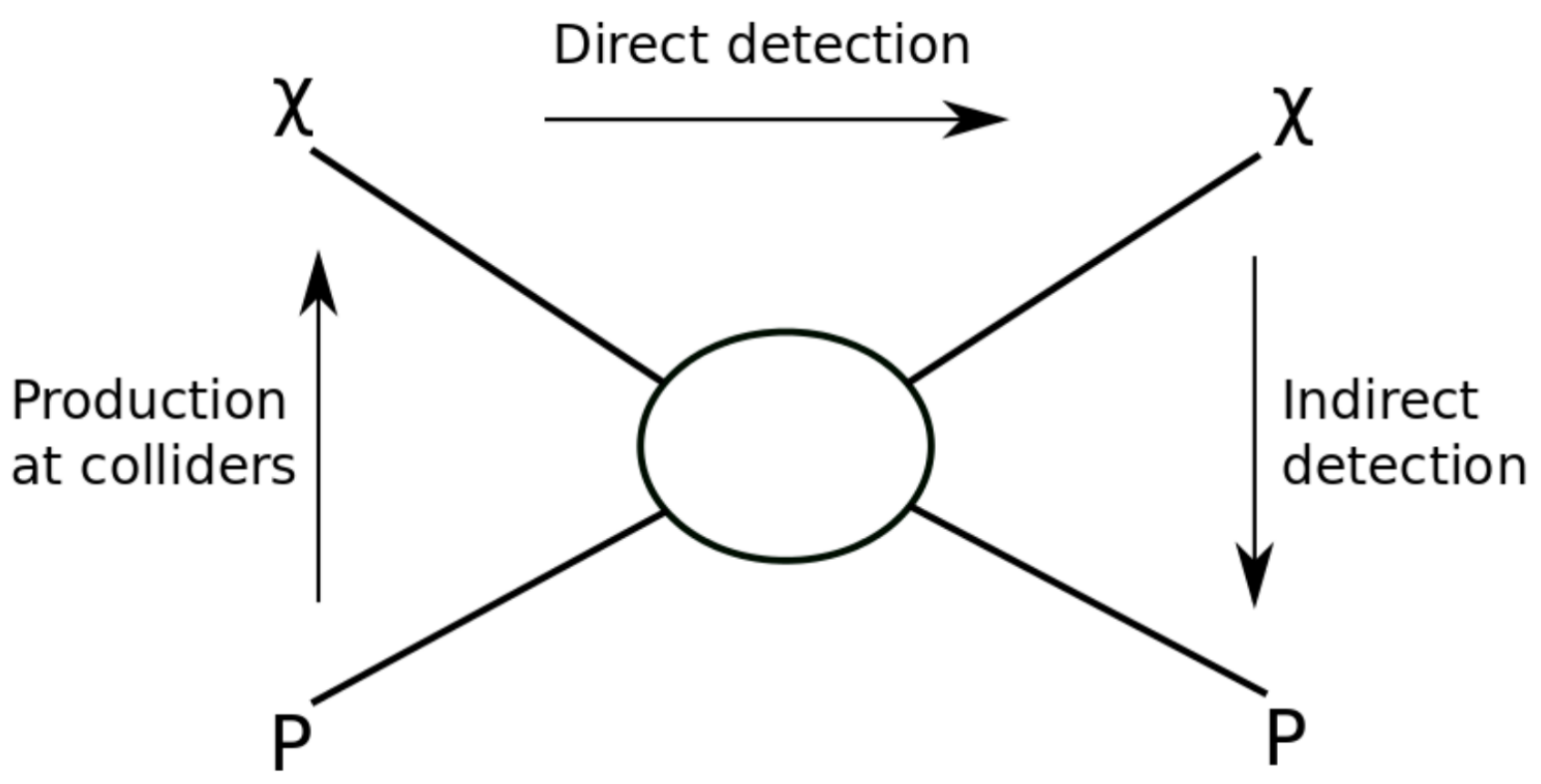

**Figure 1.6** Schematic showing the possible Dark Matter detection channels [29].

## 1.3 Experimental searches for DM

The particle Dark Matter hypothesis can be tested via three processes: production at particle accelerators [28], indirect detection by searching for signals from annihilation products, or direct detection via scattering on target nuclei [29], illustrated in the rightwards direction in the schematic representation of the possible Dark Matter couplings to a particle P of ordinary matter shown in Fig. 1.6. While the annihilation of Dark Matter particles (downwards direction) could give pairs of standard model particles, the collision of electrons or protons at colliders could produce pairs of Dark Matter particles.

### 1.3.1 Direct detection

Direct detection (DD) experiments [30] aim to detect WIMPs by observing their elastic scattering off target nuclei in the laboratory. Experiments have to measure the energy deposited by nuclear recoils from individual scattering events. Since WIMPs have a very small interaction cross section with matter, these events are rare. Moreover, the energy of the nuclear recoil is relatively small, $\sim$10-100 keV. To minimize background noise from neutrons and other particles, these experiments are conducted deep underground, using radiopure components, and shielding the detectors appropriately.
Due to the kinematics of elastic scattering, the shape of the energy spectrum depends on the mass of the target nuclei, and, for Spin-Independent interactions, the normalization of the spectrum is proportional to the square of the mass number.

Scattering of DM particles off nuclei can be detected through light (scintillation photons from excitation and later de-excitation of nuclei), charge (ionization of atoms in a target material) or heat (phonons in crystal detectors). A combination of two such discrimination techniques is often employed to disentangle a potential WIMP signal from nuclear recoils and background electron recoils. This is possible because of the different quenching factors that describe the difference between the recorded signal and the actually measured recoil energy. Different materials are used in direct detection searches depending on the technique employed. For example, phonon signals from DM-nuclei scatterings in crystal detectors provide high sensitivity to low-mass Dark Matter due to their low energy threshold. Direct Detection will be further discussed in Section 2.1

### 1.3.2 Indirect Detection

Indirect Dark Matter searches aims to detect the products of Dark Matter annihilation or decay from astrophysical sources. These searches focus on measuring the rate of reactions such as DM DM $\rightarrow$ SM SM, where DM represents the Dark Matter particle (e.g. WIMPs) and SM represents any Standard Model particle. In many cases, the SM particles produced in these reactions are unstable and decay into detectable particles. The cross section probed by indirect searches is most closely related to the process that sets the abundance of the Dark Matter in the early Universe, assuming that the DM was once in thermal equilibrium.To properly interpret the results from indirect searches, it's important to understand how Dark Matter is distributed in halos and which SM particles DM preferentially annihilates into. The most favored sources for indirect signals are the galactic center and halo, close Galaxy clusters of dwarf galaxies (dwarf spheroidals), where the density of Dark Matter is expected to be high. Enhanced self-annihilation, scattering or decay into Standard Model Particles could produce a detectable particle flux. The measurement of these secondary particles is a further detection mechanism denoted as indirect detection.

### 1.3.3 Colliders

The third classical strategy for WIMP detection involves Dark Matter particles generated directly in high-energy collisions. The Large Hadron Collider (LHC) can provide in principle an optimal instrument for pursing this experimental venue.

The vast majority of the searches for new physics at the LHC are designed to look for events that, besides the rich hadronic/leptonic activity emerging from the decay chain of the produced visible particle, are also characterized by a large amount of missing energy, which could indicate the production of neutral, stable particle (potentially WIMPs) that escape the detector without interacting. The discovery of one or more visible particles in a channel characterized by highly energetic jets or leptons, and large missing momentum, would also imply the discovery of a neutral and stable (at least within the detector bounds) particle, which could be part of the Dark Matter.
At colliders, the focus is on reactions like:

$$pp \rightarrow \chi\bar{\chi} + x \tag{1.13}$$

where $\chi$ represents Dark Matter particles, $x$ could be a hadronic jet, a photon or a leptonically decaying $Z$ or $W$ boson [31]. These missing-energy events, coupled with visible particles like jets or leptons, provide potential evidence for the existence of Dark Matter.
Collider searches are particularly useful for exploring regions of parameters space where WIMP masses are low (below $\sim$ 4 GeV for Spin-Independent interactions and below $\sim$ 700 GeV for Spin-Dependent interactions).

# 2 Dark Matter Direct detection

The search for Dark Matter is one of the greatest challenges in modern physics. As previously discussed, if Dark Matter is composed by particles which can interacts with Standard Model particles not only gravitationally, these interactions can be exploited in different ways to search a DM signature, in particular through direct detection techniques.
Direct detection focuses on measuring the interaction between Dark Matter particles and SM targets, potentially revealing information about the properties of DM. This approach typically involves detecting the energy released by a nuclear recoil (NR) that occurs when a Dark Matter particle scatters off a nucleus within the detector. While this method offers a way to constrain properties of DM particles, it faces significant limitations due to background noise and the lack of a definitive signature for DM. One major challenge is distinguishing background events, such as electron recoils (ERs), from signal events, such as nuclear recoils, particularly at the low energy scales relevant to DM searches, typically around 10 keV. At these energy levels, background and signal events can exhibit similar topological characteristics within the detector.

## 2.1 Experimental signatures of Dark Matter

The main method for direct detection of Dark Matter (DM) particles exploits a volume of material capable of measuring the recoil of ordinary matter after an interaction with a WIMP. SM particles in a material are electrons and nuclei. While DM could potentially interact with both, most attention is given to nuclear scattering. WIMPs have a much higher mass compared to electrons, which results in greater momentum transfer during nuclear recoils than in electron recoils. Since WIMPs are electrically neutral, their interactions with atomic electrons, which are much lighter are generally not expected to produce detectable signals.

### 2.1.1 DM halo

The Standard Halo Model (SHM) assumes that our Galaxy DM halo is an isotropic isothermal sphere, with a density profile $\rho \sim r^{-2}$, where $r$ is the radial distance from the galactic center. The normalized velocity distribution of DM particles in the rest frame of the Galaxy can be described by a Maxwellian distribution [32]:

$$f(v) = \begin{cases} Ne^{-\frac{v^2}{v_p^2}} & \text{if } |v| < v_{esc} \\ 0 & \text{if } |v| > v_{esc} \end{cases} \tag{2.1}$$

where $N$ is a normalization factor and $v_p$ is the velocity dispersion of the WIMPs. Assuming a perfectly flat rotation curve for the galaxy, $v_p$, can be measured at different distances and is found to be approximately 230 km/s at the Earth radius [33]. Even though the Maxwell distribution extends to infinity, the velocity distribution is typically truncated at a local escape velocity $v_{esc}$, above which particles would not be gravitationally bound to our Galaxy, posing an upper limit on the available velocities for WIMPs detection. Based on measurements from the Gaia telescope [34] and the RAVE [35] survey, $v_{esc}$ is estimated to be 544 km/s.
Another important parameter is the local DM density $\rho_0$, which is commonly taken as 0.3 GeV/cm$^3$, despite the systematic uncertainties.
To calculate the expected rate of WIMP interactions with SM particles on Earth, a coordinate transformation is needed to translate the velocity distribution into the laboratory frame. In the WIMP assumption, the DM is cold, therefore non relativistic. Since the velocities of both the Earth around the Sun and the Sun around the galactic center are also non-relativistic, a Galilean transformation suffices:

$$f(v,t) = f_{gal}(v + v_{lab}(t), t) \tag{2.2}$$

where $v_{lab}$ is the laboratory velocity relative to the Galactic rest frame. The gravitational potential of the Sun and the Earth is weak enough to affect only WIMPs with less than 40 km/s, which represent a very small fraction. In this transformation, the laboratory velocity $v_{lab}$ is the sum of the Sun's velocity relative to the Galactic rest frame and the Earth's motion around the Sun. In some cases, the direct detector is approximated as positioned in the center of Sun, neglecting Earth's motion. While this approximation affects searches for annual and daily modulation of signals, it does not significantly impact the recoil distribution or detection rate, with deviations typically less than a few percent. When only the Sun's velocity is considered, $v_{lab}$ is approximately 242 km/s.

### 2.1.2 Scattering kinematics

The expected interaction of the WIMPs with the target nuclei on Earth can be considered at rest and treated as an elastic collision, where the nucleus recoils with an energy $E_r$. Since the WIMPs velocity are non-relativistic, the momentum $q$ transferred in the collision can be written as:

$$|q| = 2\mu v \cos(\theta_r) \tag{2.3}$$

where $\mu$ is the WIMP-nucleus reduced mass defined as:

$$\mu = \frac{m_\chi m_A}{m_\chi + m_A} \tag{2.4}$$

Here, $m_\chi$ is the mass of the WIMP, $m_A$ is the mass of the target nucleus, $v$ is the mean WIMP velocity relative to the target and $\theta_r$ is the angle between the WIMP's initial direction and the recoiling nucleus in the center-of-mass-frame. The energy $E_r$ of the recoiling nucleus is then given by:

$$E_r = \frac{|q|^2}{2m_A} = 2v^2 \frac{\mu^2}{m_A} \cos^2(\theta_r) \tag{2.5}$$

The energy $E_i$ of the incoming WIMP can be written as:

$$E_i = \frac{m_\chi v^2}{2} \tag{2.6}$$

rewriting the energy recoil as a function of $E_i$, we have:

$$E_r = rE_i \frac{1 - \cos(\theta_r)}{2} \tag{2.7}$$

where r is the adimensional kinematic factor, defined as:

$$r = \frac{4 m_\chi m_A}{(m_\chi + m_A)^2} \tag{2.8}$$

This factor represents the efficiency of momentum transfer between the WIMP and the nucleus. The value of $r$ is maximal ($r$ = 1) when the two colliding particles have the same mass $m_\chi = m_A$. Depending on the angle $\theta_r$, the energy recoil varies between 0 and $rE_i$. From Eq. 2.6, the minimum WIMP velocity required to produce a recoil with energy $E_r$ is:

$$v_{min} = \sqrt{\frac{m_A E_r}{2\mu^2}} \tag{2.9}$$

### 2.1.3 Expected event rate

The event rate for WIMP-induced nuclear recoil [36; 37] is determined by integrating the time product of the incoming DM flux with the cross section $\sigma$ for the relevant WIMP-nucleus interaction. This is typically expressed in terms of the differential event rate $R$, per unit detector mass, as a function of recoil energy $E_r$ and time $t$:

$$\frac{dR}{dE_r}(E_r,t) = \frac{\rho_0}{m_\chi m_A}\int_{v>v_{min}} vf(\mathrm{v},t)\frac{d\sigma}{dE_r}(E_r,v)d^3v \tag{2.10}$$

here, $m_\chi$ is the Dark Matter mass and $\frac{d\sigma}{dE_r}(E_r,v)$ is the differential cross-section for the interaction. The WIMP cross-section $\sigma$ and $m_\chi$ are two observable of a Dark Matter experiment. The velocity $v$ of the DM particle is defined in the rest frame of the detectors and $m_A$ is the nucleus mass. The local DM density $\rho_0$ and $f(\mathrm{v},t)$ accounts for the WIMPs velocity distribution in the detector's reference frame. This velocity distribution is time dependent due to the revolution of the Earth around the Sun. The most common approach in direct-detection experiments is the attempt to measure the energy dependence of Dark Matter interactions. The integration in Eq. 2.10 is performed for WIMP speed $v$ that exceed a minimum value $v_{min}$, which is the minimum velocity required to produce a recoil of energy $E_r$:

$$v_{min}(E_r) = \sqrt{\frac{m_A E_r}{2\mu^2_{\chi A}}} \tag{2.11}$$

where $\mu_{\chi A}$ is the reduced mass of the WIMP-nucleus system.
The integral over velocities is weighted by the flux of DM particles $vf(\mathrm{v},t)$, where $f(\mathrm{v},t)$ represents the galactic DM velocity distributions. Therefore, the expected NR rate depends on the nuclear properties of the target of choice, on the assumptions about the WIMP's coupling to SM particles coming from particle physics, and the characteristics of the DM halo in our galaxy.

### 2.1.4 Cross section

The interaction between the WIMPs and the ordinary matter is unknown, the differential cross section $\frac{d\sigma}{dq}$ can be split as the product of two terms: an independent term $\sigma^{WA}$, which is unknown, the cross section, and the $F(q)$, which include the entire dependence on the transferred momentum $q$, the form factor, which is known:

$$\frac{d\sigma}{dq^2} = \frac{\sigma^{WA}F^2(q)}{4\mu^2_A v^2} \tag{2.12}$$

where $\mu_A$ is the reduced mass of the WIMP-nucleus system, previously defined (Eq. 2.8) and $v$ the velocity of the WIMP.
Note their local velocity , $\sim$ 230 km/s, the De Broglie wavelength associated to a WIMP with a mass of roughly 10 GeV/c$^2$ is:

$$\lambda_{DM} = \frac{h}{p} \sim 160 \text{ fm} \tag{2.13}$$

The typical nuclear dimension is of the order of few fm, therefore the WIMPs interacts coherently with the entire nucleus through elastic scattering with low momentum transfer. The interaction can involve both Spin-Independent (SI) and Spin-Dependent (SD) couplings. The total cross section can thus be written as the sum of these two components:

$$\sigma^{WA} = \sigma^{SI} + \sigma^{SD} \tag{2.14}$$

In the non-relativistic case, the differential cross section, can be written as:

$$\frac{d\sigma}{dE_r} = \frac{m_A}{2\mu_A^2 v^2}\left[\sigma_0^{SI} F_{SI}^2(E_r) + \sigma_0^{SD} F_{SD}^2(E_r)\right] \tag{2.15}$$

where $\sigma_0^{SI}$ and $\sigma_0^{SD}$ are the SI and SD WIMP-nucleus cross section at zero momentum, denoted by the subscript 0, and $F_{SI}$ and $F_{SD}$ are the form factors for the Spin-Independent and Spin-Dependent interaction, respectively. The form factor describes the loss of coherence at higher momentum transfer values, its effect is stronger for heavy WIMPs or heavy target nuclei.
The SI term can be written as the sum of contributions from protons and neutrons in the nucleus:

$$\sigma_0^{SI} = \frac{4\mu^2}{\pi}\left[Z f_p + (A-Z) f_n\right]^2 \tag{2.16}$$

where $f_p$ and $f_n$ are the effective Spin-Independent couplings to the proton ($p$) and neutron ($n$), respectively, while $Z$ and $A$ are the atomic and mass numbers of the target nucleus. A reasonable assumption supported by many models is that the WIMP couples equally to protons and neutrons meaning in $f_p = f_n = f$, leading to:

$$\sigma_0^{SI} = \frac{4\mu^2}{\pi} A^2 f^2 = \sigma_n^{SI} \frac{\mu_A^2}{\mu_n^2} A^2 \tag{2.17}$$

where $\mu_n$ is the reduced mass of $m_\chi$ with the mass of a nucleon and $\sigma_n^{SI}$ is the WIMP-nucleon interaction cross section. The interaction rate is thus proportional to $A^2$, making detectors with heavier target nuclei more sensitive than those with lighter one, given the same exposure.

| Nucleus | Z | Odd Nuc. | J | $\langle S_p \rangle$ | $\langle S_n \rangle$ | $\frac{4\langle S_p \rangle (J+1)}{3J}$ | $\frac{4\langle S_n \rangle (J+1)}{3J}$ |
|---|---|---|---|---|---|---|---|
| $^{1}$H | 1 | p | 1/2 | 0.500 | 0.0 | 1.0 | 0 |
| $^{19}$F | 9 | p | 1/2 | 0.447 | -0.004 | $9.1 \times 10^{-1}$ | $6.4 \times 10^{-5}$ |
| $^{73}$Ge | 32 | n | 9/2 | 0.030 | 0.378 | $1.5 \times 10^{-3}$ | $2.3 \times 10^{-1}$ |
| $^{129}$Xe | 54 | n | 1/2 | 0.028 | 0.359 | $3.1 \times 10^{-3}$ | $5.2 \times 10^{-1}$ |

**Table 2.1** The nuclear spin properties of some of the most relevant nuclei, in terms of sensitivity to Spin-Dependent (SD) interactions, are summarized [39].

The SD cross section contribution can be written as:

$$\sigma_0^{SD} = \frac{32}{\pi} G_F^2 \mu^2 \frac{J+1}{J} \left[ a_p \langle S_p \rangle + a_n \langle S_n \rangle \right]^2 \tag{2.18}$$

where $G_F$ is the Fermi constant, $J$ is the angular momentum of the target nucleus, $a_p$ and $a_n$ are the effective Spin-Dependent couplings to proton and neutron, respectively, $\langle S_p \rangle$ and $\langle S_n \rangle$ are the expectation values of proton and neutron spin inside the target nucleus. For sensitivity to SD interactions, the target nucleus must have an odd number of protons or neutrons, since the contribution of all the nucleons typically cancels out. The spin of the nucleus is carried by neutrons and protons. Example of nuclei with an odd number of neutrons ($a_p = 0$) include $^{17}$O, $^{27}$Al, $^{29}$Si, $^{37}$Ge, $^{129}$Xe, $^{131}$Xe, $^{183}$W, or in the case with odd number of protons ($a_n = 0$): $^{1}$H, $^{7}$Li, $^{19}$F, $^{23}$Na, $^{127}$I. The SD cross section of proton and neutron [38] can be written as:

$$\sigma_p = \frac{24 G_F^2 \mu_p^2 a_p^2}{\pi} \tag{2.19}$$

$$\sigma_n = \frac{24 G_F^2 \mu_n^2 a_n^2}{\pi} \tag{2.20}$$

Combining Equations 2.14, 2.17, 2.18, 2.19 and 2.20 the total cross section for a nucleus containing an odd number of protons can be written as:

$$\sigma^{WA} = \sigma_n^{SI} \frac{\mu_A^2}{\mu_p^2} A^2 + \sigma_p^{SD} \frac{\mu_A^2}{\mu_p^2} \frac{4 \langle S_p \rangle (J+1)}{3J} \tag{2.21}$$

Table 2.1 summaries some of the elements with largest expected spin value [39].

The SD term is usually less than 1, for the majority of nuclei, it makes the SD contribution negligible with respect to the SI one. Therefore experiments usually quote the results for SD and SI coupling separately.

### 2.1.5 Form factor

At high momentum transfers, the nucleus of the target would appear less and less homogeneous and the WIMP would no interact with the nucleus as a whole, but rather with its constituents. $F(q)$ is the nuclear form factor which plays a critical role in determining the dependence of the differential cross section on the momentum transfer during the collision. It depends on the structural characteristics of the nucleus, which are typically approximated using Fermi's spherical symmetry model. The nucleus is described as a sphere with a defined radius $R_n$ and a skin thickness.
For SI interaction, the form factor can be evaluated as the Fourier transformation of the nucleus radius function:

$$|F_{SI}(q)|^2 = \left| \frac{9\left[\sin(qR_n) + qR_n \cos(qR_n)\right]^2}{(qR_n)^6} \right]^2 \tag{2.22}$$

The radius is approximated as:

$$R_n \simeq \left[0.91 A^{1/3} + 0.3\right] \text{ fm} \tag{2.23}$$

where A is the mass number of the target nucleus. For SD interactions, the form factor includes additional corrections due to the spin structure of the nucleus, making the calculation more complex. At first order, the form factor for SD interactions can be expressed as:

$$|F_{SD}(q)|^2 = \left| \frac{sin(qR_n)}{qR_n} \right|^2 \tag{2.24}$$

## 2.2 Experimental signature

### 2.2.1 Energy dependence

The shape of the differential event rate depends on several factors, including the masses of the WIMPs and the target nuclei, as well as the velocity distribution of the WIMPs in the galactic rest frame. Starting from Eq. 2.5, if we assume isotropic scattering for a WIMP with kinetic energy $E_i$, then $\cos(\theta_r)$, where $\theta_r$ is the recoil angle, is uniformly distributed between -1 and 1. Therefore the recoil energy is uniformly distributed between 0 and $E_i r$. The spectrum is constrained at high energies by the maximum possible energy $E_{max}$ and in intensity by the damping effect of the form factor, especially for large A targets. On the other hand at lower energies, the distribution is limited by the energy threshold $E_{thr}$, which depends on the characteristics of the specific detector employed for the direct search.

### 2.2.2 Time dependence

The Earth's orbital motion around the Sun induces a time dependence in the WIMP recoil rate, known as annual modulation [40; 41; 42]. Because of the Earth revolution, the spreed of the Dark Matter particles in the Milky Way halo relative to the Earth is largest in June. As result, the number of particles capable of producing nuclear recoils above the detector's energy threshold is maximized in June and minimized in December.
The temporal variation of the differential event rate can be written as:

$$\frac{dR}{dE_r}(E_r,t) \approx R_0(E) + R_m(E)\cos\left[\omega\left(t-t_0\right)\right] \tag{2.25}$$

where $R_0$ is the time averaged event rate and $R_m$ is the modulated amplitude. The dependence on time is expressed by a sinusoidal function of frequency $\omega = \frac{2\pi}{year}$ and by the phase of the modulation $t_0$. A rate modulation would, in principle, enhance the ability to discriminate against background and help to confirm a Dark Matter detection.
In order to be sensitive to this signature, an experiment needs a large exposure with a highly stability in the response over an O(1) year of time.

### 2.2.3 Directional dependence

Directionality is another Dark Matter signature which can be employed for detection, as the nuclear recoils resulting from WIMP interactions exhibit strong angular dependence. The Sun's motion through the galaxy points toward a fixed direction in the sky (roughly corresponding to the Cygnus constellation), and as the Earth follows this path while orbiting the Sun, it moves through a "wind" of WIMPs. This leads to a characteristics angular distribution of nuclear recoils in the laboratory frame that change constantly during the sidereal day. This directional dependence is reflected in the differential event rate when expressed as a function of the angle $\gamma$, which defines the direction of the nuclear recoil relative to the average direction of the solar motion:

$$\frac{d^2R}{dE_r d\gamma} \propto exp\left[\frac{-\left(\left(v_E+v_S\right)\cos\left(\gamma\right)-v_{min}\right)^2}{v_c^2}\right] \tag{2.26}$$

where $v_E$ is the velocity of the Earth relative to the galactic rest frame, $v_S$ is the velocity of the Sun around the galactic center, $v_{min}$ is the minimum WIMP velocity required to produce

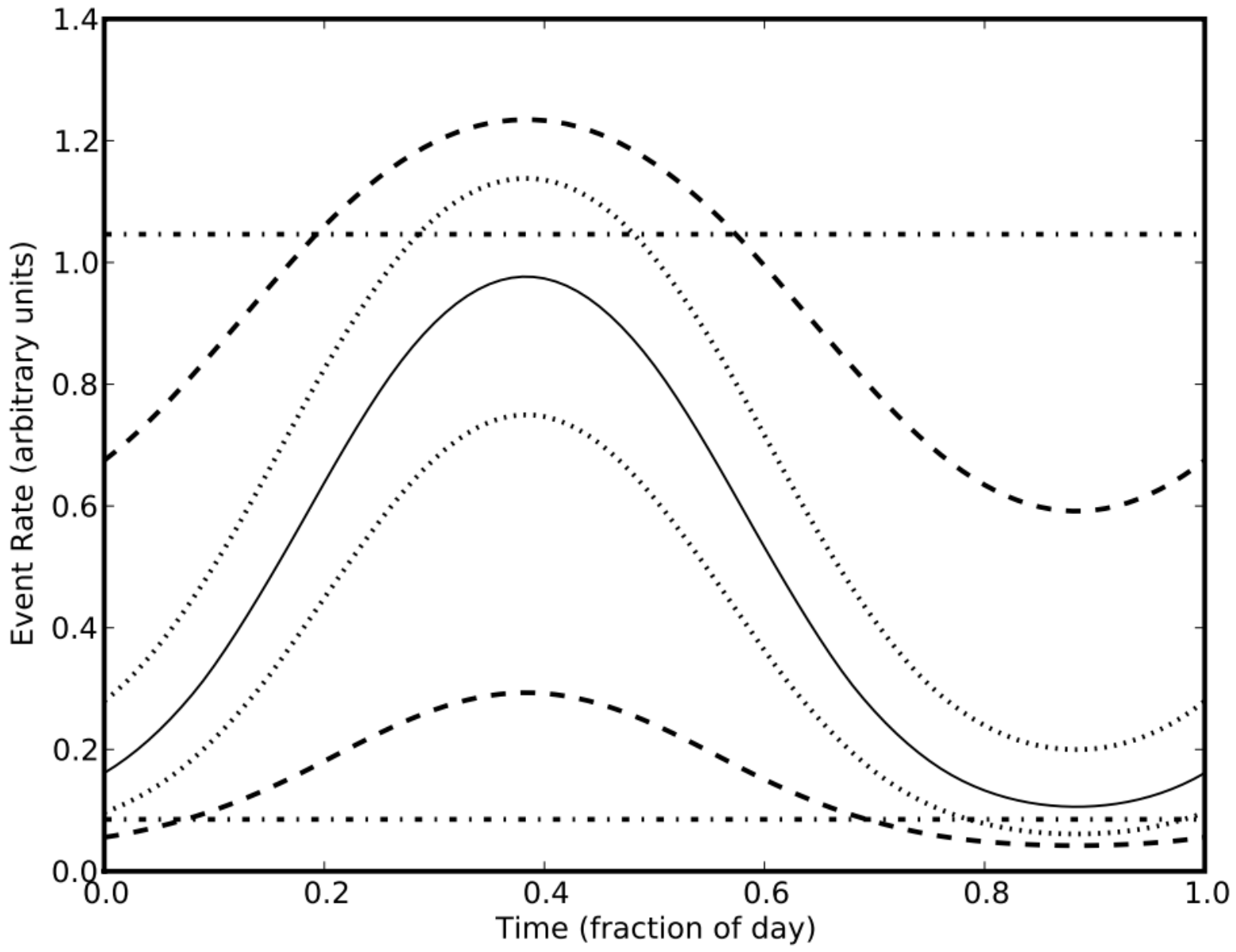

**Figure 2.1** Magnitude of the daily modulation for seven lab-fixed directions, specified as angles with respect to the Earth's equatorial plane. The solid line corresponds to zero degrees, and the dotted, dashed and dash-dot lines corresponds to $\pm$ 18°, $\pm$54° and $\pm$90°, with negative angles falling above the zero degree line and positive angles below. The $\pm$90° directions are co-aligned with the Earth's rotation axis and therefore exhibit no daily modulation. This calculation assumes a WIMP mass of 100 GeV and $CS_2$ target gas.

a nuclear recoil of an energy $E_r$ and $v_c$ is the halo circular velocity, given by:

$$v_c = \sqrt{\frac{3}{2}} v_S \tag{2.27}$$

The number of recoils along a particular direction in the laboratory frame will change throughout the day. The amplitude of this daily modulation, shown in Fig. 2.1, depends on the relative orientation between the lab-fixed direction and the spin axis of the Earth, with no modulation along directions parallel to the Earth's spin axis. Since most background is expected to be isotropic, measuring both the direction and energy of nuclear recoils provides a robust method for identifying WIMP signals and excluding other backgrounds, including neutral particles.

The dependence of the rate on the $\cos(\gamma)$ remarks the anisotropic nature of the angular distribution of the recoils, which manifests in galactic coordinates as dipole. This system, a right-handed celestial coordinate system in spherical coordinates centered on the Sun, has its x-axis pointing towards the center of the Milky Way, while the y-axis aligns roughly with the Sun's peculiar motion relative to the Galactic rest frame. The Galactic plane corresponds to the equatorial plane, with Galactic longitude $l$ and latitude $b$ resembling terrestrial coordinates.

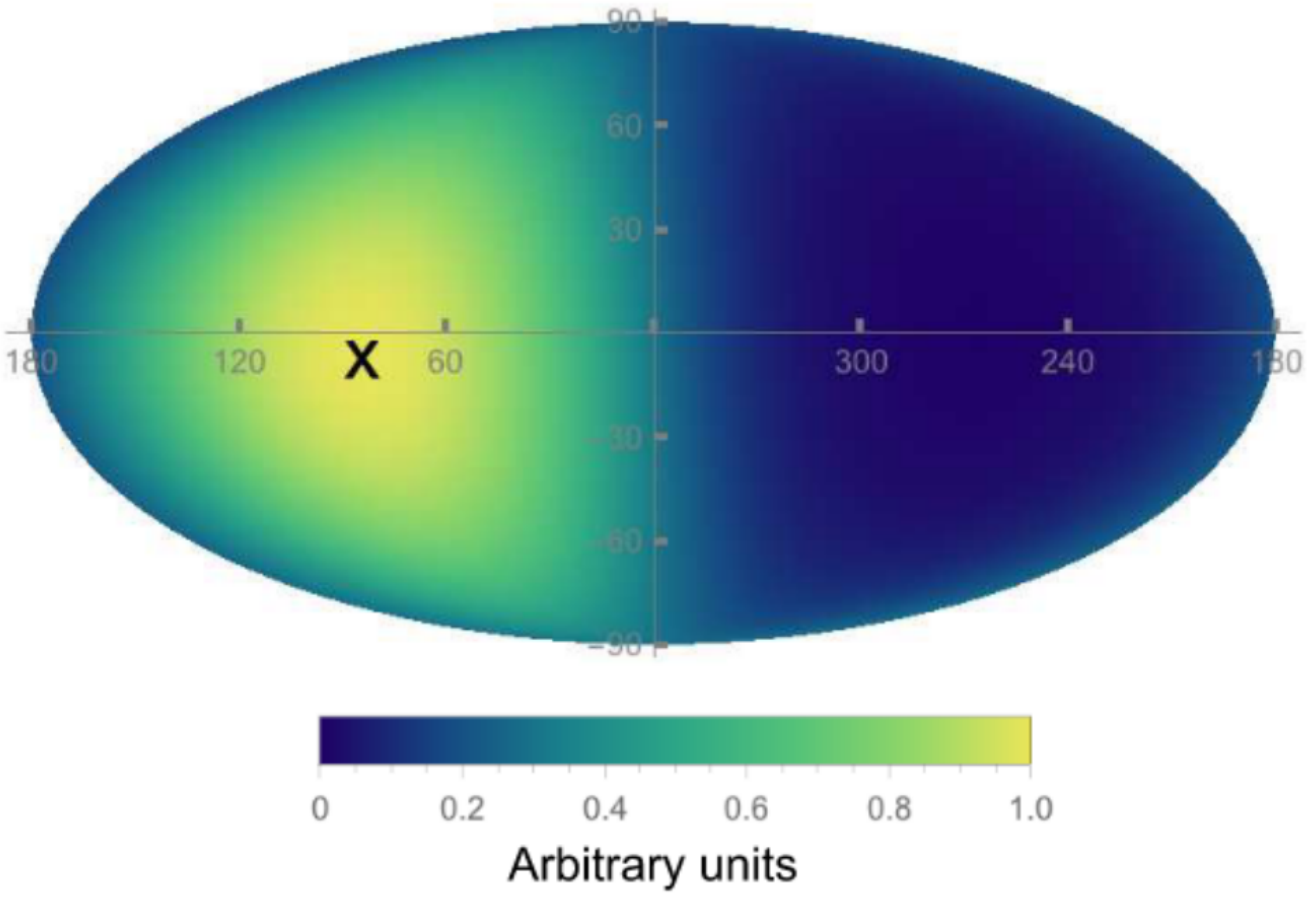

**Figure 2.2** Angular distribution of the energy differential recoil rate for 25 keV fluorine recoils from 100 GeV WIMPs [32].

In this coordinate system, the constellation of Cygnus is located at $(l, b) \simeq (81, 0.5)$, and an excess of recoil events at negative longitude, opposite to the motion of the Sun is expected. Fig. 2.2 illustrates the normalized angular distribution of 25 keV fluorine recoils generated by 100 GeV WIMPs in Galactic coordinates. The cross marks the laboratory's direction of motion. For clarity, the angular distribution is depicted based on the direction from which the recoils originate. The figure distinctly shows a dipole pattern in the distribution, which is widely regarded as a "smoking gun" signature for the detection of galactic WIMPs.

## 2.3 Background sources and reduction techniques

Direct detection experiments aim to detect signals caused by interactions between Dark Matter particles and ordinary matter, which typically result in nuclear recoils with energies on the order of a few keV, as predicted by the kinematics. However, the expected event rate is below one event per kilogram per year, making DM detection a search for rare events. Given the low expected rate, the experiments are required to keep increasing the target mass and the time exposure of the detectors.

Detecting Dark Matter is complicated by various source of background noise that can mimic DM signals. These include cosmic rays and natural radioactivity from the environment and detector materials, which produce at rates typically from $10^6$ to $10^8$ times larger than the expected DM signals. The main strategies to suppress these backgrounds involve shielding, material selection and rejection in data analysis. Neutrons and neutrinos can also induce an irreducible background of events that resemble WIMPs interactions, making it necessary to use techniques like directional or topological discrimination.

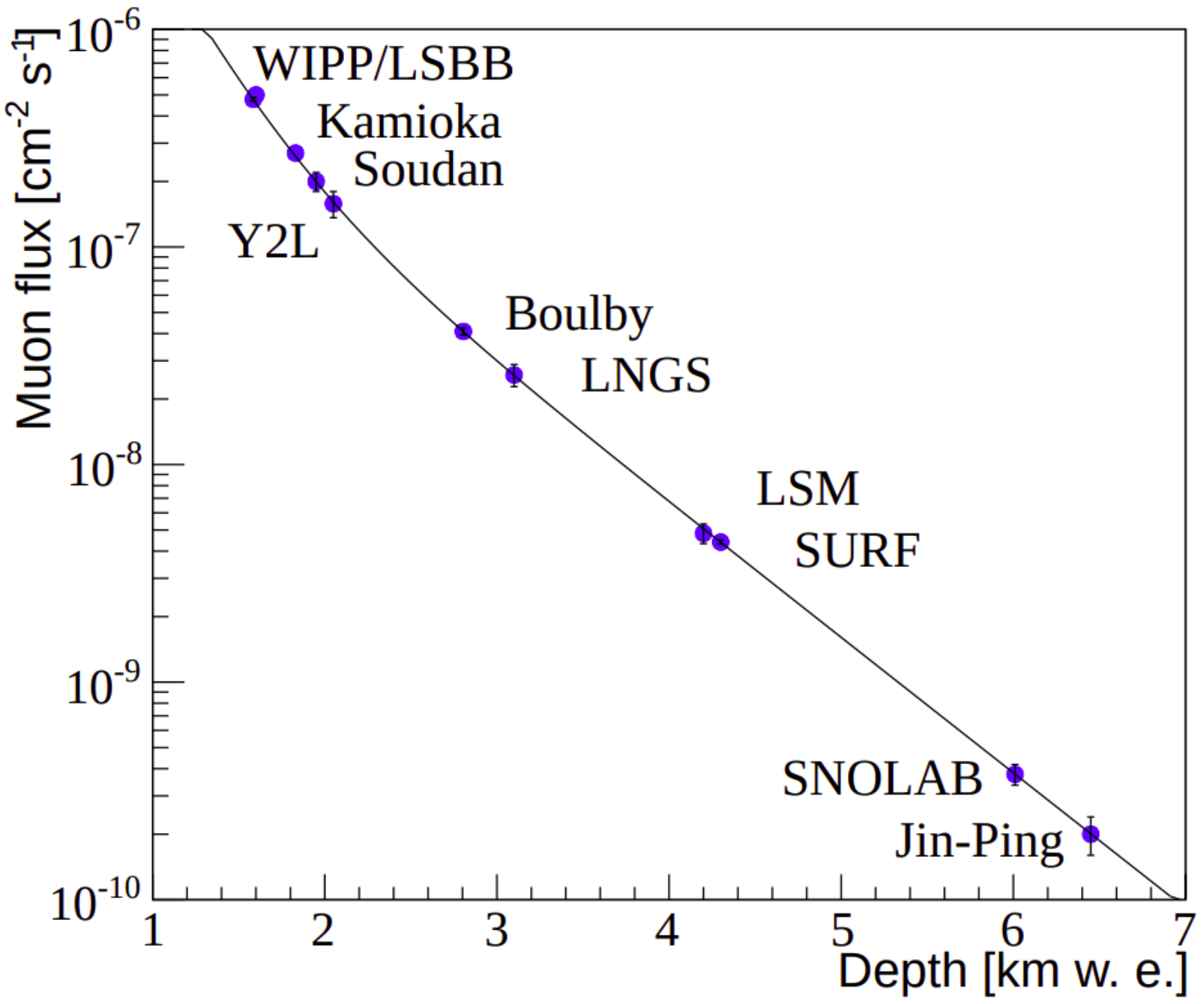

**Figure 2.3** Muon flux as function of depth in kilometers water equivalent (km w. e.) for various underground laboratories hosting Dark Matter experiments [29].

## 2.3.1 Cosmic rays

One of the main challenges in direct detection experiments arises from cosmic rays, which interact with Earth's atmosphere producing shower of high-energy particles, mainly muons, but also pions, neutrinos and neutrons. Muons, in particular, are highly penetrating, they have energies ranging from MeV to hundreds of GeV, and can generate background events within the detector's sensitive volume. To reduce cosmic ray background, experiments are conducted in underground laboratories, where the rock above acts as passive shielding. The shielding power of the rock also depends on its composition. The depth of these laboratories is usually expressed in meter water equivalent (m.w.e) that is the water that would reduce by the same factor the flux of cosmic rays as measured at the laboratory location. Fig. 2.3 shows how the muon flux decreases with depth in various underground labs. Underground laboratories are usually built by excavating tunnels below mountains, or by exploiting mining facilities. One of those is the National Laboratories of Gran Sasso (LNGS) of Istituto Nazionale di Fisica Nucleare (INFN) which hosts different direct DM search experiments, including CYGNO. Neutrons are the main responsible for the production of NRs in a detector. While hadrons produced in cosmic rays showers are suppressed by few m.w.e., muons can penetrate much more, and interact inside the rock producing neutrons via inelastic processes. These neutrons, called cosmogenic, can have energies up to several GeV and can interact with nuclei in the detector target via elastic scattering producing nuclear recoils [43]. This is a dangerous background because the type of signal is identical to that of the WIMPs.

### 2.3.2 External background

External background sources refer to those originating outside the experimental setup, typically from the surrounding environment. Natural radioactivity from the surrounding rocks remains a significant background source. The main sources of natural radioactivity are from the decay chains of $^{232}$Th, $^{238}$U, commonly found in the rocks and surrounding natural environment near the detector, $^{235}$U and the long-lived isotope $^{40}$K. These decay chain include both alpha and beta emitters, often accompanied by gamma emission. Additionally, anthropogenic and cosmogenic radioisotopes such as $^{85}$Kr, $^{137}$Cs, $^{39}$Ar, $^{110}$Ag and $^{60}$Co, present in laboratory objects, also contribute to the environmental radioactivity. These isotopes can decay in many ways with the emission of alpha particles, electrons, positrons or $\gamma$. Alphas, positrons and electrons generally are completely absorbed in the materials in which they are produced, while the $\gamma$ with energies from tens of keV up to 2.6 MeV can easily escape and potentially reach the experiment and induce ERs. $\gamma$-rays emitted from these isotopes can interact with the detector medium through the photoelectric effect, Compton scattering, or pair production. While the photoelectric effect has the highest cross-section at energies up to few hundred keV, the cross-section for pair production dominates above several MeV. For the energies in between, Compton scattering is the most probable process. All these interactions result in the emission of an electron, or electron and positron, which can deposit its energy in the target medium. Such energy depositions can be at energies of a few keV affecting the sensitivity of the experiments because this is the energy region of interest for Dark Matter searches.

Reaction such as ($\alpha$,n), where alpha particle induces neutron emission upon absorption by another nucleus, may occur. These emitted neutrons, known as radiogenic neutrons, can attain energies up to several tens of MeV and easily escape the material where they are generated. If they interact with the detector, they can produce both an ER and a NR background. Radiogenic neutrons also arise spontaneous from spontaneous fission processes, predominantly from isotopes like $^{238}$U and $^{235}$U, with energies around a few MeV. The cosmogenic and radiogenic contribution depend largely on the chosen material and their inherent alpha activity. Selecting radiopure materials, radiogenic contribution may be reduced of several order of magnitude compared to the cosmogenic flux.

A contribution to the environmental radioactivity comes from the radon isotopes produced in the natural radioactive decay chains as shown in Fig. 2.4. Radon, being a noble gas, does not chemically bind to the materials where it is produced, consequently it escapes from the rocks where U and Th are abundant and accumulate in the air. In particular, $^{222}$Rn, a daughter product in the $^{238}$U decay chain, has a relatively long half-life of 3.8 days. This allows it to diffuse over long distances in the air from its emanation site, entering inside the detector setup. $^{222}$Rn alpha decays to $^{218}$Po, which can be produced as ions and adhere to surfaces due to electrostatic attraction. This adhesion can lead to the accumulation of radioactive contaminants inside the detector, contributing to the background through alpha, beta, and gamma emissions.

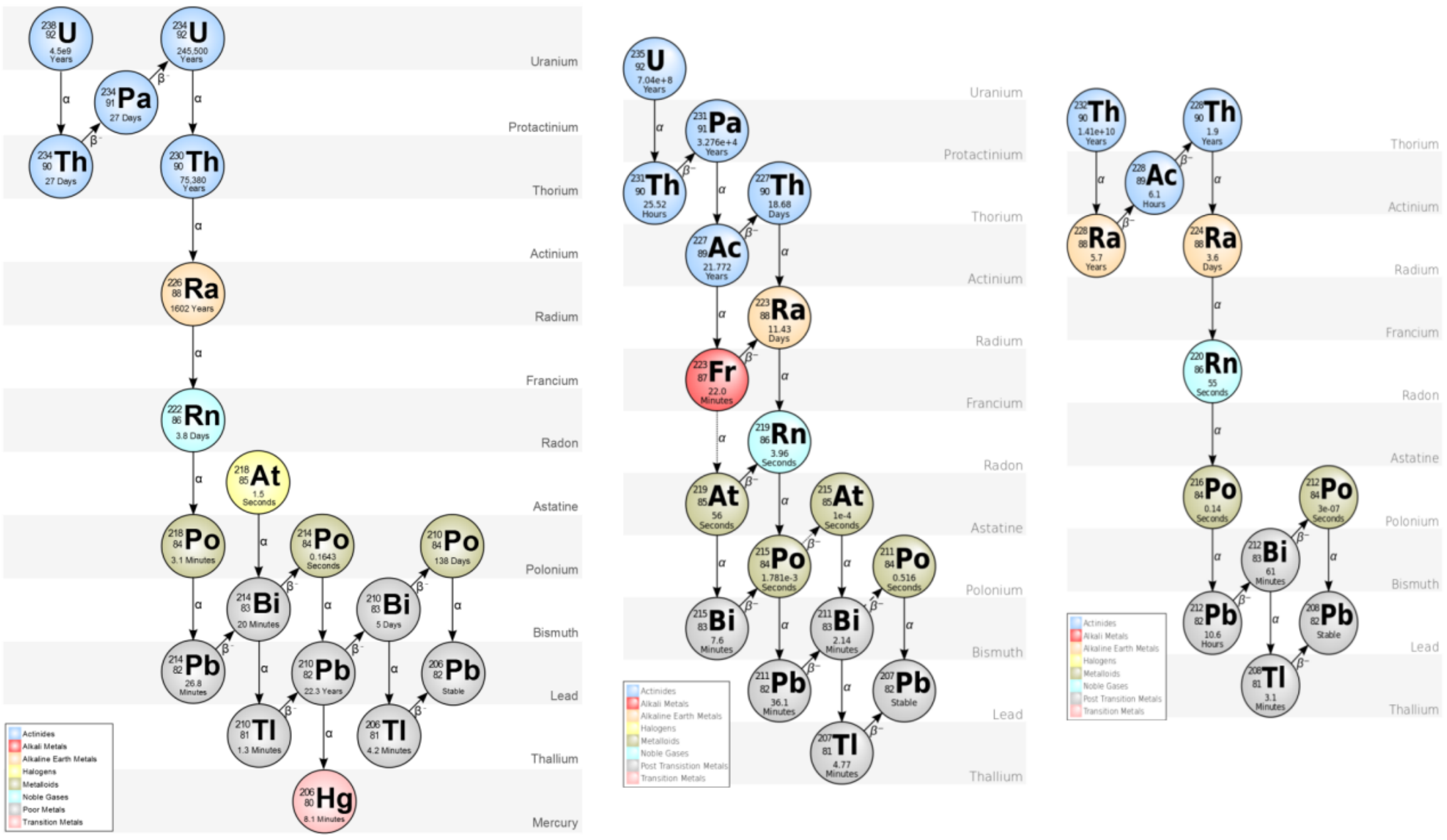

**Figure 2.4** Decay scheme of the natural radioactive families of $^{235}$U, $^{238}$U and $^{232}$Th [44].

In order to reduce the background induced by the external gamma and neutrons fluxes, experiments typically employ passive or active shieldings. The shields can be passive, used for its stopping power properties , or active, when it is instrumented and signals can be obtained from the energy deposited in the shield material to characterize the external background and actively suppress it through vetos. High Z materials are usually employed to suppress external gamma because interaction cross section scales with a power of Z. Typical materials of choice are copper (Z=29) and lead (Z=82). Lead offers superior stopping power due to its higher Z, but may introduce additional background from radioisotope $^{210}$Pb, which decay via beta emission. This decay process can produce gammas through bremsstrahlung, potentially contributing to an additional background. An innovative solution to the radioactive lead issue involves the use of archaeological lead, such as ancient Roman ingots retrieved from a sunken ship off Sardinia $\sim$ 2000 years ago. This ancient lead shows an extremely low radioactivity thanks to the long time spent underwater, protected from cosmogenic activation, and during which the activity of $^{210}$Pb was highly reduced, given its half-life of 22.3 years. Material rich in hydrogen, such as water or polyethylene, are employed to slow down and absorb neutrons thanks to the large kinematic match with H.
In addition to passive shielding, some active methods to reduce the external background are also often applied, like water Cherenkov tank surrounding the detector. The residual high energy muons reaching the underground laboratory can pass through the water and produce the emission of Cherenkov radiation, which is detected by a set of photosensors. To exclude this possible background, the muons are tagged and the events detected in a time window close to the detection of the muon are excluded. This technique is used for instance by XENONnT [45], COSINUS [46] and DarkSide-50 [47]. For the neutron veto system tanks of liquid scintillator enriched with elements with a high cross section for neutrons radiative capture (Gd, B) or with water (H) can be used. The neutrons can be tagged by detecting the light produced by the emitted gammas following the capture. This technique is used for example by LUX-ZEPLIN [48], XENONnT [49] and DarkSide-20k [50]. In large dense detecting sensitive volumes neutrons are expected to scatter more than once. Multiple neutron scattering events are typically excluded from the analysis.

### 2.3.3 Neutrino background

Another NRs background contribution for DM search experiment is given by neutrinos, in particular neutrinos emitted by the Sun, produced in the atmosphere by cosmic rays and coming from the diffuse flux of the Supernovae. Solar neutrinos are the most intense source of neutrinos on Earth, they are produced by the $pp$ cycle, which mostly induce ERs of the order of few keV. The main source of coherent elastic neutrino-nucleus scattering (CE$\nu$NS) are the neutrino emitted in the decay of $^{8}$B which generate NRs of the same energies, impossible to be singularly discriminated from WIMP induced ones. The neutrino energies and fluxes span over order of magnitude. In Fig. 2.5 the energy spectrum of the neutrinos from these different astrophysical source is shown. With increasing target masses approaching hundreds of kilograms to tons, direct Dark Matter detectors with sensitivity to keV energies start being sensitive to neutrino interactions.

### 2.3.4 Internal background

Another significant source of background arises from the intrinsic radioactivity of the materials used in the detector, target medium and shieldings. Radioactive contaminates in the detector materials, can introduce many kinds of backgrounds with the emission of alphas, electrons, positrons and gamma rays. Unlike the case of the radioactivity of the surroundings, short-range particles such as alphas emitted close to (or inside) the sensitive volume of the detector are likely to induce a background event. Therefore, ensuring the radiopurity of the detector materials is crucial for any direct Dark Matter search experiment.

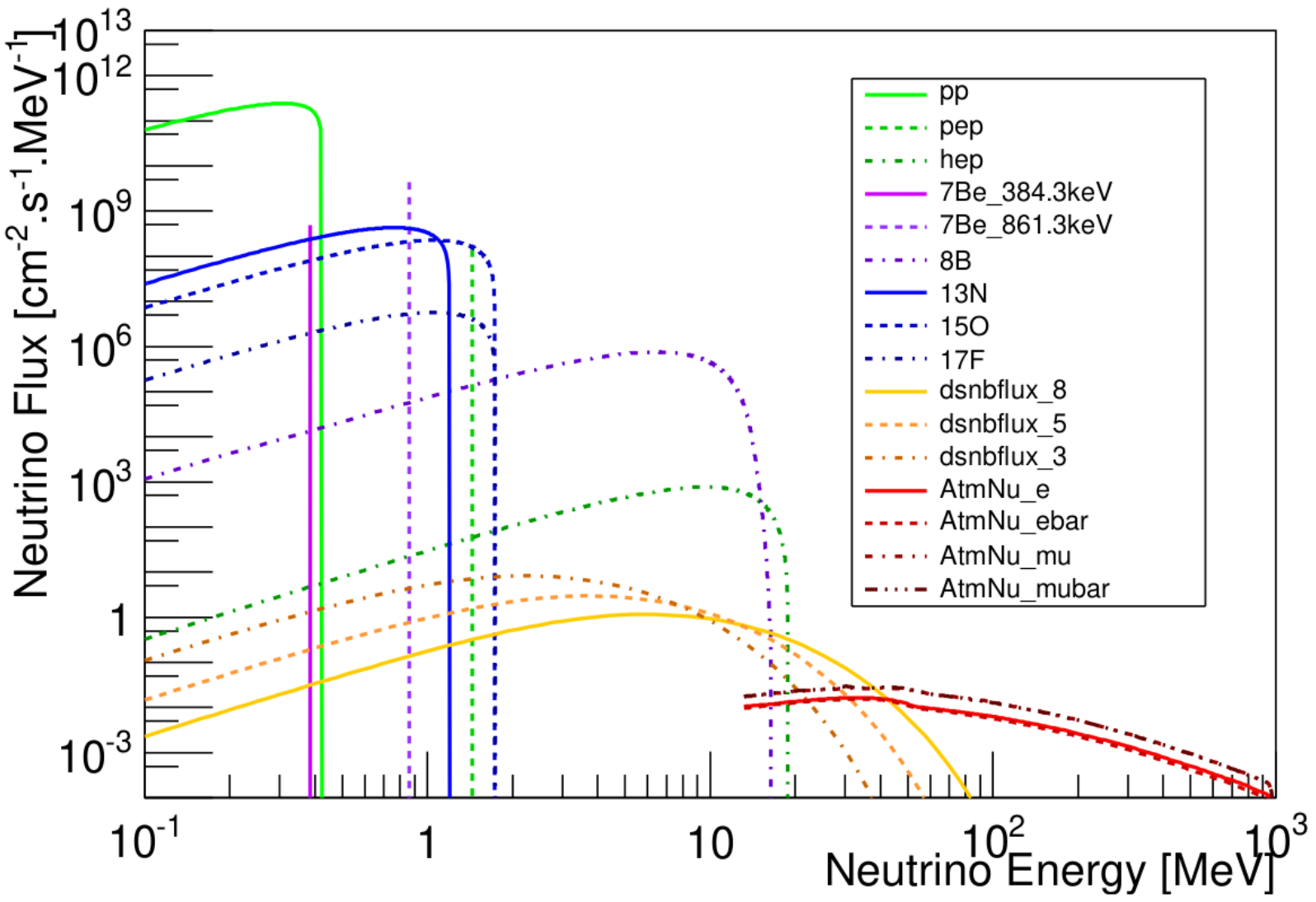

**Figure 2.5** Spectra of the most relevant neutrino sources [51].

The radioactive contamination of materials is often inevitable due to inherent production processes that introduce contaminants. To mitigate further contamination from cosmogenic activation, materials for ultra-low background experiments are preferably stored underground to limit surface exposure time. If this is unavoidable, suitable shieldings should be installed to minimize the interaction of cosmic rays.

## 2.4 Detector signals

The elastic scattering of a Dark Matter particle off a target medium induces for the case of the WIMP an energy transfer to nuclei which can be observed through three different signals, depending on the detector technology in use. These can be the production of heat (phonons in a crystal), an excitation of the target nucleus which de-excites releasing scintillation photons or by the direct ionization of the target atoms. Detection strategies focus either on one of the three signals, or on a combination of two of them. Combining multiple detection channels enhances the ability to distinguish DM signals from background events, since the response of media to an interaction is not only proportional to the deposited energy but depends on the type of particle that deposits the energy. More precisely, the relative size of two signals depends on the type of particle. This enables the discrimination of nuclear recoils from electronic recoils which is an important method to reduce the background of the experiment. To measure the ionization signal either germanium detectors or gases (low pressure, for directional searches) are employed while scintillation can be recorded for crystal and for noble-gas liquids. To detect heat, the phonons produces in crystals are collected using cryogenic bolometers at mK temperatures. The following section provides an overview of the primary techniques used for direct Dark Matter detection.

**Scintillating crystal** High-purity scintillating crystals, such as NaI(Tl) and CsI(Tl), offer high scintillation light yields, which enable them to achieve low energy thresholds. These crystals can be assembled into arrays to cover large masses and operate reliably for long durations. When a charged particle deposits energy in the crystal lattice, part of that energy is converted into visible or UV light through de-excitation. Dopants like thallium (Tl) are used to introduce additional energy levels into the lattice, increasing scintillation efficiency at room temperature. The emitted light is typically detected using photomultiplier tubes (PMTs). However, growing these crystals with high radiopurity is challenging, leading to significant background noise. Experiments using scintillating crystals include ANAIS-112 [52], COSINE-100, SABRE [53], and DAMA [54].

**Bolometers** Bolometers are cryogenic detectors that operate at extremely low temperatures, typically below 50 mK. They detect the tiny temperature changes caused by particle interactions with phonons. These low temperatures suppress thermal noise and reduce heat capacity, which is primarily due to lattice vibrations scaling as $T^3$. Both ERs and NRs primarily dissipate energy as heat, but they differ in their ionization or scintillation responses. NRs usually transfer less energy to ionization and excitation of atoms compared to ERs. Thus, bolometers often combine temperature measurements with scintillation light or ionization charge to differentiate between ERs and NRs based on their different energy loss ratios. Experiments using bolometers include EDELWEISS-II [55], CRESST-III [56], COSINUS, and Super-CDMS [57].

**Noble liquids detectos** Noble liquids like argon and xenon are easily ionized and have excellent scintillation properties. When an ionizing particle passes through, it excites atoms, leading to the emission of UV photons with varying decay times. The scintillation pulse shape, determined by the relative populations of singlet and triplet states, differs between electronic and nuclear recoils. Experiments instrument large volumes of these liquids with light detectors to measure primary scintillation. In their liquid form, these detectors can achieve masses on the order of tonnes. Examples of single-phase noble liquid detectors include DEAP-3600 [58] and XMASS [59].
In dual-phase Time Projection Chambers (TPCs), noble liquids are used to measure both scintillation light and ionization charge. The primary light (S1 signal) is collected by photosensors, while ionization electrons are drifted by an electric field towards a gas-liquid interface. Here, the electrons are accelerated, generating secondary scintillation (S2 signal) through electroluminescence, which is also detected optically. By combining S1 and S2, the energy deposited, particle identification, and interaction position can be determined. Experiments using this technology include XenonnT [60] and DarkSide [61].

**Bubble Chambers** Bubble chambers use superheated liquids maintained at temperatures just above their boiling points. Energy deposits above a set threshold trigger local phase transitions, nucleating bubbles that grow to macroscopic sizes and can be optically detected. Bubble nucleation requires a critical energy deposit within a specific stopping power range, allowing operational parameters to be tuned so that only NRs trigger bubble formation. This renders the detectors insensitive to ER backgrounds. Alpha particles create larger bubbles (about 10 $\mu$m), while NRs produce much smaller ones (about 10 nm). Stereoscopic cameras with millimeter precision are typically used to capture the position and number of bubbles. Using fluorine-rich liquids, such as $C_3F_8$ and $C_4F_{10}$, enhances sensitivity to Spin-Dependent interactions, while combining these with heavier elements like iodine improves Spin-Independent sensitivity. However, bubble chambers cannot measure the energy of deposited events, relying solely on event counts, which limits their ability to characterize DM mass. The PICO [62] experiment is a leading example of this technique.

**Gaseous Detectors** Gaseous detectors optimized for pure ionization detection are being developed with a focus on low-mass DM searches. These detectors offer low energy thresholds and can provide directional information. Directional detectors often use low-pressure TPCs to reconstruct angular distributions of events. These setups typically combine a gaseous TPC with high-resolution readouts for ionization charge or scintillation light detection. The lower density of the medium results in longer particle tracks, which facilitates directional reconstruction and enhances ER/NR discrimination based on distinct energy deposition patterns along the particle track.

### 2.4.1 Current limits

The observable rate of WIMP-induced NRs depends primarily on two unknown parameters: the total WIMP-nucleus cross section and the mass of the WIMP. If a positive detection of WIMP-induced NRs is made, it would identify a region in the parameter space defined by these two variables. In contrast, absence of detected events allows experiments to set upper bounds on the cross section.

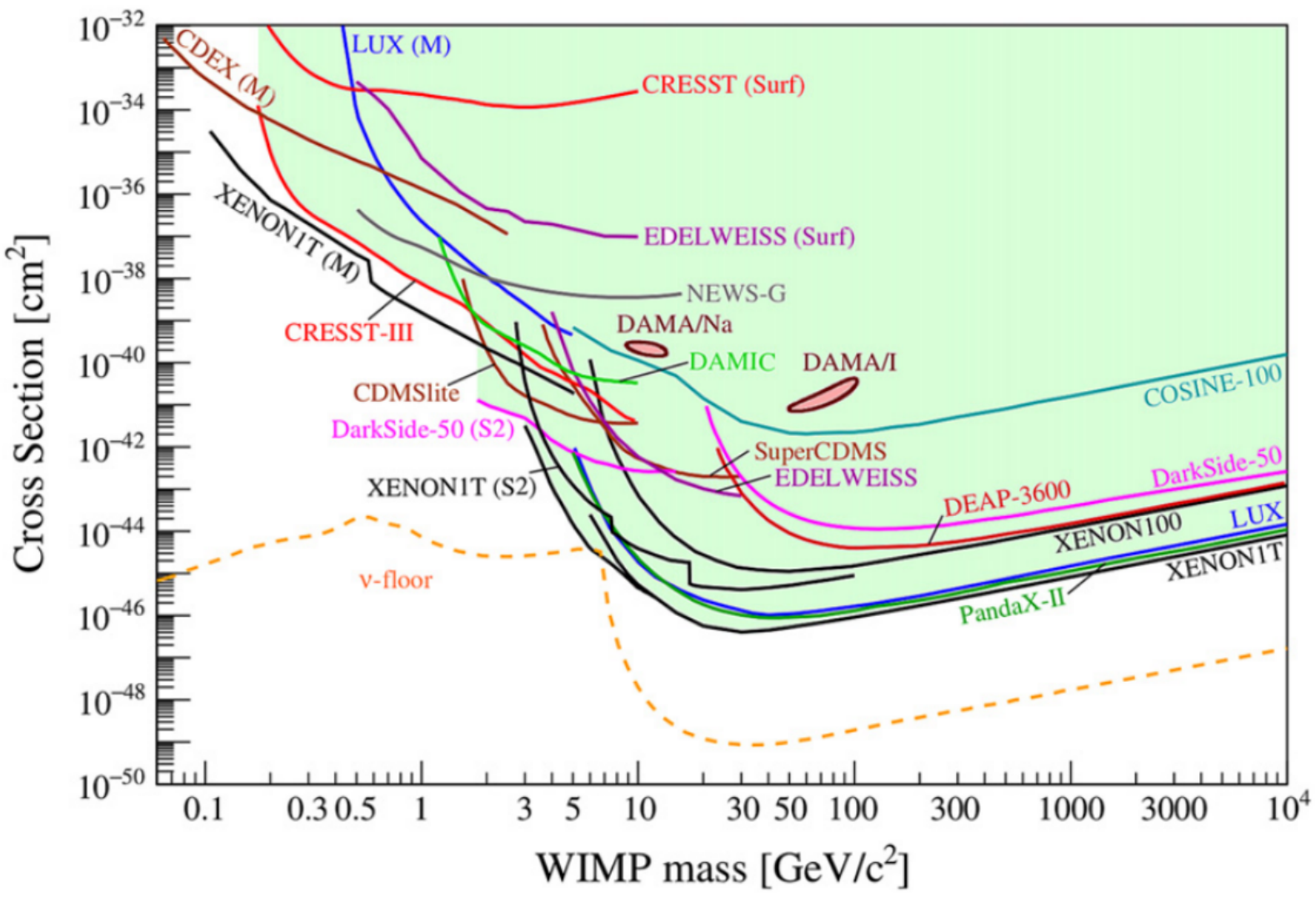

**Figure 2.6** Current status of the SI limit on the cross section WIMP [18].

In Fig. 2.6 the current status of the limits on the SI cross section as a function of the WIMP mass is shown. The DAMA/LIBRA experiment observed an annual modulation of NRs in the energy range between 2-6 keV with a 2.86 tonne y exposure and 13.7 $\sigma$ significance, consistence with the WIMP hypothesis [63]. To solve this inconsistency some experiments, such as the aforementioned ANAIS-112, COSINE, SABRE and COSINUS were specifically designed to replicate DAMA/LIBRA's results by employing the same crystal medium. These experiments aimed to either confirm or refute DAMA's claims. Recent analyses from COSINE-100 [64] and ANAIS-112 [65] have already excluded the DAMA modulation region. As seen in Fig. 2.6 , the shape of the exclusion limits shows a minimum where the WIMP mass closely matches the atomic mass of the target, for kinematic reasons. At higher masses, since the local Dark Matter density is constant, heavier WIMPs become rarer, leading to a suppression in scattering rates, scaling inversely with WIMP mass, $\propto 1/m_\chi$. Below the minimum, experimental sensitivity is restricted by the energy threshold, and thus the exclusion limits rise steeply. For WIMP masses exceeding 3 GeV, noble liquid TPC experiments, such as XENON, provide the most stringent limits on the cross section. This is due to their large, high density targets and excellent scintillation properties, which make reaching the tonne scale easier. In the 1 GeV range, bolometer detectors like CRESST-III and SuperCDMS dominate due to their exceptionally low energy thresholds, down to tens of eV. Light nuclei are ideal for probing this mass range, and the future CYGNO detector is expected to explore WIMP masses as low as 0.3 GeV by utilizing a hydrogen-enriched target.
For Spin-Dependent (SD) couplings, the sensitivity is generally about five orders of magnitude weaker than for SI couplings, due to the limited availability of spin-odd nuclei required for SD sensitivity. Future generations of experiments are expected to reach sensitivities close to the neutrino background, which will impose an inherent limit on detection capabilities. Therefore, the development of new experimental methods that can unambiguously identify Dark Matter signals, even in the presence of poorly understood backgrounds, is critical. One promising approach is directional Dark Matter detection, which seeks to address this challenge.

## 2.5 Directional Dark Matter search

As discussed in Sec. 2.2.3, WIMP-induced recoils are expected to be aligned opposite to the direction of the laboratory's motion to the galactic rest frame. In galactic coordinates, this translate into a dipole-like angular distribution, with the majority of recoils appearing from the direction of the Cygnus constellation (∼90°,0). The recoil direction, denoted by $\theta_r$, depends on several parameters including the recoil energy $E_r$, the WIMP mass $m_\chi$, the target mass $m_A$ and the relative WIMP velocity relative to the target $v$ according to:

$$cos\theta_r = \sqrt{\frac{m_A E_r}{2\mu^2 v^2}} = \frac{v_{min}}{v} \tag{2.28}$$

Here, $\mu$ is the WIMP-target reduced mass. The angle $\theta_r$ can range between 0° and 90°, depending on the target and the recoil energy. Higher recoil energies result in a stronger dipole feature, i.e. $\theta_r$ close to 0°, but the event rate decrease. On the other hand, at lower recoil energies, achieving good angular resolution becomes more challenging experimentally. For larger WIMP-target reduced mass and lower recoil energies and target masses, the maximum event rate forms a ring around the WIMP arrival direction. Additionally, since the earth's velocity relative to the galactic rest frame varies throughout the year, so does the angular distributions of recoils. The difference in angular distribution at the minimum and maximum velocities introduces an aberration pattern, which is difficult to detect and requires a large number of events to measure appreciably. As a result, the most investigated feature of WIMP-induced recoils is the dipole shape in the angular distribution, which can provide positive identification of a WIMP signal with sufficient exposure and a low energy threshold. Directional detection, specifically the dipole feature in the angular distribution of nuclear recoils (NRs), is considered one of the most promising methods for detecting galactic Dark Matter.

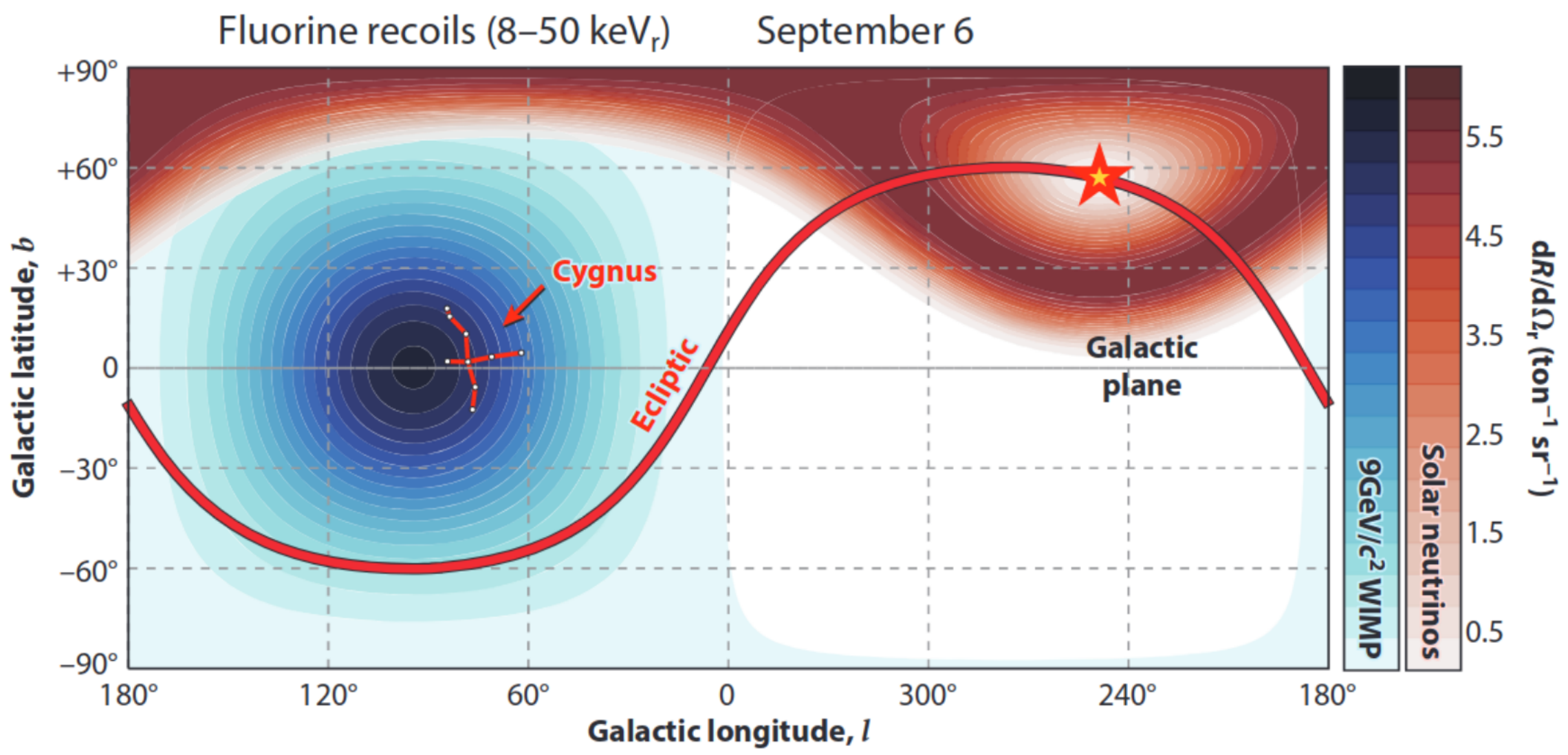

**Figure 2.7** Directional event rates from a 9 GeV WIMP (blue) and solar neutrinos (red) displayed in galactic coordinates. The red lines indicates the path of the ecliptic, showing minimal overlap between the two distribution over the course of the year [66].

Unlike non-directional detectors, directional detectors provide additional angular information for each recoil event. This extra information significantly enhances the ability to distinguish between potential Dark Matter signals and background events, especially since background signals tend to have different angular distributions compared to WIMP-induced recoils. Most internal and external backgrounds exhibit nearly isotropic distributions in galactic coordinates, as Earth's rotation and the tilt of its axis relative to the galactic plane reduce local anisotropies. This ensures that background signals are unlikely to mimic WIMP-induced events, enabling directional detectors to set stronger exclusion limits than non-directional detectors with the same exposure. This advantage arises because directional detectors can exclude backgrounds by rejecting the hypothesis of isotropic event distributions, providing a more powerful tool for probing lower WIMP-nucleon cross sections.

One of the most significanct source of directional background is coherent elastic neutrino-nucleus scattering (CE$\nu$ES) from solar neutrino, which non-directional detectors cannot exclude on an event-by-event basis. However, directional detectors with sufficient performance, could effectively exclude solar neutrino events throughout the year. This is because the direction of the Sun and the Cygnus constellation never overlap, as shown in Fig.2.7.

### 2.5.1 Directional Detection techniques

Directional detection techniques need to achieve low energy thresholds with high angular resolution while also maintaining a large exposure. Low-energy nuclear recoils typically produce sub-micrometer track lengths in solids and liquids, making directional measurement challenging. However, the use of dense liquid or solid targets allows detectors to achieve large target masses, and thus large exposures. In contrast, gaseous targets have much lower densities, which reduces the exposed mass compared to solid or liquid targets of the same volume. Nevertheless, nuclear recoils in gaseous targets can produce track lengths of approximately 1 mm, allowing for more precise measurement of both the direction and sense of the recoil. Additionally, gaseous detectors can be operated at pressures below atmospheric levels, further enhancing their directional capabilities, though this comes at the expense of reducing the total target mass.

**NEWSdm** The NEWSdm experiment [67] utilizes an emulsion film, composed of silver halide crystal dispersed in a polymer, which serves as both a target and tracking detector. The crystals are composed of a mix of elements, including H (1.6%), C (10.1%), N (2.7%), O (7.4%), Br (32.0%), Ag (44.0%) and I (1.9%). The total density is 3.2 g/cm$^3$. The presence of both light elements like C, O and N, and heavier element like Ag and Br, enables the detector to be sensitive to WIMPs of both low and high masses. To minimize external background, the sensitive volume is enclosed within a shield. Because nuclear emulsion detectors lack time resolution, the system must be installed on an equatorial telescope to account for the Earth's rotation, ensuring the detector's orientation remains fixed relative to the incoming WIMP flux. The current prototype setup features a 10 g target, housed at the underground LNGS laboratories, with the goal of testing production, calibration and analysis techniques while also measuring environmental backgrounds. The passive shielding is made of 10 cm of lead and 40 cm of polyethylene to shield the detector from environmental gammas and neutrons respectively.
Assuming 100 background events, the experiment can achieve a sensitivity of approximately $10^{-42}$ cm$^2$ for a WIMP nucleus cross section at 100 GeV with an exposure of 100 kg$\times$year. The detector's spatial resolution, which reaches O(10) nm using polarization based techniques, could allow it to probe the neutrino floor with an exposure of 10 ton$\times$year, assuming zero background.

**DRIFT** DRIFT (Directional Recoil Identification From Tracks) [68] is one of the pioneering experiments in direction Dark Matter detection. It uses a dual Time Projection Chamber (TPC) filled with low-pressure gas and equipped with Multi Wire Proportional Counters (MWPC) for charge amplification and readout. The two TPCs share a common aluminized Mylar cathode (0.9 $\mu$m thick), creating two separate sensitive volumes. Each TPC is read out by a MWPC that consists of an anode plane with 522 stainless steel wires (20 $\mu$m diameter, 2 mm pitch) for the x coordinate and grid planes with 100 $\mu$m wires (2 mm pitch) for the y-coordinate.
The detector uses a gas mixture of $CS_2$:$CF_4$:$O_2$ (30:10:1 Torr) in negative ion drift mode. It operates underground at the Boulby mine in the UK, with additional plastic shielding to reduce neutron backgrounds. The DRIFT experiment demonstrated sensitivity to the head-tail (HT) directional signature of the nuclear recoils down to 39 keV$_{nr}$ using a $^{252}$Cf neutron source.

**NEWAGE** The NEWAGE experiment [69] (NEw generation WIMP search with an Advanced Gaseous tracker Experiment) uses a micro-TPC filled with $CF_4$ gas at 100 mbar, operated underground at the Kamioka laboratory in Japan. The detector amplifies ionization signals using Gas Electron Multipliers and reads them with a low-radioactivity $\mu$-PIC (micro-pixels) plane, which also provides additional charge amplification. Signals from the anode and cathode strips are recorded by a 100 MHz fast-ADC capturing both charge and time-over threshold information, enabling 3D track reconstruction.
NEWAGE achieved an energy threshold of 50 keV$_{ee}$ with an angular resolution of approximately 40°. After 318 days of data taking, the experiment set a 90% confidence level limit on the SD WIMP-proton cross section of 25.7 pb for a WIMP mass of 150 GeV, based on forward-backward asymmetry in galactic coordinates.

**DMTPC** The DMTPC [70] (Dark MAtter Time Projection Chamber) experiment uses a low-pessure TPC filled with $CF_4$ gas (30-100 Torr), with external optical readouts (CCD, PMTs) and charge readouts. Electrons from primary ionization are amplified between closely spaced ground and anode mesh planes under a high electric field, producing secondary scintillation light detected by CCD cameras and PMTs. This setup allows for the measurement of the energy of each event (from the total light) and the 2D shape of the tracks (from the pixelated light sensor). PMTs and charge readouts provide tentative 3D track reconstruction. Various DMTPC prototypes were initially operated at MIT and later underground at WIPP in the USA. The experiment reported recoil direction measurements, achieving an HT efficiency of about 75% at 200 keV. Initial limits on the Spin-Dependent WIMP-proton cross-section were also set using a 10-liter detector.

**D$^3$** The D$^3$ [71] (Directional Dark Matter Detector) project focuses on developing TPCs with highly segmented pixelated charge readouts. The readout chips used in these detectors feature pixel sizes of 50×400 $\mu$m and 50×250 $\mu$m, with each pixel incorporating an amplifier, discriminator, shaper, and digital control. Using a double GEM stack with a gain of ap-proximately $10^4$ and low electronic noise, the detector achieves nearly $100\%$ single-electron efficiency. Measurements from a small 4.5 cm drift chamber filled with He:$CO_2$ or AR:$CO_2$ gas mixtures at atmospheric pressure demonstrated spatial resolution of $\sigma_x$ = ( 197 $\pm$ 11) $\mu$m and $\sigma_y$ = (142 $\pm$ 9) $\mu$m, as well as angular resolutions around 1 degree for alpha tracks. A larger detector, CYGNUS-HD, with a 40 liter active volume is under development, aiming to scale up to a cubic meter for future experiments.

**CYGNUS** CYGNUS [72] is a proto-collaboration aimed at constructing a network of modular, multi-site, and multi-target Galactic Observatory based on gaseous TPC using a directional approach to search for DM within the neutrino fog and to study solar neutrinos. Organizing the experiment in a modular and multi-site manner alleviates some of the issues regarding the very large volumes needed for a gas target to reach the required mass. A globally distributed experiment offers additional advantages, such as mitigating seasonal or location-specific background noise and overcoming site-specific restrictions on detector size.

Past and ongoing directional detection experiments and R&D efforts such as DRIFT, MIMAC, $D^3$, DMTPC, NEWAGE and CYGNO have demonstrated the low-density gas TPC concept, and have made many impressive technological advances over the years. In the next chapter the development and principles behind the CYGNO detector are described. This includes the choices related to detector design, gas mixture, amplification techniques and readout system.

# 3 The CYGNO Project

The CYGNO project is developing a directional detector aimed to search for Dark Matter and rare events at low energy. The CYGNO experimental setup involves a high resolution Time Projection Chamber (TPC), filled with a gas mixture of helium (He) and carbon tetrafluorine ($CF_4$) operated at atmospheric pressure and room temperature. The primary charge amplification is achieved through Gas Electron Multipliers (GEMs). A regular feature of this system is the production of scintillation light during charge multiplication, enabled by the presence of $CF_4$ in the gas mixture. The CYGNO's innovative approach lies in the use of optical readout to capture this scintillation light. Scientific CMOS (sCMOS) cameras are employed to image large areas with high granularity. By combing images from sCMOS camera with signals from photomultiplier tubes (PMTs), CYGNO aims at achieving a 3D particle tracking with directional capabilities down to O(1) keV energy.

Time Projection Chambers based on Micro Patter Gaseous Detector (MPGD) and optical readout represent ideal candidates for high resolution particle tracking. MPGDs allow to equip large surface, ensuring very good spatial and timing resolution.

Several prototypes of increasing size have been developed to study this experimental technique and evaluate its performance.

CYGNO is part of the proto-collaboration CYGNUS, a joint effort between different international groups. Together, they aim to build and operate a multi-site galactic directional recoil observatory to explore the Dark Matter hypothesis beyond the neutrino fog and detect solar and supernova neutrinos at the ton scale [72].

## 3.1 Interaction with the matter

When a charged particle traverses a medium, it primarily loses energy through two mechanisms: inelastic collisions with atomic electrons and elastic scattering off nuclei. Other processes are relevant only at high energy, such as the emission of Cherenkov radiation, nuclear reactions and bremstrahlung. For most particles, the majority of the energy is lost in the inelastic collisions with the atomic electrons except when the particle's mass is similar to that of the target nucleus, as in the case of nuclear recoils. Low energy NRs lose energy through nuclear stopping (collisions with atoms) and electronic stopping (collisions with electrons). This means that when an ion or recoiling nucleus moves through the medium, its energy is distributed between kinetic energy transferred to nearby atoms and energy used to excite or ionize electrons.

The energy partition depends on the masses and charges of the interacting particles and the ion energy. Heavier ions at lower energies tend to lose more energy through nuclear stopping, resulting in less ionization. Detectors that rely solely on ionization or scintillation are particularly affected by this, leading to a phenomenon called quenching, where the ionization signal is reduced.

Elastic collisions with nuclei dominate energy loss at lower energies, while inelastic collisions with electrons are responsible for generating a detectable signal, particularly in Time Projection Chambers (TPCs), which measure the energy lost to ionization or excitation. Since inelastic collisions are inherently statistical, calculating the exact behavior of energy loss requires quantum mechanical methods. However, in practice, this energy loss is typically averaged over the particle's path, described by the stopping power, denoted as $dE/dx$.

Low-energy electrons primarily lose energy via collisions with atomic electrons, while radiative losses like bremsstrahlung become significant only for electron energies much greater than 1 MeV. Although elastic collisions with electrons are less frequent, they still occur, causing the electron to lose energy and change direction with each collision. This results in an electron recoil (ER) track characterized by uneven energy deposition, with most of the energy deposited near the track's end, and a curly, irregular shape due to multiple trajectory deviations.

In contrast, at the low energies relevant to DM searches, NR stopping power decreases with energy, meaning most energy is deposited early in the particle's path. This behavior contrasts with higher-energy particles, which deposit most of their energy at the end of their path, forming a Bragg peak. This creates an asymmetry in the track profile that can be used to infer the direction of the particle's motion, a characteristic known as the head-tail (HT) effect.

Photons interact with matter primarily through three processes: photoelectric effect, Compton scattering and pair production. In the photoelectric effect, which is dominant at low energy range around 1 MeV, a photon transfers its full energy to an atomic electron. If the photon's energy exceeds the electron's binding energy, the electron is ejected with the remaining energy as kinetic energy. After this ejection, secondary effects can occur. For instance, an electron from an outer shell may fill the vacancy in the inner shell, releasing the energy difference as an X-ray of characteristic energy. This energy can be transferred to another electron, ejecting it via the Auger effect. The cross-section of the photoelectric effects scales approximately as $\propto Z^5$, where $Z$ is the atomic number of the material. Compton scattering, becomes dominant in the medium energy range, around 1 MeV. In this process, a photon scatters off a nearly free atomic electron, transferring a portion of its energy to the electron. The energy transferred depends on the photon's scattering angle. The recoiling electron can absorb a significant fraction of the photon's energy, resulting in electron recoils up to a few hundred keV. The cross-section for Compton scattering scales linearity with $Z$. At higher photons energies, much greater than 1 MeV, pair production becomes the dominant interaction mechanism. In this process, a photon is converted into an electron-positron pair upon interacting with the Coulomb field of a nucleus. This process requires a threshold photons energy of 1.022 MeV, corresponding to the combined rest mass of the electron and positron. The residual energy between the initial energy of the photon and the sum of the rest masses of the pair and the energy of the recoiling nucleus is available as kinetic energy for the electron and the positron, which further interact in the detector. The cross section for pair production scales as $Z^2$. In Fig. 3.1 the behavior of the photon interaction cross section various energy ranges is illustrated, showing how different processes dominate at different photon energies.

## 3.2 Detector strategy

The CYGNO collaboration is developing a detector based on a gaseous Time Projection Chamber (TPC) filled with a mixture of helium (He) and carbon tetrafluoride ($CF_4$). The detector operates at room temperature and atmospheric pressure. Since the charge generated during ionization is typically too small to be directly detected, a three stage amplification process using 50 $\mu$m Gas Electron Multipliers (GEMs) is employed. During the primary electrons amplification process through the interaction of them with $CF_4$ molecules light is generated. The signal is then optically readout using a scientific CMOS camera and photomultiplier tubes (PMTs). The detector's performance is influenced by both the properties of the gas mixture and the characteristics of the amplification and readout systems.
The CYGNO project combine the following innovative strategies:

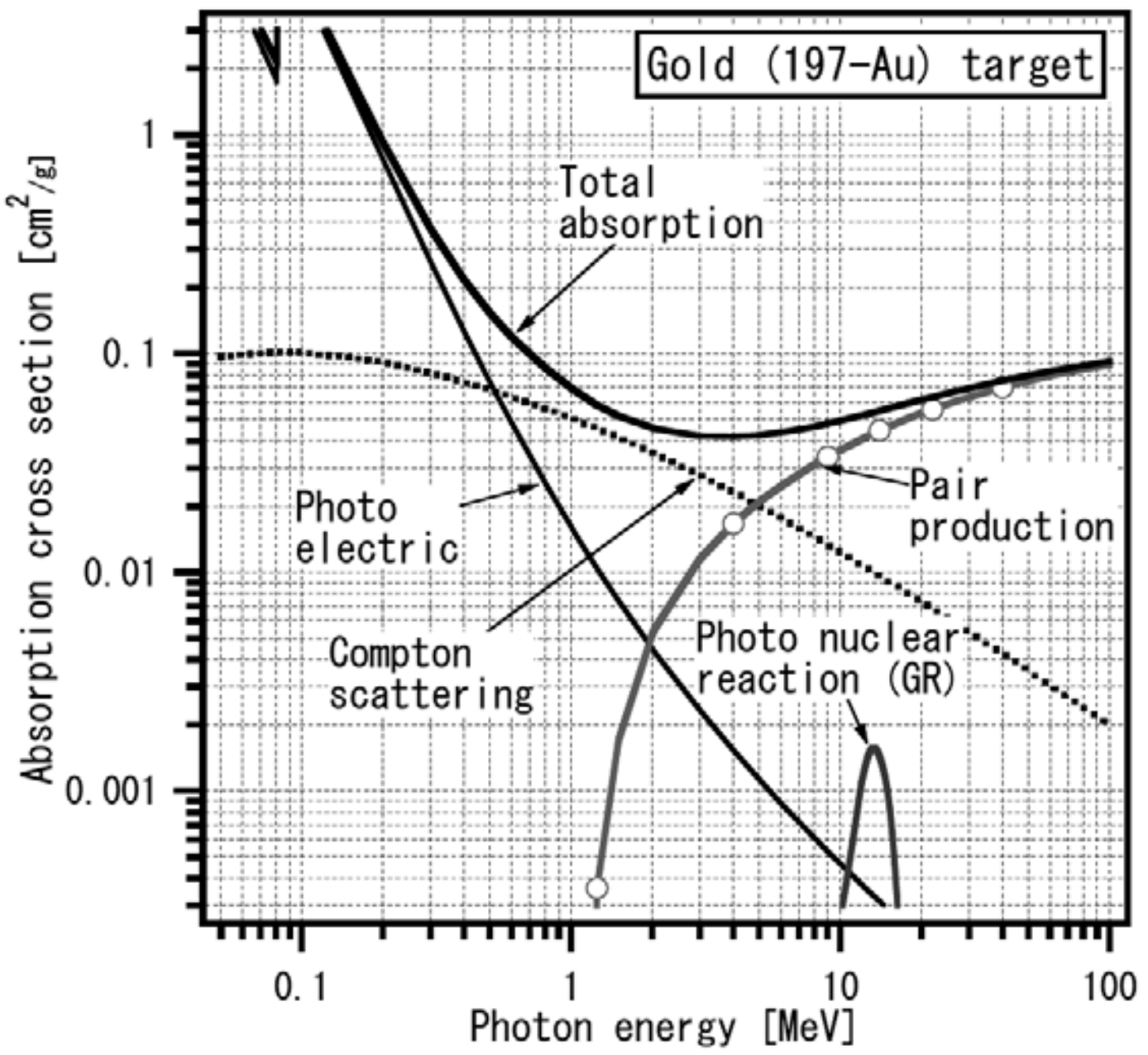

**Figure 3.1** Cross-section behavior for photon interactions as a function of photon energy and material atomic number.

- Gas mixture: He:$CF_4$ gas mixture provides sensitive to both Spin-Independent (SI) and Spin-Dependent (SD) coupling to Dark Matter, thanks to the presence of fluorine. The $CF_4$ scintillation properties are optimal for optical readout due to its high light yield and emission wavelength in the visible range. The addition of 60% helium to $CF_4$ ensures a low gas density, comparable to that of air, while also contributing to the long-term stability of the detector. This low-density environment allows both electron and nuclear recoils to travel relatively longer distances, enhancing the reconstruction of their direction;
- GEM amplification: GEMs are a type of Micro-Pattern Gas Detector (MPGD) that provide high charge amplification, enabling low energy detection thresholds. They also offer high granularity, making them well-suited for integration with the sCMOS camera sensor and optical system used in the CYGNO setup. Furthermore, GEMs are among the few MPGDs that can be manufactured in large sizes, making them an essential feature for the future stages of the experiment;
- Optical readout: The scintillation light produced by the $CF_4$ is detected using sCMOS camera and PMTs. Optical sensors offer a significantly high signal-to-noise ratio, while charge readout is accompanied by electronic noise which can be problematic for very low energy deposits. Furthermore, positioning the readout sensors outside the sensitive volume minimizes both radioactivity and gas mixture contamination. The sCMOS cameras provide high-resolution 2D imaging over large areas. Due to their long acquisition times, they effectively yield a 2D projection of all recoil tracks on the GEM plane ($x$-$y$ plane), enabling the extraction of track topology and directional information. The PMTs sample the light produced by the track as primary electrons gradually pass through the amplification plane. This allows the PMTs to measure the relative extension of the track along the drift direction (i.e., the $z$-coordinate). By combining camera data with PMT signals, full 3D reconstruction of particle recoils is achieved, along with precise energy measurements, using two distinct photosensors. The 3D topology of the recoils is particularly valuable, as it serves as a key parameter for discriminating signals from background events, specifically, nuclear recoils from electronic recoils, and enhances the directional capabilities of the detector.

- Atmospheric pressure operation: Operating at atmospheric pressure provides a favorable volume-to-target-mass ratio, making the detector more cost-effective and scalable. Among all directional detectors, CYGNO is the only directional dark matter search experiment operating at atmospheric pressure. This enhances the experiment's exposure without compromising recoil track lengths, thanks to the low gas density;

In the next Sections, each of these concepts will be examined in detail, providing insights into how they contribute to the overall effectiveness and innovation of the CYGNO project.

### 3.2.1 Time Projection Chamber (TPC)

Liquid and gaseous Time Projection Chambers have been successfully proposed and exploited in last decades for every different applications, from High Energy Physics experiments on colliders to the Dark Matter (DM) searches. TPCs are three-dimensional tracking detectors that can provide detailed information about a particle's trajectory and its specific energy loss. When a charged particle interacts in the medium, ionization occurs. The ionized electrons recombine shortly after the interaction. If an electric field is present, the primary electrons are moved, to not recombine with the ionized atoms and they drift towards the anode, where they are eventually amplified and collected (Fig. 3.2).
Gaseous TPCs constitute a promising approach to directional DM searches enabling a set of crucial features:

- The third coordinate can be evaluated from the time measurement: therefore TPCs are inherently detectors capable of acquiring large sensitive volumes with a lower amount of readout channels;
- Gaseous detectors can features very low-energy detection thresholds. A single electron cluster can be produced with energy releases of the order of few tens of eV;

- A measurement of the total ionisation indicates the energy released by the recoil and the profile of the energy deposit along the track can be measured with high precision;
- Depending on the energy and mass of the recoiling particle and on the gas density, the track itself indicates the axis of the recoil, and the charge profile along it encodes the track orientation (head-tail), providing an addition powerful observable for the DM searches;
- A large choice of gasses can be employed in TPCs, including light nuclei with an odd number of nucleons (such as fluorine), which are also sensitive to both Spin-Independent (SI) and Spin-Dependent (SD) interactions in the O(GeV) mass region;
- A room-temperature and atmospheric-pressure detector results in operational and economical advantages, with no need for cooling or vacuum sealing. These choices allow for a simpler technology and experiment realization and more straightforward scaling when compared to cryogenic solutions currently dominating the DM direct search scene;

TPCs up to 100 $m^3$ [73] of active volume have already been successfully operated, showing the feasibility of very large detectors with large active masses.
The low-pressure gas TPC is the most mature directional technology. In this scheme, the WIMP-induced recoil generated a track of ionization in the gas volume, and an electric field transports the resulting charge to an amplification and readout plane. The full three dimensions of a recoil track can be reconstructed by combining the 2D measurement of the ionization charge distribution on the readout plane, with the third dimension inferred by sampling the transported signal as a function of time. The projection of the track along this third dimension, parallel to the drift field, is found by multiplying the duration of the signal with the known drift velocity of the charge in the gas. Early works in directional detection used Multi Wire Proportional Chambers (MWPCs), which provide both gas amplification and spatial information.

### 3.2.2 Gas mixture

In the CYGNO experiment, the choice of gas mixture plays a critical role in determining the detector's performance and the sensitivity to DM coupling to SM particle. The gas mixture is composed of He:$CF_4$ in proportion 60:40, the ratio has been optimized in order to maximize the detector's performance in terms of energy resolution, photon yield spatial resolution together with the physic case, i.e. low WIMP masses for both SI and SD, as reported in [74]. Previous studies [74], indicate that small amount of $CF_4$ helps to mitigate excessive charge production during the signal amplification process, which can destabilize the detector's electrostatic configuration and lead to inefficiencies like dead time.

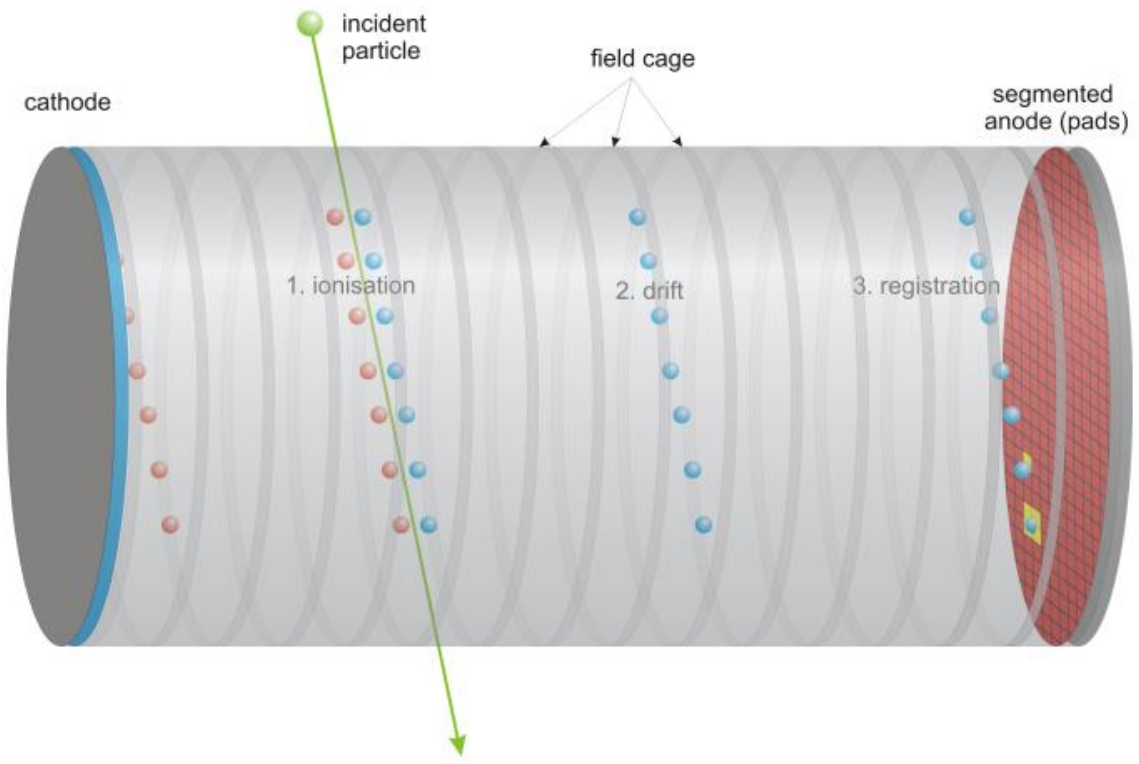

**Figure 3.2** Sketch of a TPC.

Tetrafluomethane, $CF_4$, emits scintillation light also in combination with noble gasses. The light emission spectrum actually comprises two continua, with peaks centered around 290 nm (UV region) and 620 nm (visible region), respectively [75]. The visible orange emission comes from the de-excitation of a Rydberg state of the neutral $CF_3^*$, originated from the fragmentation of $CF_4$ with an energy threshold of 12.5 eV. The UV region shows two main peaks, around 240 nm and 290 nm, coming from the different contributions of several excited ionic species that are produced by the dissociative ionization of $CF_4$, with a threshold of 15.9 eV. Thanks to the large characteristic scattering cross section of electrons, gas mixtures with $CF_4$ presents also very low diffusion during drift which improves the accuracy of 3D tracking within a TPC.

A low electron diffusion coefficient in the gas is a key factor in maintain high-quality 3D tracking in a Time Projection Chamber. Reduced diffusion enables the construction of a detector with a longer drift field, i.e., a larger volume, while preserving tracking performance. Helium is a noble gas, it is chemically stable and inert and it can provide large gains especially when coupled with other molecular gasses. Its low density allows the experiment to operate at atmospheric pressure, extending the detector's sensitivity to DM particles with masses under 10 GeV/$C^2$. In particular, the helium maximizes sensitivity to WIMP masses of few GeV/$c^2$ for SI coupling. The large presence of fluorine atoms in the $CF_4$ provides sensitivity to SD coupling also at low WIMP masses, thanks to its relative lightness. CYGNO results one of the few experiments in the field simultaneously sensitive to both SI and SD coupling at WIMP masses bellow 10 GeV/$c^2$.

Diffusion coefficients (Fig. 3.3) and drift velocities (Fig. 3.4) for the gas mixture under different electric fields were calculated using Garfield [76] simulations. The transverse diffusion coefficient is around 130 $\mu$/cm at a drift field of 500 V/cm. The energy needed for ionization (W value) is approximately 46.2 eV [77].

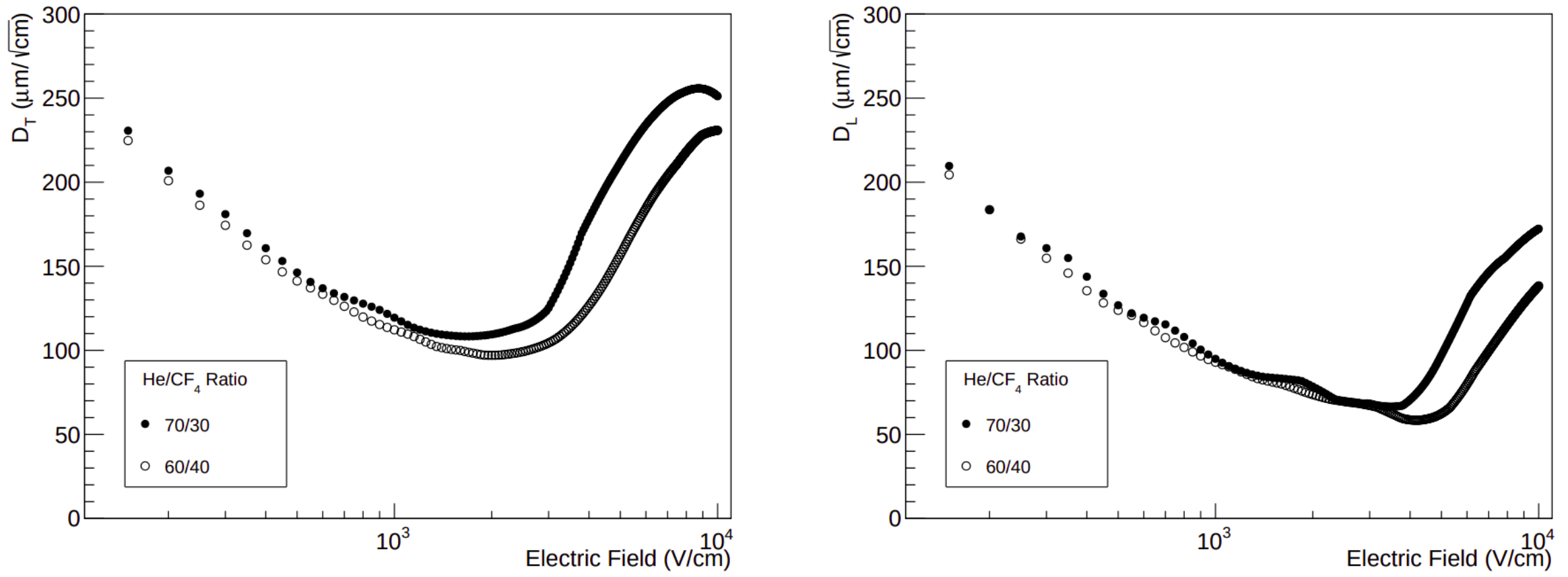

**Figure 3.3** Transverse (on the left) and longitudinal (on the right) diffusion coefficients for the mixtures as a function of the electric field [74].

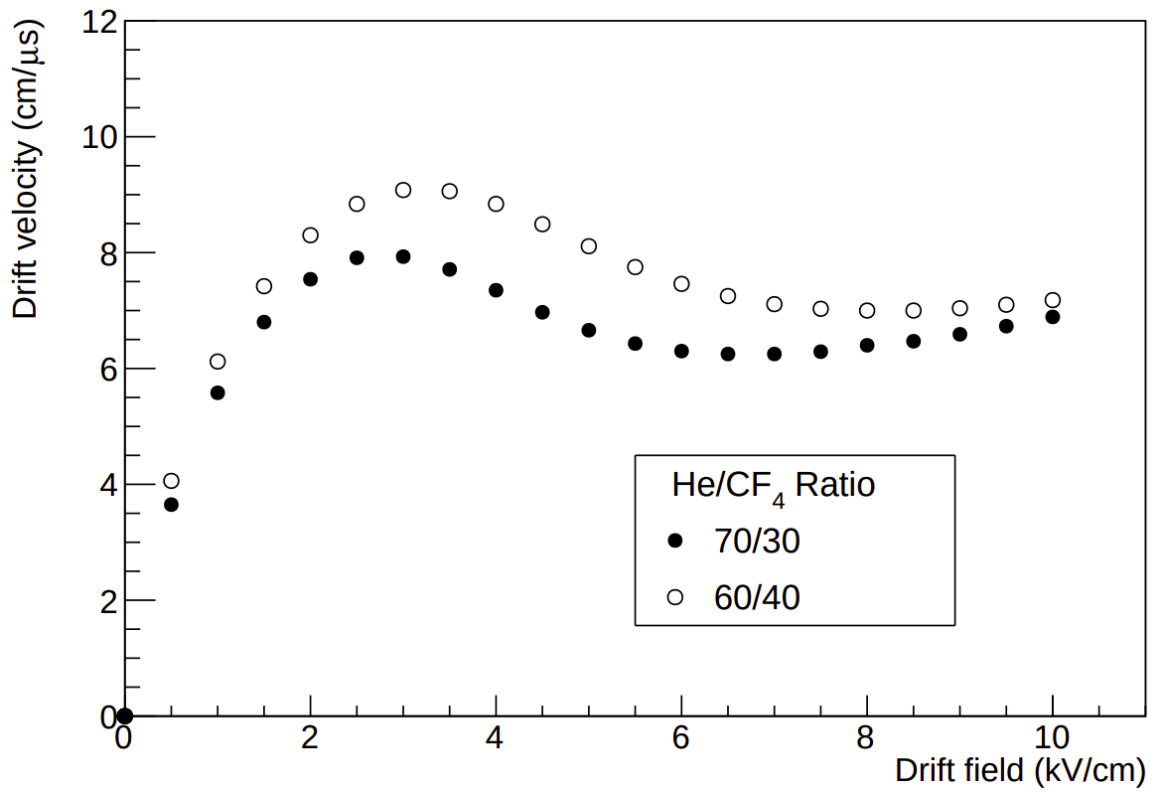

**Figure 3.4** Electron drift velocities as evaluated with Garfield [74].

Ongoing research within CYGNO is exploring other gas mixtures, including hydrogen-rich gases like isobutane or methane, and investigating the potential addition of electronegative gases for further improvements in detection modes.

### 3.2.3 GEM

The CYGNO collaboration chose Gas Electron Multipliers (GEMs), because they can reach large gains and can be stacked in multiple amplification stages while maintaining a high granularity, which prevents an excessive degradation of the track's shape. GEMs are a type of Micro Pattern Gas Detector (MPGD), initially proposed by F. Sauli in 1997 [78]; and it was widely studied and improved since then. GEM detectors are nowadays an affirmed technology, employed in many experiments worldwide.

GEMs allow to amplify the charge drifting through the channels, it is based on two metal layers separated by a thin insulator with regularly double-conical. Electrons generated in the upper gas volume drift into the channels, thanks to the high electric field, the avalanche takes place generating second electrons. The ions generated in the avalanche drift along the central field lines, avoiding charging up problems.
Unlike other gaseous detectors, the negative signal on the anode is generated only by the collection of electrons, without a contribution from the slow positive ions. This make the device potentially very fast and minimize space charge problems. Ions do not participate to the signal induction: they slowly drift back towards the top GEM electrode and the drift. The signal generated by the movement of positive charges is totally shielded by the GEM foil, therefore no ion tail is contained in the typical GEM signal.
The material employed to realise the GEM can vary, but the most common are fiberglass or polyamide, such as Kapton, for the insulator sheet, and copper for the metallic coating. GEMs can be realised with various thicknesses, from very thin 50 $\mu$m to 1 mm (Fig. 3.5). For thinner GEMs, the hole diameter is typically 70 $\mu$m wide, as smaller holes do not significantly enhance the effective gain due to the very large charge density in the hole. The pitch, i.e. the distance between holes, is typically 140 $\mu$m. Such tiny holes are usually fabricated via chemical etching, which results in a double-conical-shape hole, with a narrower diameter in the center. Holes with diameters $> 300$ $\mu$m, are fabricated through the mechanical drilling of the foil, producing cylindrical holes. Thanks to their structure, GEMs allow for a good spatial resolution, which preserve the $x$-$y$ characteristics of the electron cloud resulting from an incident particle. The secondary electrons produced by a GEM can be collected by an anode or sent to another GEM for a second amplification stage. Stacking multiple GEM layers increases the gain, achieving total gain of the order of $10^4$ to $10^6$. The advantage of a multiple structure is that the overall gain can be obtained with each foil operated at a lower voltage, reducing the risk of discharges. Foils can either be supplied with a dedicated voltage line each, or with a single voltage source and a resistive distribution chain. For these reasons, CYGNO employs a stack of three 50 $\mu$m GEMs to achieve the necessary amplification for low-energy event detection. The small size and high granularity of GEMs preserve the shape and characteristics of the electron cloud produced by an incident particle.

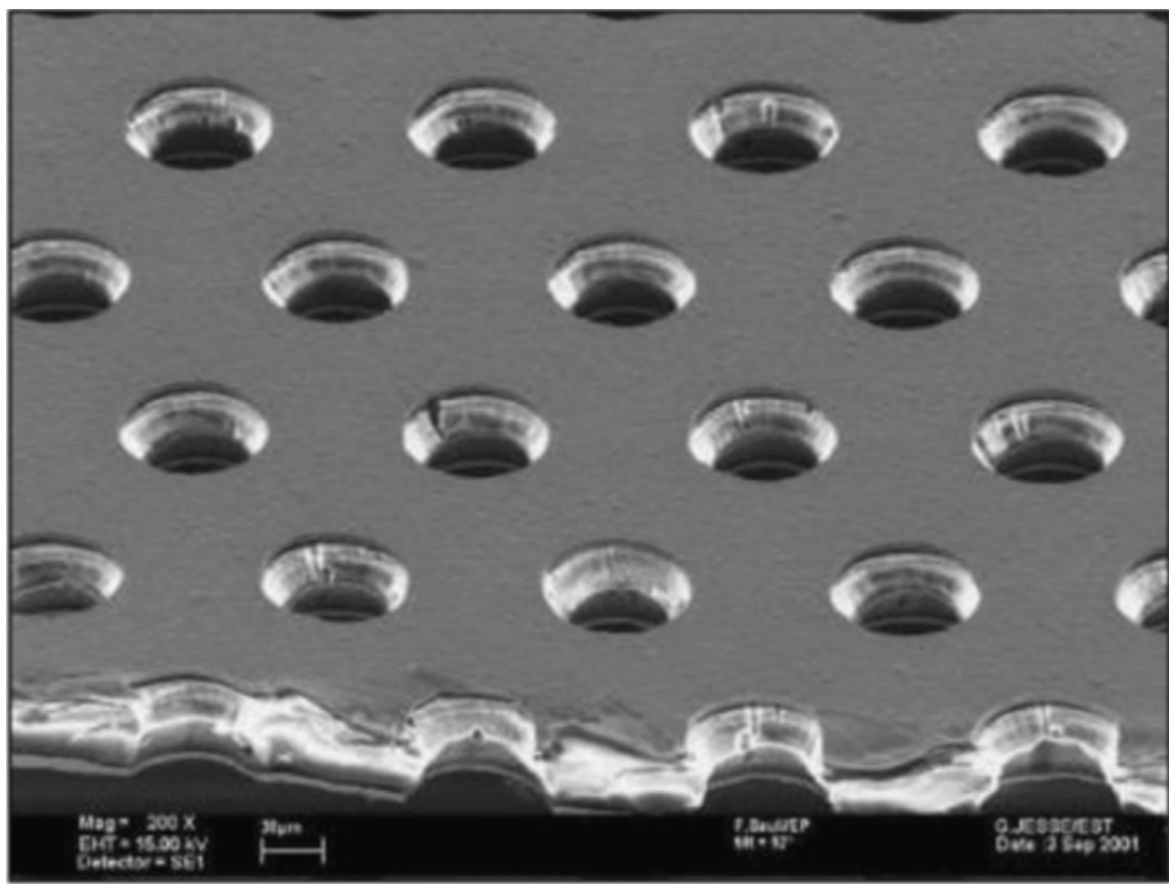

**Figure 3.5** Electron microscope picture of a section of typical GEM electrode, 50 $\mu$m thick.

### 3.2.4 Optical readout

In the search of Dark Matter, a promising approach involves optically reading out the gas electroluminescence produced during electron multiplication [79; 80; 81; 82; 83].
When a charged particle interacts with the medium, it can ionize the atoms and the molecules and it can also excite them. In the de-excitation processes, photons are emitted. The amount and spectrum of this light depend on the gas mixture, its density and the presence of an electric field. The emission spectra correspond to the second continuum spectra observed in rare gas discharges. In most common gas mixture, the number of emitted photons per avalanche electron can vary between $10^{-2}$ and $10^{-1}$.
The idea of detecting scintillation light produced during the electron multiplication process was proposed in the 90's. Using an optical readout system allows sensors to be placed outside the sensitive volume of the detector, which reduces contamination from radioactivity and the gas itself. However, this approach limits the number of photons collected due to solid angle coverage. In CYGNO, this photon collection loss is compensated by the high gain achieved using a triple GEM structure. Proper lenses are employed to capture images over large areas of about O(1) m$^2$ using a single sensor.
As a directional detector, CYGNO aims to measure the direction of particle recoils along with their energy, requiring full 3D track reconstruction. This is accomplished by using two optical sensors: a scientific CMOS (sCMOS) camera, which captures the $x$-$y$ projection of the particle tracks, and a set of PMTs that measure the integrated energy and the track development along the $z$-axis (drift direction).

#### Optics

To image a large sensitive area on a single sCMOS sensor, focusing lenses are necessary. A single camera, paired with the proper optic system, can capture images of areas as large as approximately O(1) m$^2$, maintaining high granularity. All CYGNO prototypes utilizes a Schneider Xenon lens with focal length f = 25.6 mm and an aperture ratio of 0.95, providing 0.85 transparency in the optical band was exploited. The sensor must be placed 60 cm away from the last amplification plane in order to capture images of areas measuring 33$\times$33 cm$^2$. With this setup, the LIME prototype (Section 4.1) achieves an effective granularity of

$152 \times 152$ $\mu$m$^2$.

A schematic representation of an optical system featuring a thick lens, projected onto the plane formed by the optical axis and an orthogonal axis is shown in Fig. 3.6. OP represents the object plane where the source being focused is positioned, SP the sensor plane where there is the photosensor, EP and XP the entrance and exit planes where photons enter and exit the lens, respectively. H and H$'$ represent the hyperfocal planes, corresponding to the position of a thin lens exhibiting similar optical behavior to that of a thick lens [84]. The radius of the lens opening (D) represents the area through which photons are accepted, which can be modified by manually adjusting the aperture.

The infinitesimal area of the photon-emitting object is denoted by $dA$, $u$ is the tangent of the opening angle, $s$ represents the distance between the object and the hyperfocal plane H, and $s'$ represents the distance between the sensor and the hyperfocal plane H$'$, $f_F$ is the focal distance, F is the focal point, and in the general case F = F$'$ and $f_F = f'_F = f$. The Fig. 3.6 represents the functioning of a typical thick lens, where photons within the yellow region to the left of EP reach the lens and are focused onto the sensor.

If OP, EP, XP and SP are parallel, the photons flux $\Phi$ reaching the lens from an area $dA$ of the source is given by:

$$\Phi = \pi L dA u^2 \tag{3.1}$$

where $L$ is the luminosity of the source. The fraction of solid angle $\Omega_f$ covered by the lens is defined as the flux defined in equation 3.1 per the total flux ($\Phi_{tot}$):

$$\Omega_f = \frac{\Phi}{\Phi_{tot}} = \frac{\Phi}{LdA}\frac{1}{4\pi} = \pi u^2 \frac{1}{4\pi} = \frac{(D/2)^2}{4s^2} \tag{3.2}$$

We define $s_{ep}$ as the distance between the hyperfocal plane and the entrance plane. If $s \gg s_{ep}$, which is the case of CYGNO, the relationship between the distances $s$ and $s'$ to the magnification $I$ can be considered identical to linear optics of thin lenses:

$$\frac{1}{f} = \frac{1}{s} + \frac{1}{s'} \tag{3.3}$$

$$I = \frac{y'}{y} = \frac{s'}{s} \tag{3.4}$$

where $y$ and $y'$ are the object and image dimensions respectively. Combining equations 3.1, 3.3 and 3.4 we obtain

$$\Omega_f = \frac{1}{\left(4N^2\left(\frac{1}{I}+1\right)\right)^2} \tag{3.5}$$

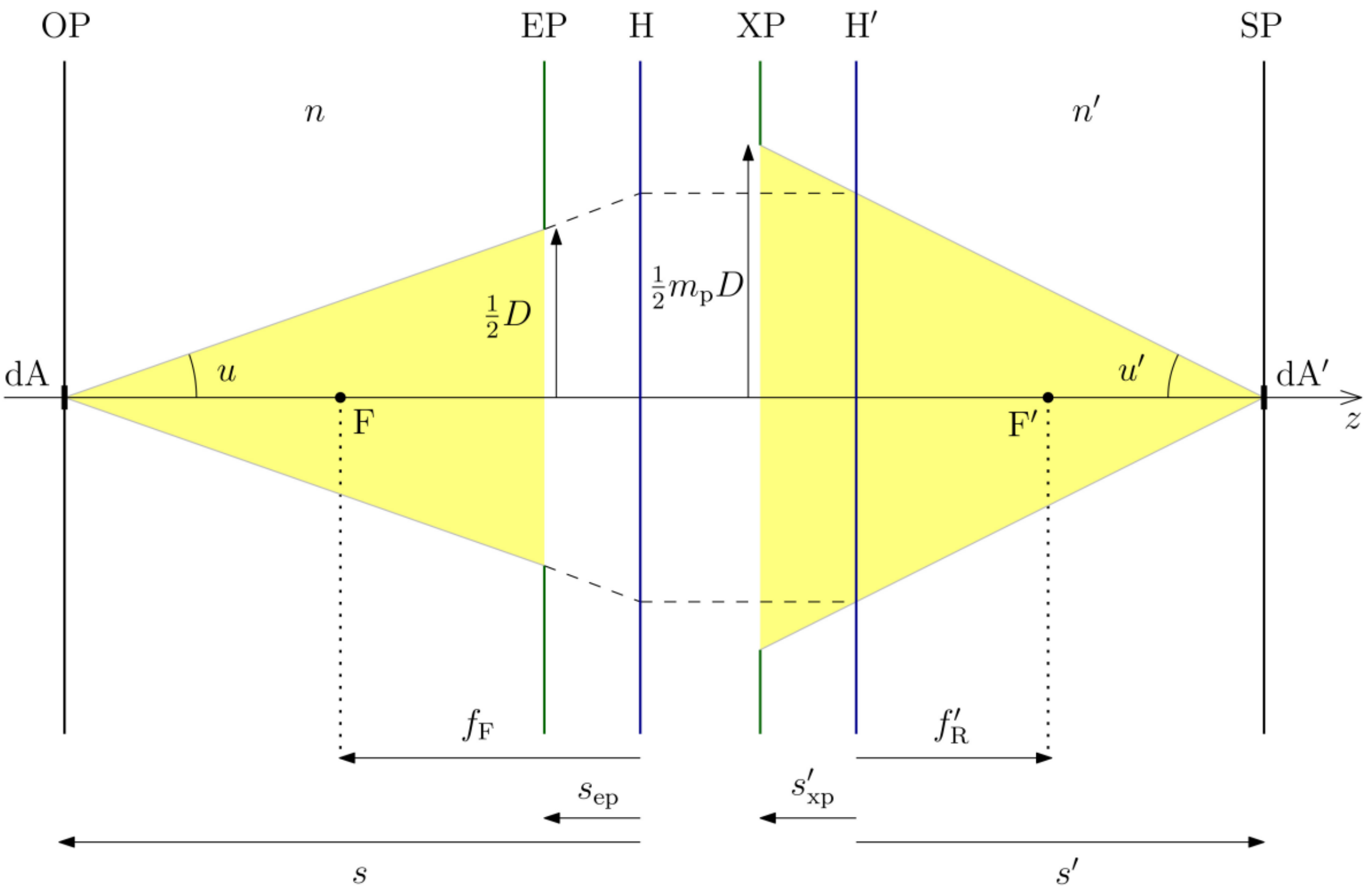

**Figure 3.6** Schematic of a lens in the Gaussian approximation. The yellow band refers to the amount of solid angle covered that emitted from the object plane (OP) is focused on the sensor plane (SP) [84].

where N = D/$f$ is f-number of the aperture ratio. The typical solid angle covered with the Schneider lens and the ORCA Fusion camera, significantly reduces the number of photons that reach the active sCMOS sensor. An extremely high gain is required from the amplification stage, which necessitates the utilization of a stack of three GEMs.
The photons flux computed was derived assuming a source placed on the optical axis; off-axis photon sources experience a decrease in flux due to the vignetting effect. The flux from a source is given by:

$$\Phi = \pi L dA u^2 cos^4 \phi \tag{3.6}$$

where $\phi$ is the angle between the optical axis and the direction connecting the source with the center of the lens entrance (EP).
This cause an $x$-$y$ non-uniformity in the light yield, called vignetting, and can be corrected in the CYGNO image analysis.

### Scientific CMOS

In recent years, significant advancement have been made in light sensor technology. Particle physics experiments, particularly those involving Time Projection Chambers (TPCs) using scintillating gas mixtures, now frequently utilize high-performance pixelated light sensors.

The most wiedly used light sensors for digital imaging are currently charged-coupled devices (CCDs) and active pixel sensors (APSs), especially those developed on complementary metaloxide semiconductors (CMOS). Both types of sensors convert light into electric charge and subsequently process it into electronic signals. In CCD sensors, each pixels consist of Metal-Oxide -Semiconductor (MOS) capacitors, biased into the depletion region where incident photons are converted into electrons. The capacitor array is placed on top of a transmission layer which acts as a shift register. After the image is captured, the charge produced in each pixel is transported to its neighbor, repeating the process until the charge reaches an amplifier, where it is converted into a voltage. This voltage can either be stored as a continuous analog signal or sampled and digitized. The CCD sensors allow nearly the entire pixel area to capture light, resulting in high output uniformity, which is critical for image quality.

CMOS sensors, on the other hand, perform charge-to-voltage conversion individually within each pixel using a pinned photodiode. These sensors often incorporate additional components like amplifiers, noise-correction circuits, and digitization features. While CMOS sensors have slightly lower uniformity due to the individual pixel processing, they support massively parallel data processing, allowing for high-speed operation and greater overall bandwidth.

Due to their low noise, high granularity, and potential for further technological improvements, CYGNO has chosen to use scientific CMOS (sCMOS) cameras. These cameras offer precise $x$-$y$ projection measurements of events occurring within the gas, along with quantitative measurements of the light emitted. This capability provides not only the location of the events but also the release density along the tracks that it is useful for particle identification. With the bulk of the photons sensitive region made of silicon and the protective window of the sensor composed of glass, sCMOS cameras have large sensitivity in the optical range. The quantum efficiency (QE) of these sensors can reach up to 80% at around 600 nm, which aligns well with the peak emission in the visible range from the He:$CF_4$ gas mixture used in CYGNO [75]. Notably, the QE of sCMOS sensors is significantly higher than that of common PMTs, and their wavelength sensitivity matches well with the emission properties of the gas mixture. To overcome the poor timing information provided by the sCMOS, a combined light readout with a fast PMT is exploited in order to provide a 3D reconstruction of the track.

Recoil ranges of both electrons and ions in the gas mixture were simulated using GEANT4 [85] and SRIM [86], respectively. Helium nuclear recoils appear as bright spots less than 1 mm across at energies up to 100 keV, while electronic recoils below 10 keV show a more diffuse, spot-like profile. At higher energies, electronic recoils can extend to several centimeters. An example of image acquired with the sCMOS camera is shown in Fig. 3.7.

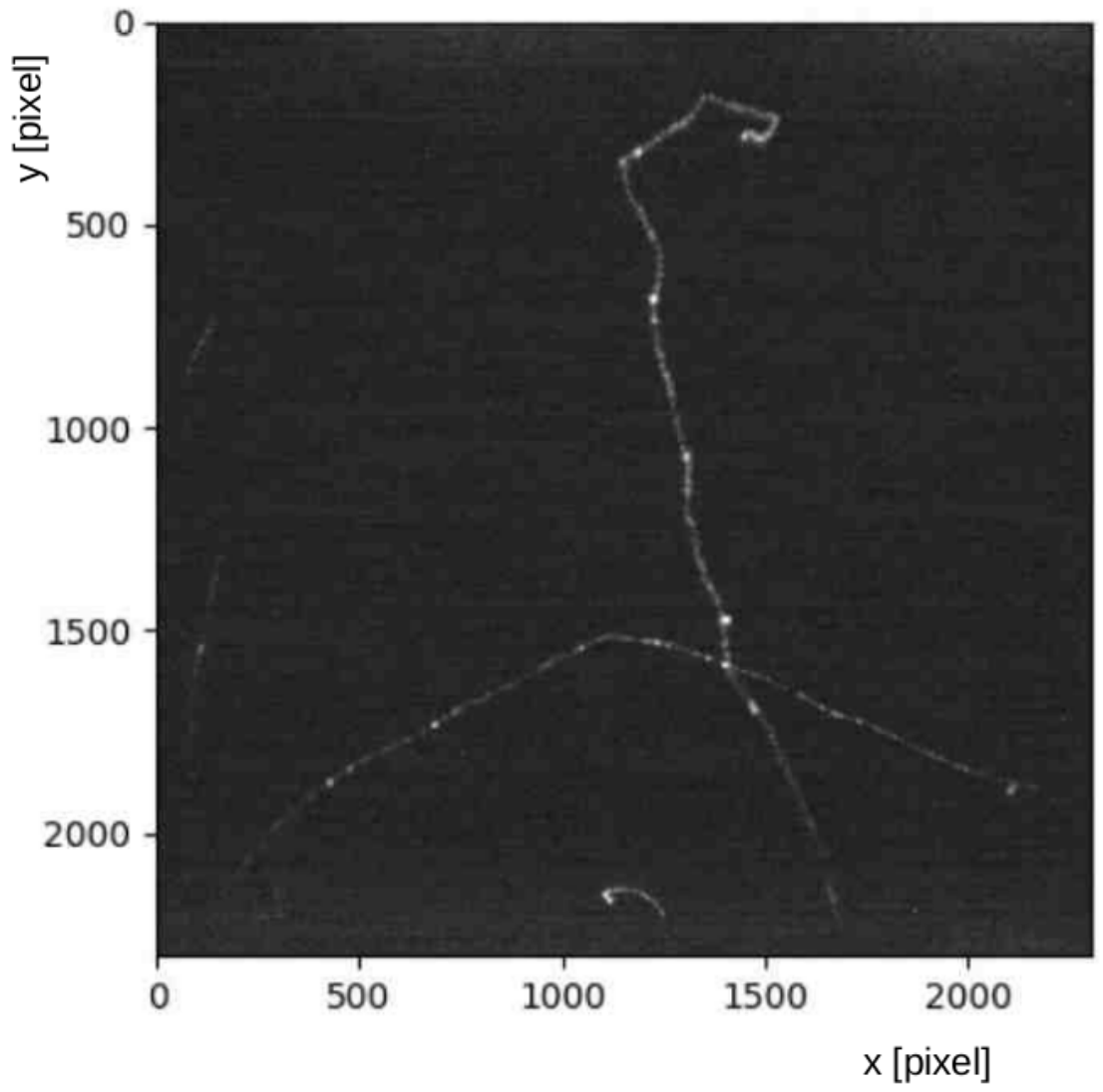

**Figure 3.7** Example of an image acquired with the sCMOS camera.

Discriminating between electronic (ER) and nuclear recoils (NR) at low energies presents a challenge, as their track shapes become similar. The key distinguishing feature is track density, the ratio of total light emitted to the size of the recoil spot. CYGNO employs measurements of specific energy loss ($dE/dx$) along the track, facilitated by high-resolution data from sCMOS cameras and PMTs, to differentiate between ERs and NRs.

**Photonmultpliers Tubes**

In TPC with optical readout, devices that can detect photons and convert them into an electrical signal at the anode play a critical role. Photmultipliers tubse (PMTs) are designed to detect very faint light, even as little as a single photon, and can handle a wide range of light intensities. It consists of a photocathode made from a photosensitive material. When a photon strikes the photocathode, an electron is emitted via the photoelectric effect. To optimize light transmission, the photosensitive material is applied as a thin layer on the inside of the PMT window, typically made of glass or quartz. The efficiency of photon-to-electron conversion is not perfect, and the fraction of photons converted to electrons is referred to as the quantum efficiency (QE).

Once an electron is emitted, it is accelerated by an applied electric field and directed towards the first electrode, called a dynode. When the electron hits the dynode, it transfers energy to the dynode material, causing multiple secondary electrons to be emitted. These secondary electrons are then accelerated towards the next dynode, where the process repeats, generating more electrons at each stage. This cascading effect produces a large number of electrons, typically ranging from $10^7$ to $10^{10}$, sufficient to represent the original scintillation event. At the anode, this electron cascade is collected, creating a current that can be amplified and analyzed. A schematic representation of the PMT and this amplification process is shown in Fig. 3.8. The output current from the PMT is directly proportional to the number of incident photons, allowing it to provide not only information about the presence of particles but also the energy deposited during the scintillation event.

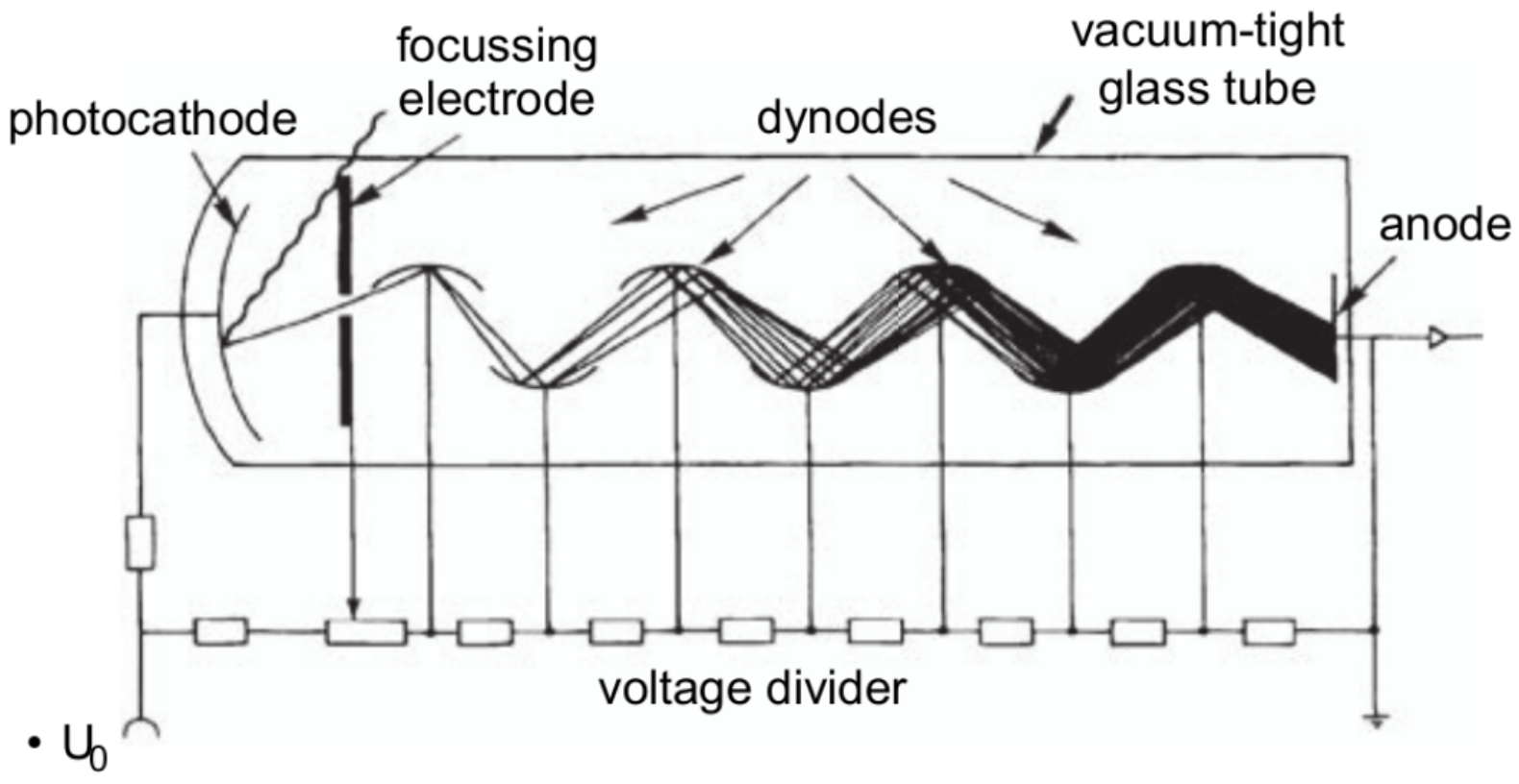

**Figure 3.8** Schematic of the photomultiplier tubes and its operation [87].

PMTs can be operated using either positive or negative high voltage, as long as the dynodes are at a positive potential relative to the photocathode. For efficient photoelectron collection, the voltage difference between the photocathode and first dynode is often several times greater than the dynode-to-dynode differences. A resistive voltage divider is used to supply these voltage differences from a single high-voltage source. The direct current through the voltage divider is determined by the ratio of applied high voltage to the total resistance of the divider string. Keeping the current small helps minimize heat dissipation issues. Even in the absence of illumination, a small current called the dark current flows through the PMT. This current arises from various sources, including thermionic emission from the cathode and dynodes, leakage currents, radioactive contamination, ionization, and other light-related phenomena.

The sensitivity of a PMT to different wavelengths of light and its quantum efficiency depend on the materials used for both the front window and the photocathode.

PMTs also preserve the timing information of the original light pulse. When exposed to a short-duration light pulse, a typical PMT will produce an electron pulse with a time width of just a few nanoseconds. The effective decay time for $CF_4$ is measured to be less than 15 ns [88]. This timing structure is directly related to the position of the electrons along the drift direction, which in turn reflects the spatial structure of the original particle track along the $z$-axis. This information complements the $x$-$y$ projection of the tracks detected by the sCMOS camera, enabling full 3D reconstruction of the events.

The integral of the PMT signal is proportional to the total light emitted by each event, providing an independent energy measurement that complements the energy information from the sCMOS camera.

## 3.3 Gain in gas detectors

When a charged particle passes through the detector, it ionizes the gas along its path. This process, known as primary ionization, produces free electrons and ions. However, only a fraction of the particle's energy is spent for ionization, and the total ionization yield can be described by the parameter $W$, representing the average energy required to create a single free electron:

$$W \langle N_I \rangle = L \left\langle \frac{dE}{dx} \right\rangle \tag{3.7}$$

Where $\langle N_I \rangle$ denotes the average number of ionization electrons created over a path length $L$ and $\left\langle \frac{dE}{dx} \right\rangle$ is the average energy loss per unit path length of the particle. The value of $W$ depends on the gas, its composition and density, and on the nature of the incident particle. Experimentally it is found that for electrons above a few keV and alpha particles above a few MeV, $W$ is generally independent of the initial energy.
In the case of a Time Projetion Chamber, an electric field is applied between the cathode and the anode, guiding the primary electrons toward the anode. During the drift, electrons can be neutralized by ions, absorbed in the walls or attached to electronegative molecules. Therefore, as the electrons drift, some may be absorbed by the gas. The number of electrons is influenced by the gas's absorption length, defined as:

$$N = N_0 e^{-d/l_{abs}} \tag{3.8}$$

where $N_0$ is the number of primary electrons, $d$ is the drift length and $l_{abs}$ is the absorption length.
As electrons traverse the gas, they undergo frequent collisions with gas molecules, which limit their average velocity, called drift velocity. Due to these random collisions, an initial electron cluster will spread over time into a Gaussian distribution with the spread given by:

$$\sigma_x = \sqrt{2Dt} \tag{3.9}$$

where $D$ is the diffusion coefficient and $t$ is the drift time in a given direction $x$. While electron diffusion is typically considered symmetric, high electric fields can lead to an asymmetry where the longitudinal diffusion coefficient, $D_L$, in the drift direction becomes smaller than the transverse diffusion coefficient, $D_T$ [89]. This discrepancy arises because electron energy gains between collisions vary depending on the alignment of their motion with respect to the electric field, particularly in gases where the mean free path is long. Therefore, the drift field's asymmetry in a TPC causes differing diffusion between the drift and transverse directions, resulting in an electron cloud with a 3D Gaussian distribution. The spreads are:

$$\sigma_{x,y} = \sqrt{2D_T t} \tag{3.10}$$

for the transverse plane and

$$\sigma_z = \sqrt{2D_L t} \tag{3.11}$$

for the drift direction.

### 3.3.1 GEM gain

When electrons enter a GEM hole, they undergo localized multiplication, resulting in an electron avalanche. With a suitable gas mixture, the excited gas species can also emit photons over a very short distance. Due to the GEM's optical transparency, these photons can exit through the channel and be detected, contributing to the signal. The high gain of the GEM enables a large number of photons to be emitted per incoming electron [90].

GEMs can be cascaded to increase gain, this configuration effectively suppresses ion feedback, preventing the accumulation of positive charges in the interaction volume and maintaining stability [91].

Multiple GEMs achieve higher gains compared to a single GEM because the charge between stages is distributed across several holes, each acting as an independent amplifier. What matters is therefore the amount of charge in the hole, or, equivalently, the charge density. The charge exiting from the considered volume is composed by the same amount of charge entering from the neighboring volume. This is valid when several holes are uniformly affected by the see amount of charge, as it happens for the high flux measurement.

The electric field increases in front of the ion cloud, upstream of the avalanche region, before all the primary electrons have been fully amplified. In this intensified region, new incoming electrons generate additional electron-ion pairs, further aggravating the electric field distortion. Some electrons in the highly ionized region diffuse against the drift, populating the high-field region ahead of the ion cloud, thereby worsening the field distortion even more. This distortion drives ionization towards the cathode, increasing the ionization rate over time. As the electrons drift through the field cage, they diffuse and reach the GEM, where they trigger avalanche multiplication over a larger area than the original ionization region.

Thanks to the high electric field inside the GEM holes, an electron avalanche develops with a typical factor on the order of $10^2$. The gain per GEM, $G_{GEM}$, can be expressed as a function of the applied voltage $V_{GEM}$:

$$G_{GEM} = ae^{bV_{GEM}} \tag{3.12}$$

Where $a$ and $b$ are parameters dependent on the specific detector setup and gas mixture.
As the electron avalanche progress in the GEM, photons in the visible spectrum are emitted via the gas molecules de-excitation process. Studies with a He:$CF_4$ gas mixture at a 60:40 ratio [75] report around 0.07 photons per secondary electron above 400 nm, which was also confirmed by recent measurements done within the CYGNO collaboration. However, only a portion of these emitted photons reach the sensor, as some are lost due to the lens optics, which vary depending on the event's position on the GEM plane.

### 3.3.2 Gas gain

To ensure a detectable signal in a small gas volume, it is desirable the gas gain not to be too low. Operating in high-gain mode offers certain advantages, such as simplifying the task of the amplification electronics. This optimization requires a knowledge of distortion due to space charge effects which can be estimated by modeling gas avalanches. In particular, this modeling relies on knowledge of the Townsend coefficient as a function of the electric field. The mean free path for ionization describes the average distance an electron travels between ionizing collisions. Its inverse, the number of collisions per cm, is the first Townsend coefficient, denoted as $\alpha$. This coefficient is a fundamental parameter that determines the gas gain. The value of $\alpha$ depends on several factors, with the main ones being the gas composition, the electric field strength and the gas pressure. The addition of polyatomic molecules to noble gas mixtures raises the required electric field for achieving similar gains, due to the presence of various excitation, vibrational and rotational levels [92].
The presence of contaminants such as $O_2$, $N_2$ and $H_2O$, which are commonly present during experimental operation, can influence the performance of GEM detectors, especially in terms of signal efficiency and amplification gain. Common sources of impurities include detector components and the gas system itself, and these contaminants can be mitigated by using gas purifier modules.
Increasing gas flow significantly improves GEM gain. It has been observed that low gas flows lead to a performance drop due to impurity buildup in the detector. This accumulation decreases exponentially as gas flow increases, allowing for higher gains as the same applied voltage. Oxygen, being a common impurity, has a particularly strong effect because of its high electron attachment coefficient, which leads to electron capture during the avalanche process. The presence of $O_2$ limits both primary ionization and the development of the electron avalanche [93]. Even small amounts of electronegative gases in the mixture can significantly alter the drift properties and overall detector performance. Oxygen, due to its high electron attachment probability (attachment coefficient of $h = 2.5 \times 10^{-5}$), can inhibit the avalanche multiplication process and lead to the formation of ozone $O_3$ during the avalanche. Ozone, in turn, acts as a strong absorber of ultraviolet photons and can quench discharges, limiting the charge multiplication process.
Another contaminant like $H_2O$, being electronegative molecule, tend to capture free electrons in the avalanche, reducing the overall gain. Increasing the gas flow reduces the concentration of these molecules, helping maintain higher detector performance.
The gain, charge multiplication factor, is given by the Townsend equation:

$$G = e^{\alpha x} \tag{3.13}$$

where $x$ is the distance over which the multiplication avalanche occurs and $\alpha$ is the Townsend coefficient, which depends on the ratio $E/\rho$, with $\rho$ being the gas density. From the ideal gas law:

$$\alpha \propto 1/\rho \propto T/P \tag{3.14}$$

The gain dependence on temperature ($T$) and pressure ($P$) is then expressed as:

$$G(T/P) = Ae^{B\frac{T}{P}} \tag{3.15}$$

where $A$ and $B$ are parameters that are experimentally determined. In controlled environments with small fluctuations, a first order approximation of 3.15 is possible, giving a good behavior of the gain:

$$G(T/P) = A + B\frac{T}{P} \tag{3.16}$$

In this thesis (Chapter 5), the impact of impurities and pressure variations on the detector's gain is analyed for the LIME prototype.

## 3.4 TimeLine

The CYGNO project aims at developing a O(30) $m^3$ detector using an innovative approach in astroparticle physics, particularly for directional DM searches. The experiment involves an underground Time Projection Chamber with high-resolution optical readout, exploiting all TPC features. It employs Gas Electron Multipliers at atmospheric pressure with He:$CF_4$ gas mixture, designed for detecting rare processes with energy releases ranging from hundreds of eV to tens of keV.

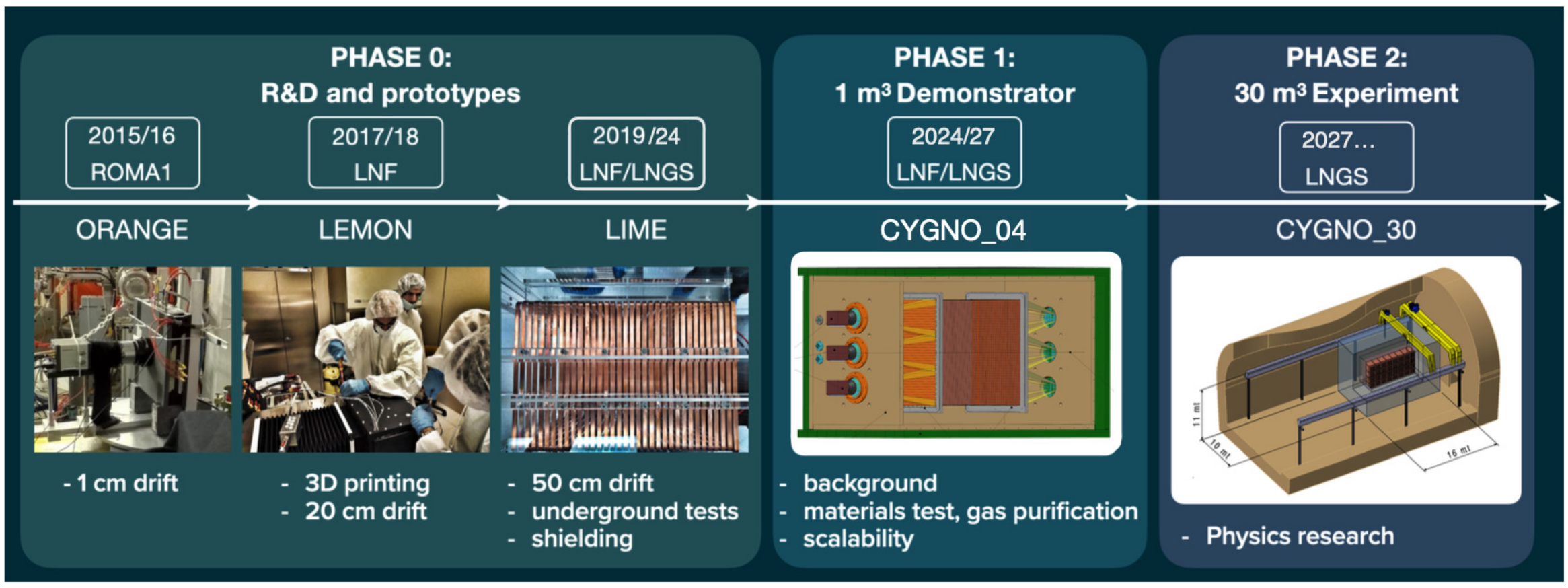

**Figure 3.9** The CYGNO experiment timeline.

To achieve this, the experiment is progressing through prototypes of increasing size. The CYGNO timeline consists of three main phases, as illustrated in the Fig. 3.9. Currently, the experiment is nearing the completion of PHASE_0, the R&D stage. During this phase, multiple prototypes were constructed to evaluate and optimize CYGNO's performance, leading to the installation of the LIME detector underground. The next stage, PHASE_1, will focus on scaling up the technology by deploying a 0.4 $m^3$ demonstrator, CYGNO-04, with a modular readout system. This phase will also serve as a test for material radiopurity. The final goal is PHASE_2, which involves constructing a large-scale detector with a sensitive volume of tens of cubic meters.
It is important to stress here that the project was discussed and approved only up to PHASE_1 which is currently starting.

### 3.4.1 PHASE_0

To optimize the experimental technique, three main prototypes were developed. The first, ORANGE (Optically Readout GEM), explored the optical readout capabilities of GEMs for 3D tracking. It featured a triple 10 $\times$ 10 $cm^2$ GEM stack with 2 mm spacing and a 1 cm drift gap, resulting in a sensitive volume of 100 $cm^3$. The emitted light was detected by a Hamamatsu ORCA-Flash 4.0 camera, positioned 20 cm away from the last GEM layer with a geometrical acceptance of $3.46 \times 10^{-2}$ (Fig. 3.10). ORANGE provided the first test for track image reconstruction and assessed light yield performance, demonstrating 3D track measurement using PMTs [94].
Subsequently, to explore the feasibility of using TPCs for directional Dark Matter searches and to test CYGNO experimental approach with longer drift length, the LEMOn (Long Ellipticall MOdule) detector has been developed [95]. A 7 liters active volume TPC with a

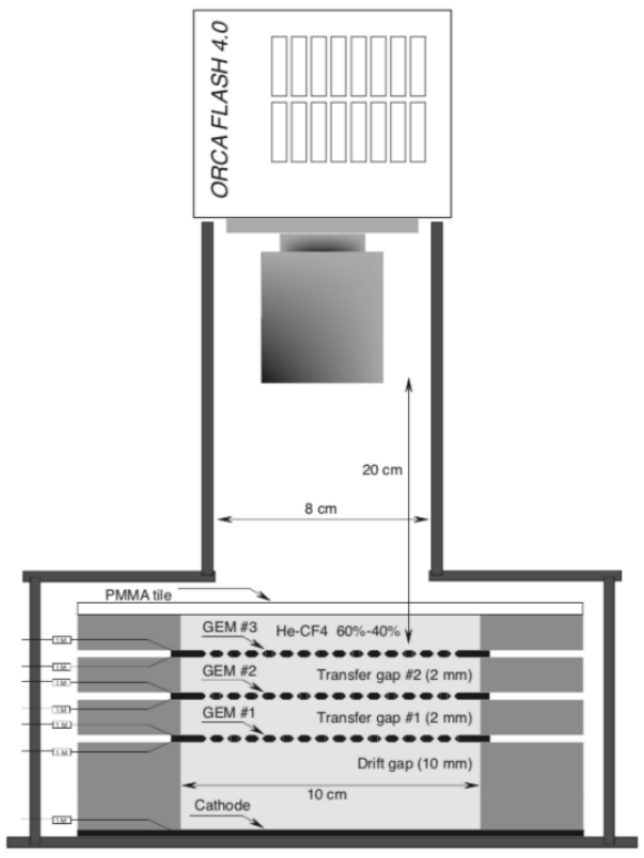

**Figure 3.10** Schematic of the ORANGE prototype.

20 cm drift length and a 24 $\times$ 20 cm$^2$ readout area is enclosed in a gas-tight acrylic vessel and operated in continuous gas flux mode. Inside the chamber, an ellipsoidal field cage made of silver wire held by 3D-printed plastic supports with a 1 cm pitch ensures the uniformity of the drift field in the 20 cm gap. The cathode is constructed using a mesh with a wire diameter of 30 $\mu$m and a pitch of 50 $\mu$m. The amplification stage consists of three 24 $\times$ 20 cm$^2$ GEMs spaced 2 mm apart. The LEMOn detector is optically coupled to a Hamamatsu sCMOS camera via a TEDLAR transparent window. The camera has a calibrated response rate of 0.9 counts per photons. On the opposite side, behind the transparent cathode, a HZC Photonic XP3392 PMT is placed.

LEMOn validated the detector concept on a medium scale, demonstrating 100% detection efficiency for 5.9 keV electron recoils with an energy resolution of 12%. It also showed the background rejection capabilities, with 96.5% of 5.9 keV electrons being rejected while retaining 50% of the neutron recoil signal [96].

The final prototype for PHASE_0 is LIME, which is the prototype studied in this work; it will be detailed in Section 4.1.

### 3.4.2 PHASE_1: CYGNO-04

The primary goal of PHASE_1 is to build the CYGNO-04 demonstrator, focusing on reducing material radioactivity and developing the modular optical readout system and data acquisition (DAQ) for a full-scale experiment. The technical design report for CYGNO-04 was submitted to the LNGS administration and funding agencies and it was approved in July 2022. CYGNO-04 will be installed in the underground Hall F (Fig. 3.12) of LNGS, located between Hall A and Hall B.

As shown in Fig. 3.13, CYGNO-04 will be composed of two TPC volumes back-to-back,

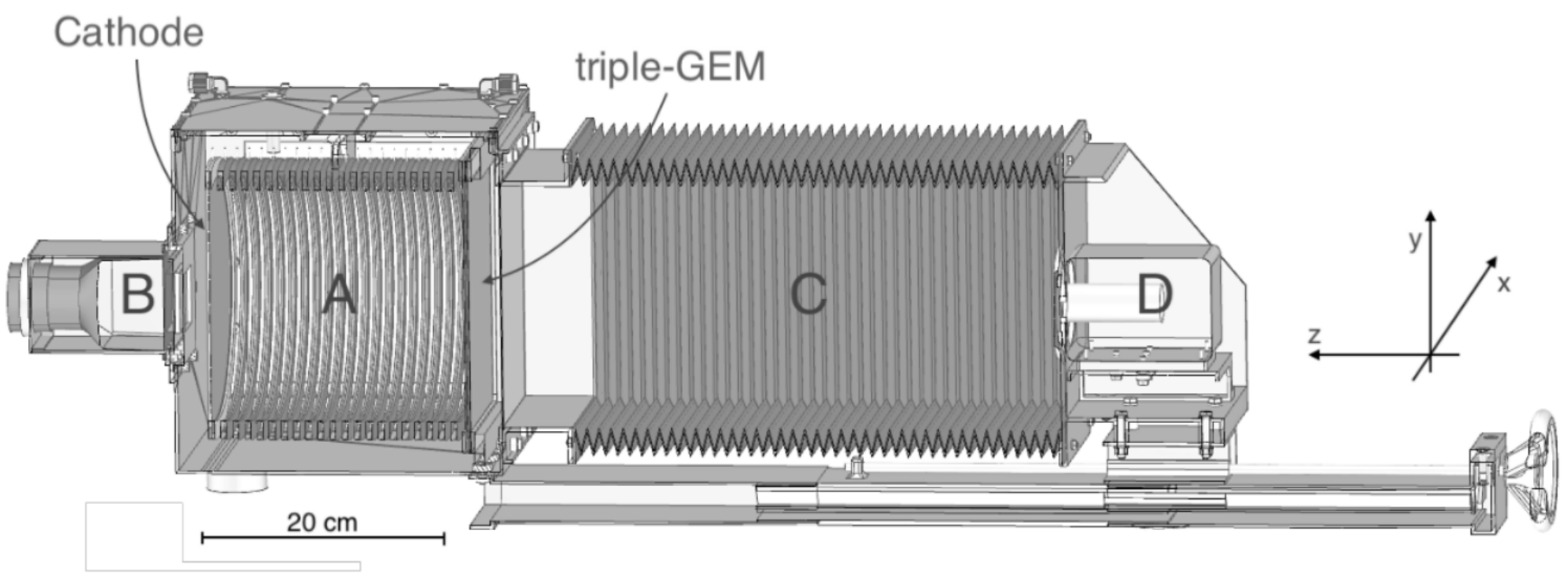

**Figure 3.11** The schematic of the LEMOn prototype: B is the PMT holder, A the elliptical field cage ring, C the optical bellow and D the sCMOS camera holder.

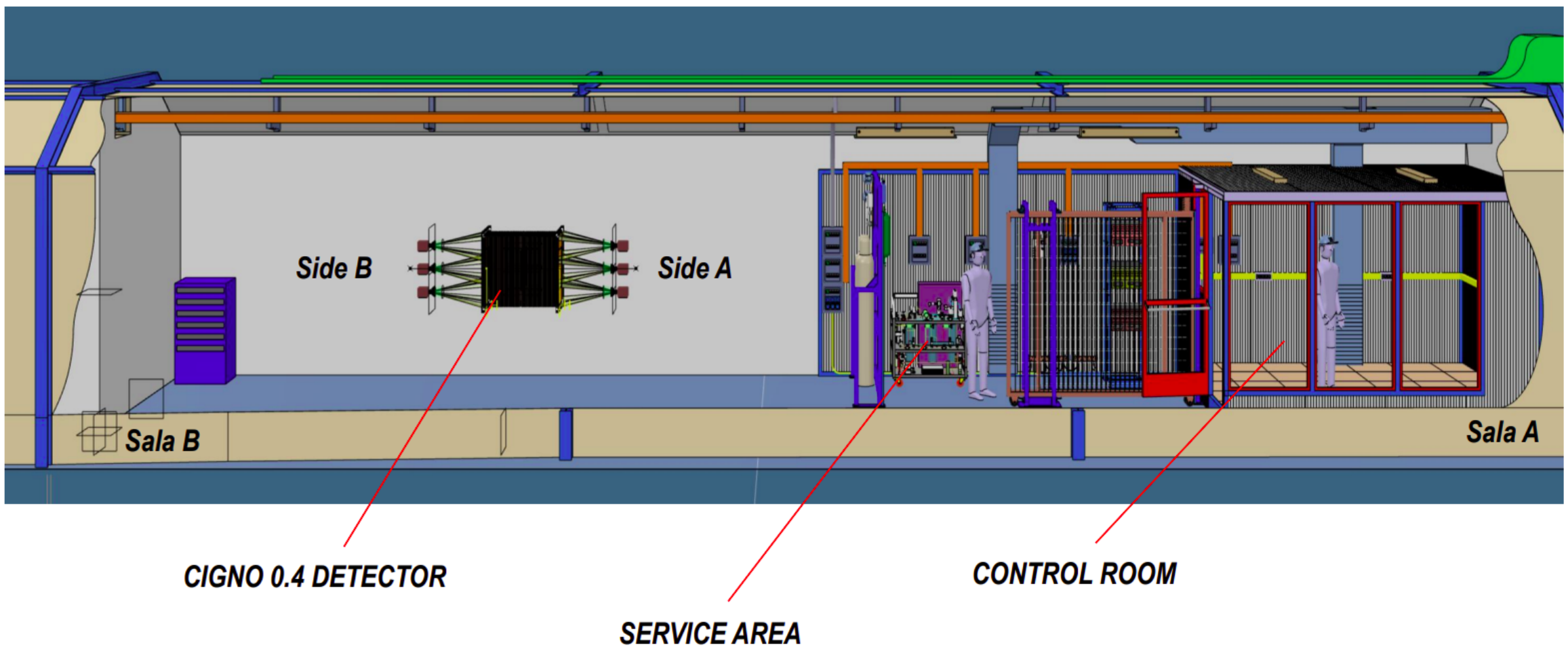

**Figure 3.12** General setup of the Hall F, where the CYGNO-04 demonstrator will be hosted.

with a common cathode at the center of the volume. The amplification stage will employ triple thin GEMs, with an area of 50 $\times$ 80 cm$^2$. The optical system, as updated, includes 3 sCMOS ORCA-Quest camera and 8 Hamamatsu R7378 PMTs per side. Each sCMOS camera will provide an effective pixel granularity of 130 $\times$130 $\mu$m$^2$, offering improved resolution compared to the LIME prototype. The PMTs, which are the same model used in LIME, will be positioned at the corners of each camera facing the GEMs. These PMTs have a QE of about 15% at 500 nm and a typical gain of 2 $\times$ 10$^6$. A GEANT4 MC simulation was defined to optimized the shielding materials used in CYGNO-04. The most cost-effective solution is shown in Fig. 3.14 and involves an internal thickness of 4 cm of radiopure copper, surrounded by 6 cm of refurbished copper from the OPERA experiment. The radioactivity of the materials used in the setup is still being studied to ensure minimal

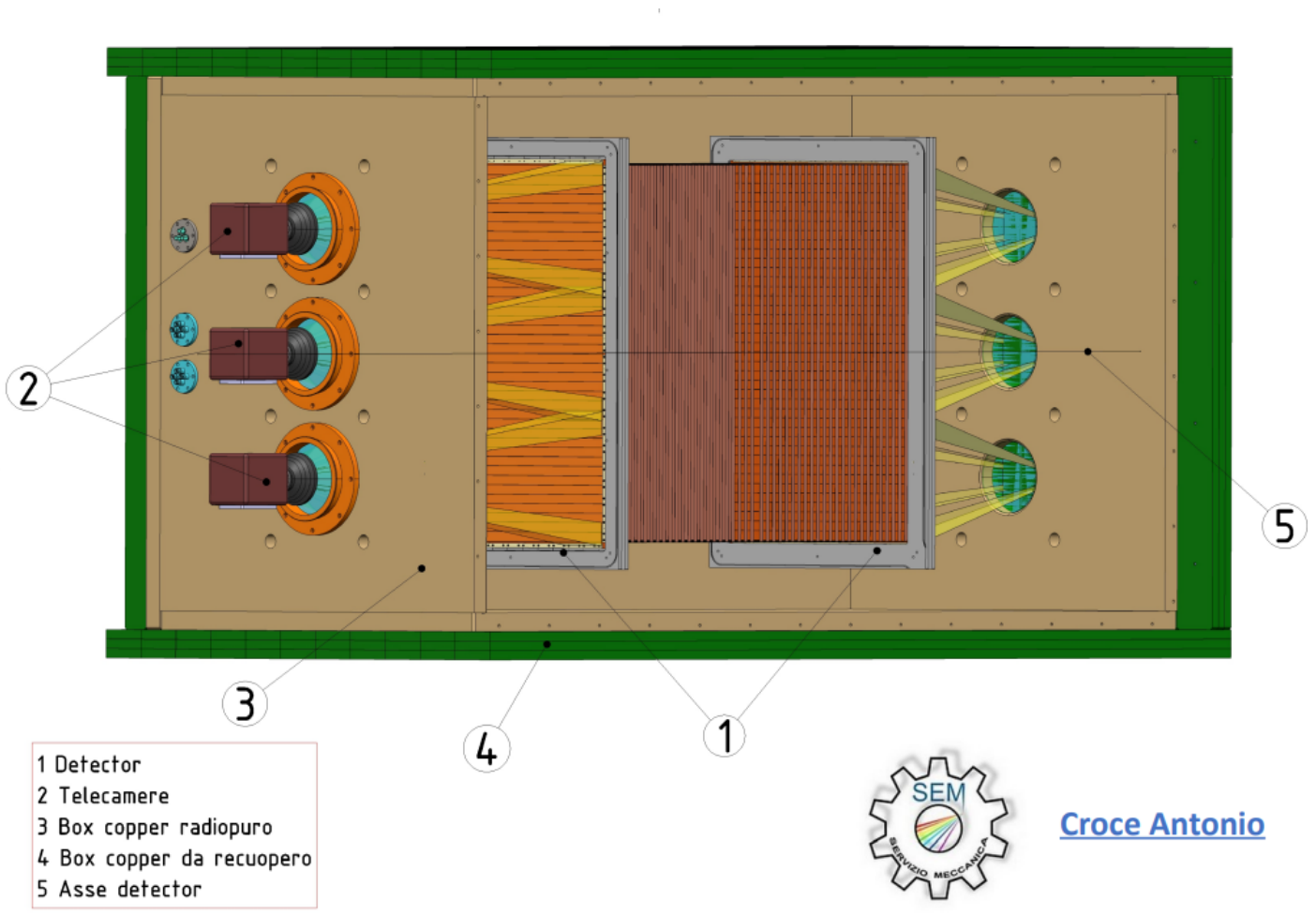

**Figure 3.13** Schematic of the CYGNO-04 experimental setup.

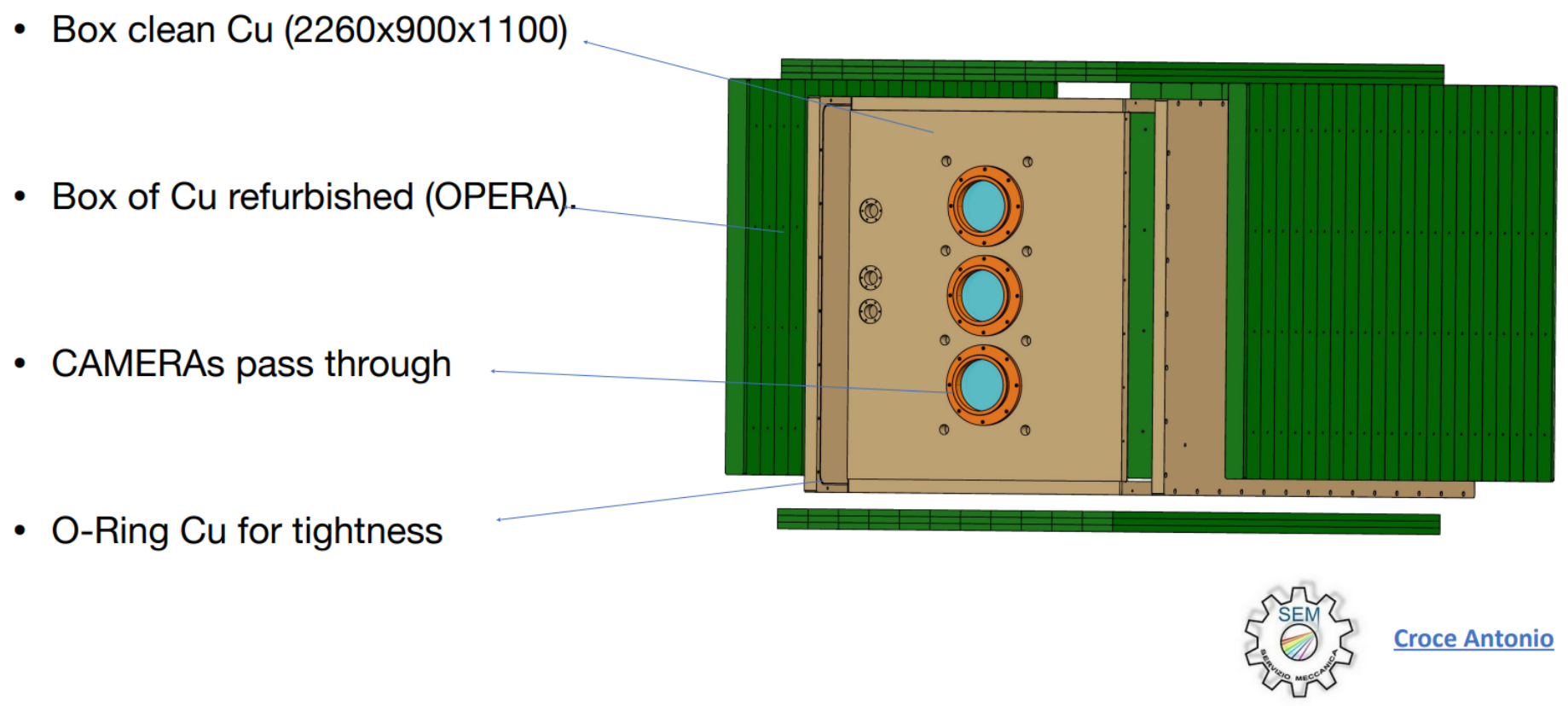

**Figure 3.14** Schematic of the optimized copper shielding fro CYGNO-04.

background interference.

### 3.4.3 PHASE_2: CYGNO-30

The final stage, CYGNO-30 [97], envisions a large-scale detector with a volume of approximately 30 m³, created by stacking multiple CYGNO-04-like detectors. This setup is expected to significantly contribute to DM searches, particularly in the mass range below 10 GeV, covering both Spin-Independent (SI) and Spin-Dependent (SD) coupling scenarios. The detector would be hosted in LNGS, with passive shielding to reduce external background. In the event of a DM discovery by other experiments, CYGNO's directional capability would be essential to confirm the galactic origin of the observed signal.
A Bayesian statistical analysis was performed to evaluate the sensitivity of the CYGNO-30

detector for WIMP searches in the presence of background noise. The lower limit on the detectable WIMP mass depends on the target of choice and the energy threshold, which in turn depends on the ionization quenching factor. The likelhood assumes that both background events and DM events are present in the sample, each following their probability distribution. The background events are assumed to be isotropic, since any local background source which is not inherently isotropic with respect to the detector, would be smeared in galactic coordinated due to the motion of the Earth around its own axis, and would have a mostly flat distribution. The WIMP recoils were assumed to follow the angular distribution expected under the Standard Halo Model (SHM). The sampled data were then smeared with a Gaussian to account for the angular resolution and were fitted to extract the 90% CI on the number of DM events.

The number of DM events is closely related to the Spin-Independent (SI) or Spin-Dependent (SD) cross sections through the expected recoil rate. This rate considers various characteristics of CYGNO, such as gas mixture properties, total mass, relative percentage of each element, exposure time and energy threshold. Additionally, the rate is based on assumptions regarding the properties of the DM halo, such as the WIMP velocity distribution and the local DM density. Setting the true number of DM events equal to the 90% CI value found from the analysis, the only dependencies on unknown parameters are the mass of the WIMPs and the interaction cross section for the SI and SD cases, which are plotted in Fig. 3.15. Together with limits from other experiments, the computed constraints for the SI and SD coupling in Fig. 3.15 are shown, considering a 30 $m^3$ detector with a 3 year exposure and a 1 keV electron equivalent energy threshold. An angular resolution of 30° is assumed across the entire energy spectrum, with a presumed 100% efficiency in recognizing head-tail (HT) events, indicative of the ability to accurately measure the full direction of all events. This choice of angular resolution is supported by measurement conducted by the NEWAGE collaboration in the 50-400 $keV_{nr}$ energy range and by recent simulations within the CYGNUS context. While acknowledging that a 100% HT efficiency may be optimistic for energies below 10 $keV_{nr}$. Conversely, the angular resolution of 30° is considered conservative for higher energy NRs in the context of CYGNO. For the SI limit, the plot in the top left of Fig. 3.15 suggests that a CYGNO-30 detector could explore previously unexplored regions of the cross-section-mass range. Although anticipated future sensitives of other experiments such as SuperCDMs [98], CRESST [99] and DarkSide [100] low-mass may also cover these regions, CYGNO's ability to leverage the directional properties of NRs represents a crucial and definitive test for positively identifying a DM signal. The plot in the bottom left quadrant indicated that by potentially lowering the energy threshold to 0.5 KeV, as suggested by results from LIME, CYGNO probe the lower mass region down to approximately 0.7 GeV, extending to approximately 0.5 GeV if a small percentage of hydrogen rich elements such as isobutane is incorporated into the CYGNO mixture. CYGNO has been testing addition of isobutane and methane for this scope and obtained a configuration with methane that can maintain same LY up to 5% methane concentration [97]. The top right plot in Fig. 3.15 illustrates the limit on the SD cross-section achievable by CYGNO-30. With a 1 keV energy threshold, the sensitivity approaches levels comparable to PICO, which currently exhibits the best sensitivity among experiments exploring SD coupling. In the lowest background scenario, lowering the threshold to 0.5 keV, as depicted in the bottom right plot, would enable CYGNO to explore regions inaccessible to PICO. Notably, PICO operates on an energy threshold approach and lacks the capability, unlike CYGNO, to confirm the galactic origin of a detected signal.

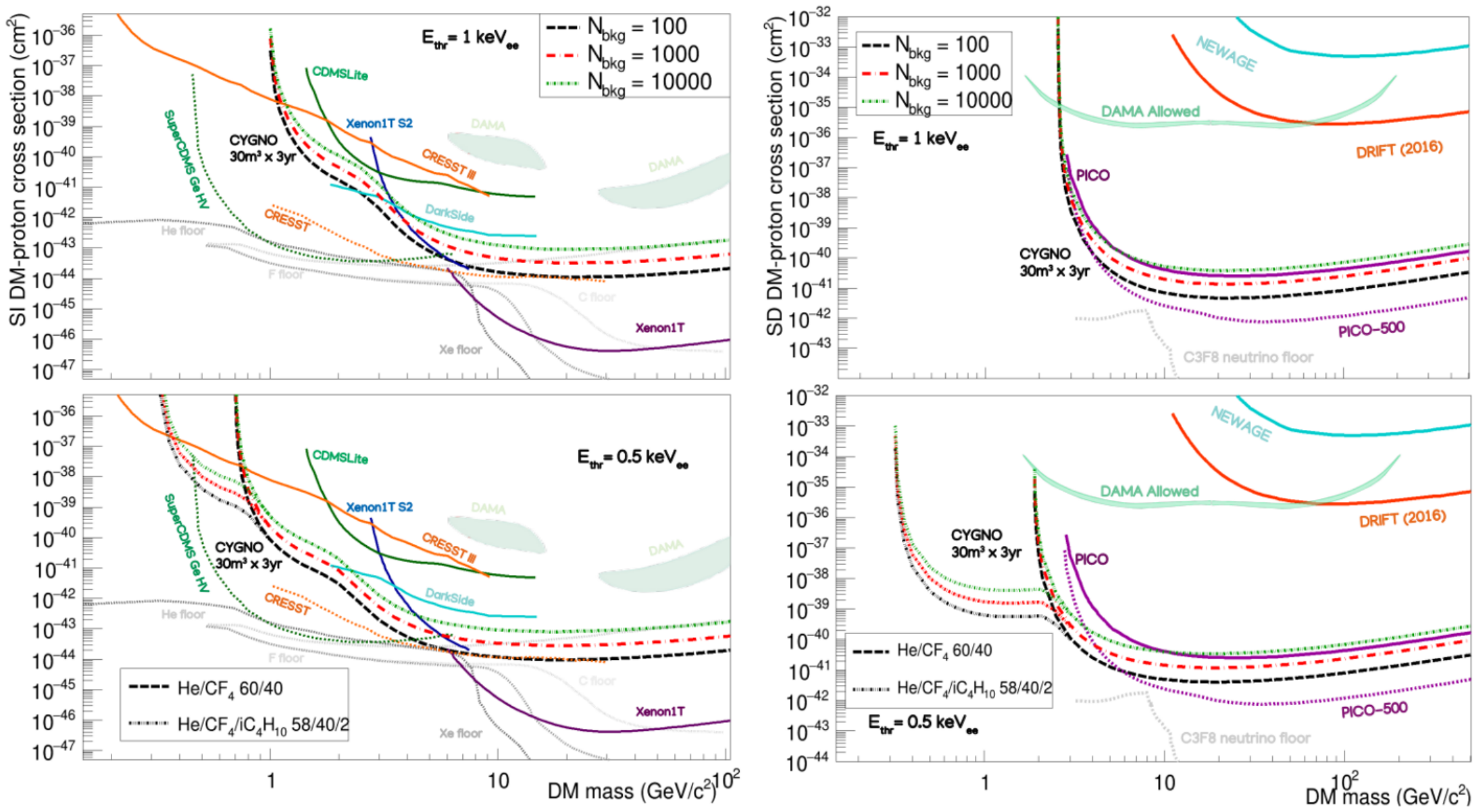

**Figure 3.15** Spin-independent (left) and spin-dependent(right) 90% CI limits for CYGNO-30 with 3 years of exposure, considering different background levels and operative threshold 1 keV$_{ee}$ (top) and 0.5 keV$_{ee}$ (bottom). The dashed curves represent a HE:$CF_4$ (60:40) gas mixture with 100 (black), 1000 (red) and 10000 (dark green) background events. The dotted curves show the sensitivity of a He:CF: 4:isobutane (58/40/2) mixture. SI limits are depicted on the left, while SD limits are on the right.

With an estimated event rate of 1 event/year/m³, CYGNO-30 could be relevant not only for DM searches but also for neutrino physics studies. CYGNO-30 could provide the first directional measurement of electron scattering from solar neutrinos in the pp chain.

# 4 LIME: The largest prototype

The CYGNO projects aims to develop a high-resolution gaseous Time Projection Chamber (TPC) with optical readout, designed to operate at atmospheric pressure and room temperature. Several prototypes, which are described in Section 3.4.1, have been built to study the performances of the approach and its scalability in order to build a detector with a sensitive volume of tens of cubic meters. The R&D phase is concluding with the LIME (Long Imaging ModulE) prototype.
The energy response of LIME was fully characterized in a range from few keV to tens of keV electron kinetic energy using different photon sources. After initial commissioning at the Laboratori Nazionali di Frascati (LNF), where its energy response and operational stability were tested using X-ray sources and cosmic rays, LIME was installed at Laboratori Nazionali del Gran Sasso (LNGS) in February 2022. Its primary goals include validating the Monte Carlo (MC) background simulations and providing spectral and directional measurements of the environmental neutron flux at LNGS. As each of these goals necessitates different configurations and background levels, the underground LIME program has been implemented in a phased manner, with additional shielding layers progressively added around the detector. In Section 4.1, the LIME prototype is introduced, with detailed descriptions of the overground studies conducted at LNF (Section 4.2) and the underground tests at LNGS (Section 4.3). The development and description of the algorithm for track recognition in the detector are provided in Section 4.4. The various sources of background noise are detailed in Section 4.5. Finally, Section 4.6 presents the LIME underground program.

## 4.1 The LIME prototype

The LIME (Long Imaging ModulE) [101] prototype (Fig. 4.1) is a 50 L gaseous TPC designed to operate at atmospheric pressure and room temperature. The gas vessel, made of 10 mm thick transparent acrylic (PMMA) maintains an internal overpressure of about 3 mbar compared to the external atmospheric pressure. The sensitive part of the gas volume is about

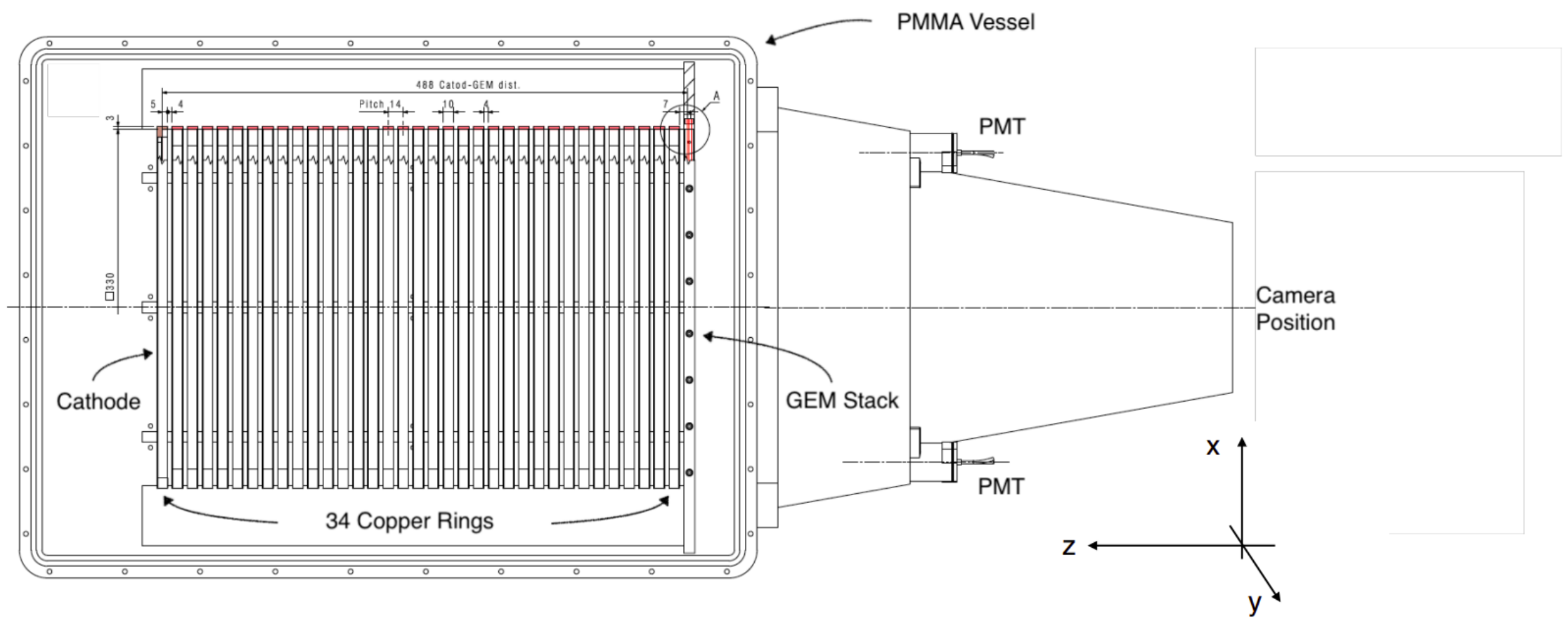

**Figure 4.1** The side view of the LIME detector shows the acrylic vessel housing the 50 cm drift region, which is defined by 34 copper rings. This drift region is enclosed between a copper cathode at one end and a triple GEM stack at the other. The positions of the PMTs and the sCMOS camera are also indicated, highlighting their placement for optimal detection and readout of the signals generated within the detector [101].

50 liters with a 50 cm long electric field, closed by a 33×33 $cm^2$ triple-GEM stack.

Inside the vessel, a uniform electric field is created to drift ionization electrons generated by the interaction of charged particles with the gas molecules. This field is maintained by a series of 34 copper rings, each 10 mm wide, spaced 4 mm apart and connected via 100 MΩ resistors. These rings establish a controlled potential gradient from the cathode to the GEMs, ensuring a uniform drift field orthogonal to the cathode plane. The cathode is a 0.5 mm thin copper plate, matching the dimensions of the copper rings.

Thanks to the high electric field inside the GEMs hole, an avalanche of secondary electrons and ions is produced. Interactions of secondary electrons with gas molecules produce also photons. The light emitted is collected by an optical system comprising Orca-Fusion sCMOS camera and four Hamamatsu R7378 photomultiplier tubes [102]. The PMTs, each with a 22 mm diameter bialkali photocathode, are symmetrically arranged around the camera for fast photon detection. The sCMOS camera [103], with a resolution of 2304 × 2304 pixels, each pixel with an active area of 6.5 × 6.5 $\mu m^2$, is equipped with a Schneider lens with 25 mm focal length and 0.95 aperture, positioned 623 mm away from the GEMs. In this configuration, the sensor covers a surface area of 34.9×34.9 $cm^2$, resulting in each pixel covering an area of 152×152 $\mu m^2$. The camera's quantum efficiency (QE) is optimized for wavelengths between 450 nm and 630 nm, matching with the light emitted by the He:$CF_4$ gas mixture, which has two main emission peaks at 300 nm and 620 nm as shown in Fig. 4.2. In order to allow efficient transmission of light to the outside, the vessel's window in front of the GEM stack is 1 mm. The PMMA window induces reflection of alpha particle tracks onto the GEMs plane. In Fig. 4.3, a portion of the image acquired by the camera is shown, the brighter track corresponds to the real event caused by an alpha particle, while the fainter track, known to as the shadow, is the result of the reflection.

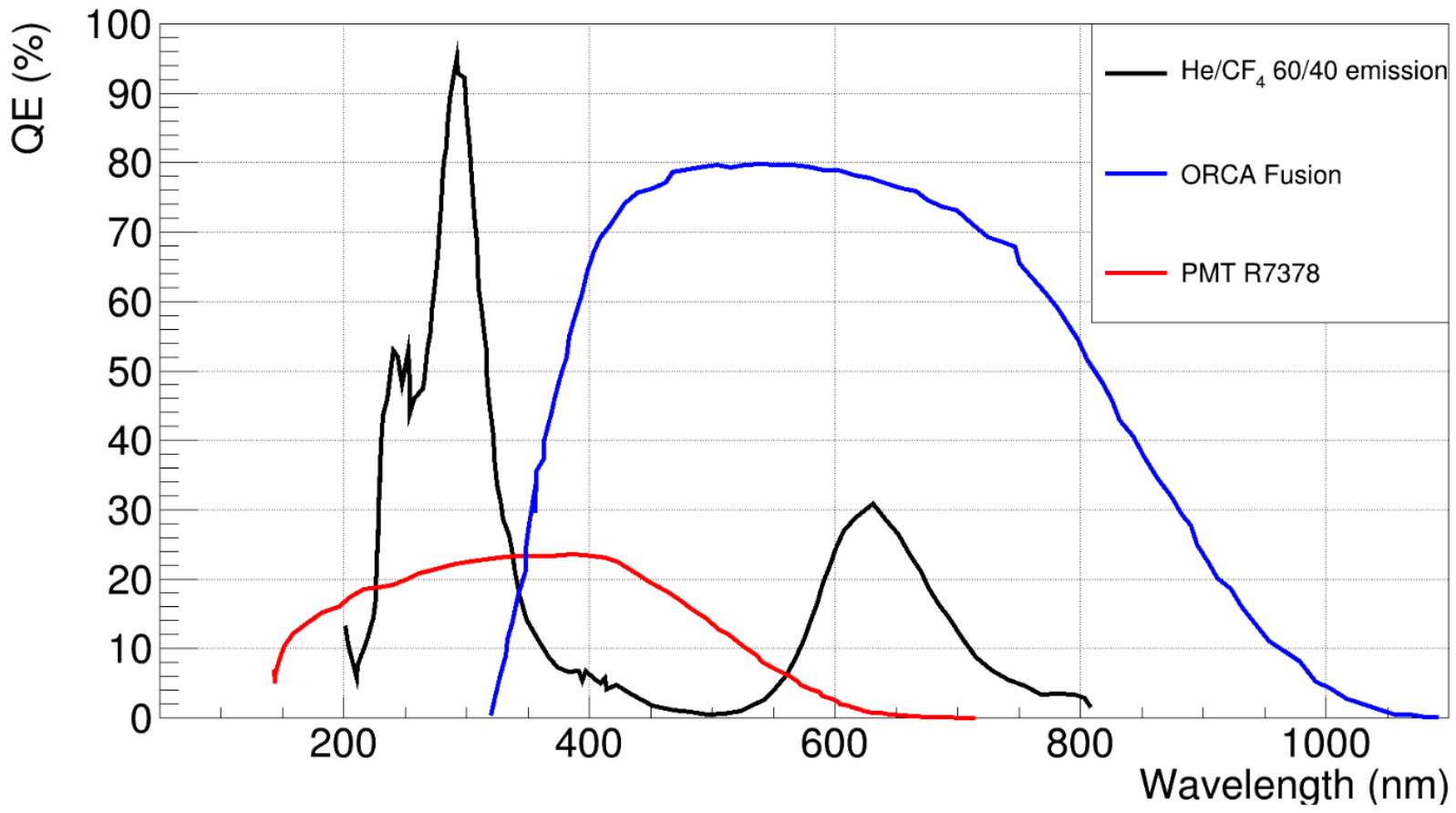

**Figure 4.2** The He:$CF_4$ (60:40) emission spectrum is superimposesd with the quantum efficiency (QE) of the ORCA-Fusion camera [103] and the R7378 PMT [102].

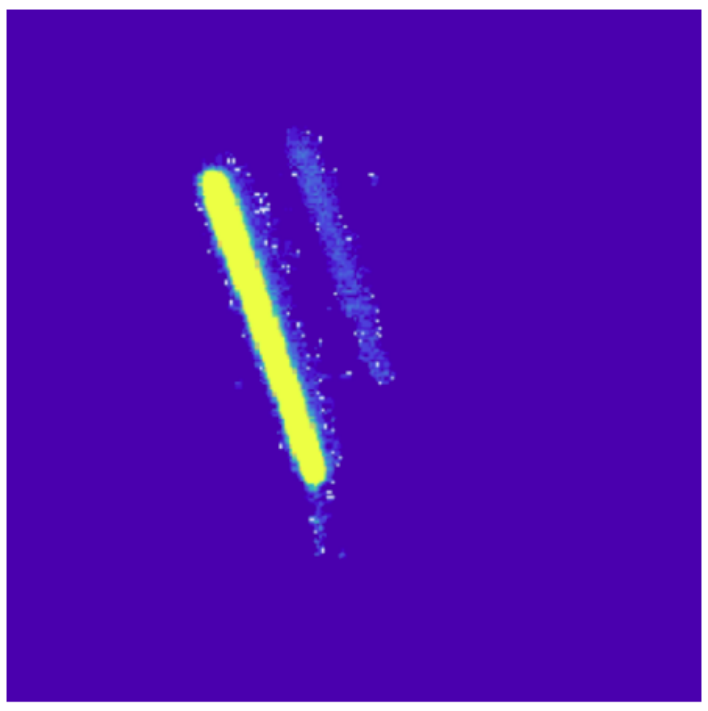

**Figure 4.3** Example of an alpha particle track and its reflection captured by the sCMOS camera.

For calibration purposes, a 5 cm wide and 50 cm long window, sealed with a 150 µm thick layer of ethylene-tetrafluoroethylene (ETFE), is installed on the upper face of the vessel, as shown in Fig. 4.4. This window allows low-energy photons to enter the gas volume from an external source. A remotely controlled trolley moves along a track 18 cm above the sensitive volume, holding a radioactive source that can be positioned at distances ranging from 5 cm to 45 cm from the GEMs. The trolley has a 5 mm diameter hole to collimate the photon beam for precise calibration.

The entire detector is enclosed within a 3 mm thick aluminum Faraday cage to prevent electromagnetic interference and ensure light-tightness. The detector is operated with voltages provided by two different power supplies: an ISEG HPn 500 generator for the cathode, providing up to 50 kV with a ripple of less than 0.2%, and a CAEN A1515TG board for the GEM stack, delivering up to 1 kV with a precision of 20 mV.
The radioactivity of the components in LIME, which includes cathode, copper shielding, field rings, resistors, GEM, acrylic vessel, camera lens and body, has been measured using a high-purity germanium detector at the LNGS low radioactivity laboratory. These results were used to simulate the internal radioactivity of the detector. While most of the materials and the detection elements used in LIME are not at the radiopurity level required for a DM search, however they can be produced in a radiopure version without affecting the detector performance of the 1 $m^3$ CYGNO demonstrator.

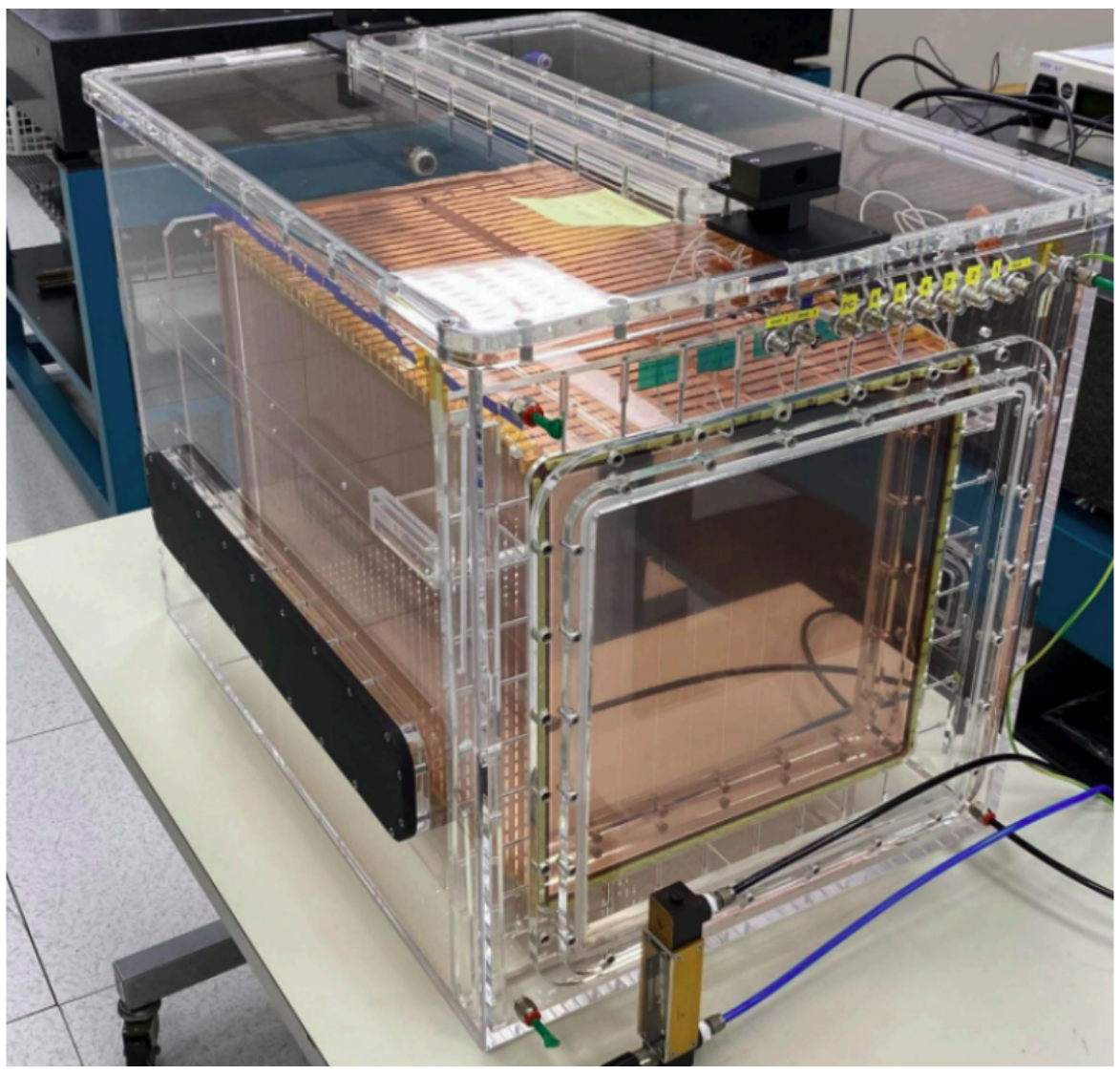

**Figure 4.4** A picture of the LIME detector inside the acrylic vessel, on the top also the ETFE window is shown.

## 4.2 Overground test at LNF

The Lime prototype operated for several months overground at Laboratori Nazionali di Frascati (LNF) in a long campaign of data-taking. The primary goals of this campaign were to assess the long-term operational stability of the detector, collect data for analysis, develop image processing techniques, evaluate particle energy reconstruction performance, and validate the feasibility of the detector's technological choices.
The data acquisition system was implemented through a C++-based integrated framework within the MIDAS [104] software system. Signals from the PMTs were processed by a discriminator and logic module, which generated a trigger signal for image acquisition based on the coincidence of at least two out of four PMTs.
The gas mixture, maintained at an overpressure of about 3 mbar relative to atmospheric pressure, flowed through the vessel at a rate of 200 cc/min to ensure optimal operating conditions. The drift field was set at 0.8 kV/cm, with a fixed voltage of 440 V applied to each GEM. Several test and measurements were done, providing a wide characterization of the detector's performances, which are described in details in [101]
For the results presented in the following sections, the sCMOS exposure time was set to 50 ms in order to minimize the pile-up from natural radioactivity events.

### 4.2.1 Sensor electronic noise

The fluctuations of the dark offset of the optical sensor arise mainly from two different contribution: readout noise, which is the electronic noise generate by the amplifiers in each pixel (less than 0.7 electrons RMS), and a dark current, a small current flowing through each camera photo-diode, amounting to approximately 0.5 electrons/pixel/s [105]. To isolate this noise contributions, dedicated runs, referred to as pedestal runs, were taken through the data taking period with the values of $V_{GEM}$ set to 220 V. Under these conditions, the counts recorded by the camera pixels were only due to the electronic noise of the sensor itself.
During each pedestal run, 100 images were recorded, and for each pixel the average value ($pix_{ped}$) and the standard deviation ($pix_{rms}$) of its response has been evaluated. The distribution of $pix_{rms}$ for all pixels in a typical pedestal run is show in Fig. 4.5. A long tail above the most probable value can be observed, corresponding to pixels located at the top and bottom boundaries of the sensor, which tend to be slightly noisier than those in the central region, To mitigate this effect, a fiducial cut is applied, excluding these noisier pixels from the analysis, more details in Section 5.3.
The stability of both the pedestal value and the electronic noise was monitored examining the mean values of the $pix_{ped}$ and $pix_{rms}$ distributions from regular pedestal runs. Fig. 4.6 shows the distribution of these two quantities over a period of approximately two weeks, demonstrating that the sensor maintains a very good stability throughout the data-taking period.

### 4.2.2 Light Yield

The energy calibration process in the detector involves converting the camera sensor counts into energy measurements using a $^{55}$Fe source as a standard reference. The 5.9 keV X-rays emitted by this source interact with the gas inside the detector through the photoelectric effect, producing spot-like tracks in the camera images. These spots typically measure around 20 mm$^2$, accounting for diffusion and the sensor's granularity. The images are then subjected to a reconstruction analysis, during which the tracks are identified, and key properties are extracted for further examination.
The light yield, typically on the order of $10^3$ counts per keV, may vary over time due to environmental factors such as changes in pressure, temperature, and gas quality. Despite these variations, the calibration process ensures accurate energy measurements, which are essential for the detector's overall performance in particle identification and event reconstruction.
In Chapter 5, a description of the spectrum analysis and the procedure to evaluate the iron peak is reported.

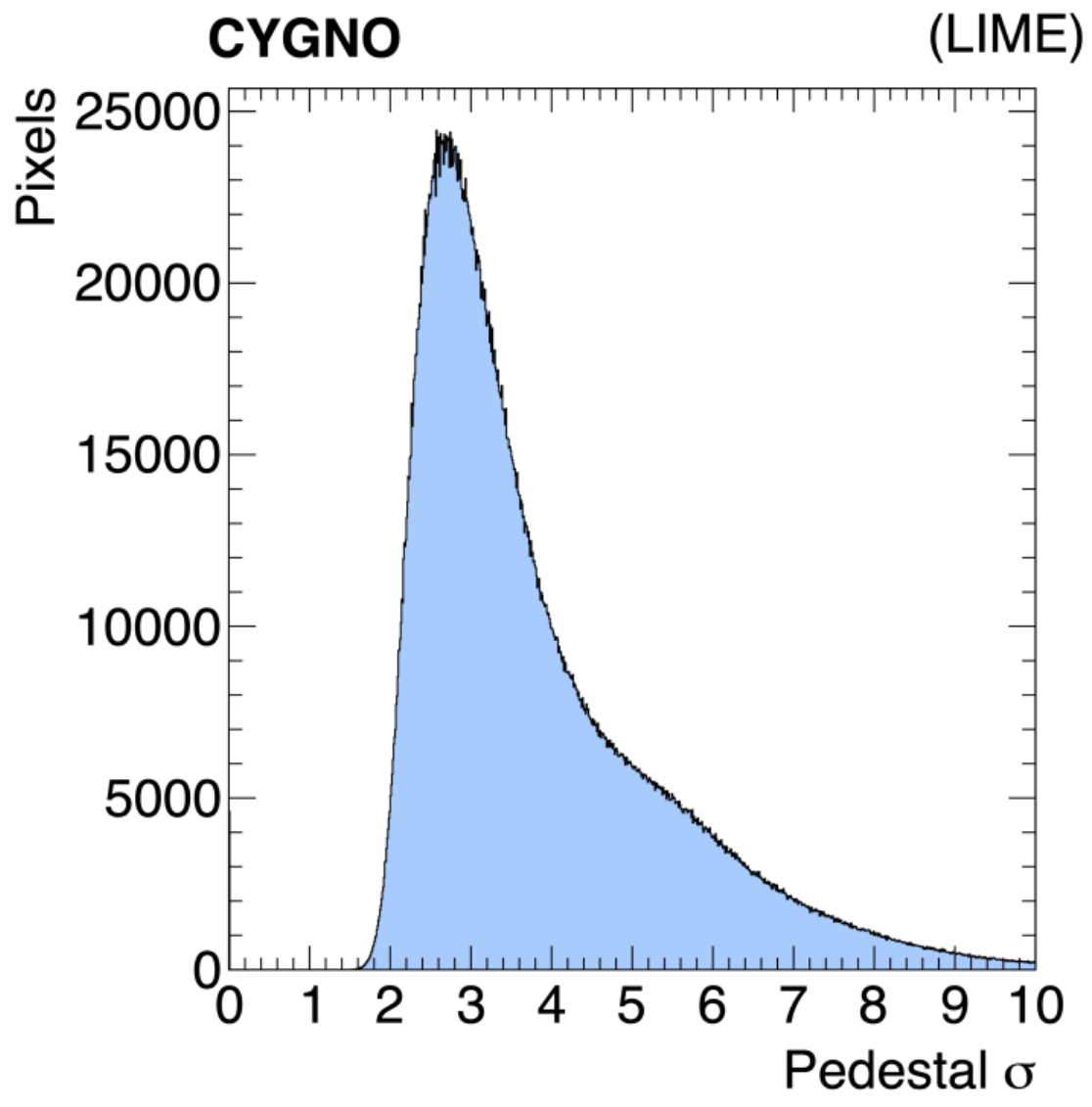

**Figure 4.5** Distribution of pix$_{rms}$ in one pedestal run [101].

### 4.2.3 Gain saturation

The large charge generated during electron avalanches can significantly reduce the electric field inside the GEM holes, particularly in the last GEM, which in turn diminishes the effective gain. This saturation effect is influenced by both the total charge entering the GEM and the area over which the ionization electrons are spread (i.e., the specific energy loss, $dE/dx$).
This phenomenon was observed in LIME with a $^{55}$Fe source placed at various distances from the GEMs. As the source gets closer to the GEMs, the saturation effect becomes more pronounced, resulting in a lower light yield (Fig. 4.7). This occurs because diffusion in the drift region distributes the primary charge over a larger area, leading to a less dense electron cloud reaching the GEMs.
Additionally, saturation is expected to have a lesser impact on higher energy electromagnetic recoils (ERs), where the energy loss per unit length is lower, causing in a more sparse ionization charge distribution compared to denser, spot-like tracks. Consequently, nuclear recoils (NRs) are anticipated to be more strongly affected by GEM gain saturation, as their energy deposition is consistently denser than that of ERs across all energy levels.

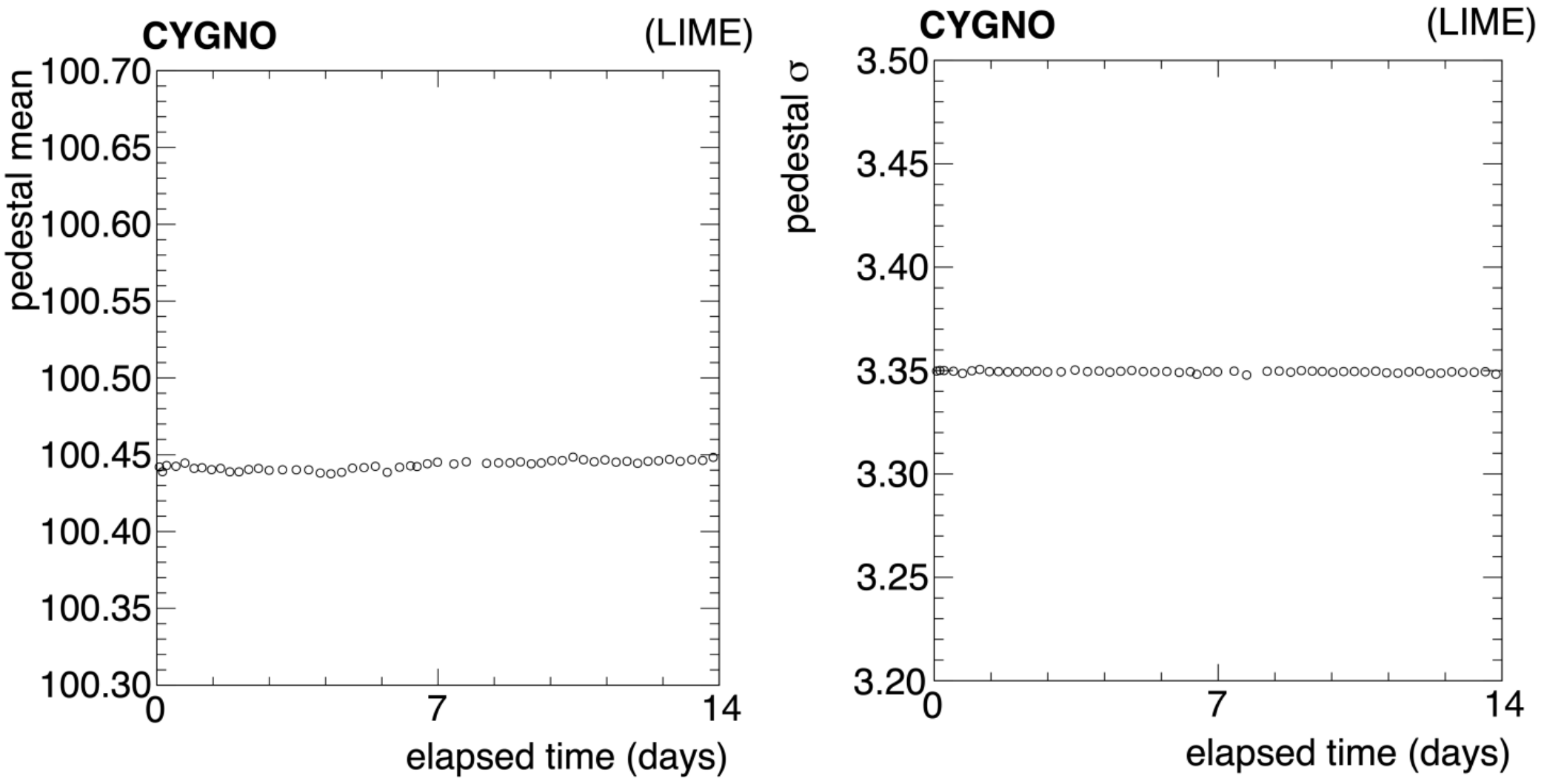

**Figure 4.6** On the left the average $pix_{ped}$ and on the right the $pix_{rms}$ as a function of time for a period of two weeks of data taking [101].

In LIME, the diffusion of the primary ionization electrons over the 50 cm drift path can almost quadruple the area involved in the multiplication, as shown in Fig. 4.8, which presents two iron spot at different distances. Thus reducing the charge density and therefore the effect of a gain decrease. The light yield for spots originated by interactions farther from the GEM is larger than for spots closer to the GEM. Thus, the overall trend of light yield as a function of the $z$ position of the ionization site initially increases, followed by an almost plateau region, as shown in Fig. 4.7.

### 4.2.4 Response to low energy recoils

The energy response of the detector as a function of the impinging X-ray energy is studied by selecting clusters reconstructed from various radioactive sources. A multi-target X-ray source utilizing a $^{241}$Am primary source was employed to test the detector's response to low-energy electron recoils (ERs). The $^{241}$Am source produce $\sim$5 MeV alpha particles directed onto disks composed of different materials placed on a rotating support. The material used include copper (Cu), rubidium (Rb), molybdenum (Mo), silver (Ag), barium (Ba) and terbium (Tb), which can be selected rotating a wheel. This wheel rotation positions the disks in alignment with an aperture for X-ray passage. A schematic of the source setup is shown in Fig. 4.9.

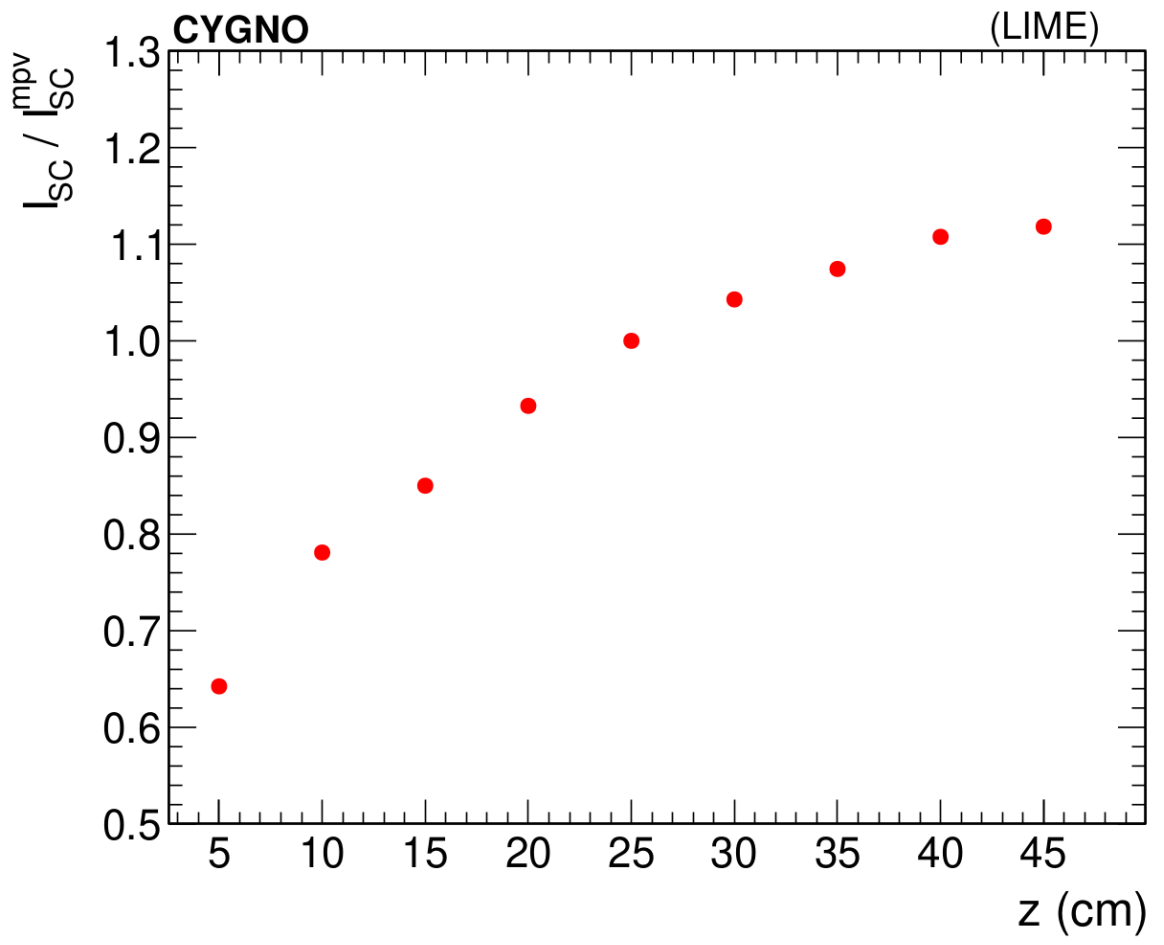

**Figure 4.7** Average light integral, normalized to its most probable vale ($I_{SC}^{mpv}$) for cluster reconstructed in presence of the $^{55}$Fe source as a function of the distance from the GEM plane [101].

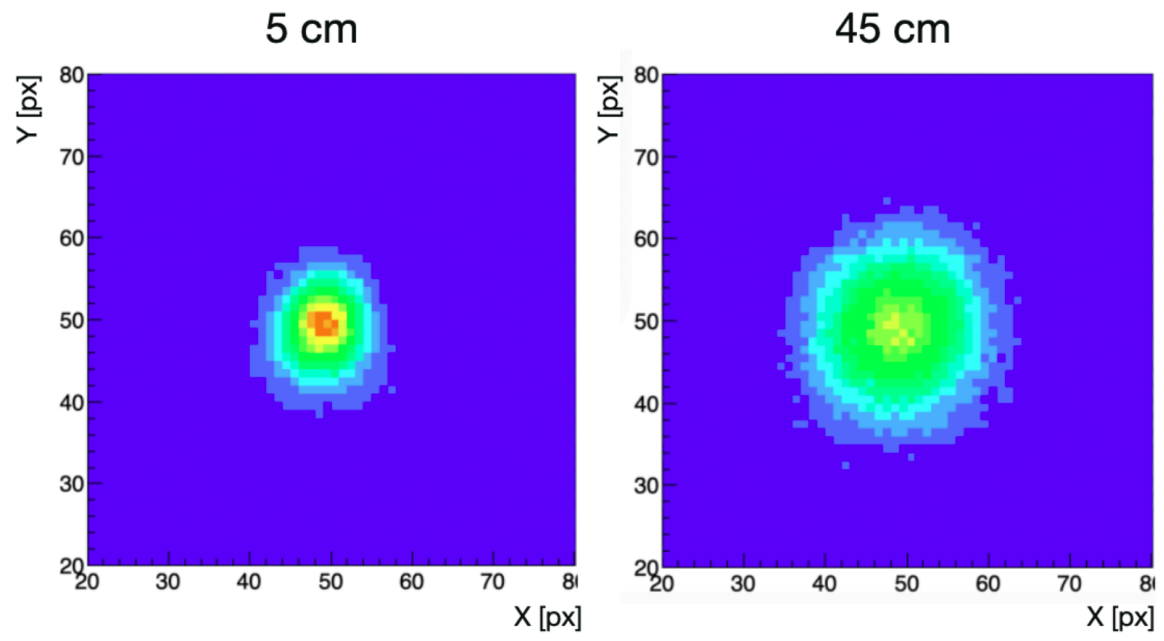

**Figure 4.8** Two iron spots acquired with the source placed 5 cm and 45 cm distance from the GEM plane.

Upon interaction with the material, the alpha particles, through ionization, cause the ejection of an electron from the atom's inner shell, K-shell. An outer-shell electron from either the M-shell or L-shell fills the vacancy, leading to the production of a monochromatic X-ray, referred to as $K_\alpha$ and $K_\beta$ emission. The gammas rays interact with the detector through the photoelectric effect, producing electron recoils. Fig. 4.10 displays images of the ERs from X-ray interactions at different energies. It can be observed that the tracks appear as a spot at an the energy of 8 keV. This phenomenon occurs because, at that energy, the distance covered by the electron traveling in the gas is considerably smaller than the size of the diffused tracks. Tracks starts to exhibit significant extension for energies above 15 keV.
Additionally, a custom set-up with a $^{55}$Fe X-ray source was used to induce the emission of X-rays below 6 keV from a calcium (Ca) and titanium (Ti) target. The energies of the tested X-rays are summarized in Table 4.1. Data collection involved positioning the radioactive source 25 cm away from the GEM plane, corresponding to the center of the drift region. The average energy response from the $^{55}$Fe source was utilized to derive the absolute energy scale calibration constant. The detector's light clusters, which represent the energy deposited by incoming X-rays, were reconstructed and analyzed for their total energy distribution. The background, predominantly due to natural radioactivity, exhibited an exponential decay. In contrast, the X-ray interactions resulted in distinct peaks in the energy spectrum, with peak positions indicating the mean energy deposit and peak widths reflecting the detector's energy resolution. Each spectrum from the radioactive sources was fitted using a combination of two Cruijff functions to represent the $K_\alpha$ and $K_\beta$ lines, along with an exponential function to model the background. The spectra of the calibrated energy E for data collected in presence of the radioactive sources is shown in Fig. 4.11 The energy peaks are calibrated using the light yield computed from the $^{55}$Fe measurement. As shown in Fig. 4.12 the linear fit between the estimated and expected energies demonstrates good linearity from 3.5 keV to around 37 keV. The uncertainties on each point represent the statistical contribution and the systematic uncertainty arising from the knowledge of the $z$ position.

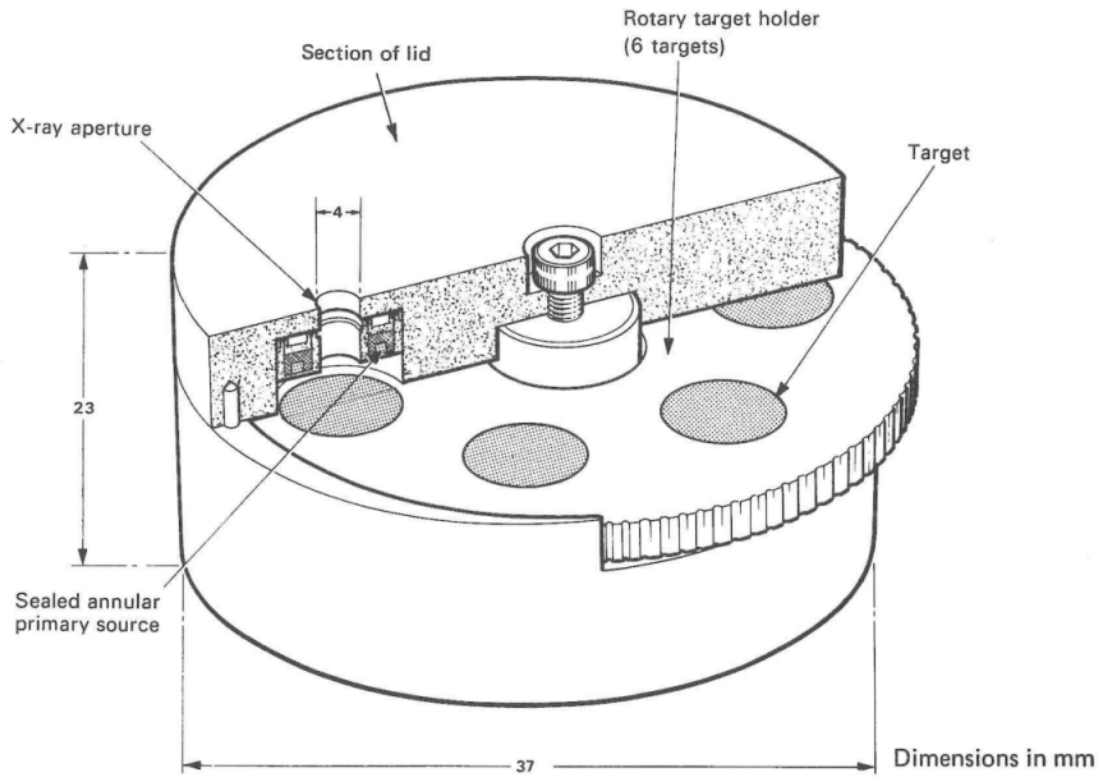

**Figure 4.9** Scheme of the multi-target X-ray.

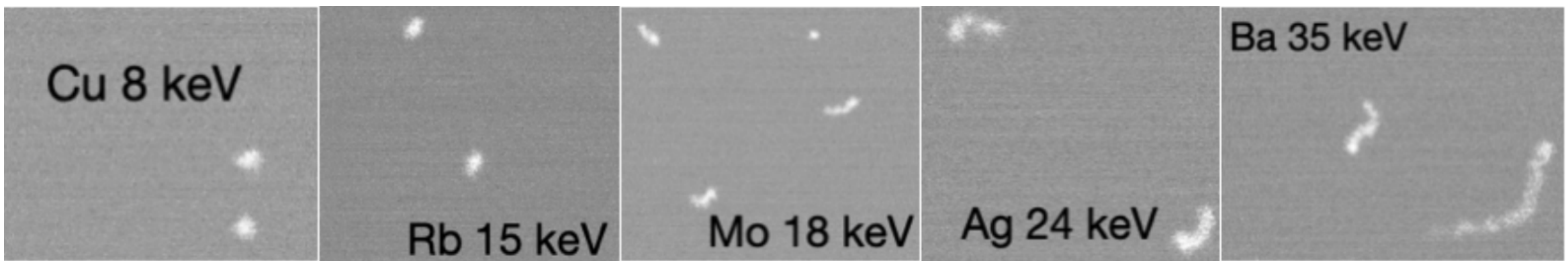

**Figure 4.10** The ER produced by gammas at different energies.

| Target | Energy $K_\alpha$ [keV] | Energy $K_\beta$ [keV] |
|---|---|---|
| Calcium | 3.69 | 4.01 |
| Titanium | 4.51 | 4.93 |
| Copper | 8.04 | 8.91 |
| Rubidium | 13.37 | 14.97 |
| Molibdenum | 17.44 | 19.63 |
| Silver | 22.10 | 24.99 |
| Barium | 32.06 | 36.55 |

**Table 4.1** X-ray energies for various target material used.

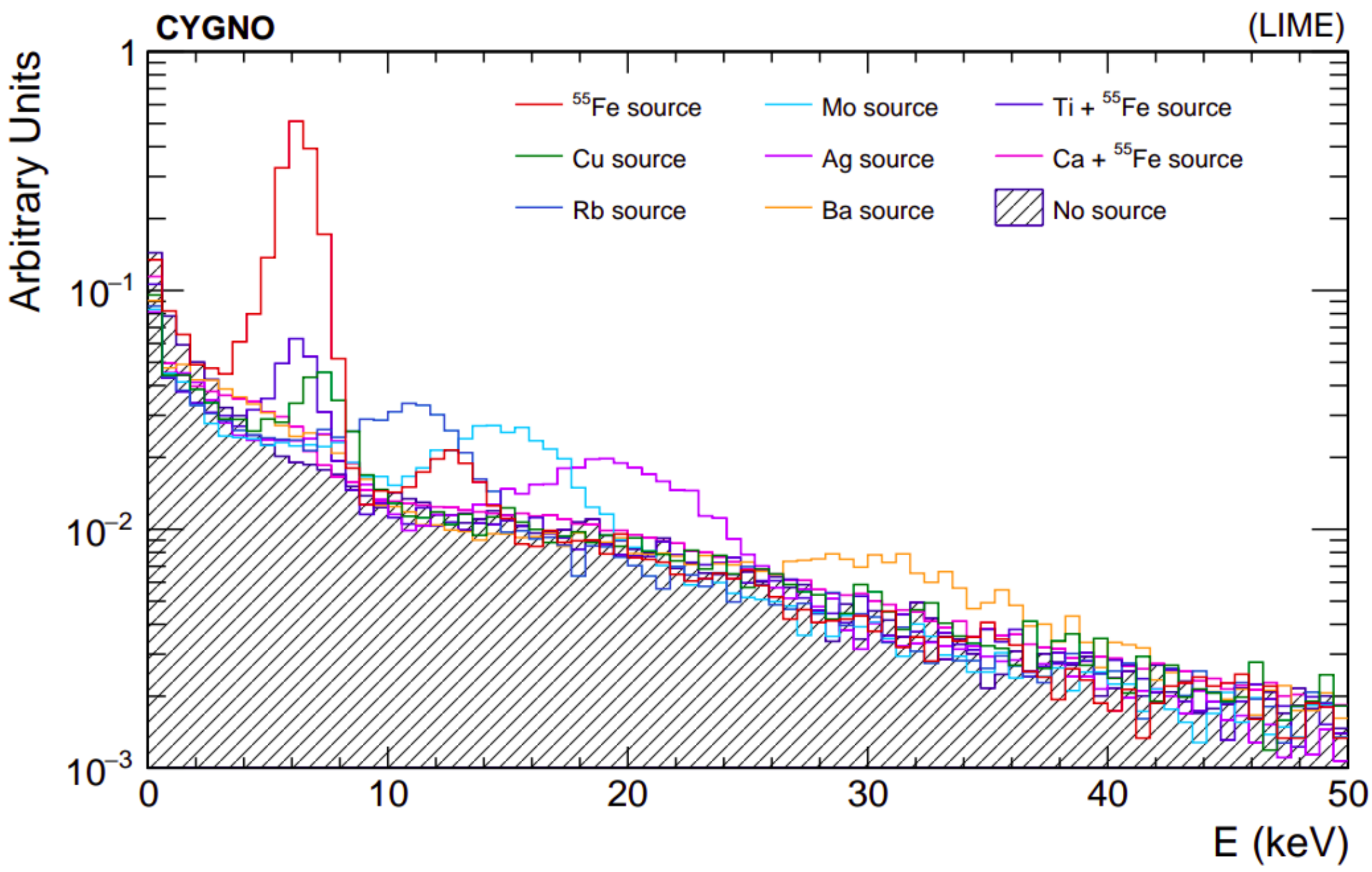

**Figure 4.11** The calibrated energy spectra E for data collected with the radioactive sources positioned at $z$ = 25 cm [101].

### 4.2.5 Energy threshold

The image reconstruction code occasionally clusters pixels corresponding to noise over-fluctuations that are not completely eliminated by the zero suppression process. However, based on the analysis of pedestal runs, where no real signals from particle tracks are expected, the contribution of these false clusters becomes negligible above 400 counts, equivalent to roughly 0.5 keV. Setting a threshold of 1 keV a limit of 10 false events per year has been established.

### 4.2.6 Determination of absolute z

Due to gas diffusion, the shape of the primary ionization electron cloud that reaches the GEMs is influenced by the drift distance, which in turn alters the shape of the light spot generated by the GEM and captured by the CMOS sensor. The $z$-reconstruction performance associated with the $^{55}$Fe source has been examined by introducing the variable $\zeta$, defined as the product of the Gaussian sigma fitted to the transverse profile of the spots, $\sigma_T$, and the standard deviation of the counts per pixel inside the spots, $I_{rms}$. The distribution of $\zeta$ for all reconstructed spots as a function of distance is illustrated in the Fig. 4.13. It is evident that $\zeta$ exhibits a clear dependence on $z$ position of the event. Despite this correlation, the $\zeta$ distribution features extended tails in all instances. The accuracy of the absolute $z$ estimation deteriorates with increasing distance from the GEMs, resulting in a resolution between 4 cm for shorter drift distances to 8 cm for larger ones.

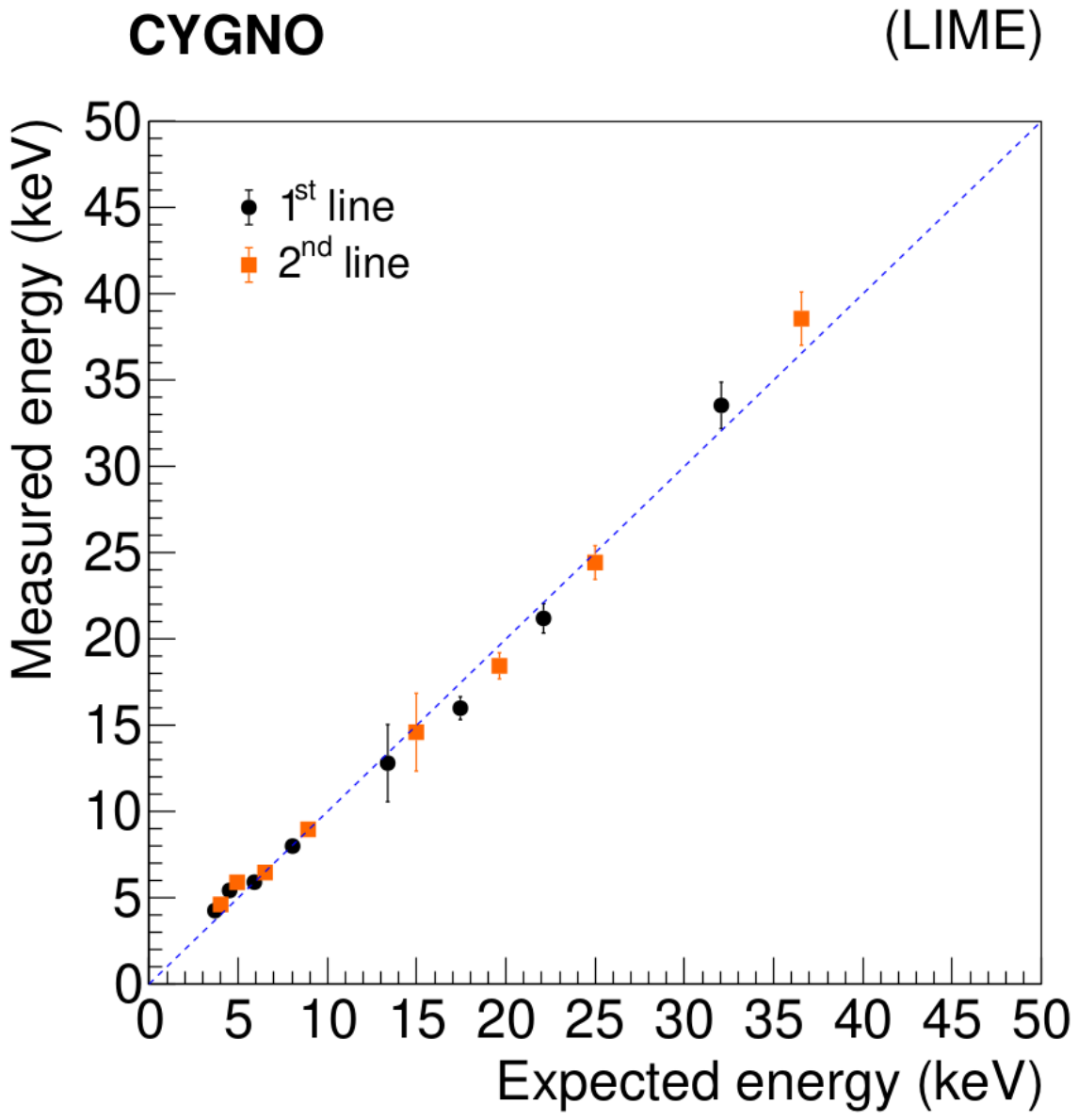

**Figure 4.12** Estimated average energy response against expected average energy response from $K_\alpha$ (black points) and $K_\beta$ (orange squares) lines contributions. The dotted line represents a linear response of the detector [101].

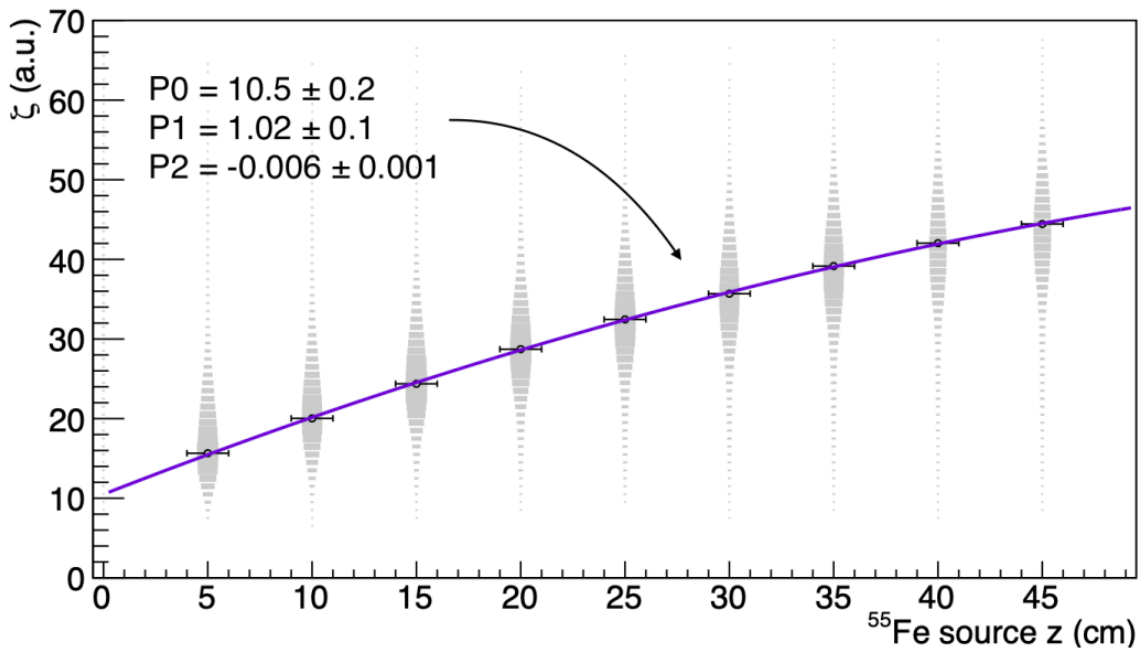

**Figure 4.13** The distribution of the values of $\zeta$ in the runs with $^{55}$Fe source at different distances from the GEMs [101].

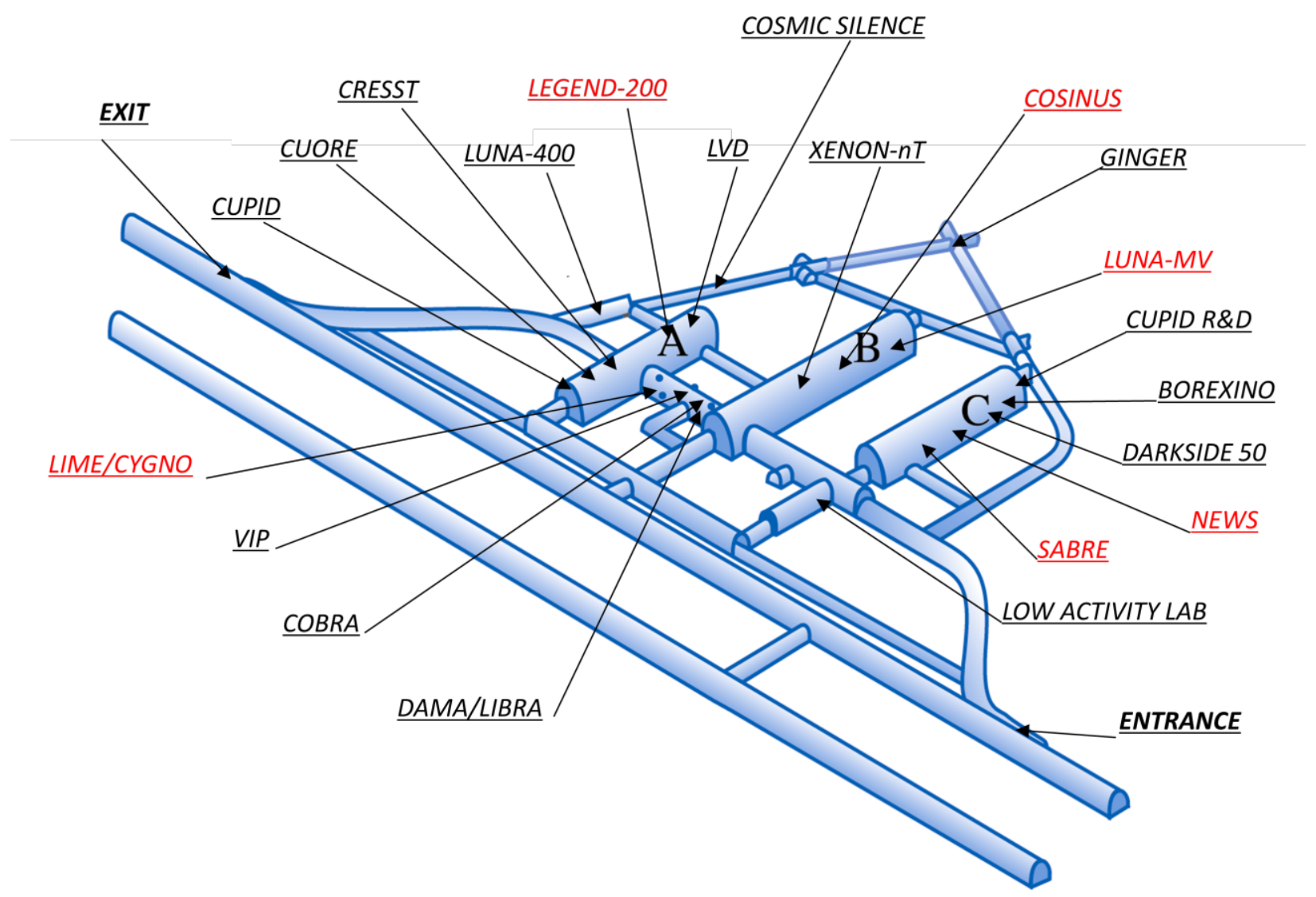

**Figure 4.14** A schematic view of the experiments currently installed in the underground laboratories of Gran Sasso (LNGS), showing the LIME detector positioned in a connecting gallery between Hall A and Hall B.

## 4.3 Underground installation at LNGS

The primary objective of conducting LIME operations at LNGS is to evaluate the feasibility of the detector's technique in an underground environment, demonstrating also the long-term functionality of the detector and its auxiliary systems. Additionally, the aims is to quantify the effective background arising from radioactivity and external sources, while also assessing the potential for reducing these background levels through the use of passive shielding. The prototype was installed at its current locations at LNGS in February 2022. The experimental setup is housed in a container located in the TIR gallery between Hall B and Hall A as shown in Fig. 4.14, with the container spanning two floors (Fig. 4.15). The lower floor contains the detector, while the upper floor serves as the control room, housing the DAQ servers, modules and high voltage supplies. The gas system, developed by the CYGNO experiment in collaboration with Air Liquide, is located just outside the experimental area.
After an initial commissioning phase, various data-taking phases have been carried out and are still ongoing today.

### 4.3.1 Gas System

The gas system in LIME serves four key functions: providing the correct gas mixture of He:$CF_4$ in a 60:40 ratio, chemically removing impurities, recirculating the gas, and recovering. The system consists of a cart and a ramp with four cylinders: two supply cylinders for helium (He) and $CF_4$, and two recovery cylinders, which operate at a maximum pressure of 40 bar. The entire gas mixture is fully recovered within the system, ensuring no atmospheric discharge, thus supporting environmentally responsible gas disposal.

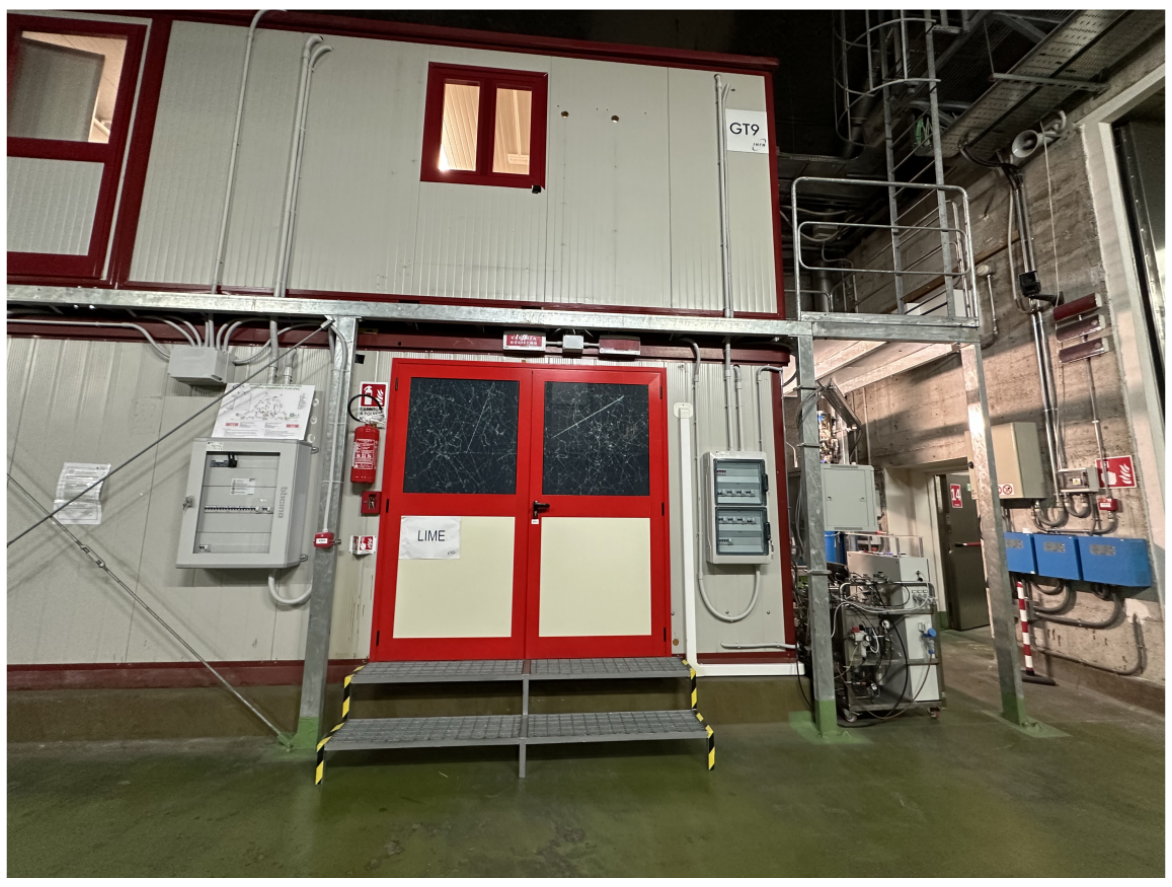

**Figure 4.15** Two floors barrack, underground at LNGS. LIME is installed on the bottom floor, while the control room is on the upper floor.

The gas system also allows for the purification and recirculation of gas within the detector. Initially, before July 2023, the gas flowed directly from the supply bottles through the LIME detector and was retrieved in the recovery bottles. After this, the recirculation system was employed. The recovery process begins at the LIME chamber's output, where a pump raises the gas pressure to 1-2 bar above atmospheric pressure within a buffer volume. The pressurized gas is then directed through a flux meter to a purification system, which contains commercial filters.

Once purified, the gas is mixed with a small percentage of fresh gas O(20%) from the supply bottles and recirculated back into the LIME detector. As fresh gas is introduced, the pressure in the buffer volume gradually increases, and a pressure booster periodically pumps this gas into the recovery bottles.

Additionally, the gas system is equipped with sensors to monitor critical parameters such as pressure, temperature, humidity, and oxygen levels inside the detector. The purification system is set to be upgraded with low-radioactivity 5 Å molecular sieves, currently under development by collaborators at the University of Sheffield (UK), which will help reduce radon contamination during future phases of the CYGNO experiment.

The same gas system will serve the CYGNO-04 demonstrator.

At the end of 2021, I took part in the underground installation of the gas system and in verifying its functionality.

### 4.3.2 Data acquisition system (DAQ)

The Data Acquisition System (DAQ) is responsible for the collection and processing of the acquired data. For the LIME prototype, the signal of the PMTs are sent to a digitizer and to a discriminator and a logic module to produce a trigger signal based on a coincidence of at least two of four PMTs with signal above 2 mV.
LIME uses a sCMOS ORCA-Fusion camera, which operates in an external trigger mode, meaning the exposure of the sensor is initiated via software to avoid capturing images without tracks. The camera employs a rolling shutter mechanism, where each line of the sensor is exposed sequentially with a time delay to ensure uniform exposure time across all rows. Operating in its lowest noise mode, the camera takes 80 $\mu$s to activate each row, from top to bottom, resulting in a total activation time of 184 ms. Afterward, each row remains open for a predetermined exposure period, set to 300 ms in LIME, before being read out. This exposure time was chosen to minimize event pile-up while maximizing track acquisition, making the sensor open for a total of 484 ms. During this period, the image is saved only if there is a trigger from the PMTs; otherwise, it is discarded. Once an image is read or discarded, an external software trigger restarts the row exposure cycle.
The data acquisition of all the sensors was developed and integrated in the MIDAS software. The acquisition is divided into runs, pedestal runs consist of 100 images and standard runs contain 400 images. The runtime of a standard run varies depending on the shielding configuration, ranging from 5 to 10 minutes as the shielding increases.

### 4.3.3 Slow control

The LIME detector's slow control system is designed for the setup, monitoring, and management of various hardware and environmental parameters that are not time-critical, operating at a lower priority. The system is built on the MIDAS framework, an open-source C++ software capable of interfacing with a wide range of sensors and electronic boards. This versatile software controls the high-voltage supplies for the PMTs, cathode, and GEMs, and records critical data such as the currents drawn by the GEMs. If any instability is detected, recovery procedures are automatically triggered to maintain stable operation.
In addition to controlling the high-voltage supplies, the slow control system continuously monitors environmental conditions such as temperature, pressure, and humidity both inside the detector and in the ground floor room where it is housed. Examples of humidity (Fig 4.18) and pressure (Fig. 4.17) monitor are shown. In particular, the system is able to follow the environmental pressure and adjust the internal pressure to maintain an overpressure between 4 and 6 mbar. This system helps ensure optimal operating conditions and can detect potential gas leaks or other issues with the gas system. The gas system itself, including gas flow rates within the chamber, recirculation flow and gas levels in the supply bottles, is also constantly monitored. In case of any malfunction within the gas system, the slow control manages error messages and alarms to prevent further complications.
In particular, I developed the Gas System page, illustrated in Fig. 4.16. This interface displays an alarm notification in the event of a malfunction in the gas system. The second section of the page provides controls to turn the gas system on and off, as well as to set the total gas and recirculation flow rates. The pressure levels of each gas are visually represented, with colors indicating the gas level status: orange for levels below 50% and red for levels below 15%. Additionally, a box displays the readings from various sensors for easy monitoring. This page allows for real-time monitoring and adjustments to the gas system status during data-taking periods.

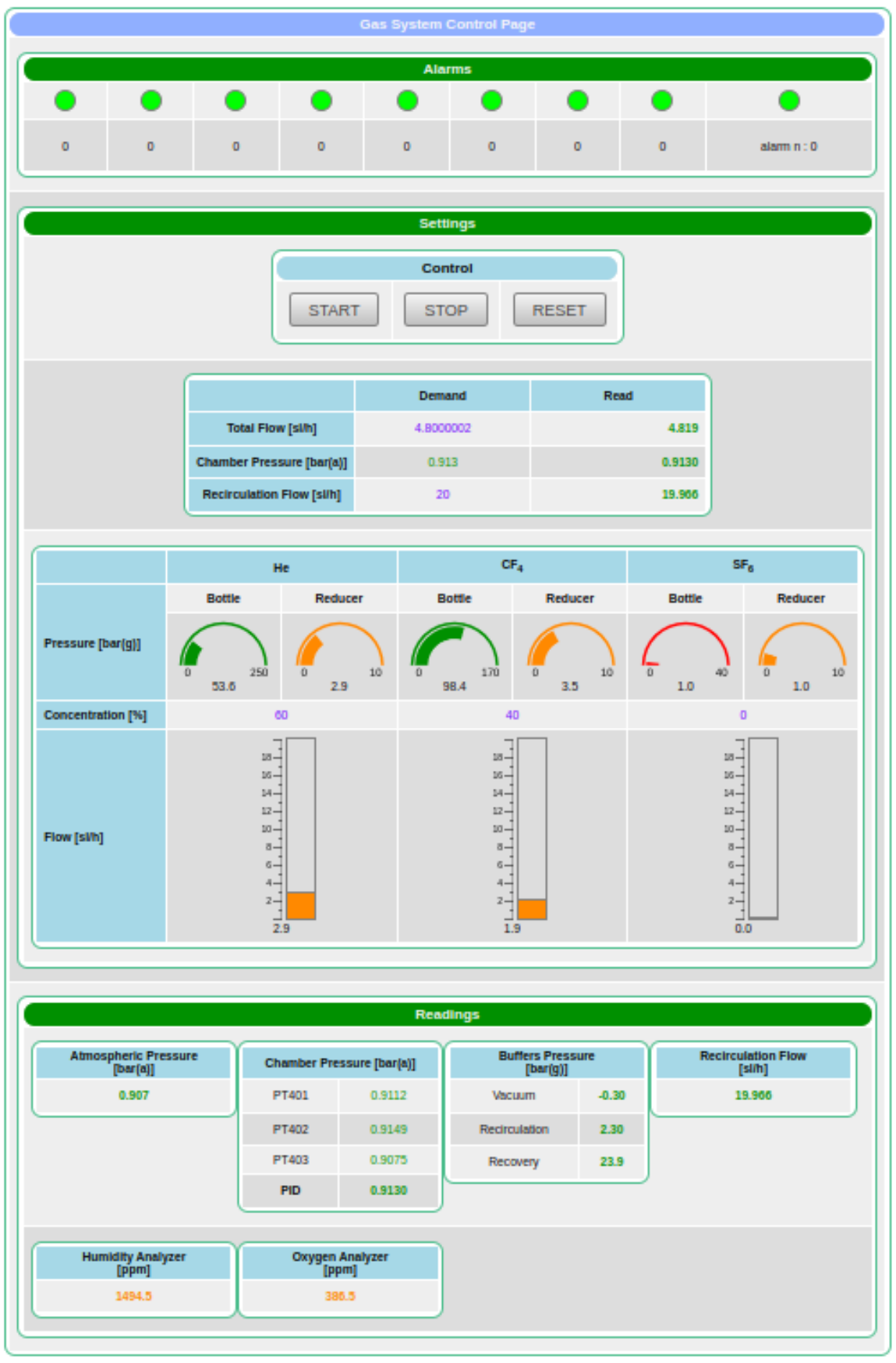

**Figure 4.16** A picture of the Gas System page.

Using a custom system developed by the collaboration, acquired runs are automatically uploaded to the cloud storage for processing and reconstruction. Grafana [106], an open-source platform, enables real-time monitoring, visualization and analysis of these runs. The status of the reconstructed runs is displayed (Fig. 4.19), reporting the last acquired run, the last reconstructed run and the last running run. Additionally, Grafana allows the monitoring of variables such as pressure and humidity as seen in Fig. 4.20. Another important feature of the slow control system is its capacity to monitor the PMT trigger rate, providing an ongoing check of the data collection process. It also performs an initial raw analysis of PMT waveforms and sCMOS images during data acquisition. This raw analysis helps monitor the quality of the data and ensures correct gas purity and electronic stability. Indicators such as the average number of clusters detected in images and the total photon count are used to assess system performance. Specifically, for each calibration run, the light integral distribution is fitted, and the most probable value is shown (Fig. 4.20). Thus, the slow control system plays a crucial role in ensuring the smooth operation of LIME.

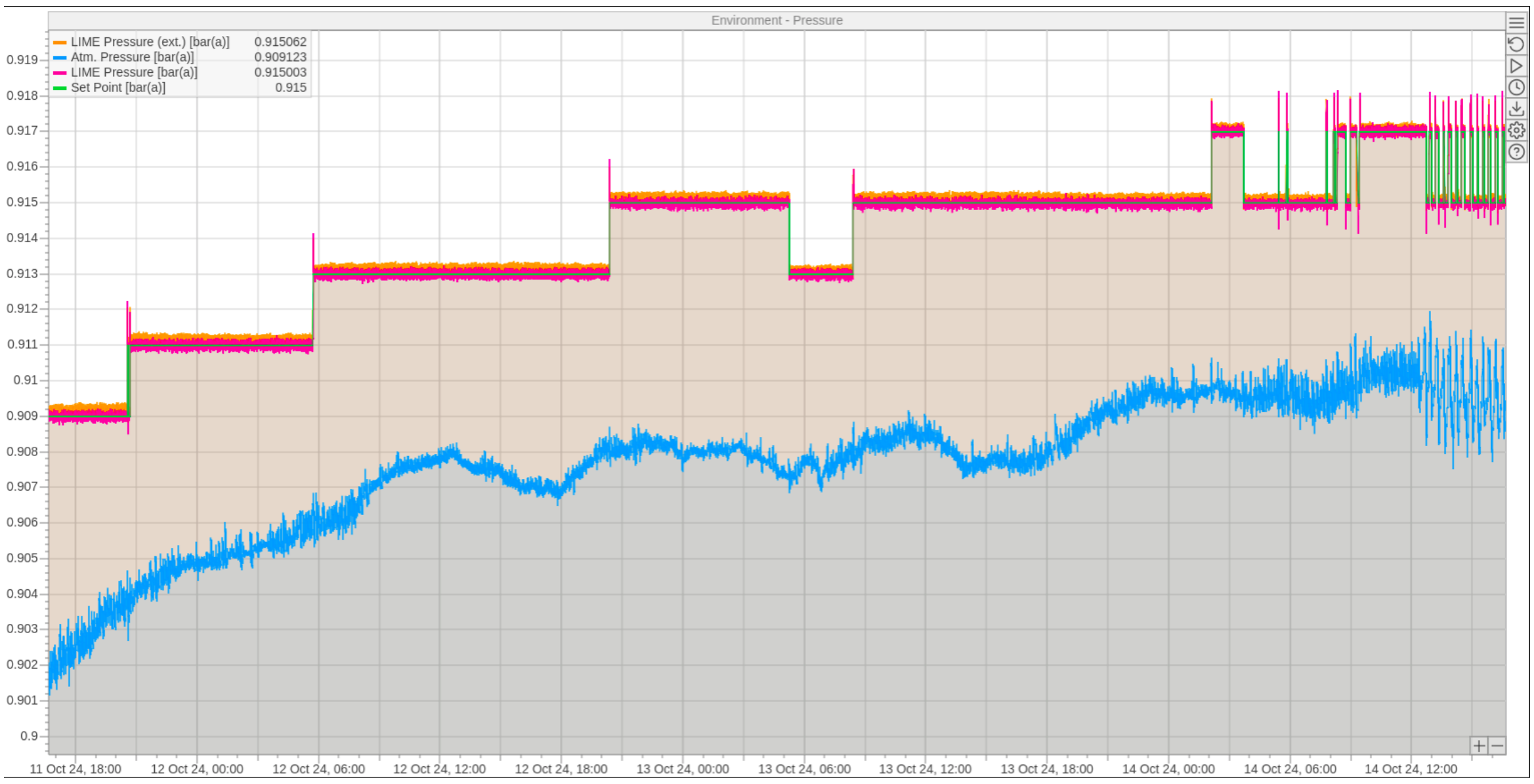

**Figure 4.17** Three-day pressure history from the MIDAS page, showing external pressure in blue, internal set pressure in green, and measured internal pressure in red.

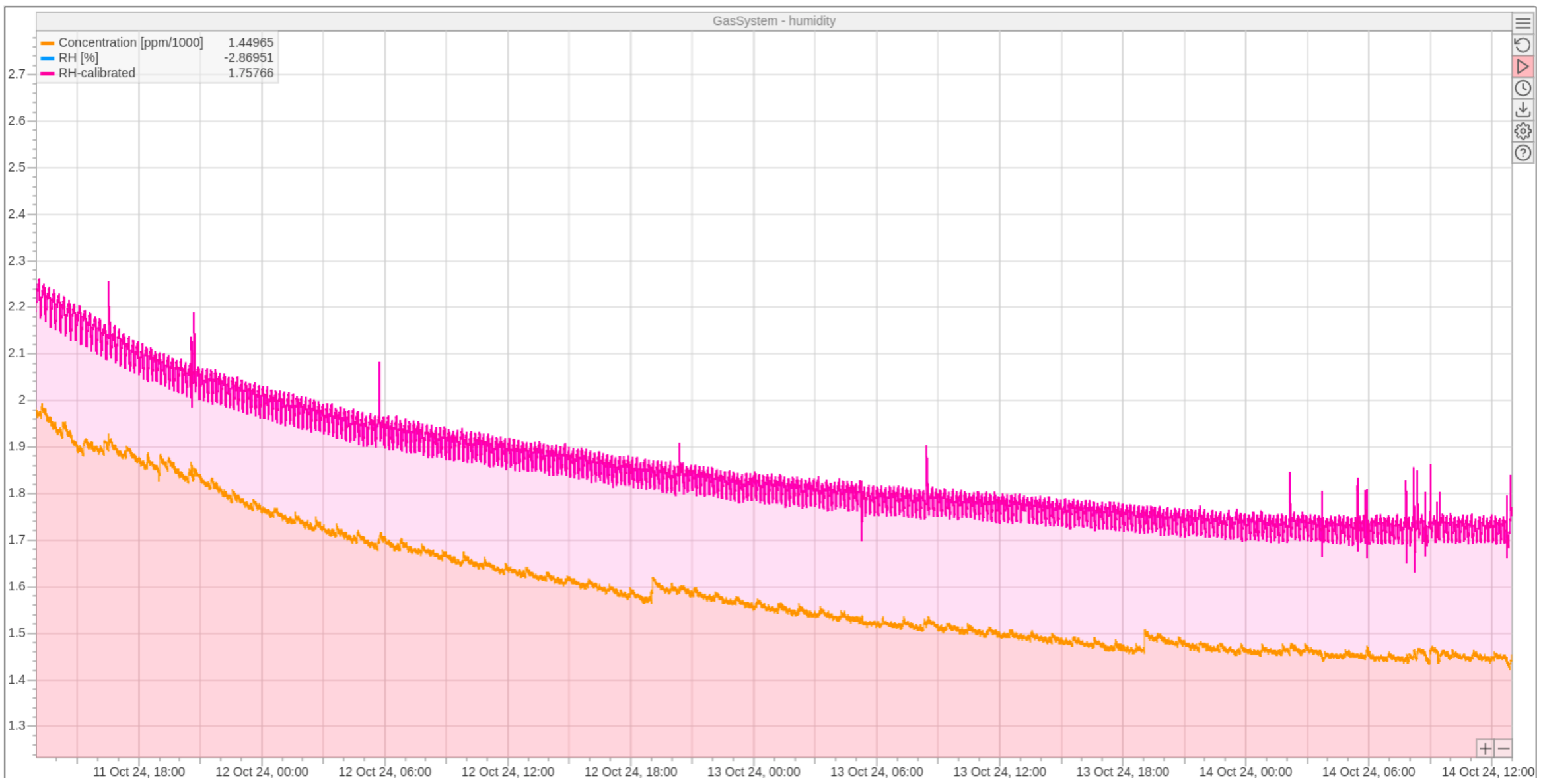

**Figure 4.18** Display of three-day humidity level history captured from the MIDAS page, in magenta the calibrated relative humidity (RH[%]).

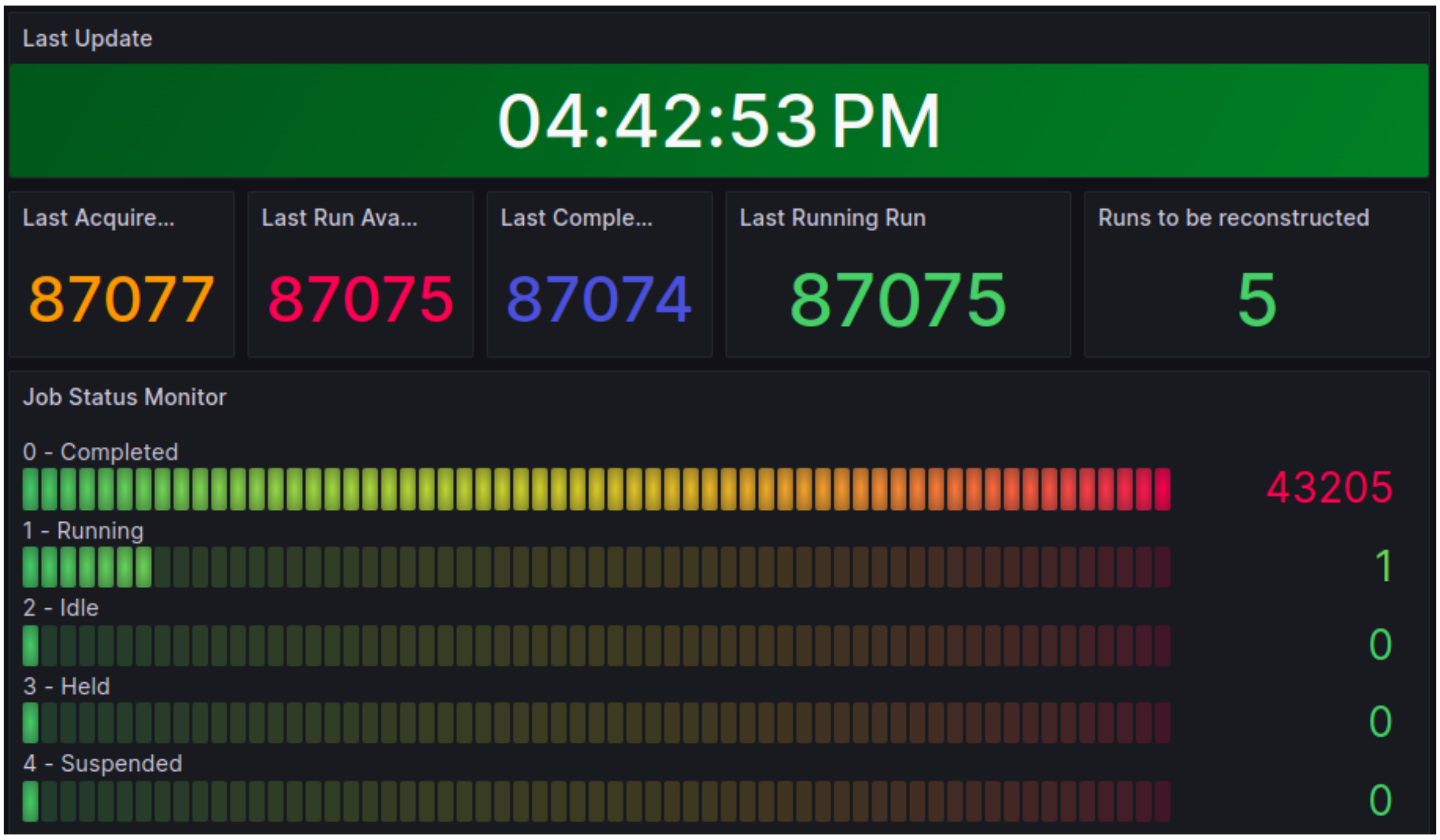

**Figure 4.19** The status of the running and reconstructed runs.

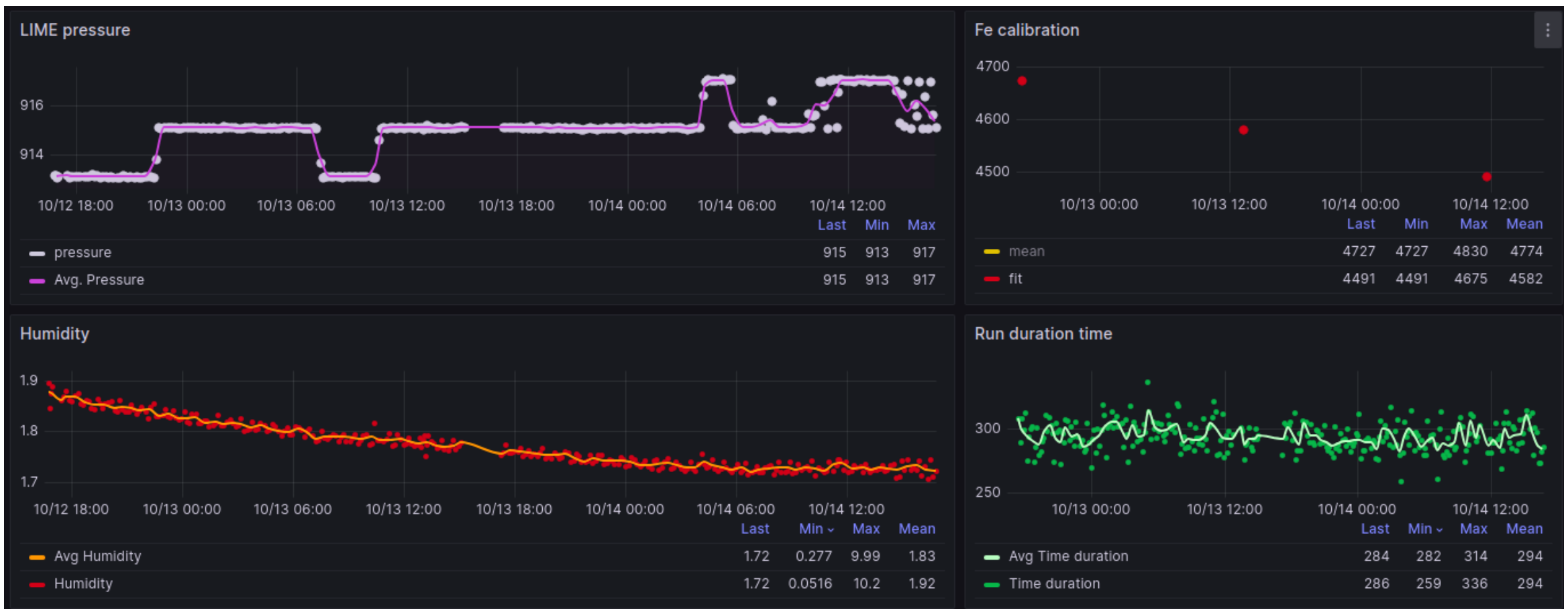

**Figure 4.20** Monitoring of the variables of interest.

## 4.4 Tracks Reconstruction

The CYGNO experiment employs an innovative method for reconstructing particle tracks by using optical detection of light generated during electron avalanches in the GEM amplification stage. The setup utilizes PMTs, which are sensitive to the development of particle tracks along the drift direction ($z$-axis), and sCMOS cameras that capture the tracks' projection on the $x$-$y$ plane. While the analysis of PMT waveforms is still being refined, a specific algorithm has been developed for reconstructing particle tracks from the sCMOS camera data.
To analyze both real and simulated images, the CYGNO collaboration developed a Python-based clustering algorithm based on DBSCAN (Density-Based Spatial Clustering of Applications with Noise) [107], which is designed to discover clusters of arbitrary shape.
Building on this, the collaboration introduced an enhanced version of the algorithm called iDBSCAN (intensity-based DBSCAN), [108] which incorporates pixel intensity as a weight when identifying clusters. Each pixel's intensity is proportional to the number of collected photons, with a baseline (called the pedestal) that represents the intensity corresponding to zero photons. This pedestal varies between pixels and is subtracted during the analysis to ensure accurate reconstruction.
Particles interacting with the active gas volume of the detector produce ionization patterns with different characteristics depending on their energy and type. Electrons in the tens of keV range leave curved tracks that become denser near their stopping points, while nuclear recoils result in short, straight, and dense tracks due to their high energy release over a small area. The reconstruction algorithm must be versatile enough to handle these diverse patterns accurately. Moreover, data collected above ground often include many overlapping tracks caused by cosmic rays and environmental radioactivity, requiring the algorithm to not only identify individual tracks but also separate them effectively.

#### Noise reduction and pedestal subtraction

Each picture captured by the sCMOS camera is a matrix of 2304 $\times$ 2304 integers representing the raw ADC counts for each pixel. These images contain intrinsic noise, which must be corrected before data analysis. To accomplish this, pedestal runs are performed at low GEM voltage, ensuring that no amplification occurs and no real events are visible, leaving only sensor noise in the image.
For each pixel in the image, the average noise level $\mu_{ij}$ and standard deviation $\sigma_{ij}$ are calculated. Where $i$ detonates the row and $j$ denotes the column. These values are stored in a pedestal map. During data analysis, the noise correction is applied by subtracting $\mu_{ij}$ from each pixel's intensity $I_{ij}$, resulting in a corrected value $I_{sub,ij}$.

### Image pre-processing

The first step in pre-processing is zero suppression, where all pixels with an intensity lower than $n_{sigma} \times \sigma_{ij}$ are set to zero, the standard deviation $\sigma_{ij}$ has been evaluated on the pedestal run. This threshold is chosen to be sightly above the noise level, and the value of $n_{sigma}$ is a free parameter that can be adjusted for each data-taking campaign depending on the camera's noise conditions.
After the noise suppression, the image is scaled grouping 4$\times$4 pixels into a single macro-pixel and assigning an intensity value corresponding to the average intensity of the 16 individual pixels occupying that area of the sensor. To further suppress noise, the intensity of each macro-pixel is replaced by the median intensity of its neighboring macro-pixels.

### Vignetting correction

The rebinned matrix of pixel is further corrected to compensate for the vignetting effect, which is a natural geometrical reduction of the light intensity of an object imaged by a lens. The light acceptance is maximal at the center of the lens and decreases as $1/R^4$, where $R$ is the distance from the focal plane's center. This reduction occurs due to the inclination of the lens with respect to the emitted light cone, effectively reducing the solid angle covered by the lens.
To correct for this distortion, a vignetting correction map is applied. This map (Fig. 4.21) is generated from uniformly illuminated images of a white surface, the camera was also rotated to avoid any preferential direction of the light impinging on the sensor. Each pixel's light intensity is normalized to the value of the central pixel. A drop of the collected light as a function of the radial distance from the center, down to 20%, with respect the center of the image is shown (Fig 4.22).
The correction is applied after noise reduction to avoid amplifying the noise contribution. The final, corrected image is then ready for analysis using the clustering algorithm.

### Clustering Algorithm: iDBSCAN

After pre-processing, the full-resolution image is restored and analyzed. All pixels with non-zero intensity are passed to the iDBSCAN clustering algorithm, a modified version of the classical DBSCAN method, designed to handle particle track data from the CYGNO experiment. This algorithm uses two key parameters: the radius $\epsilon$ and the minimum number of pixels $N_{min}$ required to form a cluster. Starting from a pixel, if the number of neighboring pixels within a radius $\epsilon$ exceeds $N_{min}$, the algorithm activates these pixels, initiating the formation of a cluster.

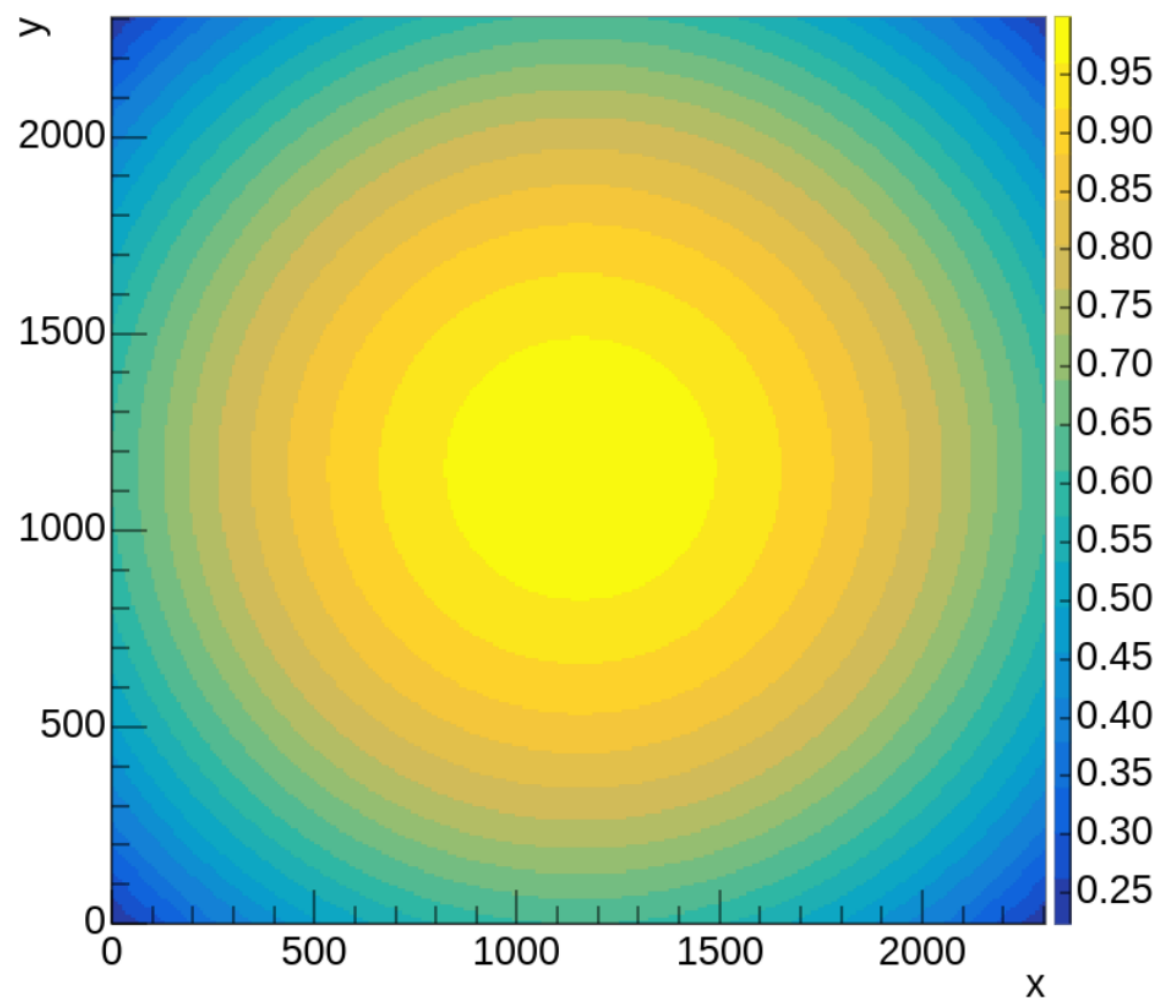

**Figure 4.21** The vignetting map generated by averaging multiple images captured in front of a uniformly illuminated smooth surface. To eliminate any bias in the light's incident direction on the sensor, the camera was rotated. The pixel intensity of each reconstructed cluster is then divided by the corresponding value in the map based on its $x$-$y$ position, compensating for any light loss.

In the iDBSCAN implementation, the sum of pixel intensities $I_{ij}$ is used instead of just counting the number of pixels, making it more sensitive to actual signal intensities. The parameters $\epsilon$ and $N_{min}$ are carefully optimized to minimize noise clustering while maximizing the efficiency of reconstructing real events. A smaller radius $\epsilon$ can isolate false clusters, while a larger value may capture too much noise. Similarly, a low $N_{min}$ accepts low-intensity noise as genuine events, whereas a high $N_{min}$ could miss real events. Therefore, these parameters are adjusted at the start of each data-taking session based on noise levels and track density in the images. Typical values for $\epsilon$ range from 1 to 10, and for $N_{min}$, from 4 to 30.

The iDBSCAN clustering algorithm works iteratively, accommodating different energy deposits. The first iteration aims to detect clusters with dense, high-intensity distributions, typically corresponding to high-energy particles like nuclear recoils or alpha particles. Once identified, these pixels are removed, and the process is repeated with more relaxed clustering criteria to capture softer electron tracks or lower-intensity segments. A third iteration further relaxes the parameters to target faint recoil track fragments. This multi-step process helps to separate overlapping tracks and recognize a broad range of particle interactions.

Once the clusters (or seeds) are identified, they undergo a process called superclustering to merge related clusters into a single track. While increasing the search radius significantly could merge noise into the tracks, it would also degrade energy resolution and fail to differentiate closely spaced tracks. To avoid this, the superclustering method starts with seeds found by iDBSCAN as a basis. Initially, Geodesic Active Contour (GAC) [109] algorithms were employed, which trace track paths by calculating intensity gradients across pixels and identifying cluster boundaries by detecting gradient variations. This approach works well for images with sharp edges but struggles with low-energy, blurred tracks.
To address this limitation, the Chan-Vese [110] algorithm was introduced. Unlike GAC, Chan-Vese identifies track boundaries by distinguishing between average noise levels and pixel intensity values, making it particularly effective for handling blurred or low-intensity features. This algorithm, based on binary minimization techniques, allowed for improved reconstruction of both electron recoil (ER) and nuclear recoil (NR) tracks, contributing to more accurate particle identification and event classification.
A picture of the clustering procedure is shown in Fig. 4.23.

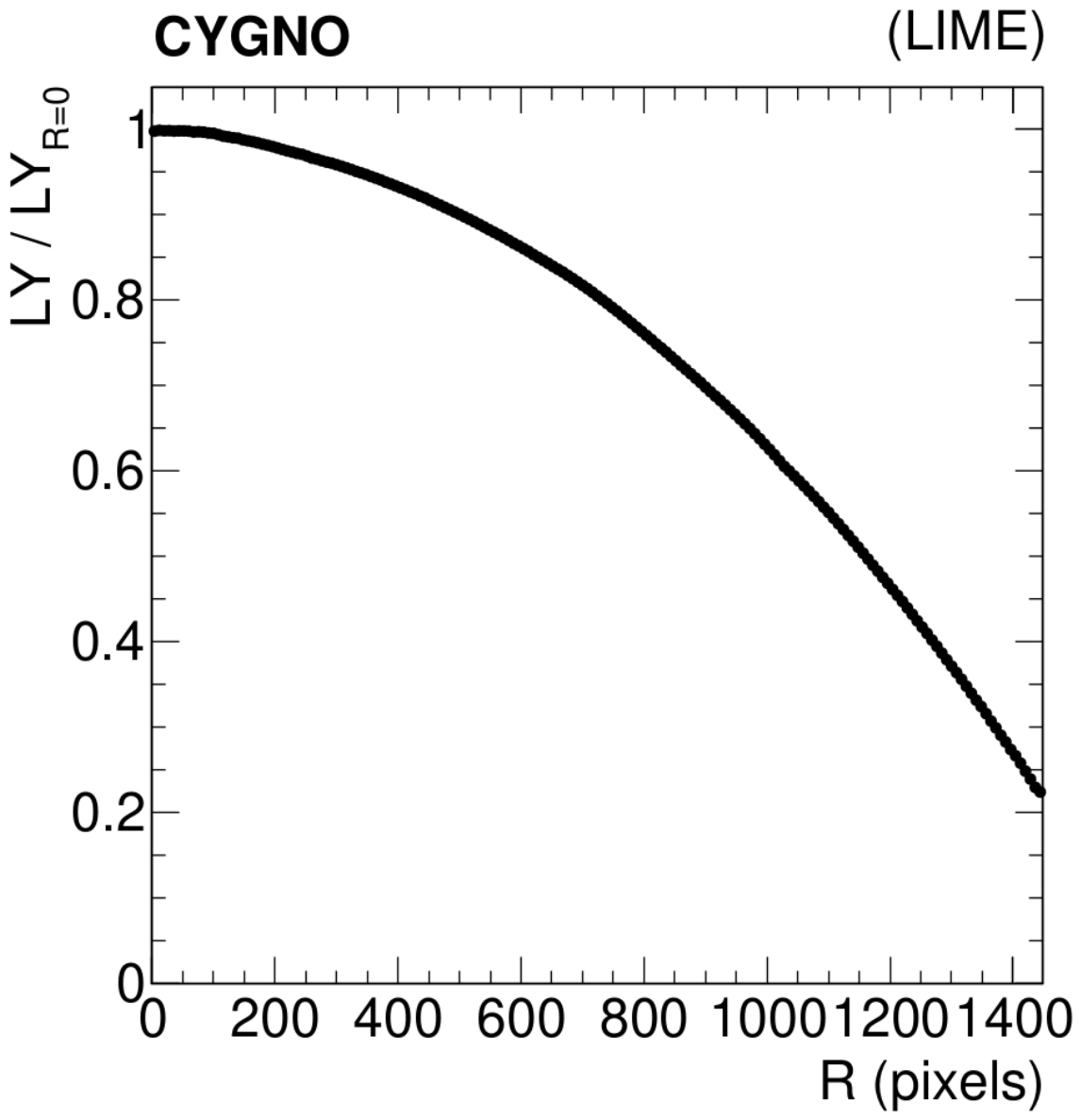

**Figure 4.22** Pixel count as a function of the radial distance from the center of the sensor, normalized to the one at the center, using pictures of a uniformly illuminated white surface [101].

### Directional iDBSCAN (iDDBSCAN)

The CYGNO experiment employs an advanced method for reconstructing particle tracks, utilizing the iDDBSCAN (intensity-Directional DBSCAN) [111] algorithm to improve the accuracy of clustering and track identification. This approach addresses the challenges posed by complex particle tracks, particularly those generated by events like electron recoils (ERs) and cosmic rays, which can result in long or overlapping tracks within the detector. The primary objective of the iDDBSCAN algorithm is to ensure that these long tracks, often split into separate clusters by simpler algorithms, are accurately reconstructed and merged where necessary.

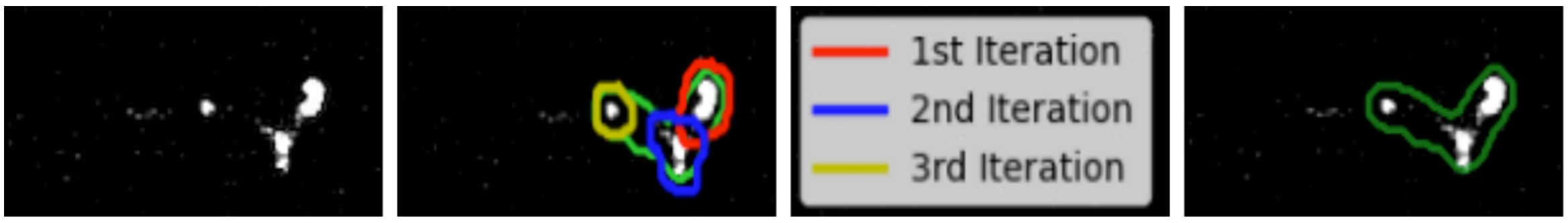

**Figure 4.23** The flow reconstruction algorithm applied on a simulated ER track. The left plot displays the original track after pedestal subtraction and and zero noise suppression. The middle plot show the clusters identified by the iDBSCAN algorithm during different iteration represented by different colors. The right plot presents the supercluster found by GAC algorithm seeded by the iDBSCAN clusters.

Once the iDBSCAN algorithm has been applied to the data and clusters have been identified, a second phase is introduced: the directional iDBSCAN (iDDBSCAN). This algorithm applies a linear RANSAC [112] procedure multiple times to the clusters identified by iDBSCAN, aiming to detect and merge long, straight, or low-density clusters that might represent cosmic rays or high-energy electrons. In cases where no directional clusters are found, iDDBSCAN retains the original clusters identified by iDBSCAN. However, when directional clusters are detected, the algorithm performs a more sophisticated search, fitting clusters to polynomial curves of varying complexity—ranging from straight lines to third-order polynomials.

The fitting process is iterative. When a good polynomial fit is found, pixels within a defined width ($\omega$) around the fit curve, as well as those within a radius $\epsilon_{\text{dir}}$ around a pixel, are included in the cluster. This iterative approach continues until no further pixels can be added to the cluster. This process is particularly effective at disentangling overlapping tracks, a common challenge in high-exposure or high-event-rate scenarios (e.g., when the event rate exceeds 10 Hz). Once all possible superclusters have been identified, the remaining pixels are analyzed in the next step of the algorithm.

The next phase of iDDBSCAN involves Small Cluster Identification, which focuses on detecting residual, small deposits of energy that may have been missed in the earlier stages. These deposits are often indicative of low-energy particle interactions or the terminal points of longer tracks. The clustering radius is kept intentionally small to ensure the accurate identification of these smaller clusters.

Additionally, iDDBSCAN incorporates a feature known as Isolation Seeding, which prevents the formation of new clusters near existing cosmic tracks, helping to reduce false positives and maintain energy resolution.

The success of iDDBSCAN is measured by its ability to improve background rejection, a crucial task for rare event searches like those in the CYGNO experiment. By more accurately reconstructing tracks from natural background radiation (e.g., cosmic rays), the algorithm enhances the experiment's signal detection efficiency without compromising the accuracy of energy estimation. This is vital for identifying and distinguishing between background noise and potential signals of interest, which is key in the search for new physical phenomena. Consequently, all Monte Carlo simulations and detector data in the CYGNO experiment are reconstructed using the iDDBSCAN algorithm to ensure the highest possible precision in track reconstruction and event identification.
The processing time for the DBSCAN algorithm, as well as the total computation time, was evaluated for each image. Fig. 4.24 displays the two distribution, based on data collected over a 30 days run.
After clustering and applying corrections, various features of the reconstructed tracks are calculated, including track length, width and total light intensity, which is proportional to particle energy. These features are stored in an a nTuple structure, implemented through ROOT [113] as a TTree, allowing detailed analysis of particle behavior within the detector. The extracted track properties are essential for differentiating between signal and background events, which is crucial in rare event searches.

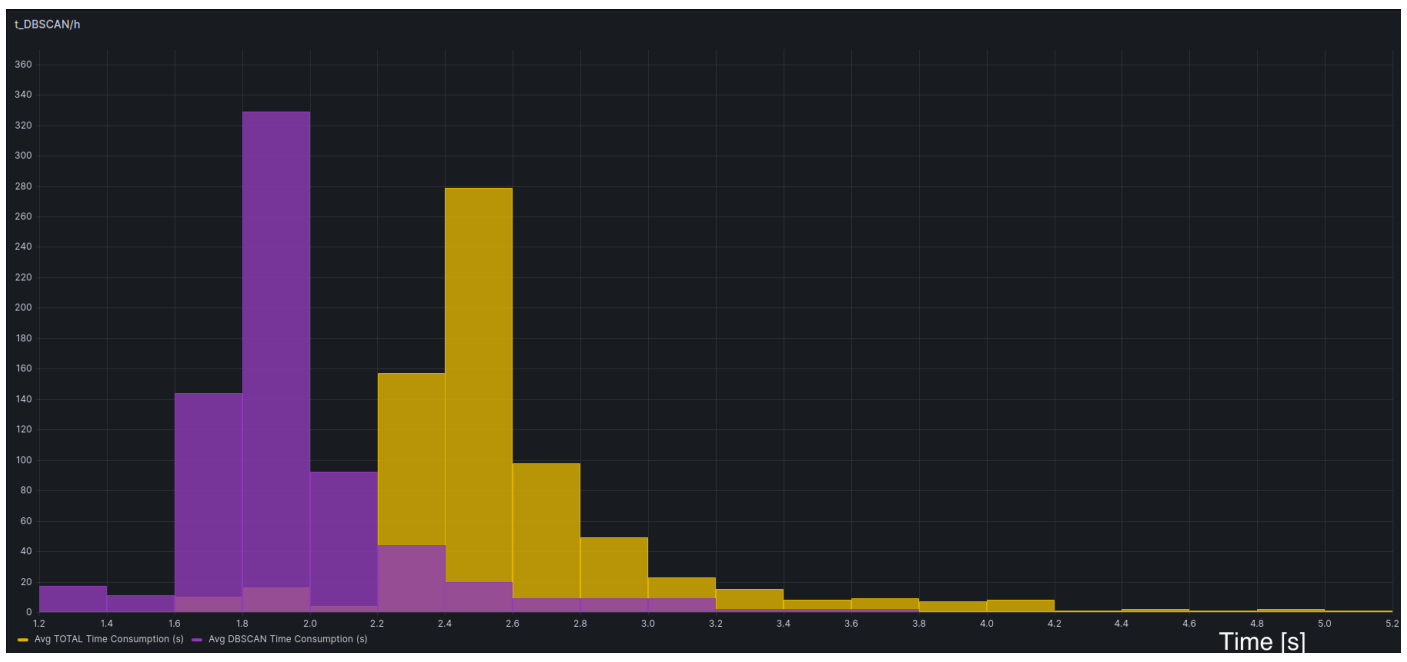

**Figure 4.24** Processing time distribution of the DBSCAN algorithm (in magenta) and the total time (in yellow) for 30 days of acquired data.

## 4.5 LIME background

A thorough characterization and understanding of radioactive background are essential for experiments involving rare event searches. A comprehensive Monte Carlo (MC) simulation of the expected background in LIME was performed using the GEANT4 [85] package. The simulation included the CAD engineering model of the prototype and various shielding configurations. External background radiation, originating from neutrons and gamma rays, was simulated based on flux spectra measured at LNGS by other experiments. While the intrinsic radioactivity components used in the LIME construction, such as the cathode, copper field rings, resistors, GEMs acrylic vessel, the camera lens and the body, was measured using HPGE detectors, facilitated by the special services division at LNGS, and incorporated into the internal radioactivity simulation. The expected spectrum of the internal background for different components is shown in Fig. 4.25.

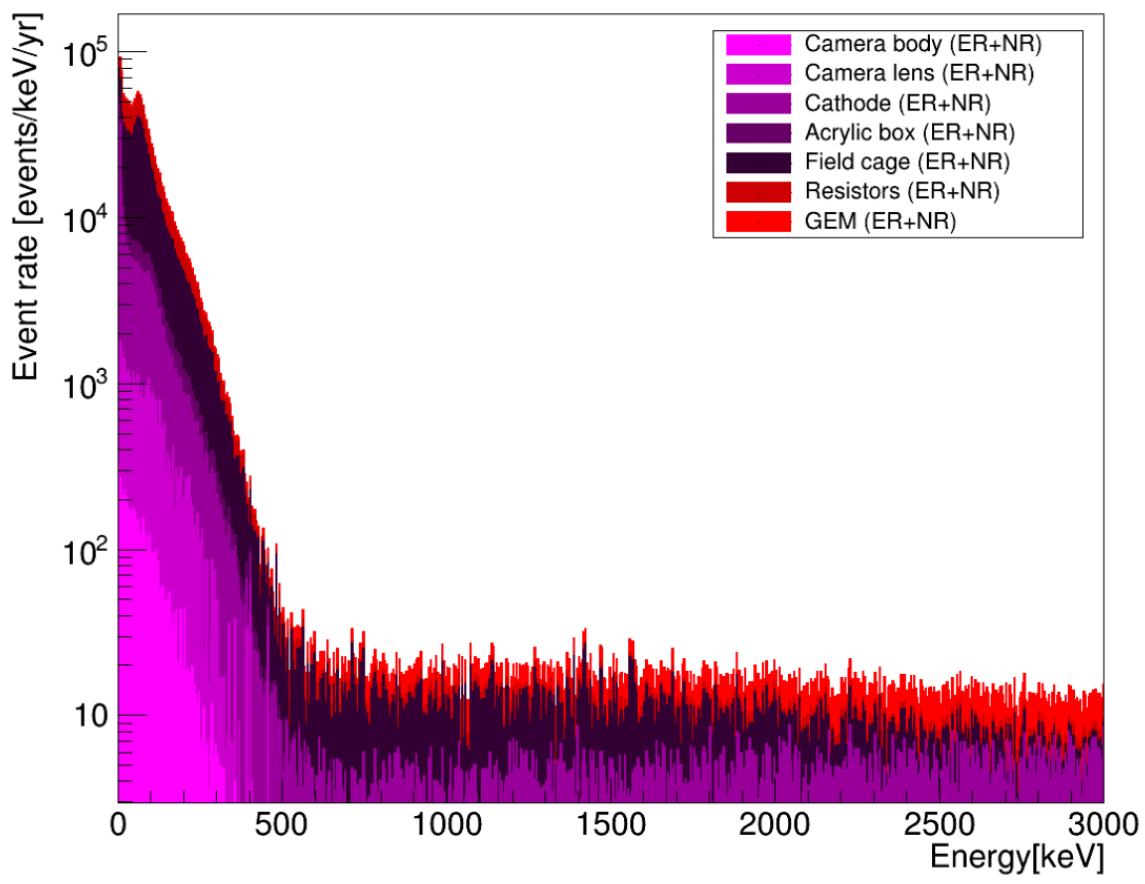

**Figure 4.25** Energy deposition spectra in LIME obtained from Monte Carlo simulations of the primary radioactive components of the detector.

Radioactivity was detected from isotopes in the $^{238}$U, $^{235}$U and $^{232}$Th decay chains, as well as $^{40}$K, with significant $^{234}$Th contamination observed in various components like the resistors, cathode, field rings, camera body, lens and GEMs. Together, these sources contributed up to 20 Bq/kg. Elevated levels of $^{234}$Pa, $^{226}$Ra and $^{210}$Pb were also observed in the resistors, and activation isotopes such as $^{58}$Co were detected in the field rings and cathode. Additionally, high contamination of $^{40}$K was found in the camera sensor, lens, GEMs and resistors. It is important to note that LIME was constructed using standard materials not optimized for low intrinsic radioactivity.

In the unshielded configuration, where only a Faraday cage was used, external background dominated, leading to an estimated $10^9$ events per year above 1 keV, corresponding to a rate of approximately 30 Hz. A cost-benefit analysis was conducted to optimize the shielding setup, aiming to maximize the suppression of external background while minimizing costs, space, and the additional radioactive contribution introduced by the shielding materials. The most cost-effective solution involved a 10 cm thick copper shield encased in a 40 cm thick water tank. Copper blocks all beta emissions and significantly reduces gamma rays, while the water, rich in hydrogen, decelerates and captures neutrons.

This shielding configuration is expected to reduce the number of neutron recoils (NRs) above 1 keV to less than 2 per year, while electron recoils (ERs) are reduced to about $10^5$ per year in the 1 to 20 keV range. These background levels are low enough for LIME's purposes, given that internal radioactivity is predicted to produce about $1.5 \times 10^6$ events per year in the sensitive detector volume, making it the primary source of background in the fully shielded setup.
In future GYGNO detectors the radioactive materials such as the copper and the PMMA could be replaced by more radiopure ones, while the GEMs, the lens and the camera body are more difficult to substitute. For the reduction of the GEM contribution, some studied showed that a cleaning procedure in a deionized water bath can significantly reduce radioactive contamination, in particular coming from $^{40}$K.
Dedicated measurements have been done within the CYGNO collaboration in order to identify which specific elements of the camera setup produce the larger radioactivity contributions. It was found that the protective glass of the sCMOS sensor is the most radioactive part of the camera body, while the lens resulted to be highly contaminated with $^{40}$K. The camera-induced background reduction strategy is still under study.
The expected background is almost unaffected by the increasing copper shielding while it becomes completely dominated by the internal contribution once the water shielding is added. The ER background is completely dominated by the external gamma contribution in the unshielding configuration, which is then progressively suppressed below the internal one with the increasing shielding thickness. With the full 10 cm thick copper shielding the internal background becomes dominating with respect to the external one, especially near the borders of the sensitive volume.

## 4.6 LIME underground program

The goal of operating the LIME prototype at the Laboratori Nazionali del Gran Sasso (LNGS) is to test its performance in an underground environment, to study the background, to validate the MC simulations and to perform a measurement of the neutron flux. The data collection strategy was refined based on MC simulation results and is implemented in a staged approach with progressively increasing shielding thickness to study different background sources. These stages are outlined below:

#### Run0: Commissioning phase

The first phase involves commissioning LIME after its underground installation, marking the first evaluation of its performance in a low background environment, typical of a Dark Matter experiment. Unlike overground operations where cosmic rays dominate, this phase aims to test LIME's response under minimal background interference.

### Run1: No shielding

It is dedicated to collecting data without any added shielding. LIME was housed in the same aluminum Faraday cage used during overground tests at the LNF. During this phase, the dominant background is from the natural gamma radiation in the laboratory. Run1 concluded in the autumn of 2022.

### Run2: 4 cm copper shielding

In this phase, a 4 cm thick copper shielding was installed around LIME on all sides, reducing external gamma radiation by a factor of about 50. This significantly decreased the event rate and track occupancy in the images. While internal backgrounds became more prominent, they did not surpass the contribution from external sources. Run2 was completed in the winter of 2023.

### Run3: 10 cm copper shielding

For Run3, additional copper bars were installed, providing a total of 10 cm of shielding. This configuration suppressed external backgrounds to the level of internal detector material radioactivity. More than one million background images were collected from May to November 2023, with detector settings optimized for maximum efficiency. Following this phase, a measurement of the environmental neutron flux is planned to characterize neutron-induced nuclear recoil backgrounds, crucial for future experiments, including CYGNO. Additionally, during Run3, a neutron measurement using an AmBe source was performed, providing a sample of nuclear recoils (NRs) to optimize signal identification and background rejection algorithms

### Run4: 40 cm water shielding

Run4 involves the installation of water tanks around LIME, creating a 40 cm layer of hydrogen-rich shielding. This water shielding moderates environmental neutrons, minimizing nuclear recoil backgrounds from external sources. The installation took place in December 2023, with data collection continuing until April 2024.

### Run5

The same shielding configuration as in Run3 has been installed. To assess the effect of gain, and consequently the saturation effect, the $V_{GEM}$ was set to 440 V, 20 V lower than in the previous Runs. An other goal is to perform a measurement of the environmental neutron flux. Data taking began in May 2024 and is still ongoing as of December 2024.

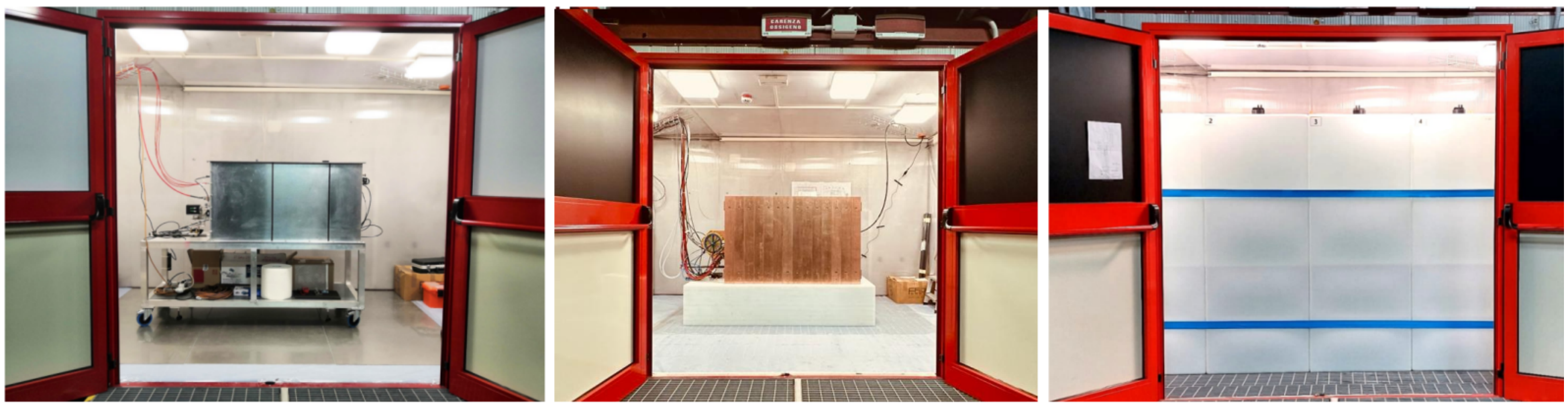

**Figure 4.26** Pictures of the LIME prototype with different shielding configuration. On the left without shielding, in the middle with 10 cm of copper and on the right with an additional 40 cm of water.

In Fig. 4.26, a picture of LIME with the different shielding configuration is shown. The total number of expected events during the different phases of data collection, from both external and internal sources, is summarized in Table 4.2. External sources include gamma rays and neutrons from outside the detector, while internal sources refer to the radioactivity of the main LIME internal components as well as the shielding material itself. As anticipated, the neutron recoil background remains largely unaffected by the increasing thickness of copper shielding. However, with the addition of water shielding, the internal sources become the dominant contributors to the NR background. For the electronic recoil (ER) background, external gamma radiation dominates in the unshielded configuration. As shielding is progressively added, external gamma radiation is increasingly suppressed, falling below the internal background levels. Once the full 10 cm copper shielding is in place, the internal background becomes the dominant factor over external sources.

| Phase | External | | Internal | | Total | |
|---|---|---|---|---|---|---|
| | ER[$10^6$yr$^{-1}$] | NR[yr$^{-1}$] | ER[$10^6$yr$^{-1}$] | NR[yr$^{-1}$] | ER[$10^6$yr$^{-1}$] | NR[yr$^{-1}$] |
| Run1 | $1140 \pm 35$ | $1480 \pm 90$ | $7.34 \pm 0.01$ | $79000 \pm 470$ | $1140 \pm 35$ | $80480 \pm 480$ |
| Run2 | $26.6 \pm 0.6$ | $870 \pm 10$ | $7.87 \pm 0.3$ | $79000 \pm 470$ | $34.5 \pm 0.7$ | $79870 \pm 470$ |
| Run3 | $1.49 \pm 0.04$ | $930 \pm 25$ | $7.88 \pm 0.3$ | $79000 \pm 470$ | $9.37 \pm 0.3$ | $79930 \pm 470$ |
| Run4 | $0.5 \pm 0.2$ | $2.0 \pm 0.2$ | $7.88 \pm 0.3$ | $79000 \pm 470$ | $8.38 \pm 0.4$ | $79000 \pm 470$ |

**Table 4.2** Expected events rates in LIME across different data-taking phases, distinguishing between ERs and NRs events. External components account for environmental gamma rays and neutrons, while internal components refer to the radioactivity of LIME's internal parts as well as the radioactivity from the shielding materials.

LIME has been successfully operated underground for two years, with promising results. The challenges encountered during these months of data collection significantly expanded the understanding of CYGNO detector operations and yielded encouraging results to move towards the future of the experiment.

### 4.6.1 Data/MC comparison

For each phase, the energy spectra from experimental data and Monte Carlo (MC) simulations were compared [114], focusing on the energy range up to 100 keV, which is relevant in the region of interest for WIMP-nuclei scattering searches. Fig 4.27a shows the comparison between data and MC for Run1. Around 8 keV a peak is visible both in the data and the MC energy spectra, which originates from interactions in the gas caused by photons from the copper X-ray fluorescence ($K_\alpha$ = 8.05 keV). This fluorescence occurs in materials close to the gas, such as the field cage, cathode, and GEMs. Above 10 keV, the data and MC spectra are consistent within $1\sigma$ up to 100 eV, although the MC simulation predicts a slightly lower event rate. For Run2 (Fig. 4.27b) the data and MC spectra show a similar shape, while for Run3 (Fig. 4.27c) the MC simulation underestimates the measured spectrum, especially at lower energy, with the discrepancy decreasing as the energy increases.
The discrepancy between data and MC with increasing copper shielding thickness can be attributed to internal background components that were unforeseen and not simulated. These components become relatively more significant as the external background is reduced.

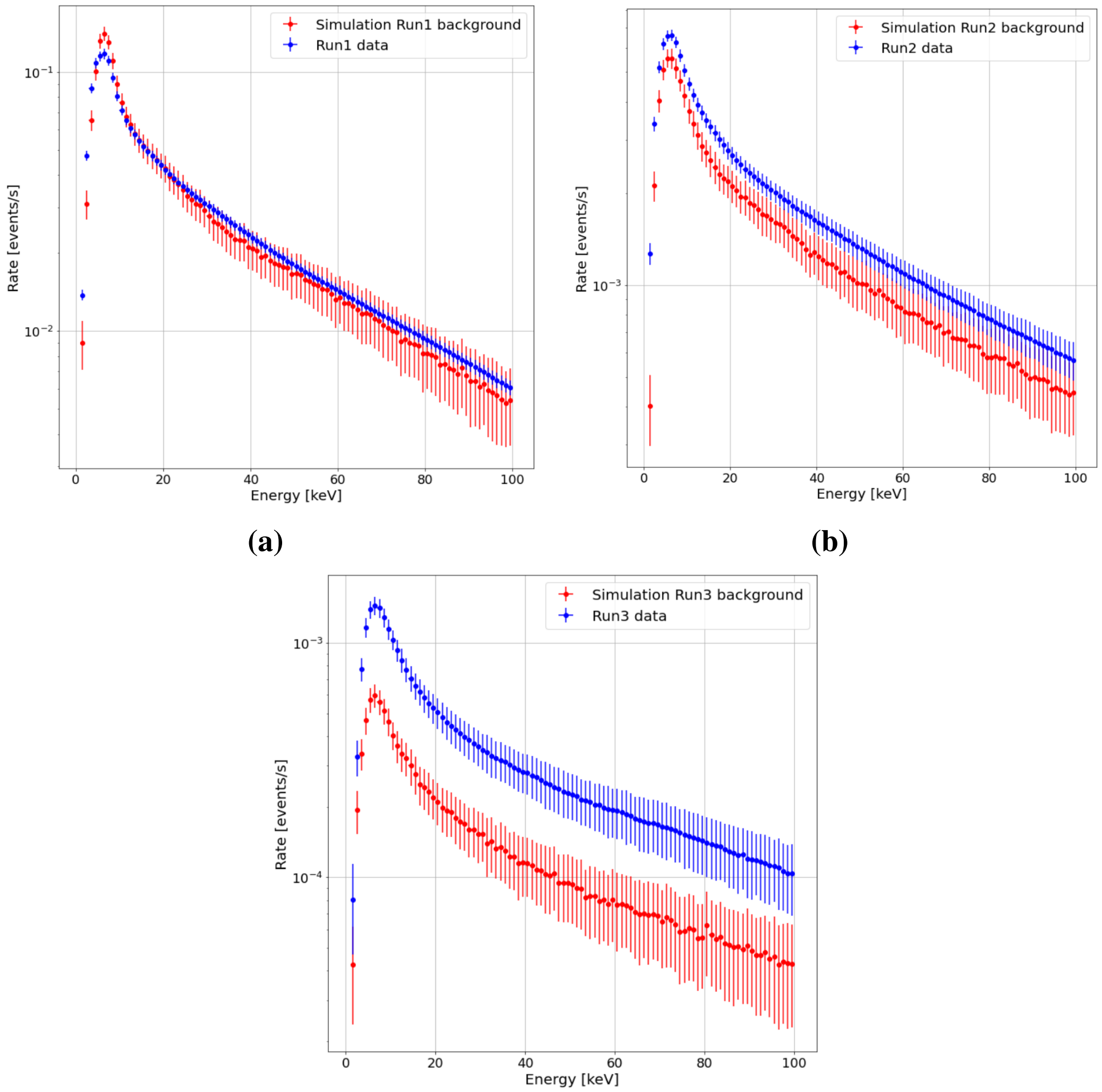

(a) (b)

(c)

**Figure 4.27** Comparison of the measured (blue) and simulated (red) energy spectra for Run1 (a), Run2 (b) and Run3 (c) LIME configurations between 0 keV and 100 keV.

# 5 Light yield stability

Variations in environmental conditions, such as pressure and humidity, can significantly influence the detector's gain, causing fluctuations in the light yield over time. In this chapter, the performance of the LIME detector is thoroughly examined in relation to these variations, with a focus on analyzing its behavior as a function of both pressure and humidity changes.

## 5.1 Datasets

During the data-taking campaign, conducted both overground at the Laboratori Nazionali di Frascati (LNF) and underground at the Laboratori Nazionali del Gran Sasso (LNGS), two distinct types of datasets were collected: background runs and calibration runs.
Background runs were aimed at measuring the natural or environmental background noise present during the experiment, while calibration runs were used to adjust and fine-tune the detector's response, involving a known radiation source.

### Background dataset

The background runs record images without external radiation sources, allowing for a measure of the natural background levels in LIME. Through the data taking campaign, with the different shielding configurations, multiple background runs were collected. After optimizing gas flow to maximize the LY, a series of so-called golden runs were collected, where detector operation and response were stable. These runs are selected for the data analysis and the comparison with simulated samples.

### Calibration dataset

Calibration runs are conducted periodically to monitor the light yield with a $^{55}$Fe X-ray source placed on top of the upper side of the acrylic vessel of the detector, which allows X-rays to penetrate a thin window and enter the sensitive gas volume. There, they interact primarily through the photoelectric effect, producing detectable 5.9 keV energy deposits (ERs). Initially, the source's position along the drift direction could only be adjusted manually, this applied during commissioning overground and at the beginning of the underground phase. Starting from Run2, a pulley system was introduced, enabling the source to be moved to various distances from the GEMs for calibration at different points along drift path through the rotation of a wheel. Starting from Run3, a motorized system was implemented to allow remote control of the source's position. This upgrade enable precise adjustments to the source's location along the drift direction without manual intervention. The motor system allows the repositioning of the $^{55}$Fe source to multiple predetermined positions along the drift length, ensuring calibration coverage throughout the detector's active volume.
The $^{55}$Fe source is housed in a lead container, placed 18 cm from the sensitive gas volume, which blocks X-rays except those directed through a 1 cm opening towards the gas. The geometry of the container ensures that ER events are spread along $z$ between 4 cm ( in the upper part of the volume, closer to the source) and 11 cm (in the bottom part) from the central position of the source along the drift volume. Due to the X-ray absorption in the gas, these events are mainly concentrated in the upper part of the detector, closer to the source. Therefore, this configuration results in a conical distribution of the source's photons, which aligns with observed spreads in both the $x$ and $y$ distributions, as shown in Fig. 5.1. For Run1, the source was fixed at a single position, 25 cm from the GEM plane, at the center of the detector. Starting from Run2, daily calibration runs were conducted with the source at five specific positions on top of the LIME drift region. These positions, reported in Tab. 5.1, spanning from 4.75  cm to 45.75 cm from the GEMs, were chosen to ensure full drift coverage. The ER events resulting from the $^{55}$Fe events are identified using image reconstruction code already described in Section 4.4, revealing the ER signals as bright and round spots. While the sub-millimeter topology of these events is obscured by diffusion and the camera's granularity, the total light detected provides a reliable measure for calibrating LY in terms of counts per keV. This calibration is not only essential for accurate energy scaling of background data but also allows for continuous monitoring of LY stability throughout the data-taking period.

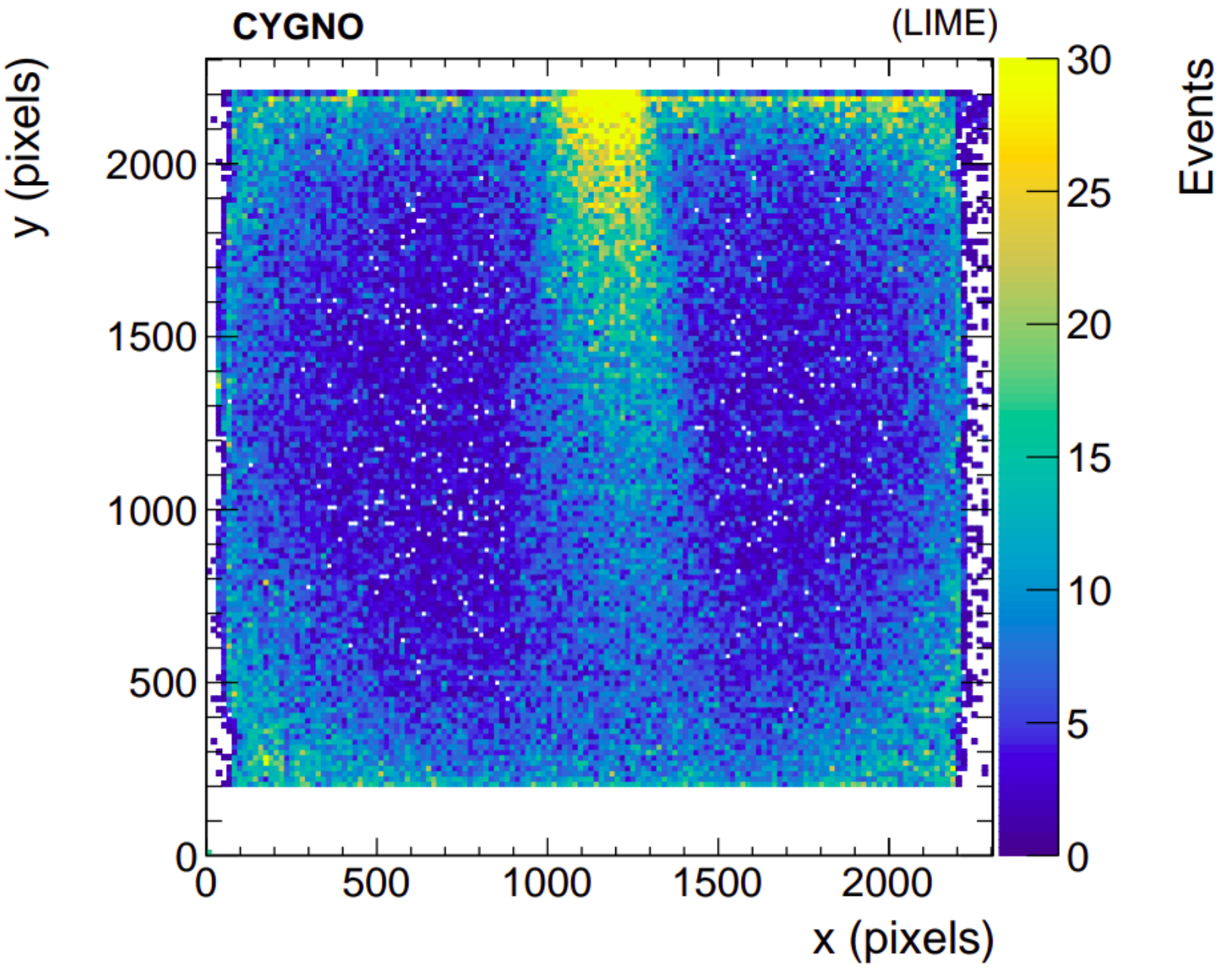

**Figure 5.1** Spatial distribution of the reconstructed clusters in data collected with $^{55}$Fe source placed outside the detector at high values of $y$ [101].

| Step | Between rings | Distance [cm] |
|---|---|---|
| 1 | $3^{rd}$ - $4^{th}$ | 4.75 |
| 2 | $10^{th}$ - $11^{th}$ | 14.75 |
| 3 | $17^{th}$ - $18^{th}$ | 24.75 |
| 4 | $24^{th}$ - $25^{th}$ | 34.75 |
| 5 | $32^{rd}$ - $33^{th}$ | 45.75 |

**Table 5.1** The five positions used for the calibration were defined in terms of specific rings of the field cage and their corresponding distances from the GEMs.

## 5.2 Light yield monitoring

The light yield (LY) can fluctuate over time due to variations in operational parameters such as GEM voltages, total gas flux and humidity levels. Additionally, LY depends on the $z$-position along the drift direction, as the primary ionization electrons in the gas may be re-absorbed based on their distance from the GEMs, the applied drift field, and any contaminants present in the gas-mixture. This re-absorption can reduce the number of primary electrons that reach the amplification stage, effectively lowering the LY.

The diffusion of electrons also impacts the LY. A significant spread of the primary electron cloud over long drift distances can result in counts at the edges of the reconstructed cluster falling below the threshold used for zero suppression, meaning they are not included in the energy estimation. Additionally, events occurring closer to the GEMs can experience gain saturation, which also reduces the effective LY. These factors make the LY sensitive to both the $z$-position of the event and the gas mixture's purity.

Understanding the relationship between LY and the position along the drift direction is critical for characterizing the detector's response. During Run3, humidity and gas system issues were noted, particularly due to a malfunction in the pump responsible for gas recirculation.

Consequently, several adjustments were made to address these issues, which may have altered the gas conditions and introduced impurities into the mixture.

## 5.3 Data analysis

The images acquired by the LIME prototype are analyzed using the reconstruction algorithm called iDBSCAN, described in Section 4.4, which identifies and isolates clusters of pixels that are associated with events in the gas. The algorithm relies on external parameters that can be adjusted to enhance reconstruction efficiency, and these parameters are optimized for each data taking phase. The reconstruction code outputs a set of identified tracks and calculates various variables based on the pixels corresponding to each track, including the light integral intensity, the track dimensions and other shape-related characteristics. The relevant variables used in the analysis are:

- **sc_integral**: The total light intensity across all pixels within a reconstructed cluster. Since the light emission correlates with the energy deposited in the gas, this variable is used to measure the energy of the event;
- **sc_nhits**: The number of pixels associated with the track that exhibit non-zero intensity after applying zero suppression. This variable is proportional to the spatial extent of the original energy deposition, which is smeared by diffusion during the drift and amplification stages;
- **sc_xmin, sc_xmax, sc_ymin, sc_ymax**: These variables represent the pixel coordinates marking the extreme boundaries of each cluster in the $x$, $y$ coordinates. Each cluster is enclosed within a rectangle defined by these coordinates;
- **sc_length, sc_width**: The length of the major and minor axis of each cluster. The axes are determined through principal component analysis (PCA) applied to the pixels associated with the track projected onto the $x$-$y$ plane;
- **sc_tgausssigma**: The sigma of the Gaussian transverse profile, obtained by fitting the intensity distribution of the pixels along the minor axis of the track. This variable measures how much the energy deposition spreads relative to the original path of the event in the gas;
- **sc_rms**: The standard deviation of pixel intensity values within a cluster, providing a measure of how evenly the intensity is distributed. Lower values indicate more uniform intensity across the track.

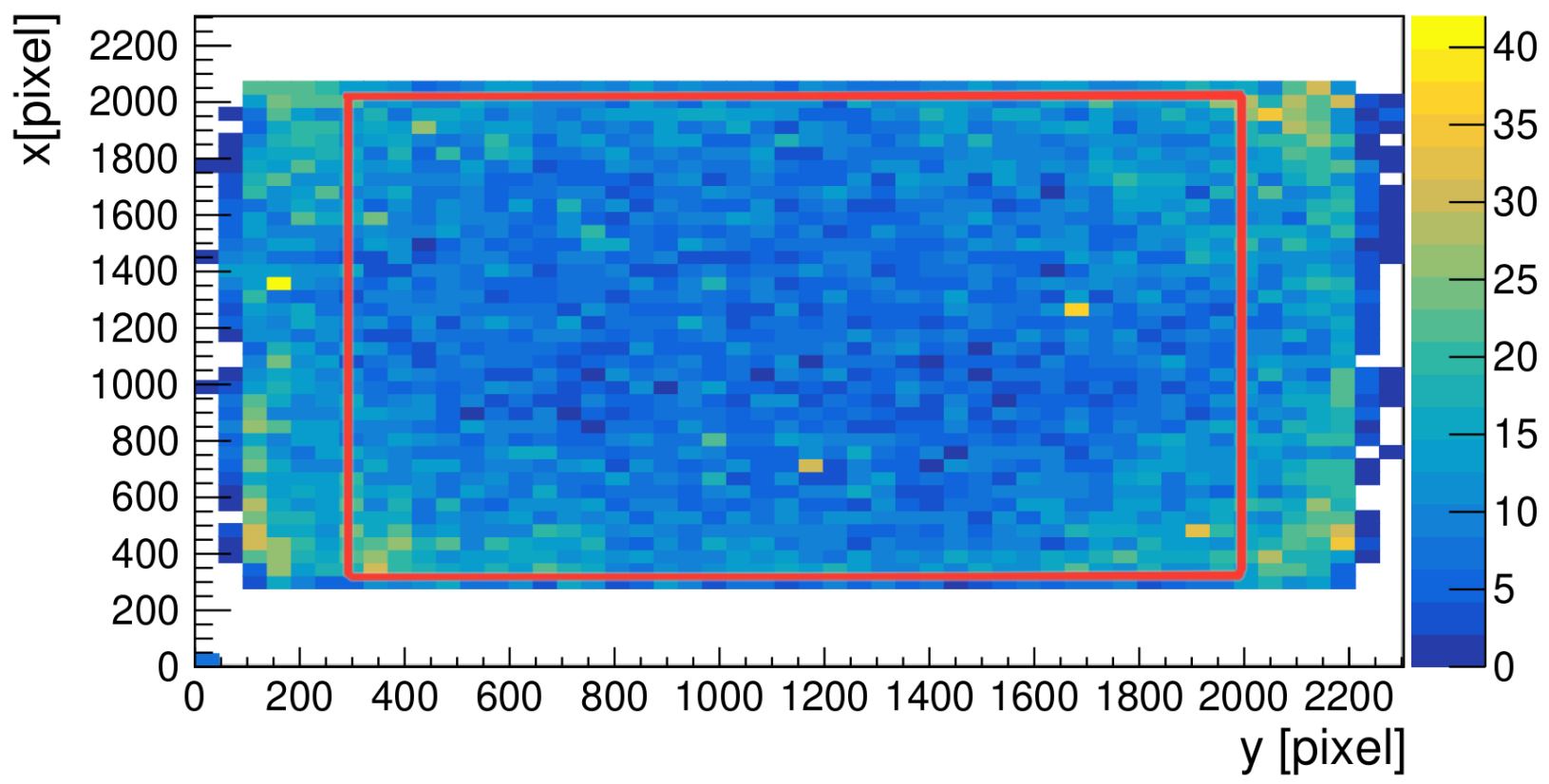

**Figure 5.2** The occupancy of the reconstructed cluster, the cluster taken in consideration for the analysis are inside the red rectangular.

These variables play a critical role in characterizing tracks and help in determining the energy deposition and spatial features of events detected by LIME.

### 5.3.1 Fiducial and quality cuts

Several selection criteria were applied to remove the sCMOS sensor noise, mis-reconstructed tracks, border noise and undesired event type such as alpha-like tracks, minimum ionizing particle (MIP)-like events, and others from the integral distribution.
A fiducial cut in the $x$ and $y$ coordinates is applied because camera noise is concentrate near the image borders, especially along the top and bottom edges. Additionally, the light yield in the outer image regions is reduced by lens vignetting. Therefore, the outer borders of the image are excluded by applying the following cuts, isolating only tracks in the central square where the detector's response is more uniform (Fig. 5.2):

$$\text{sc_xmin[pixel]} > 300 \quad (5.1)$$

$$\text{sc_ymin[pixel]} > 300 \quad (5.2)$$

$$\text{sc_xmax[pixel]} < 2000 \quad (5.3)$$

$$\text{sc_ymax[pixel]} < 2000 \quad (5.4)$$

As shown in Fig. 5.2, a cut at the top and bottom of the acquired image is applied directly within the reconstruction algorithm.
Despite these cuts, some noise may still persist in the image center, leading the reconstruction algorithm to falsely identify noise as real tracks. To address this, an analysis of pedestal runs was performed by the CYGNO collaboration to identify distinguishing variables from the reconstruction code's output. The primary source of fake events is the sCMOS sensor noise. Unlike real tracks, noise is not affected by diffusion in the gas, so false clusters tend to appear more uniform in their intensity.
Real events, on the other hand, undergo diffusion in the gas as they drift and additional smearing occurs during amplification in GEMs. This results in a track image with a bright central "core" and gradually decreasing intensity in surrounding pixels. Among all the reconstructed variables, sc_rms and sc_tgausssigma were identified as the most effective for distinguishing real tracks from noise.
To distinguish signal tracks from noise, a cut is applied to the root mean square (sc_rms) of the pixels counts within the superclusters. Noise-induced fake events are typically sharp, with pixel intensity dropping off rapidly from the center, resulting in a very small sigma. The following quality cuts were optimized and applied to all the events:

$$\text{sc_rms[counts]} > 6 \tag{5.5}$$

Tracks below this threshold are mainly due to noisy pixels in the sCMOS sensor.

$$\text{sc_tgausssigma[pixel]} \times 0.152[\text{mm/pixel}] > 0.5 \tag{5.6}$$

This ensures the tracks are wide enough to correspond to real events, as fake tracks tend to have very sharp, narrow profiles.

### Analysis of the $^{55}$Fe spot shape

The absorption of the 5.9 keV X-ray produces an electron recoil that travels a few hundreds of $\mu$m before being stopped in the CYGNO gas mixture. The diffusion of the primary charges while drifting for tens of cm in the CYGNO gas mixture results in a standard deviation of a few hundred $\mu$m, as described in Section 3.2.2. The amplification stage introduces additional diffusion of a similar magnitude, which is added in quadrature to the original diffusion. As a result, the $^{55}$Fe spot, as imaged by sCMOS camera, appears as a round spot with intensity following a two-dimensional Gaussian distribution when projected onto the amplification plane. The shape of this spot is dominated by the diffusion of the electron cloud. Therefore, to measure diffusion from the $^{55}$Fe data it is necessary to accurately determine the $x$-$y$ dimensions of the spot.

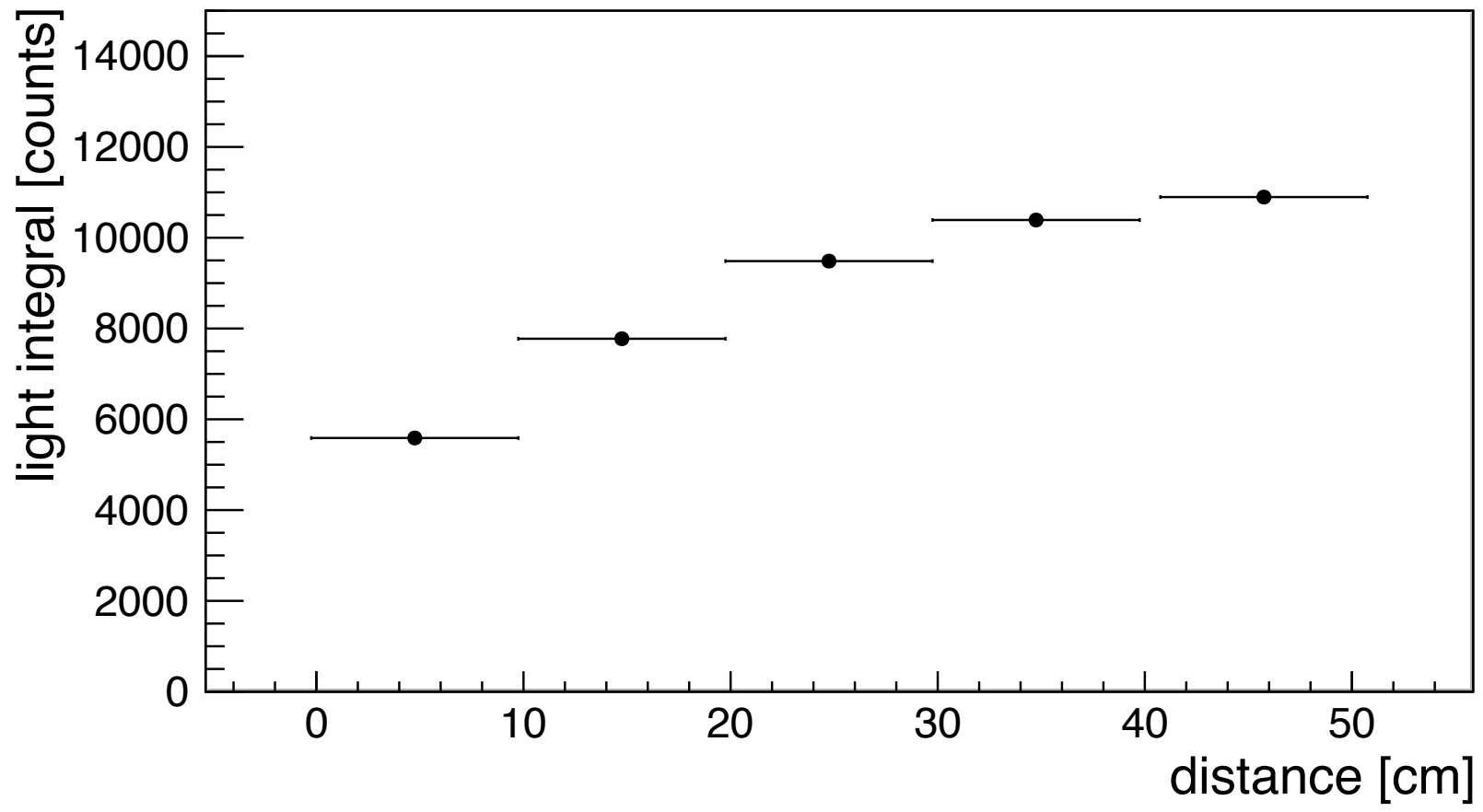

**Figure 5.3** Energy response as a function of drift distance.

## 5.4 Energy response as a function of distance

Energy linearity in a gaseous TPC is crucial for understanding how accurately the detector can measure particle interactions over a range of energies. Energy linearity ensures that the detector's response is proportional to the actual energy deposited in the gas, which is critical for precise particle identification and energy measurement.

To study the energy response as a function of drift distance, data from Run3, where $^{55}Fe$ source was positioned at five different distances from the GEMs are shown in Fig. 5.3. The observed trend shows that the light integral increases as the source moves further from the GEMs.

This can be explained as follows: during the amplification process, the GEM foil channels are filled with ion and electrons generated in avalanches. These particles, having small size, can rapidly evacuate. However, at high gain ($10^6$ – $10^7$), the charge accumulation from a single avalanche is sufficient to locally alter the electric field, leading to a saturation effect that reduces the effective gain of the GEM. As a result, the GEM system's response become sensitive to the charge density entering the GEM holes, particularly in cases with a large number of primary electrons from gas ionization, and to the area over which the electrons are dispersed. In the LIME detector, primary electrons diffuse over the 50 cm drift distance, increasing the surface area for multiplication and reducing the charge density, thereby minimizing gain reduction.

This behavior could explain why $^{55}Fe$ X-ray events at greater distances from the GEM exhibit larger light integrals compared to those closer to the GEM.

### Energy calibration

Energy calibration is carried out using periodic data collected with a $^{55}$Fe source, which serves as a standard candle. These data are used to estimate the light yield and subsequently calibrate the energy of the reconstructed tracks because variations in the light integral counts have been observed and attributed to operational fluctuations in the LIME detector, such as changes in pressure, humidity and gas flow.
All the calibration runs are analyzed with the reconstruction code to extract the properties of the $^{55}$Fe-induced tracks. An additional selection is applied to obtain a clean sample of $^{55}$Fe tracks. Since the original length of a 5.9 keV ER track is only a few hundred $\mu$m, after diffusion in the gas and amplification stages, these tracks appear as bright, almost perfectly round spots. A selection based on the slimness of the track, defined as:

$$s = \frac{\text{sc_width}}{\text{sc_length}} \tag{5.7}$$

is applied to retain only clusters with $s > 0.8$, which correspond to nearly round tracks.
The distribution of the light integral for $^{55}$Fe events passing this selection is fitted using a Cruijff function. This function incorporates separate parameters to describe the width of the distribution on both sides of the peaks, capturing any asymmetry present. The function is defined as:

$$f(x) \propto exp\left(-\frac{(x-\mu)^2}{2\sigma_{L,R}^2 + \alpha_{L,R}(x-\mu)}\right) \tag{5.8}$$

where $\mu$ represents the mean of the distribution, $\sigma_{L,R}$ are the widths on the left and right sides and $\alpha_{L,R}$ control the asymmetry in the tails on either side of the distribution.
The fitted mean $\mu_{Fe}$ is then used for energy calibration and the light yield is calculated as:

$$LY = \frac{\mu_{Fe}[counts]}{5.9[keV]} \tag{5.9}$$

## 5.5 LNF

During the summer and the autumn of 2021, the detector operated overground for an extended period inside an experimental hall with controlled environmental conditions.
The detector was continuously flushed with a He:$CF_4$ (60:40) gas mixture, supplied from cylinders of pure gases at a flow rate of 12 l/h. The exhaust gas was directed to an external environment via a water-filled bubbler, ensuring an overpressure of 3 mbar inside the system.
Data acquisition for the LIME detector was handled thought the MIDAS framework which,

allowing automated run collection without the need for human intervention. A series of pedestal and signals runs was acquired.
The signals from the four photomultiplier tubes (PMTs) were processed through a discriminator and logic module to generate a trigger, requiring a coincidence between at least two PMTs. Table 5.2 outlines the typical working conditions during this data-taking period.

| Parameter | Typical value |
|---|---|
| Drift Field | 0.9 kV/cm |
| GEM Voltage | 440 V |
| Transfer Field | 2.5 kV/cm |
| Gas Flow | 12 l/h |
| PMT Threshold | 15 mV |
| Exposure Time | 50 ms |

**Table 5.2** Summary of the operating condition for LIME during the data taking.

### 5.5.1 Long term stability of detector operation

Dark Matter searches usually require extended data-taking period, spanning months or even years. This impose the capability to monitor the stability of detector performance. The LIME prototype operated continuously for two weeks, with pedestal and $^{55}$Fe source runs being automatically recorded. The iron source was placed 25 cm from the GEMs. Thanks to air conditioning, the temperature remained stable with an average of (298.7 $\pm$ 0.9 K), while atmospheric pressure varied between 970 mbar to 1000 mbar.
For each signal run, the integral distribution is built applying the cut previously described and a fit with the Cruijff function is performed in order to estimate the average energy response. The light integral has been normalized to the value of the first run acquired. Since the temperature remained nearly constant, variations in light yield were studied as a function of pressure, as shown in Fig 5.4. Base on gas gain considerations described in Section 3.3.2, a fit using the following function:

$$y = a + \frac{b}{x} \tag{5.10}$$

is performed.
As reported in [101], a decrease in light yield of approximately 0.6$\%$ per millibar is observed, consistent with the expected reduction in gas gain as gas density increases.

During data acquisition, the electric current of the third GEM is continuously monitored. If the GEM voltage remains stable and the current exceeds 1 nA, the event is classified as a spike. The time interval between two consecutive spikes is recorded and as reported in

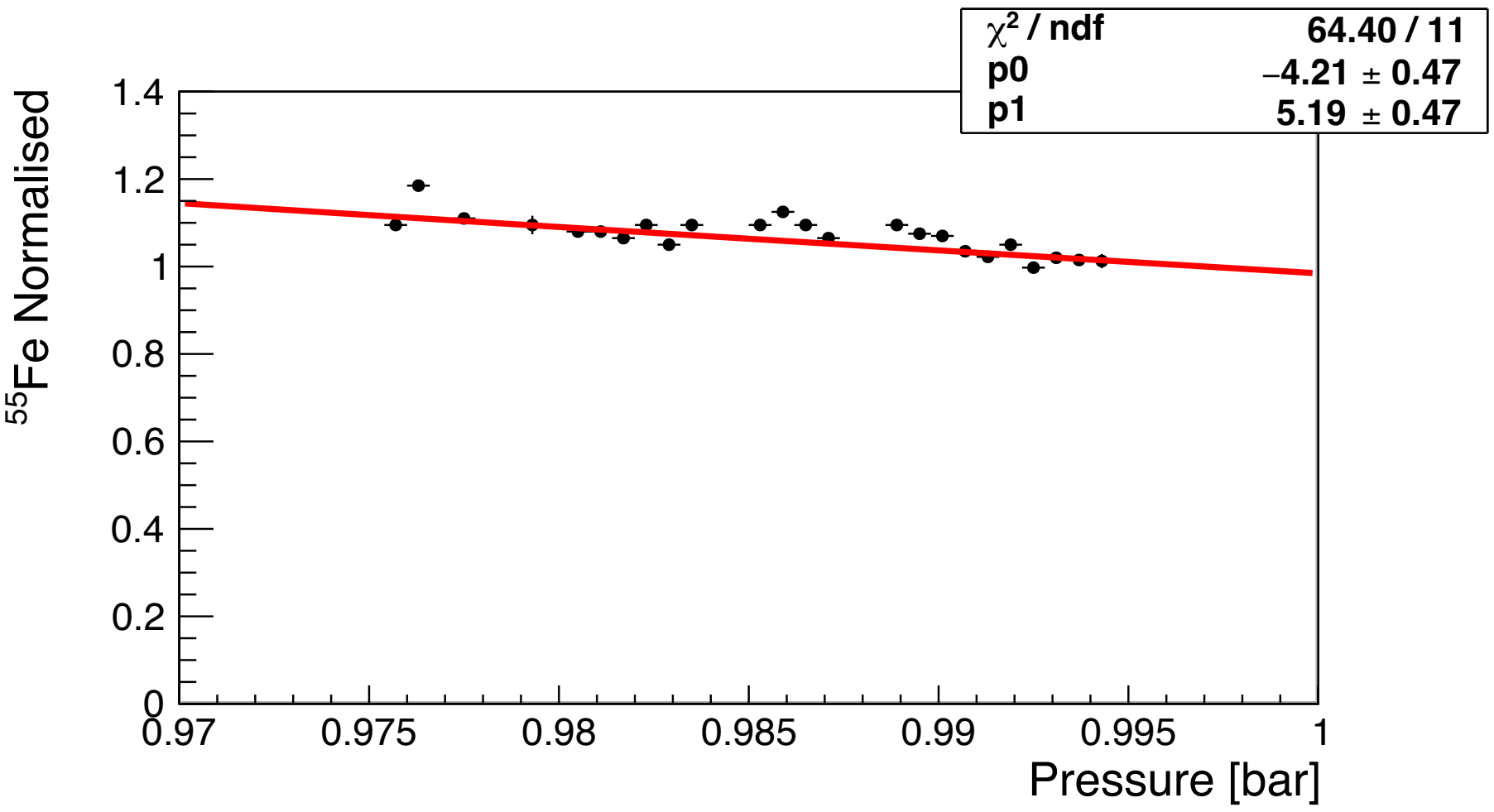

**Figure 5.4** Behavior of the $^{55}Fe$ peak normalized to the first acquired tun as function of the atmospheric pressure.

[115], less than 0.8 spikes per hour occur. This data helps monitor and maintain the stability and performance of the detector during operations.

## 5.6 LNGS

The LIME prototype, was installed at the underground Laboratori Nazionali del Gran Sasso (LNGS) in February 2022. This underground operation marks a significant milestone in the CYGNO collaboration's roadmap toward constructing a large-scale Time Projection Chamber (TPC) for directional Dark Matter (DM) searches. It represents the first test of a CYGNO prototype in a low-background environment, providing critical insights into the detector's performance under conditions essential for DM detection.

### 5.6.1 Detector setup and operational parameters

The underground data-taking campaign began in November 2022 and it is still ongoing as of December 2024. Two types of datasets have been collected: background runs and calibration runs. The detector setup was configured with a drift field of 800 V/cm to optimize primary electron transfer efficiency. During Run1, each GEM layer was set at 420 V, but in subsequent Runs (Run2, Run3 and Run4), this was increased to 440 V in order to enhance the overall light yield (LY). In Run1, LIME could not operated at a higher gain because increasing the voltage above 420 led to frequent discharges in the GEMs, affecting detector stability.
The gas system includes a recirculation and purification setup that became operational in July 2023, during Run3. During Run1 and Run2, fresh gas was continuously flushed through the detector, with exhaust collected in dedicated bottles. During Run1, the gas flow was varied between 1 and 20 l/h, while in Run2 and early Run3, it was increased to between 5 and 20 l/h. This adjustment aimed to maximize light yield and optimize fresh gas consumption since impurities in the gas reduce the total detected light. A higher gas flow rate was needed to maintain gas purity.
In July 2023, during Run3, to limit fresh gas usage, the recirculation system was activated. In this configuration, most of the gas was purified through commercial oxygen taps and recirculated, while fresh gas was introduced at a reduced flow rate. The fresh gas flow was initially set at 1 l/h in July 2023 but increased to 5 l/h in August, just before the AmBe source campaign. During Run3, several interventions were made to the gas system to set up the new gas system configuration, and some failures in a vacuum pump and a pressure booster were encountered and fixed.
The DAQ and data storage systems, initially tested during commissioning at LNF, were validated under low-background conditions at LNGS during underground data-taking. Thanks to the MIDAS framework, the system includes an automatic data acquisition procedure where, periodically, the GEM voltage is reduced to 200 V per layer, preventing any secondary scintillation light detection. At this reduced voltage, 100 pedestal images are acquired hourly to monitor camera noise for subsequent noise subtraction during image reconstruction. The camera exposure time, $T_{exp}$, is set at 300 ms to maximize light capture while minimizing the probability of pile-up events, especially during the higher-rate Run1. A rolling shutter mechanism exposes each camera sensor row sequentially, with a total activation time, $T_{act}$, of 184 ms.

| Run | Number of images | Time [s] |
|---|---|---|
| 1 | 253600 | 152160 |
| 2 | 526400 | 315840 |
| 3 | 703600 | 422160 |
| 4 | 149200 | 89520 |

**Table 5.3** Number of images and interval time taken in exam for each set of data.

Data acquisition is triggered by an external signal from the PMTs. An event is recorded if at least two PMTs register signals above a 2 mV threshold within the window $T_{exp} + T_{act}$. This setup captures the image and the PMT waveforms, storing them locally. The trigger logic, optimized through a dedicated campaign carried out during the commissioning of LIME in LNF, reduces false positives and enhances detection efficiency by requiring coincidences from multiple PMTs. While this setup minimizes fake triggers from isolated PMT signals, random coincidences can occasionally cause inefficiencies. As a result, the trigger rate may not directly match the actual event rate in the gas volume but still correlates with detectable tracks.

Finally, acquired data are transferred to a remote, cloud-based storage system where automated image reconstruction is applied to the sCMOS images for further analysis.

The number of images and interval time taken in exam for this work are summarized in Table 5.3.

### 5.6.2 Run 1

The Run1 focused on data collection without any added shielding and concluded in autumn 2022. The primary background source was from the natural gamma radiation within the laboratory environment. Due to frequent discharges, each GEM layer was operated at a voltage of 420 V.

Fresh gas was continuously flushed into the detector, and data were collected at different gas flow rates: 1 l/h, 3 l/h and 20 l/h. The stability of the detector was evaluated by analyzing events induced by the $^{55}Fe$ source. Instabilities were observed at the lowest flow rate of 1 l/h, affecting the data because of impurities. While, for the gas flow rates of 3 l/h and 20 l/h, the

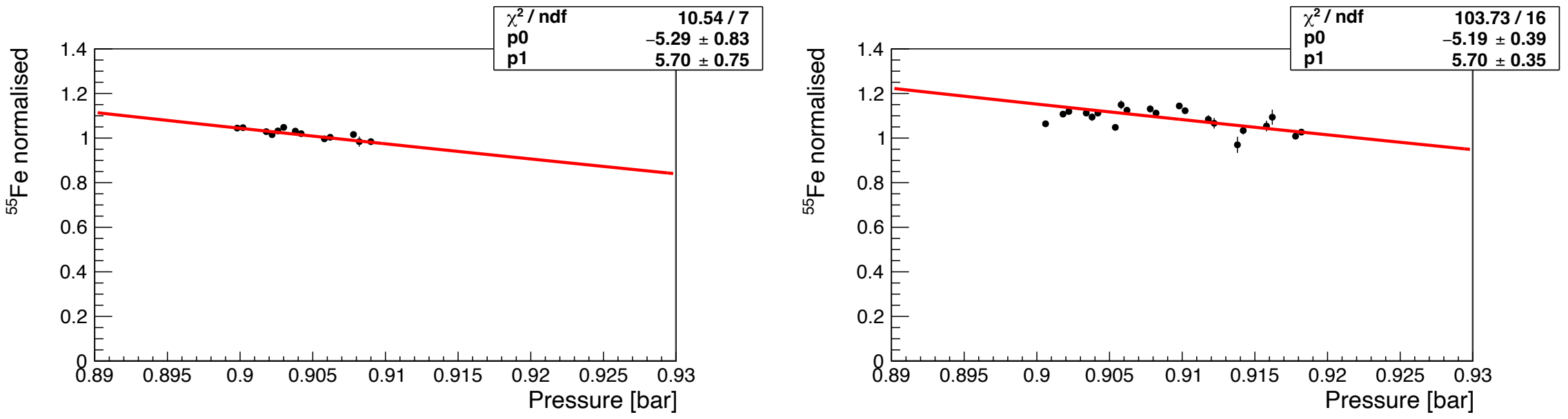

**(a)** Fresh gas flow 3 l/h **(b)** Fresh gas flow 20 l/h

**Figure 5.5** RUN1: $^{55}$Fe normalized light yield as function of gas pressure for two different fresh gas flow.

normalized light yield was studied as a function of pressure. A fit with the function 5.10 is superimposed on the data, demonstrating a behavior consistent with the results showed in Section 5.5.1 during the LNF data taking campaign.

A reduction in light of (0.57 $\pm$ 0.08) $\%$ per millibar for a fresh gas flow rate of 3 l/h and (0.57$\pm$ 0.04) $\%$ per millibar for a fresh gas flow rate of 20 l/h was observed.

### 5.6.3 Run 2

During Run2, 4 cm thick copper shielding was installed around the LIME prototype on all sides. This shielding reduced external gamma radiation by a factor of approximately 50, allowing for an increase in GEM voltage without causing discharges. The GEM voltage was raised to 440 V.

Based on the issues observed with gas impurities at low flow rates during Run1, the fresh gas flow was varied between 6 and 20 l/h. After a period of stability, the fresh gas flow has been set to 20 l/h, the normalized light yield is analyzed, showing a (0.68 $\pm$ 0.04) $\%$ decrease per millibar with increasing the gas pressure, as illustrated in Fig. 5.6.

### 5.6.4 Run 3

During Run3, additional copper was installed around the LIME prototype, increasing the total shielding thickness to 10 cm. This level of shielding suppressed the external background to the level of radioactivity originating from the internal detector material radioactivity.

The gas recirculation system was activated, allowing most of the gas to be purified trough commercial filters. In addition to the purified gas, fresh gas flow ranging from 1 l/h to 20 l/h was flushed. The pressure during this period remained nearly constant.

A humidity sensor has been installed in order to measure the Relative Humidity (RH[$\%$]).

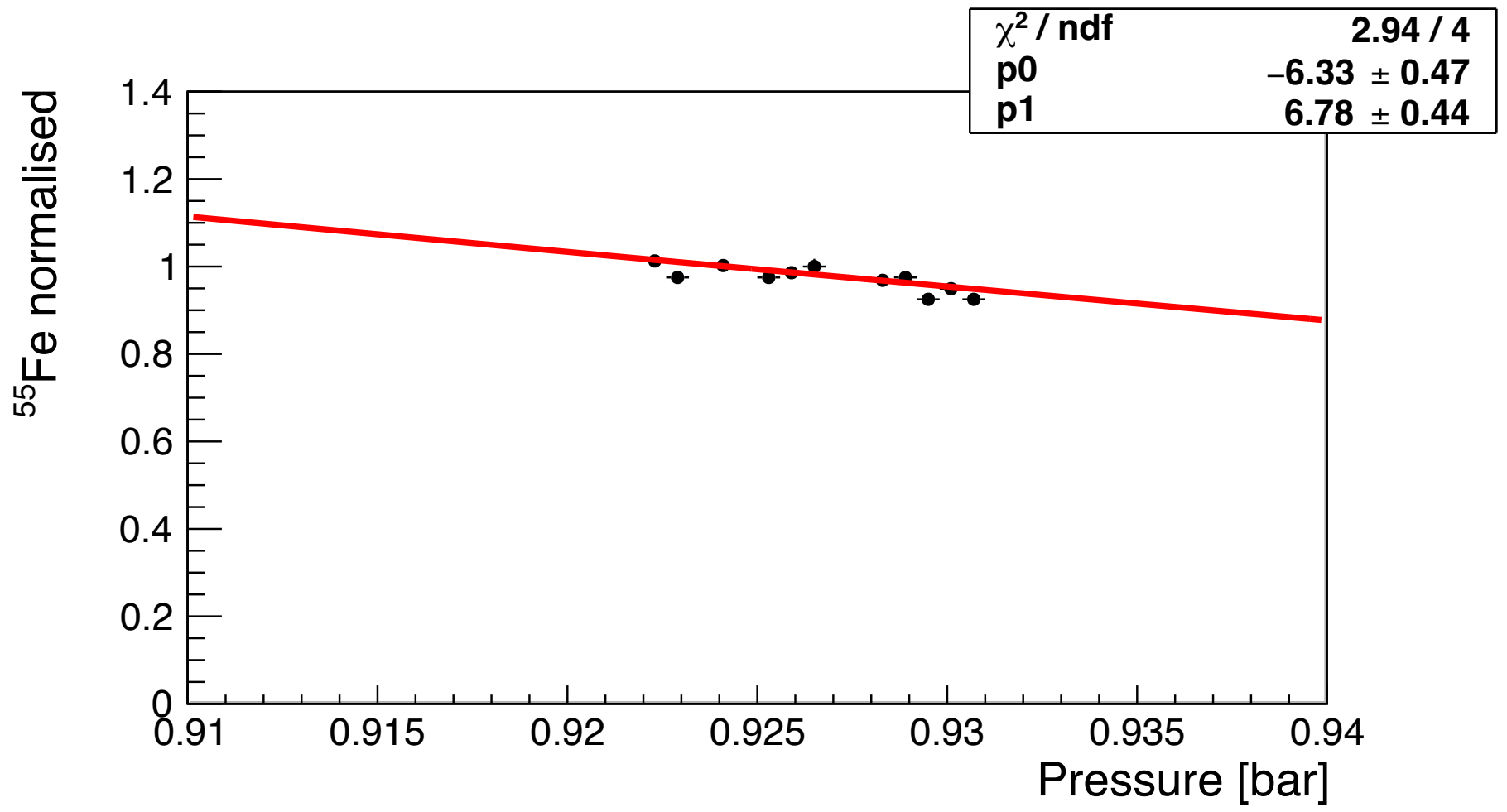

**Figure 5.6** RUN2: $^{55}$Fe normalized light yield as function of gas pressure, with a fresh gas flow of 20 l/h.

The presence of water vapor in the gas mixture affects the detector's gain. As electrons drift through the gas, they may be absorbed by forming negative ions. While noble gases and most organic molecules can only form stable negative ions at collision energies of several electronvolts, these energies are much higher than what electrons typically encounter during their drift in gas detectors. However, certain molecules, often present as impurities, such as water ($H_2O$), are capable of attaching electrons even at much lower collision energies. This electron attachment process reduces the number of free electrons available for detection, resulting in a reduction of signal efficiency and detector gain.

The normalized light yield has been studied as a function of relative humidity. To quantify this effect an exponential fit was performed, observing a 20% decrease in light yield per unit increase in RH, as shown for different fresh gas flows in Fig. 5.7.

### 5.6.5 Run4

During Run4, water tanks were installed around LIME, providing a 40 cm layer of hydrogen-rich shielding. The fresh gas flow rate was maintained at 5 l/h and the pressure remained stable throughout data collection. Humidity was well-controlled, showing only minor fluctuations. As shown in Fig. 5.8, the normalized light integral is analysed as a function of humidity. An exponential fit is performed showing a $(25 \pm 5)$ % decrease in light yield per unit increase in RH, consistent with the results observed during Run3.

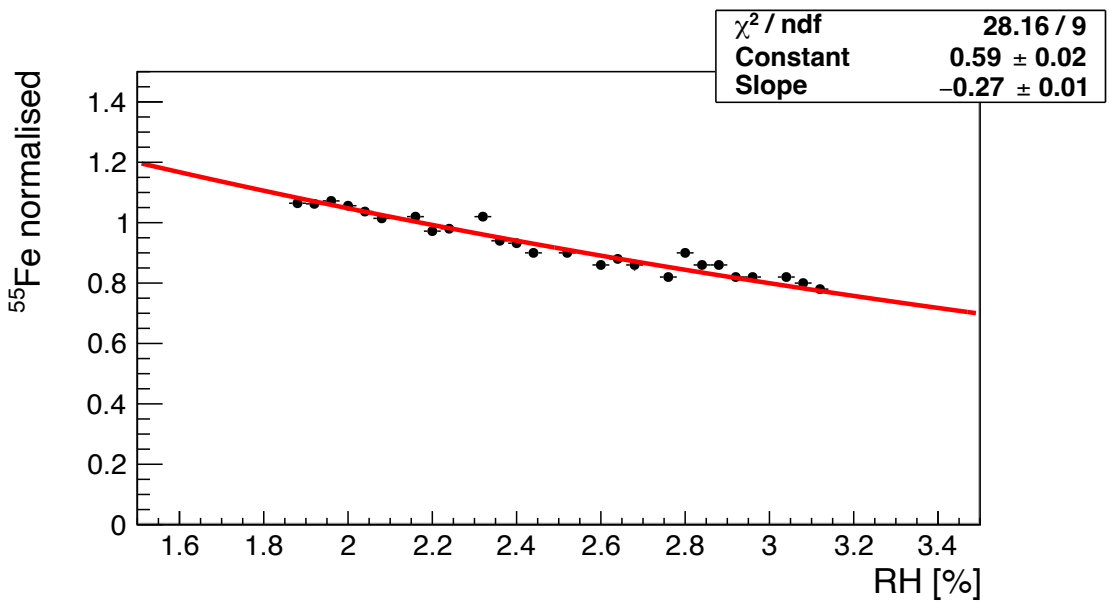

**(a)** Fresh gas flow: 1 l/h

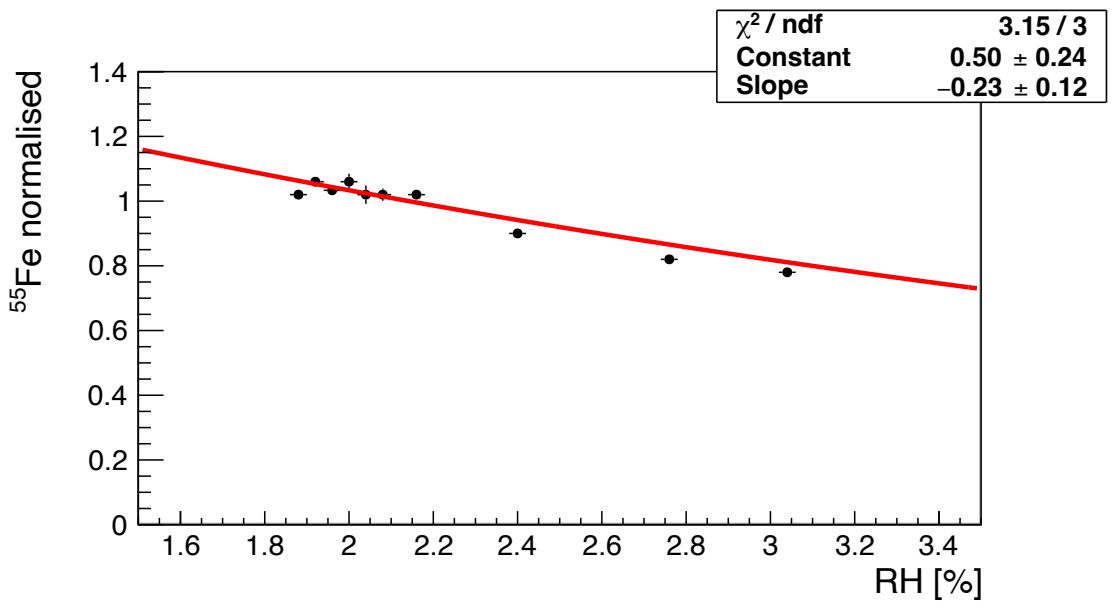

**(b)** Fresh gas flow: 2 l/h

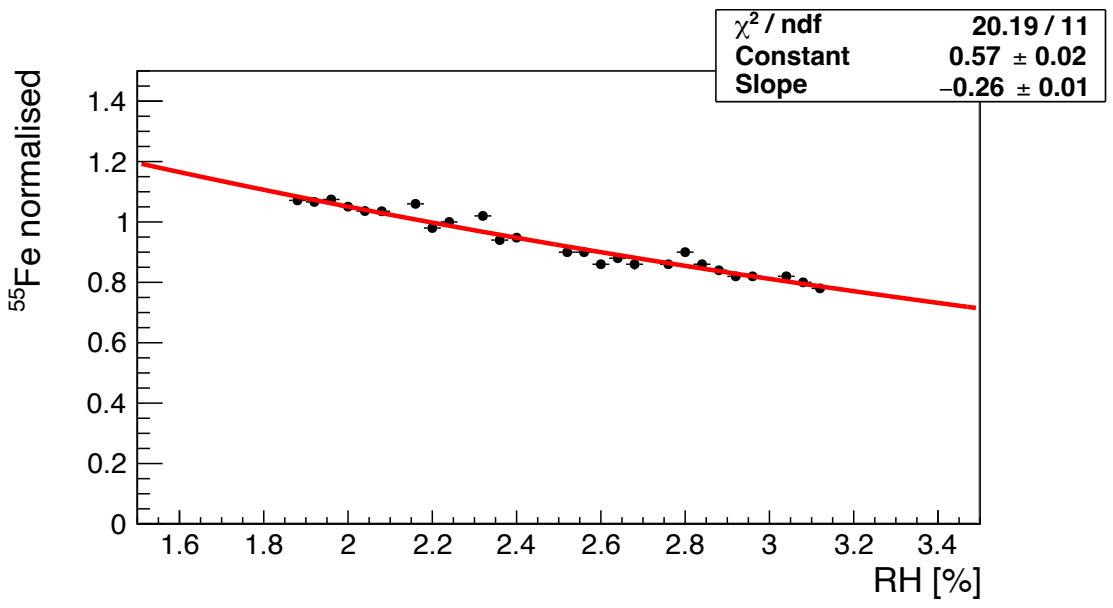

**(c)** Fresh gas flow: 5 l/h

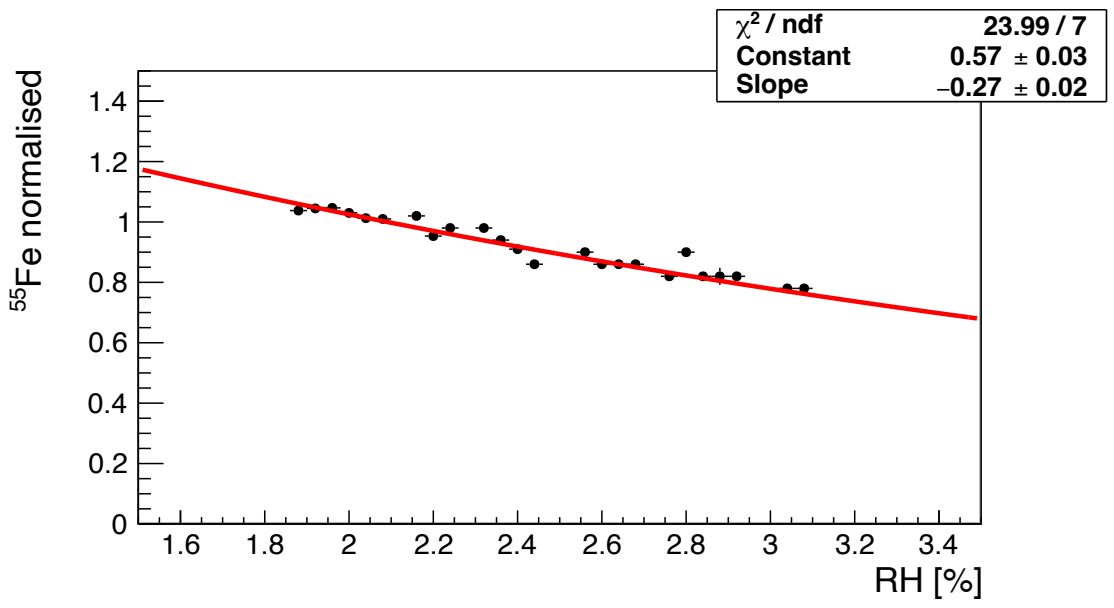

**(d)** Fresh gas flow: 6 l/h

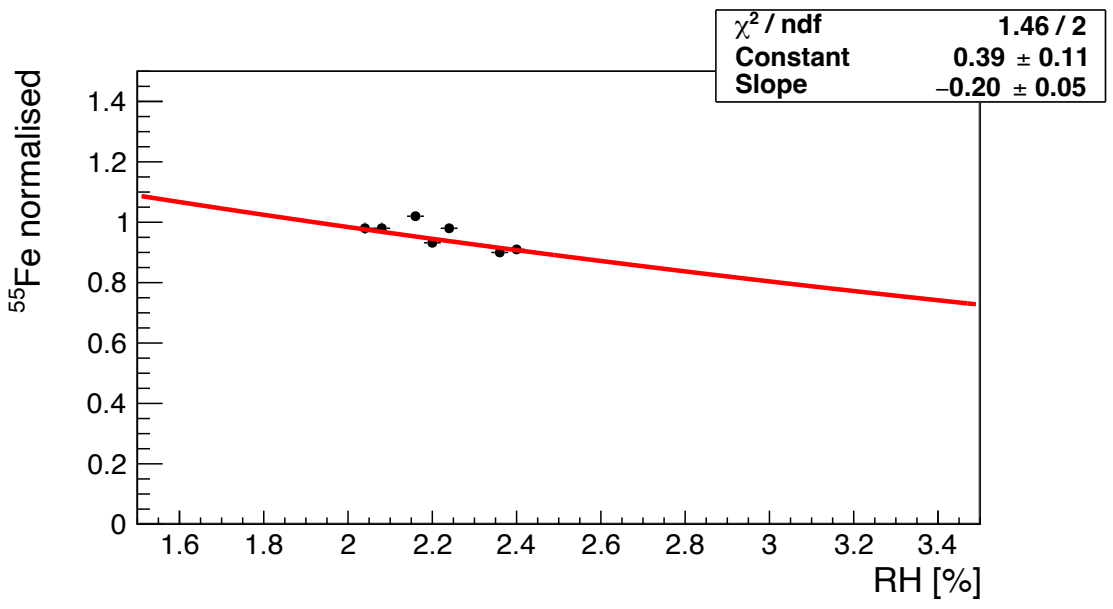

**(e)** Fresh gas flow: 10 l/h

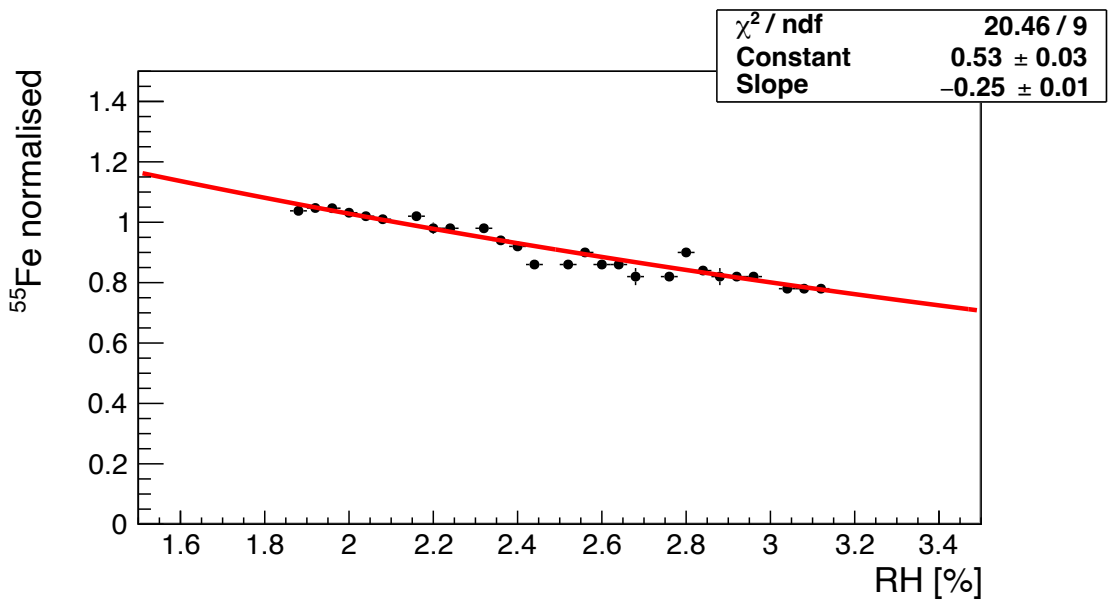

**(f)** Fresh gas flow: 20 l/h

**Figure 5.7** RUN3: $^{55}$Fe normalized light yield as function of relative humidity for different fresh gas flow.

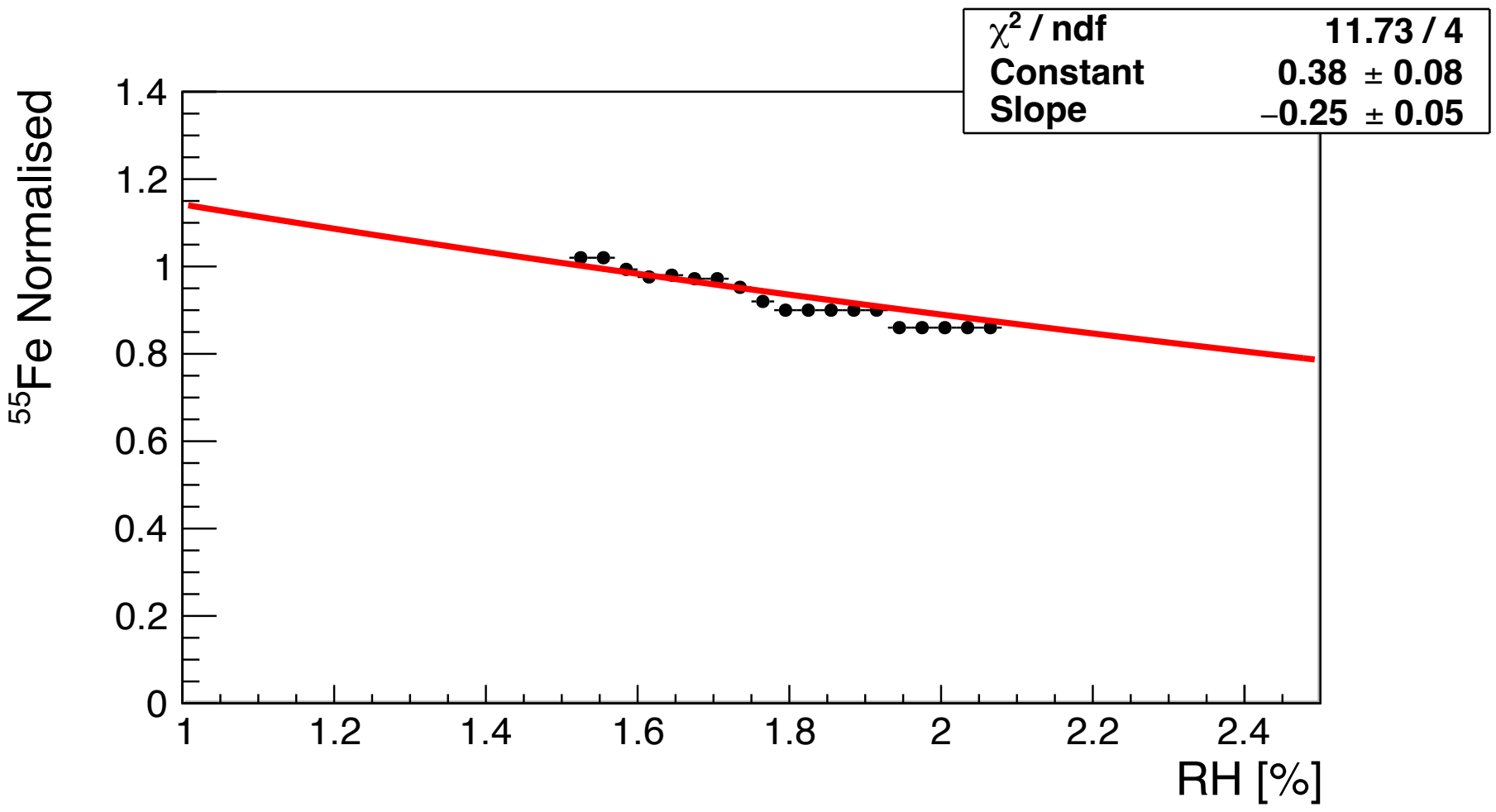

**Figure 5.8** RUN4: $^{55}$Fe normalized light yield as function of relative humidity, with a fresh gas flow of 5 l/h.

# 6 Light yield calibration

A Dark Matter detector must operate over extended periods, requiring stability during data taking. Data have to be merged, carefully considering the calibration in the detector response and its equalization over time. As described and studied in the previous chapter, several factor can influence the detector's gain, such as pressure fluctuation, humidity and impurities, all of which affect its performance. Although a consistent parametrization of the detector response with operating conditions was presented, in the end another technique resulted more effective in correcting the data.
In this chapter a novel method for calibrating and equalizing the data across all runs is introduced, ensuring uniformity in the detector's response despite environmental and operational variations.

## 6.1 Calibration procedure

In order to equalize all the runs, a new variable, LY_30, is introduced. This variable represents the average light yield of high-energy tracks. Only the fiducial area cut is applied, as explained in Section 5.3.1 . The dataset used to illustrate this procedure correspond to Run3, during which both background and $^{55}$Fe source runs, utilized as standard candle, were recorded. The clusters are reconstructed using the algorithm described in Section 4.4. For the runs where the iron source is present, after applying the fiducial and noise cut described in Section 5.3.1, the light integral distribution is fit using the Cruijff function to estimate the average energy of the signal induced by the $^{55}$Fe X-ray source in terms of counts. The iron peak typically corresponds to around 10 k counts. Occasionally, double spots are incorrectly recognized as single ones, to avoid tail contribution, the lower bound for LY_30 is set to 30 k counts. While the upper limit could be infinite, but to avoid to include miss-reconstructed tracks, it is set equals to 300 k counts. By using this variable, the expected value of $^{55}$Fe can be evaluated even when the source is not present with the procedure described in the following. This enables the equalization of all the runs to the same condition and help monitor the stability.

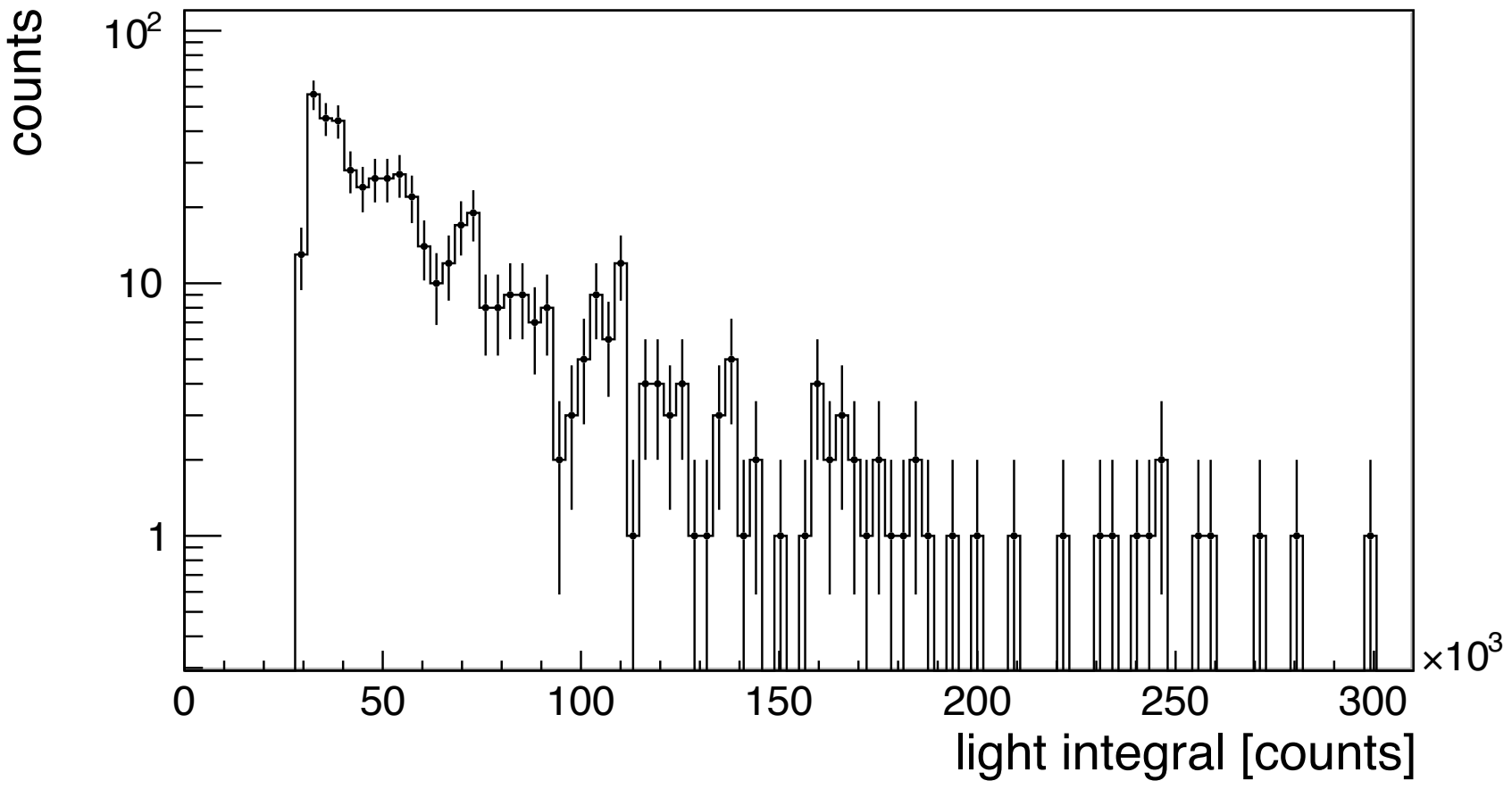

**Figure 6.1** Example of one distribution used to evaluate the LY_30 value.

For each run, the light integral distribution is built and the LY_30 is calculated as the average between 30 k and 300 k, using only the fiducial area cut, as shown in Fig. 6.1.

For calibration runs, where the iron source is placed 25 cm from the GEMs, additional cuts on noise and track shape are applied to select events due to X-ray source. The light integral distribution for each run is fitted with the Cruijff function, as shown in Fig. 6.2. The parameter $p_0$ represents the mean, while $p_1$ and $p_3$ the left and the right sigma, and $p_2$ and $p_4$ describe the left and the right tails.

The $^{55}$Fe value for each run where the source is placed 25 cm far from the GEMs, is shown in Fig. 6.3 and the LY_30 normalized to be shown in the same scale is superimposed. As can be seen, they shows similar behavior, suggesting the possibility to use the LY_30 to predict the $^{55}$Fe value.

To evaluate the parameters for calculating the expected value of the iron peak, LY_30 versus the $^{55}$Fe peak is plotted. A linear fit is performed on the profile, as shown in Fig. 6.4. The error on each point is evaluated as the standard deviation.

The expected value of the $^{55}$Fe peak, named $^{55}\text{Fe}_{eq}$, is calculated inverting the linear relation. The $^{55}\text{Fe}_{eq}$ is computed for each run, with or without the iron source. Fig. 6.5 shows $^{55}\text{Fe}_{eq}$ for the entire Run3 dataset, with black dots representing the average of 20 runs to reduce

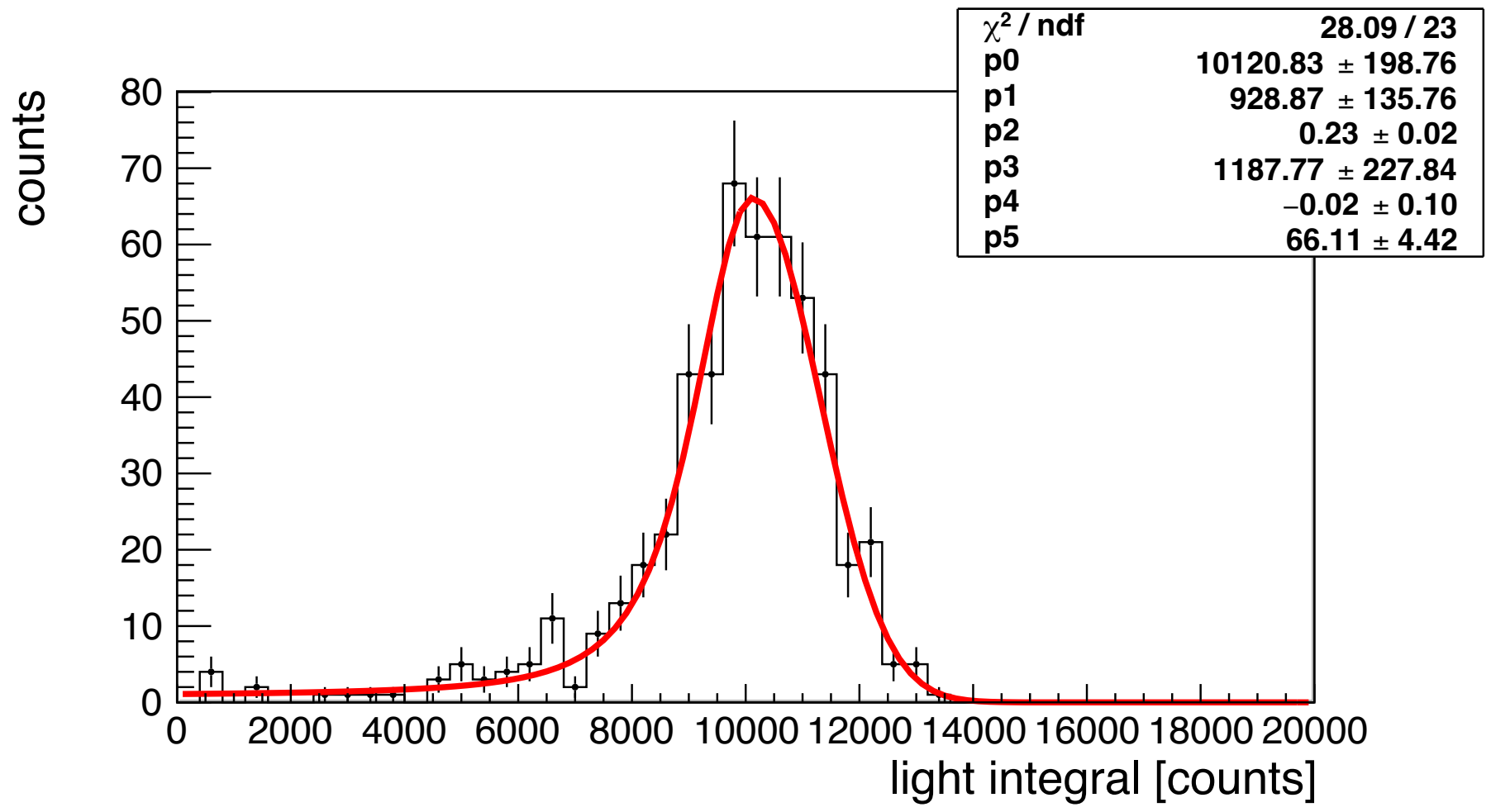

**Figure 6.2** Example of the light integral distribution for a run with the $^{55}$Fe source at 25 cm from the GEMs. A fit with the Cruijff function is superimposed.

statistical fluctuations.

The comparison between the $^{55}\mathrm{Fe}_{eq}$ and the $^{55}$Fe in the full dataset is shown, using the 20 runs average, in Fig. 6.6 red dots representing the $^{55}$Fe peak are present only for the calibration runs. Quantitative comparison (Fig. 6.7) is obtained by the relative difference between $^{55}$Fe and $^{55}\mathrm{Fe}_{eq}$ in the data points where the source is present. A Gaussian fit is performed: a 13 $\%$ dispersion and null offset is found. This 13$\%$ is the assumed error in using this method to compute the calibration peak.

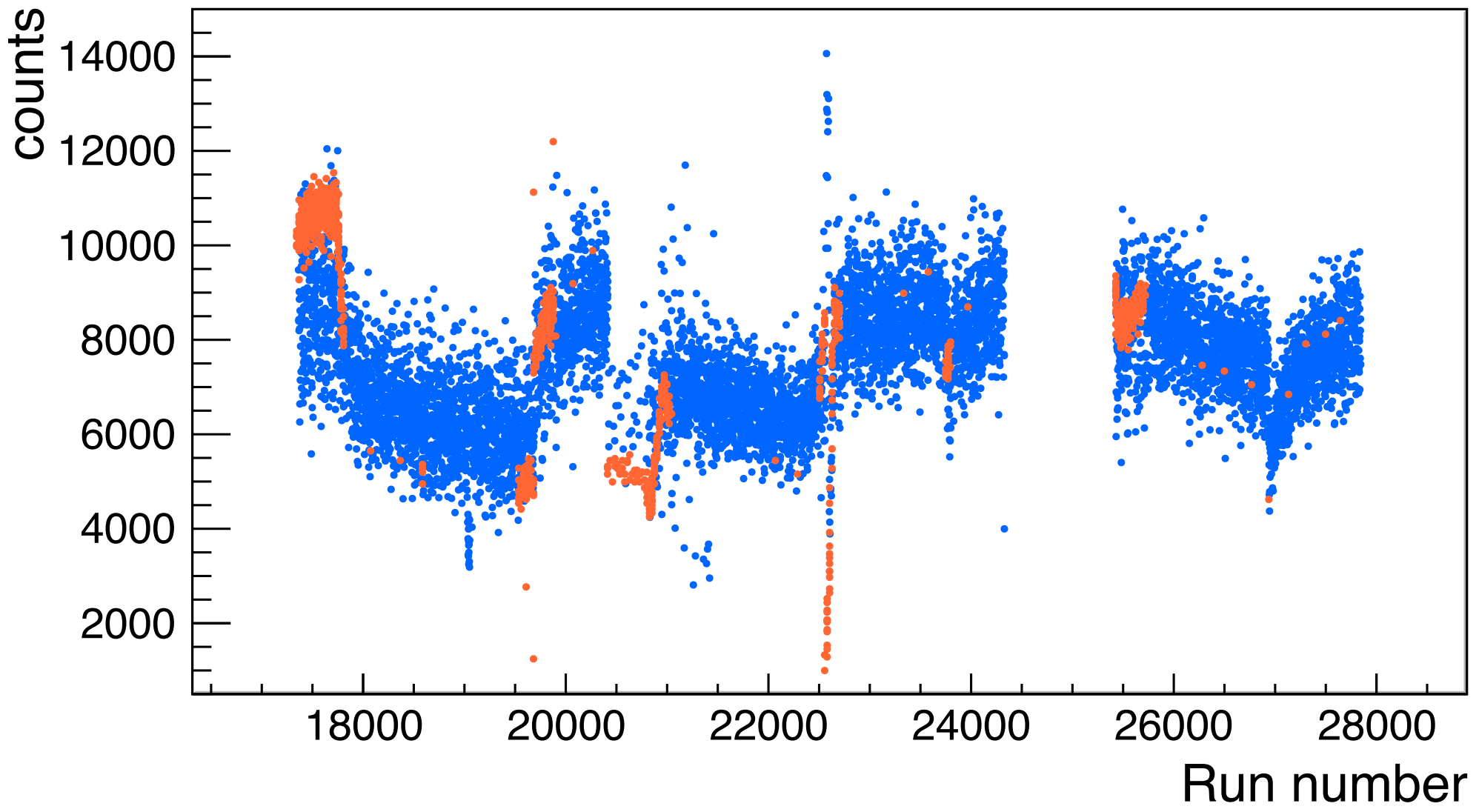

**Figure 6.3** In blue the normalized LY_30 and in orange the $^{55}$Fe peak evaluated for each run where the source is placed 25 cm far from the GEMs.

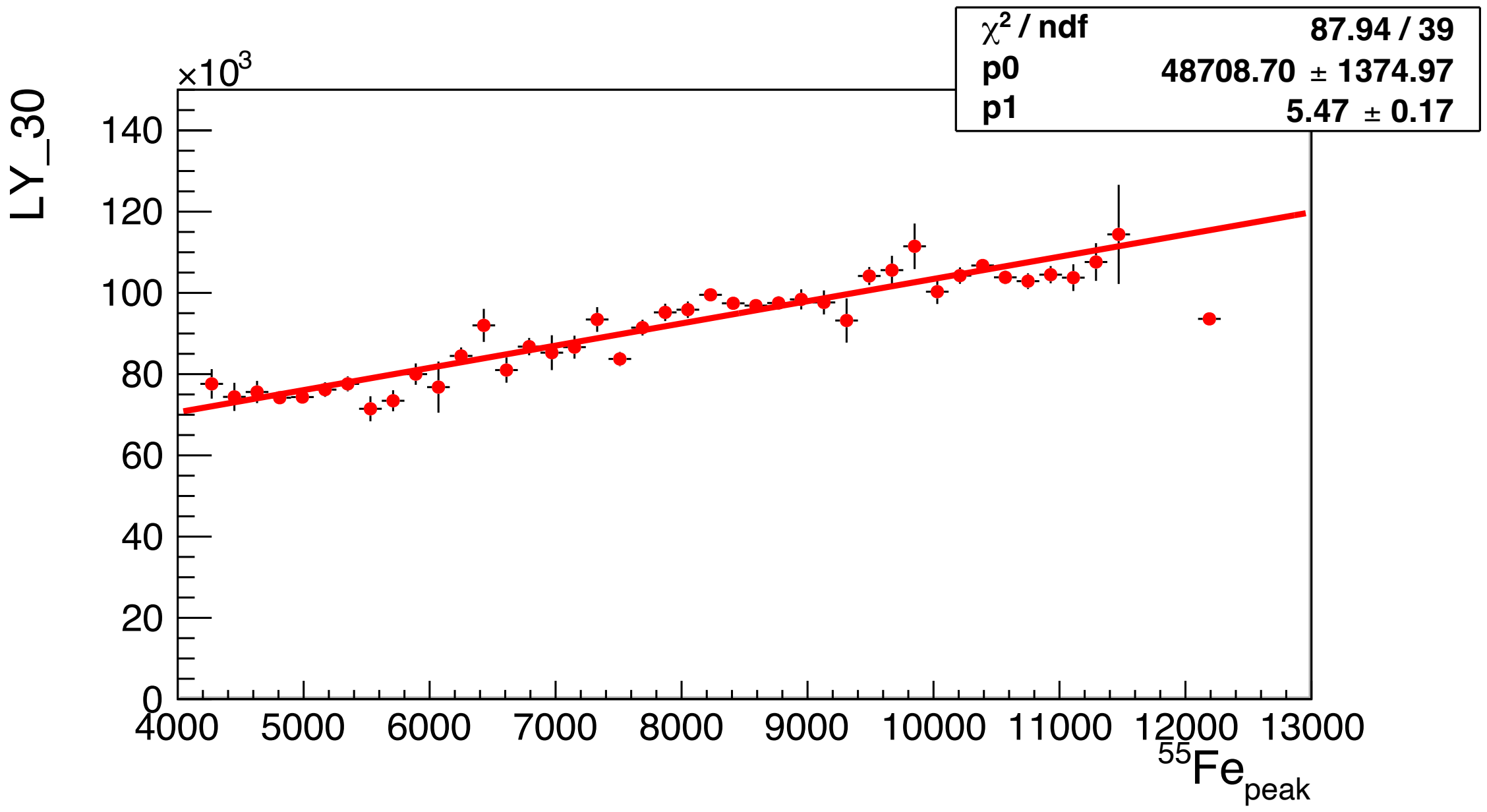

**Figure 6.4** Plot of the average LY_30 versus the $^{55}$Fe peak, with a linear fit superimposed.

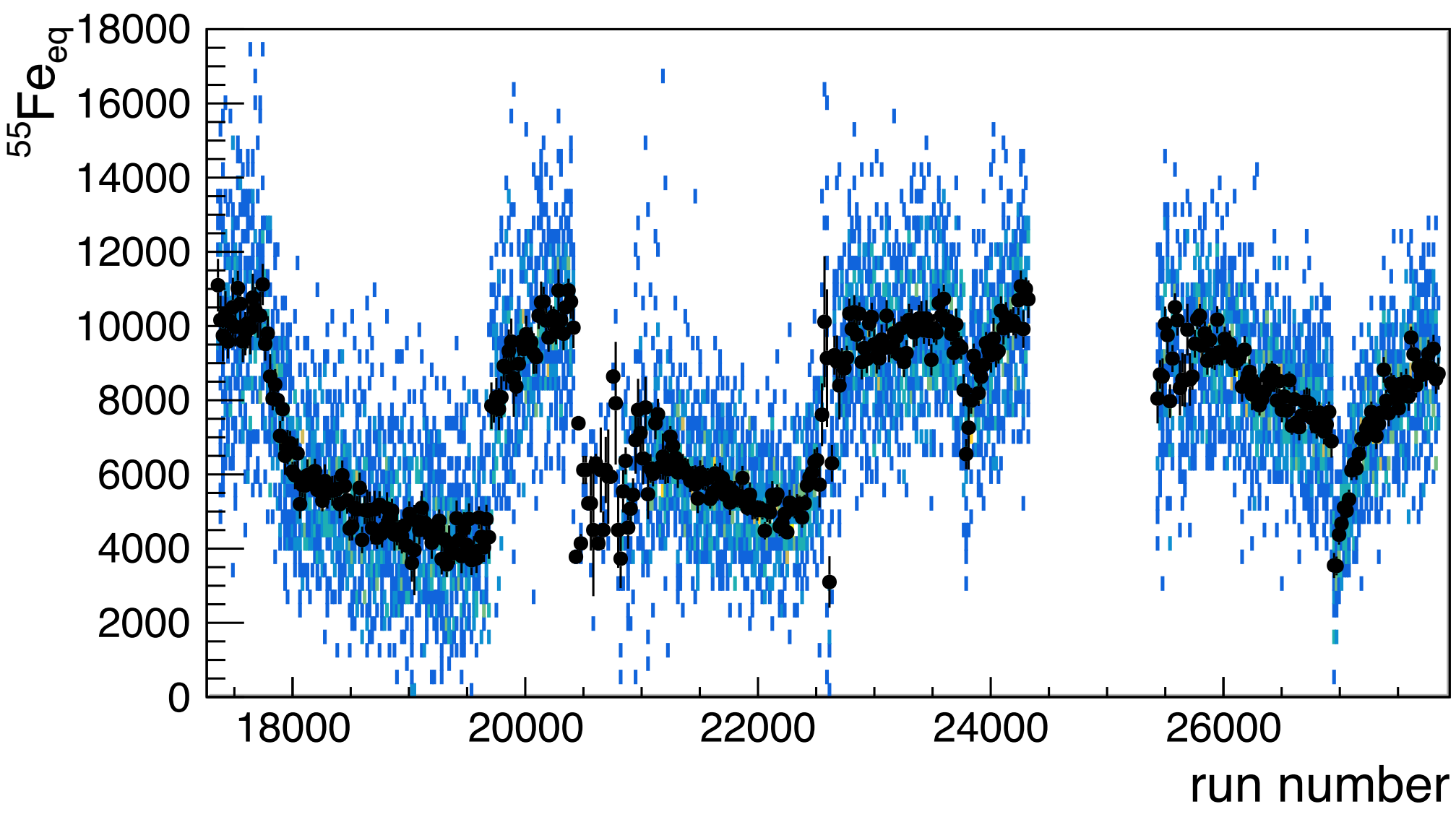

**Figure 6.5** The $^{55}Fe_{eq}$ value of each run in Run3. Black points represent 20 run average.

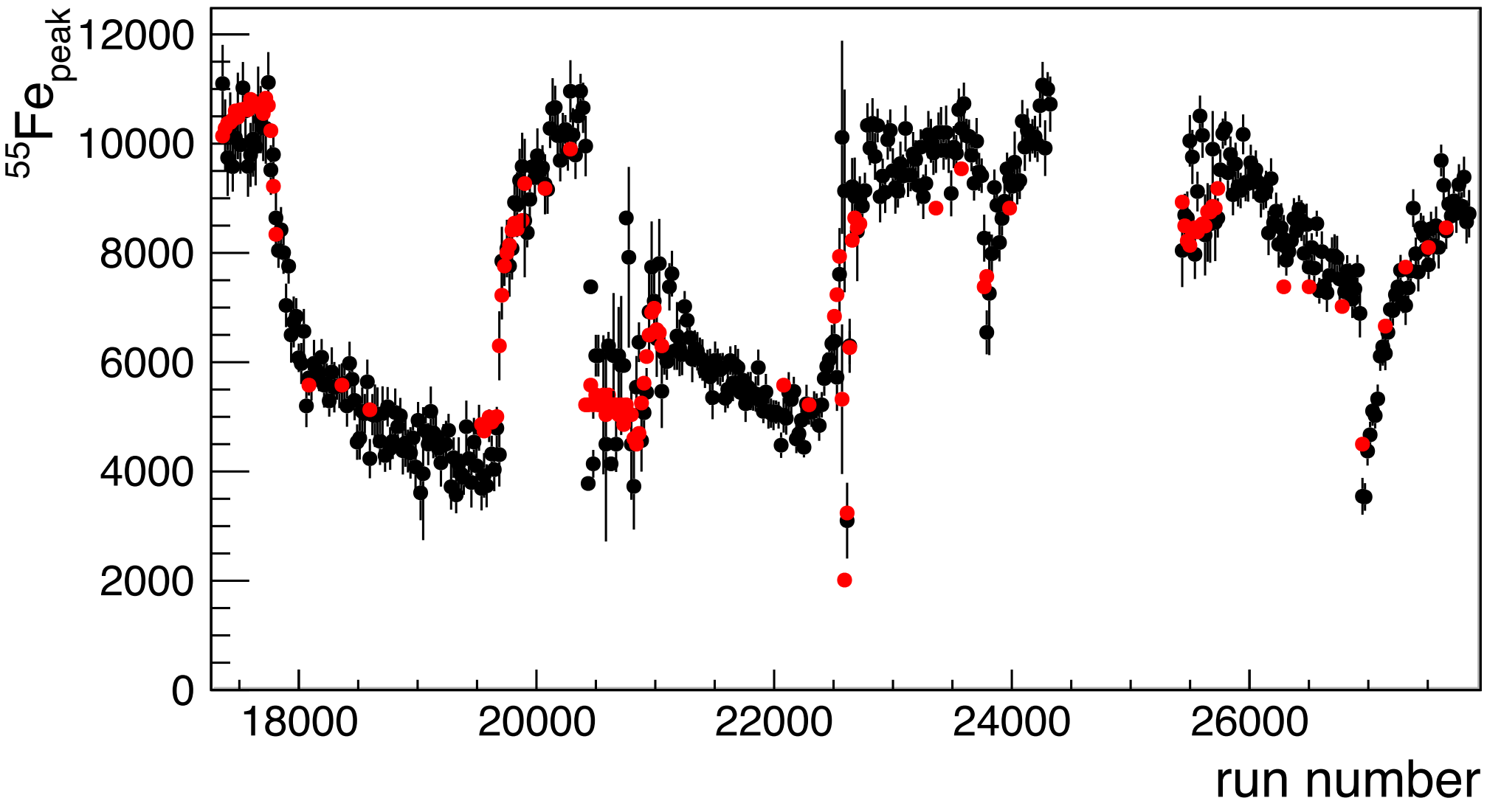

**Figure 6.6** The black points represents the averaged $^{55}Fe_{eq}$, while the red one the averaged $^{55}Fe$.

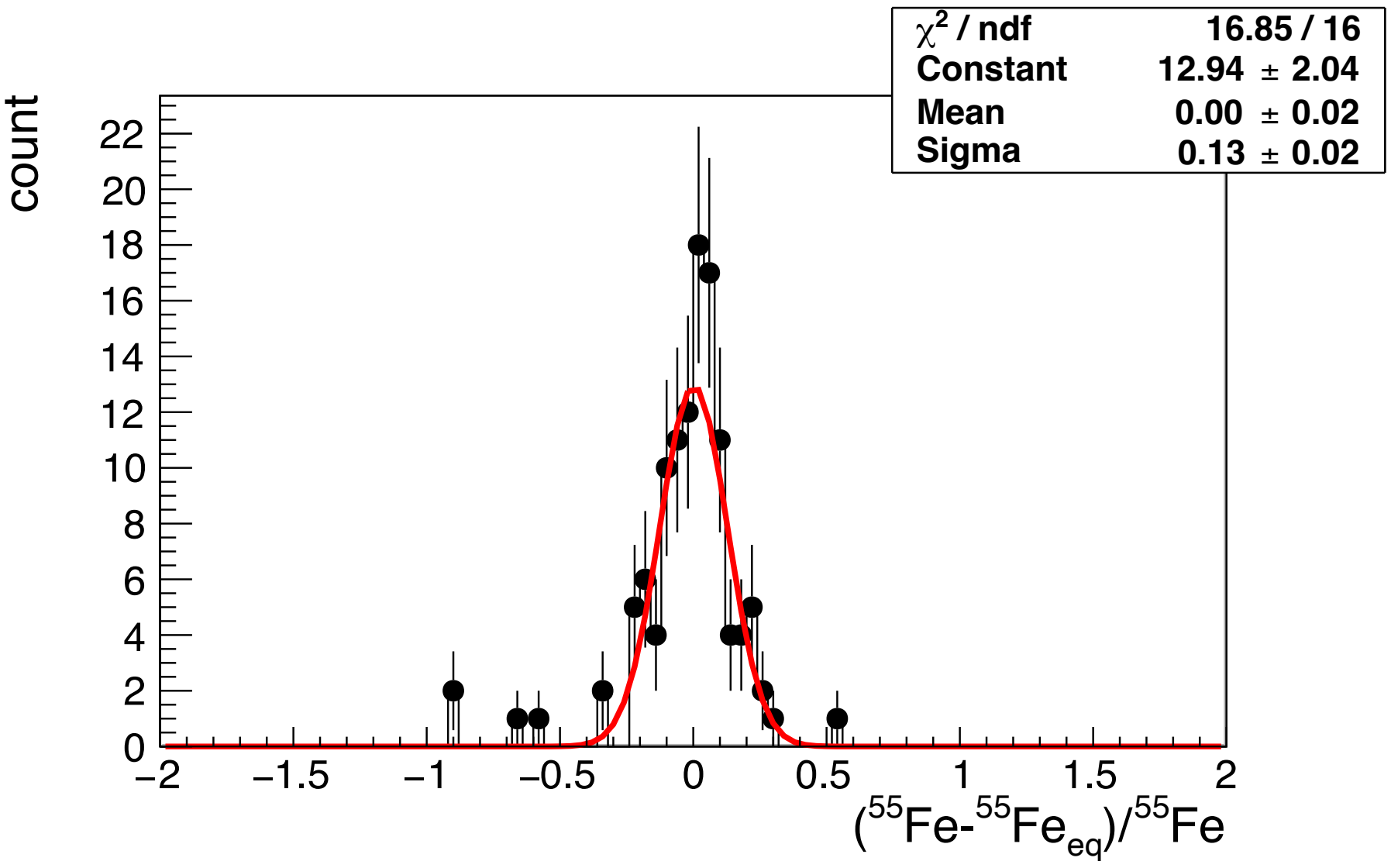

**Figure 6.7** The distribution given by the difference between the averaged $^{55}$Fe$_{eq}$ and the averaged $^{55}$Fe with respect to the averaged calculated iron peak. A Gaussian fit is superimposed.

### 6.1.1 Dataset equalization

In order to equalize all the data, a table is created counting for each run the correction value, determined through the following procedure:

- if the iron source is placed 25 cm far from the GEMs, the measured $^{55}$Fe value is reported;
- in the other configurations, the last calibration run is used and:
  - if the $^{55}$Fe$_{eq}$ differs less than 13 %, the last $^{55}$Fe value is reported;
  - if the $^{55}$Fe$_{eq}$ differs more than 13 %, the averaged $^{55}$Fe$_{eq}$ is reported until the next calibration. This last step is implemented to properly equalize runs where the light yield varied too much with respect to the last available calibration.

The 13% threshold is established based on the fit of the dispersion distribution, as shown in Fig. 6.7. The correction value is used to evaluate a scaling factor that equalizes all runs, ensuring that for each run with the iron source placed 25 cm from the GEMs, the light count of the acquired tracks has an integral of 10 k.

## 6.2 Application of the equalization

During the calibration process, runs are taken with the $^{55}$Fe source placed at different distances from the GEMs, as listed in Table 6.1.

After equalizing all runs, the resulting light integral distribution for each step is reported in Fig. 6.8 and a Gaussian fit is performed. The fitting range is defined as one standard deviation (RMS) around the mean value. Getting closer to the GEMs, the value of the $^{55}$Fe peak decreases, due to saturation effects. More precisely, the $^{55}$Fe peak has a fixed value of $\sim 10^4$ counts for steps 3 to 5; while saturation effect becomes visible at step 2 where the peak is reduced by $\sim 15\%$ and is large at step 1 where the peak is $\sim 35\%$ lower. It is also observed that the resolution of the $^{55}$Fe peak smoothly decreases as the distance from the GEMs increases, as shown in Fig 6.9.

| **Step** | **Distance from the GEMs [cm]** |
|---|---|
| 1 | 4.75 |
| 2 | 14.75 |
| 3 | 24.75 |
| 4 | 34.75 |
| 5 | 45.75 |

**Table 6.1** Calibration steps corresponding to the different distances between the $^{55}$Fe source and the GEMs.

Summing together all the calibration runs, as shown in Fig. 6.10, two contributions are visible. The first peak arises mainly from the contribution of the runs in step 1, because of the saturation effects, while the second peak is attributed to the other steps.

To study the effect of the $z$ dependence of the saturation on all the data, not on the calibration runs, we aim at finding a parametrization of this effect. To this extent, a toy Monte Carlo was used.

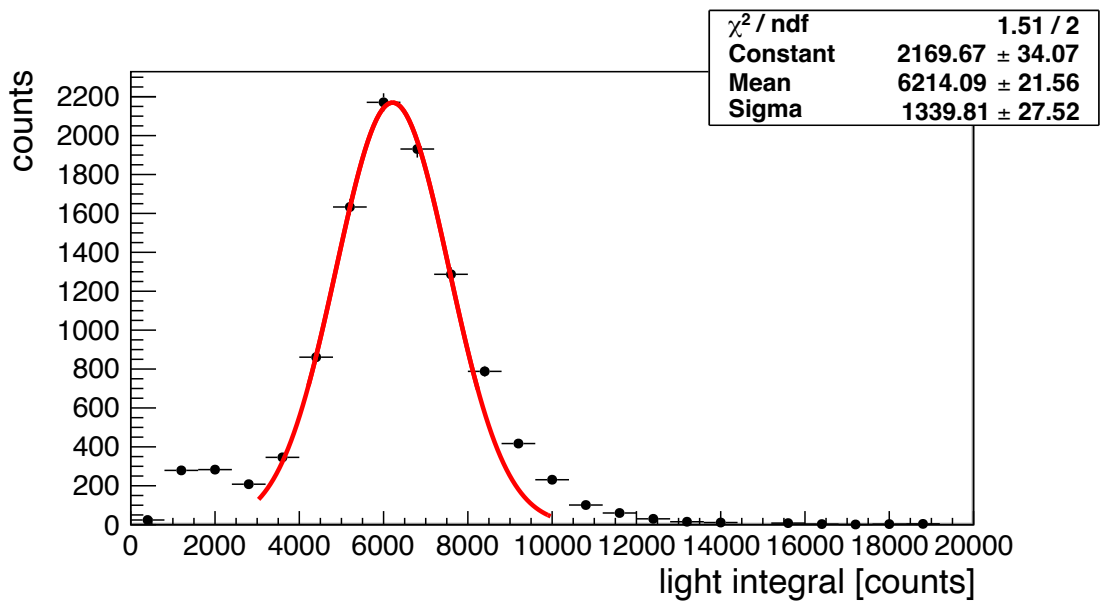

**(a)** step 1

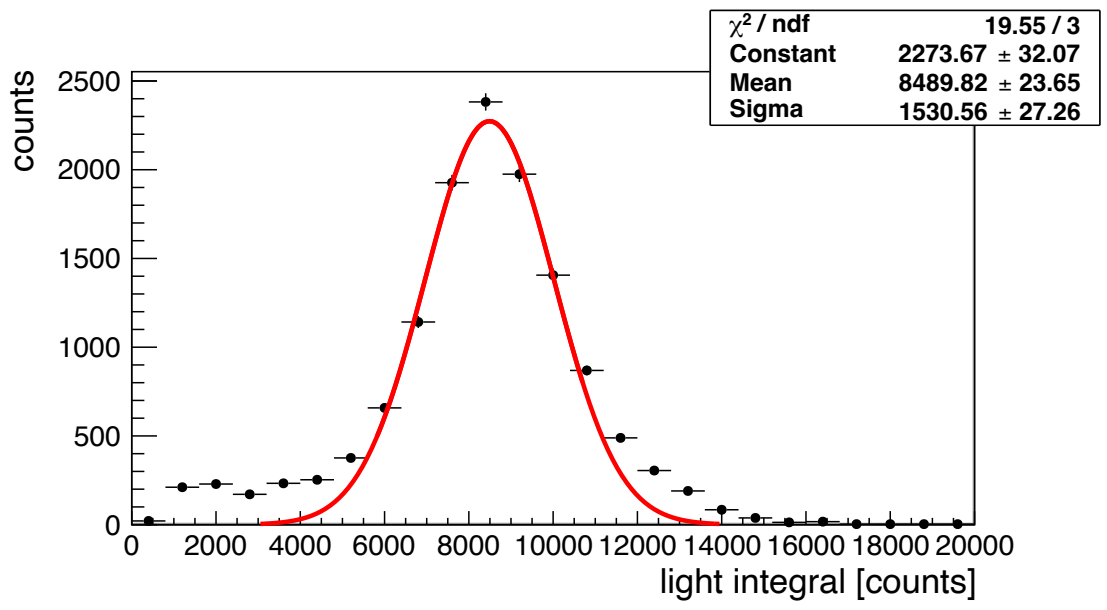

**(b)** step 2

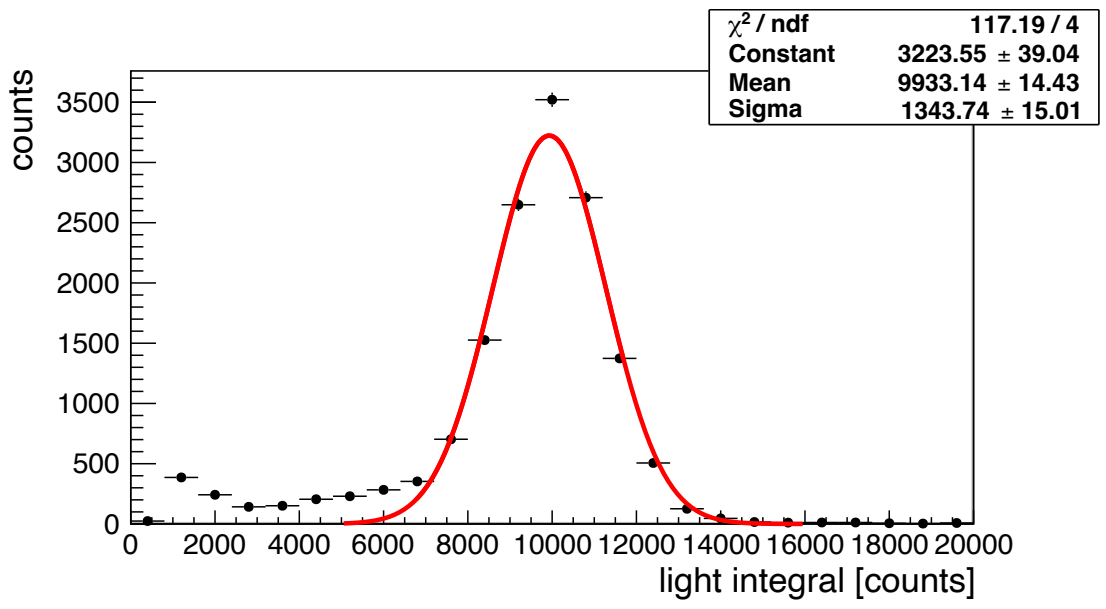

**(c)** step 3

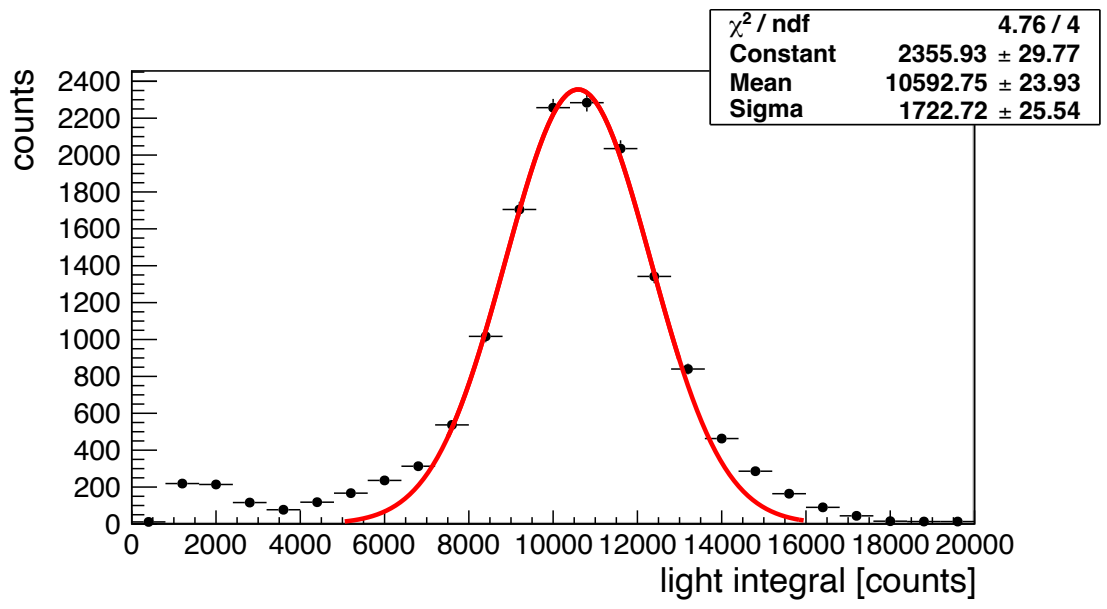

**(d)** step 4

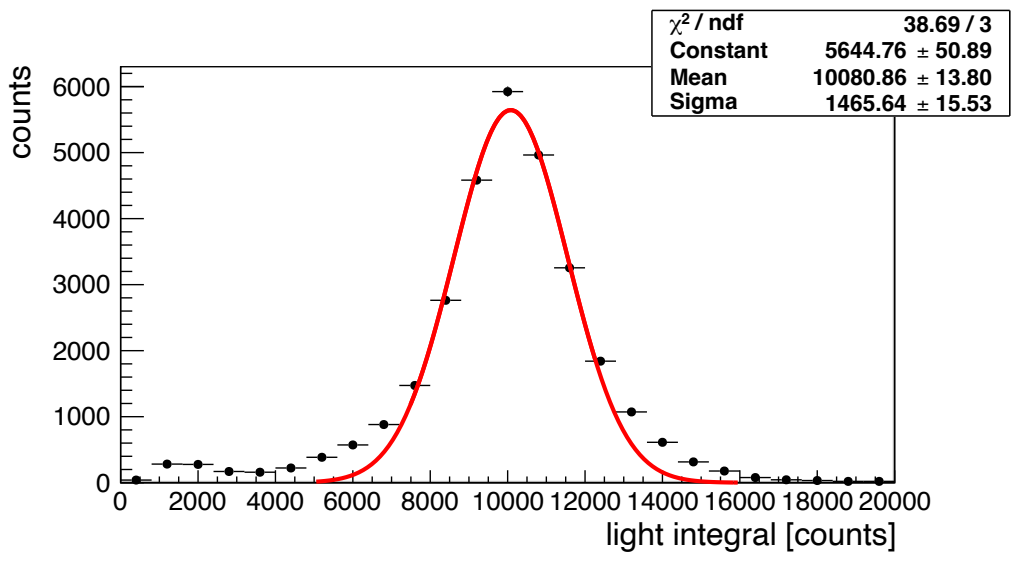

**(e)** step 5

**Figure 6.8** The light integral distribution for each calibration step, showing the Gaussian fits for each step's data.

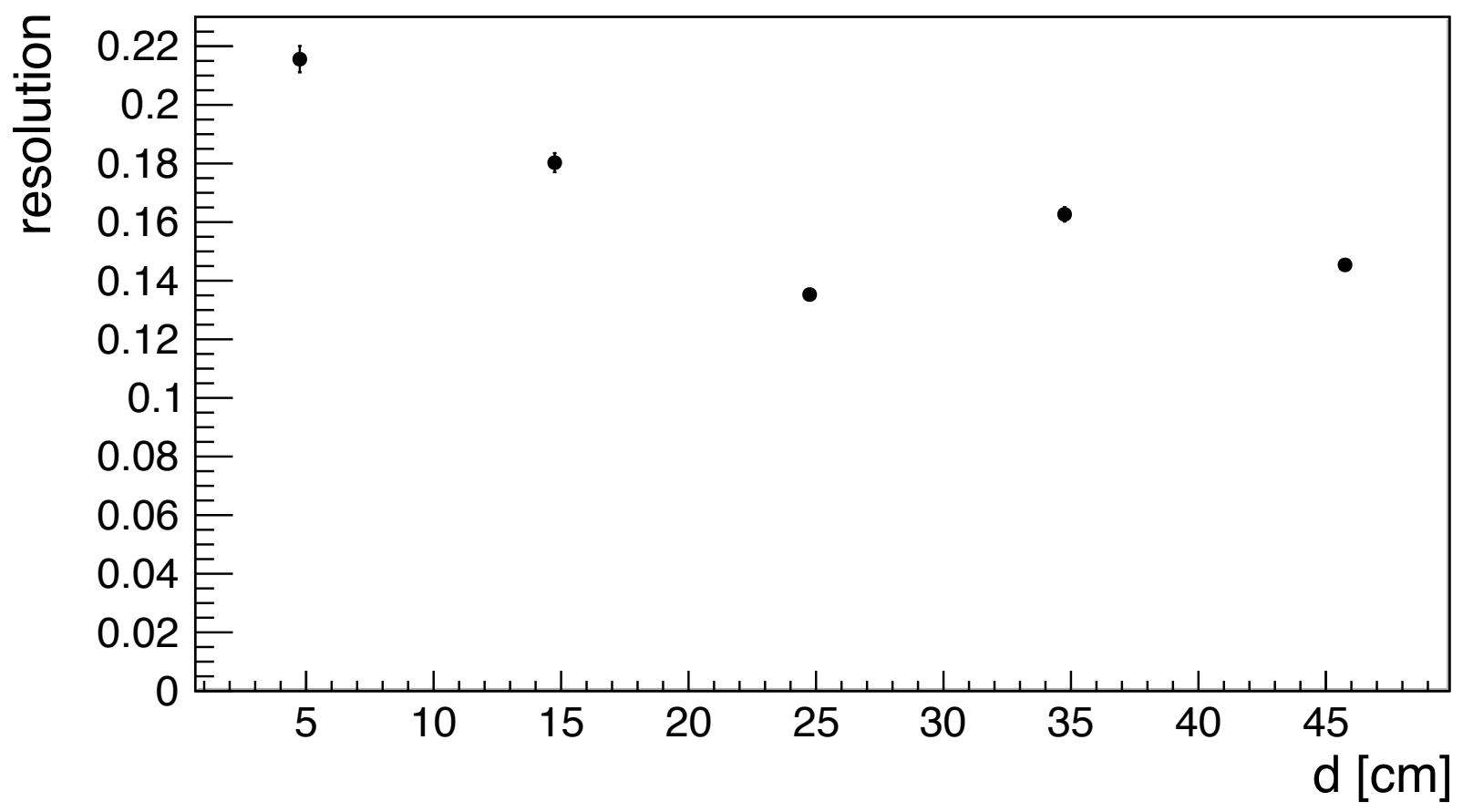

**Figure 6.9** Light integral resolution for each step.

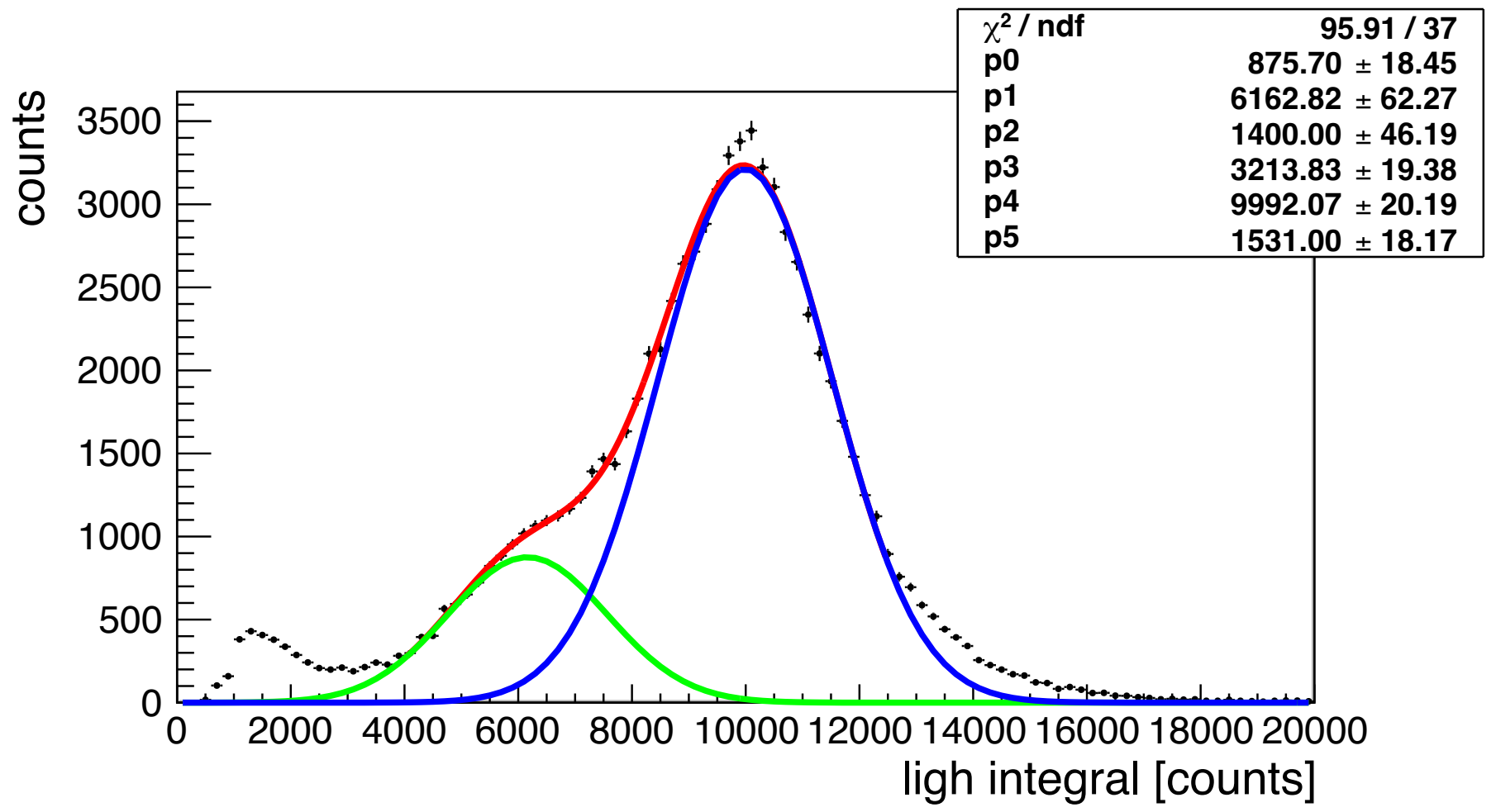

**Figure 6.10** The integral light distribution built summing all the calibration runs. A double Gaussian fit is superimposed, the red line, with the green curve representing the first peak contribution (saturation) and the blue curve representing the second peak (non-saturated steps).

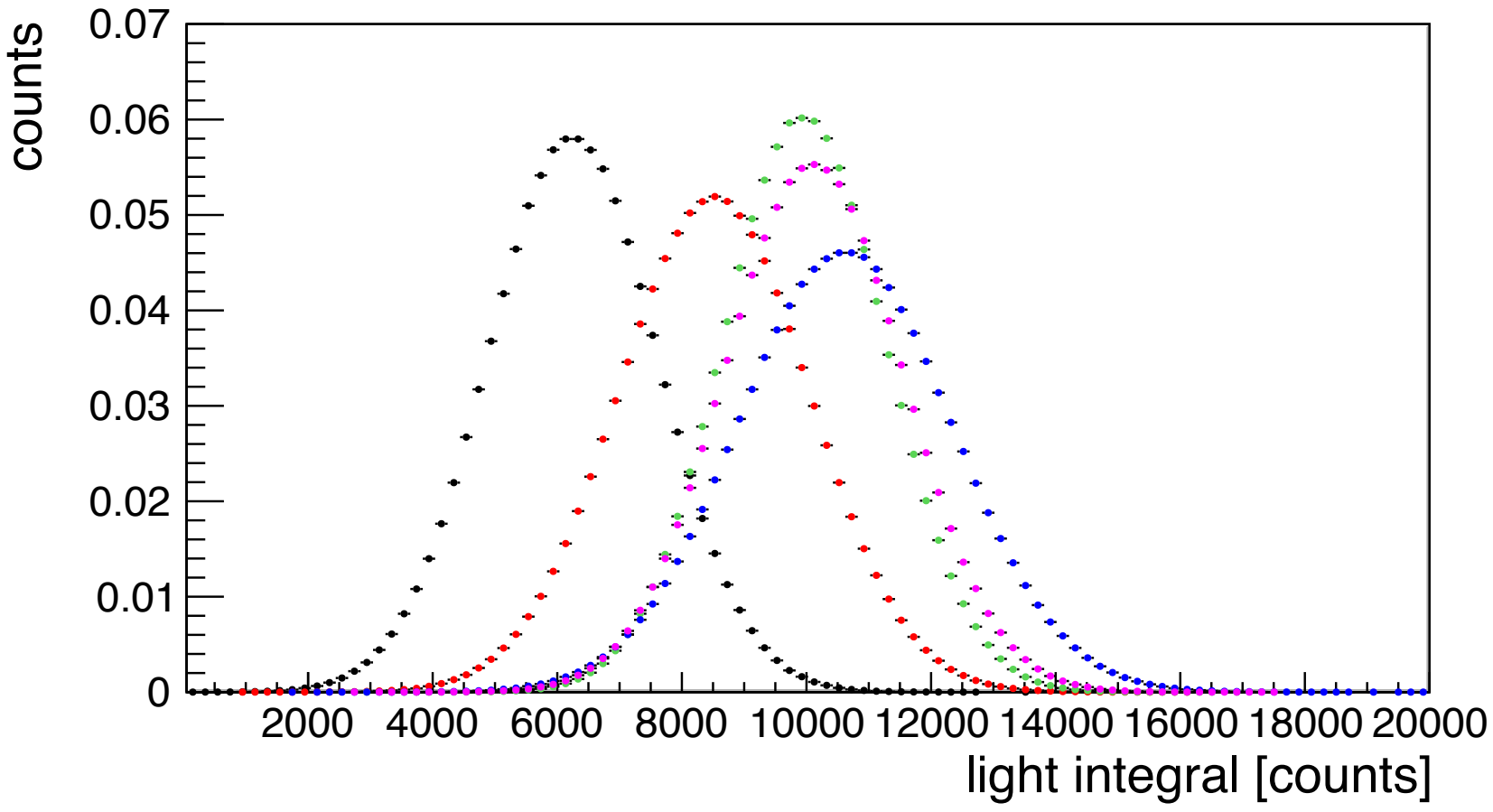

**Figure 6.11** Normalized integral light distribution generated: black represents spots generated at step1, red at step2, green at step3, blue at step 4 and purple at step 5.

### 6.2.1 Toy Monte Carlo simulation

The first step is to reproduce the double peak distribution on Fig. 6.10.
The light integral distributions at each step are fitted with a Gaussian function, as shown in Fig. 6.8. Subsequently, the parameters are used to randomly extract the light integral of the spots. The process is repeated $10^4$ times per step, as shown in Fig. 6.11.

After generating the light integral distribution for each step the total light integral distribution is built by summing the contribution from all five steps. This summed distribution is shown in Fig. 6.12. A double Gaussian fit is performed in order to evaluate the two main contributions. The agreement between this generated distribution (Fig. 6.12) and the acquired data (Fig. 6.10) confirms the toy Monte Carlo procedure's validity.

To simulate random interactions along the $z$, $10^5$ points are uniformly extracted between 0 and 46 cm, which represents the distance from the GEMs. This allows for an investigation of the saturation effect as a function of $z$. To model this, the mean, $\mu(z)$ (Fig. 6.13a), and the sigma, $\sigma(z)$ (Fig. 6.13b) of the light integral distribution are plotted versus the distance from the GEMs. The uncertainty on the mean and sigma is given by the error on the fit parameters. A parabolic and a linear fit is performed on the mean and on the sigma, respectively:

$$\mu = p_{00} + p_{01}z + p_{02}z^2 \tag{6.1}$$

$$\sigma = p_{10} + p_{11}z \tag{6.2}$$

where $p_{n0}$, $p_{n1}$ and $p_2$ are the parameters given by the fit, with $n$ equal to 0 for the fit on the mean plot and equal to 1 for that on the sigma. These functions allow for the evaluation of $\mu$

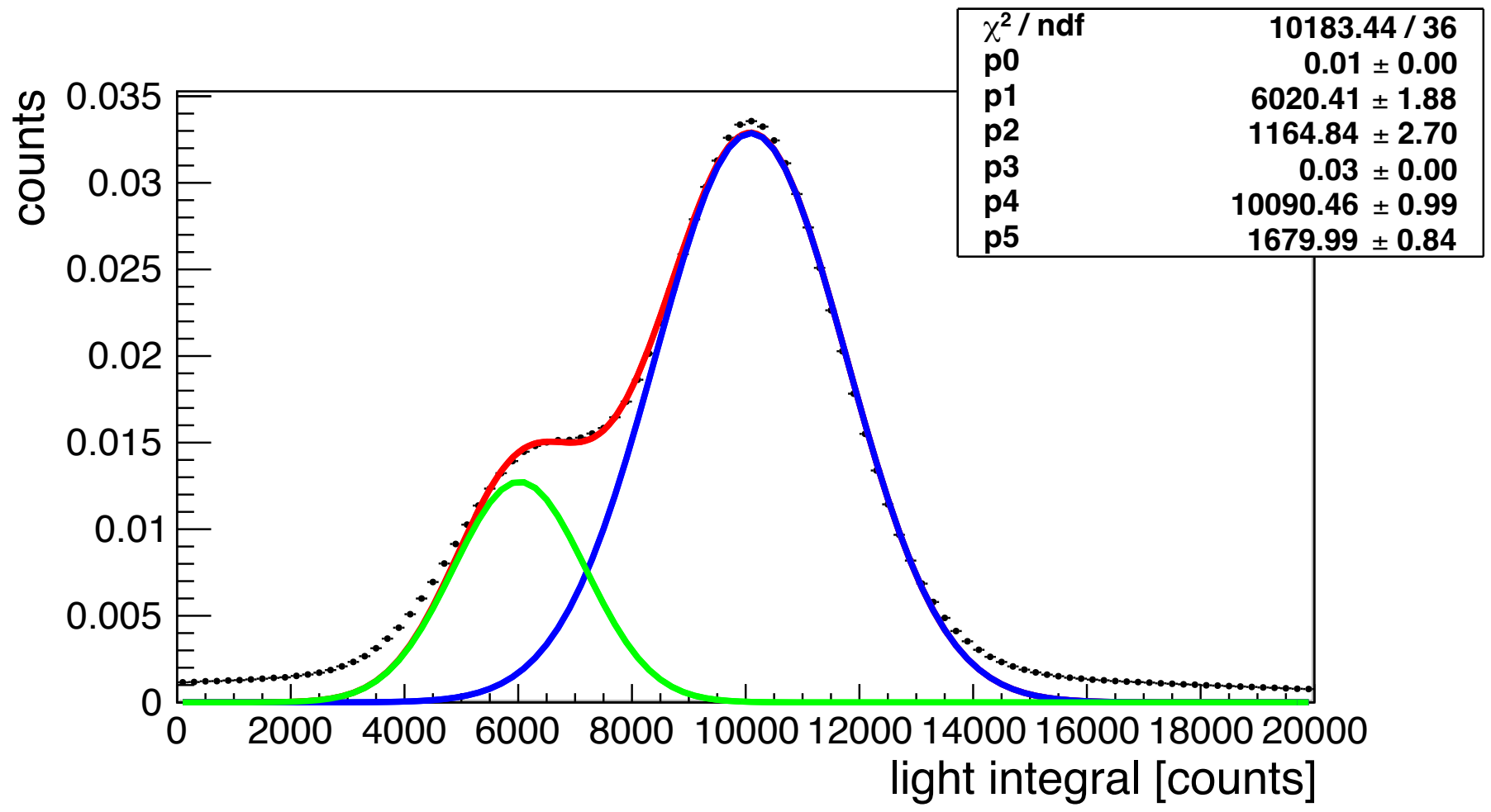

**Figure 6.12** The normalized integral distribution built summing all the contribution given by the five steps. In red the double Gaussian fit, in green the contribution given by the first peak, in blue the contribution given by the second one.

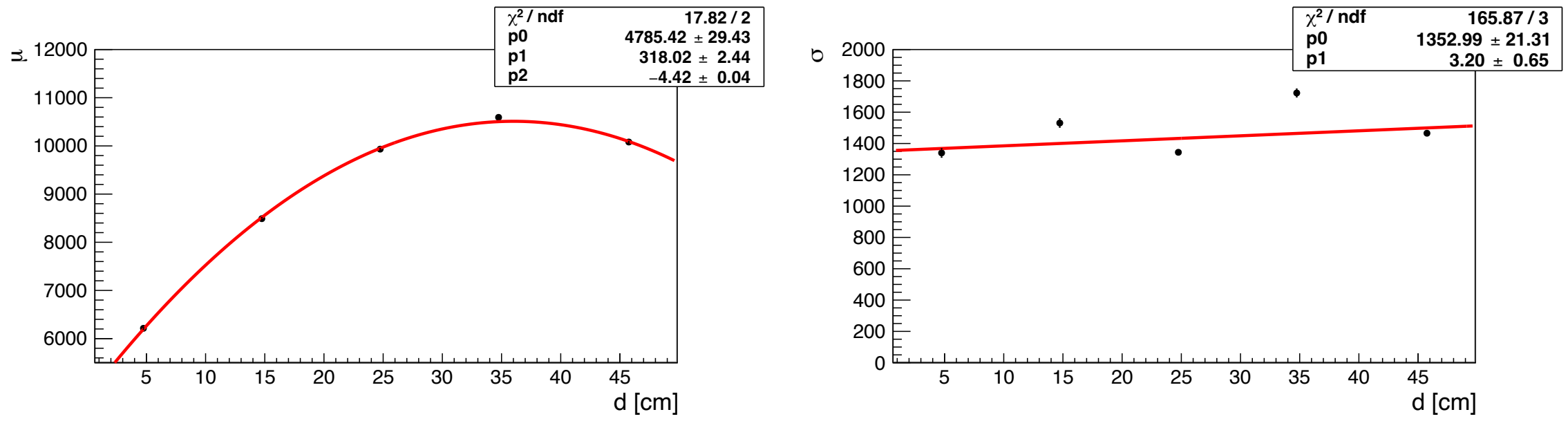

**(a)** **(b)**

**Figure 6.13** In 6.13a the mean given by the gaussian fit on the light integral distribution for the five steps, while in 6.13b the sigma. The uncertanty are given by the standard daviation.

and $\sigma$ at any given $z$.

Using the $z$-dependent $\mu$ and $\sigma$ values, the light integral of the spots is extracted from a Gaussian distribution for each $z$. The resulting normalized distribution of the light integral is shown in Fig. 6.14. To account for both the saturated and non-saturated components, a fit using a double Gaussian function is applied. The non-saturated component is found to have a width of 15% with a mean of 10 k counts.

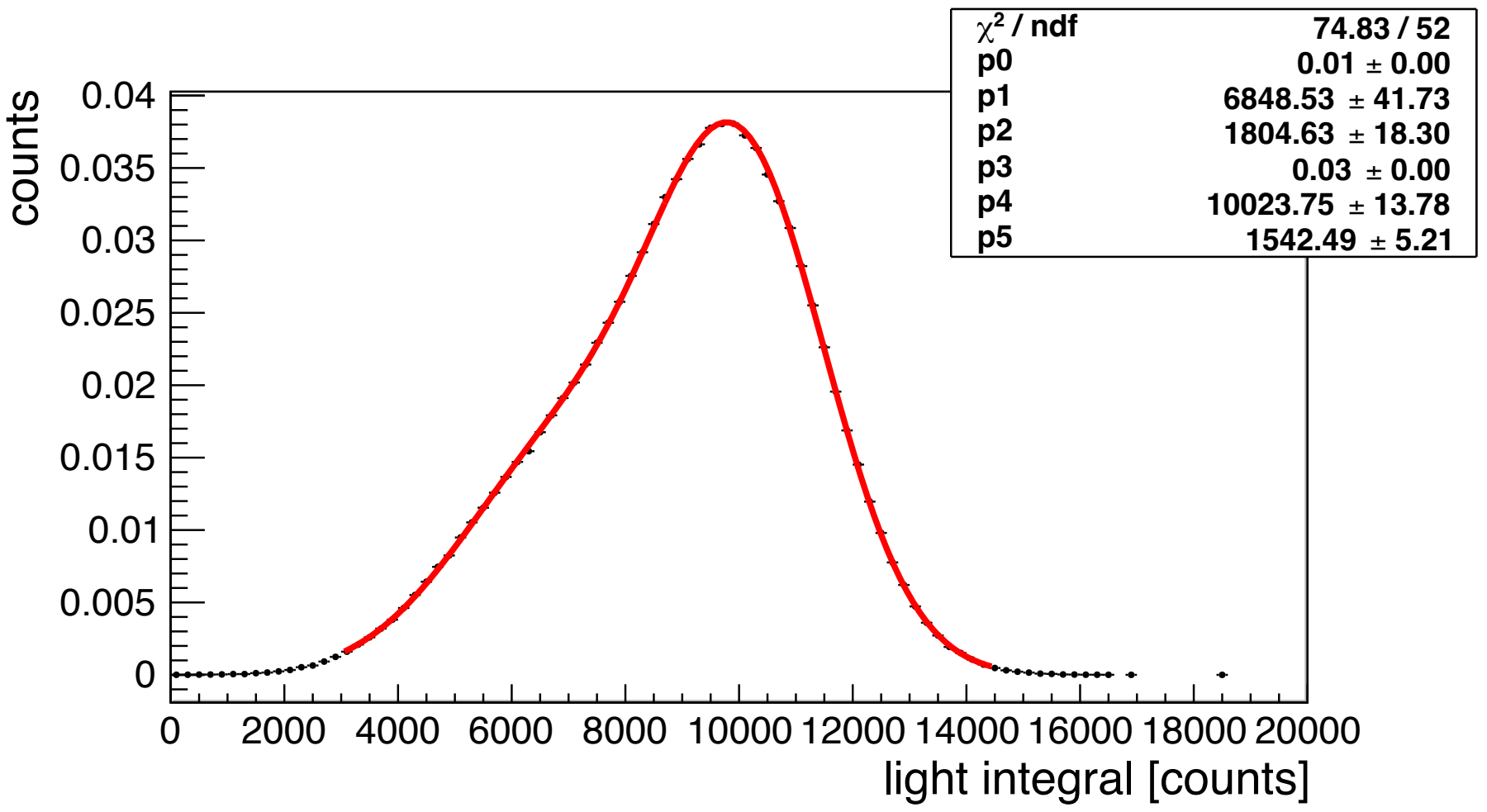

**Figure 6.14** The normalised generated light integral distribution.The fit with a double Gaussian is superimposed to account for both the saturated and non-saturated components.

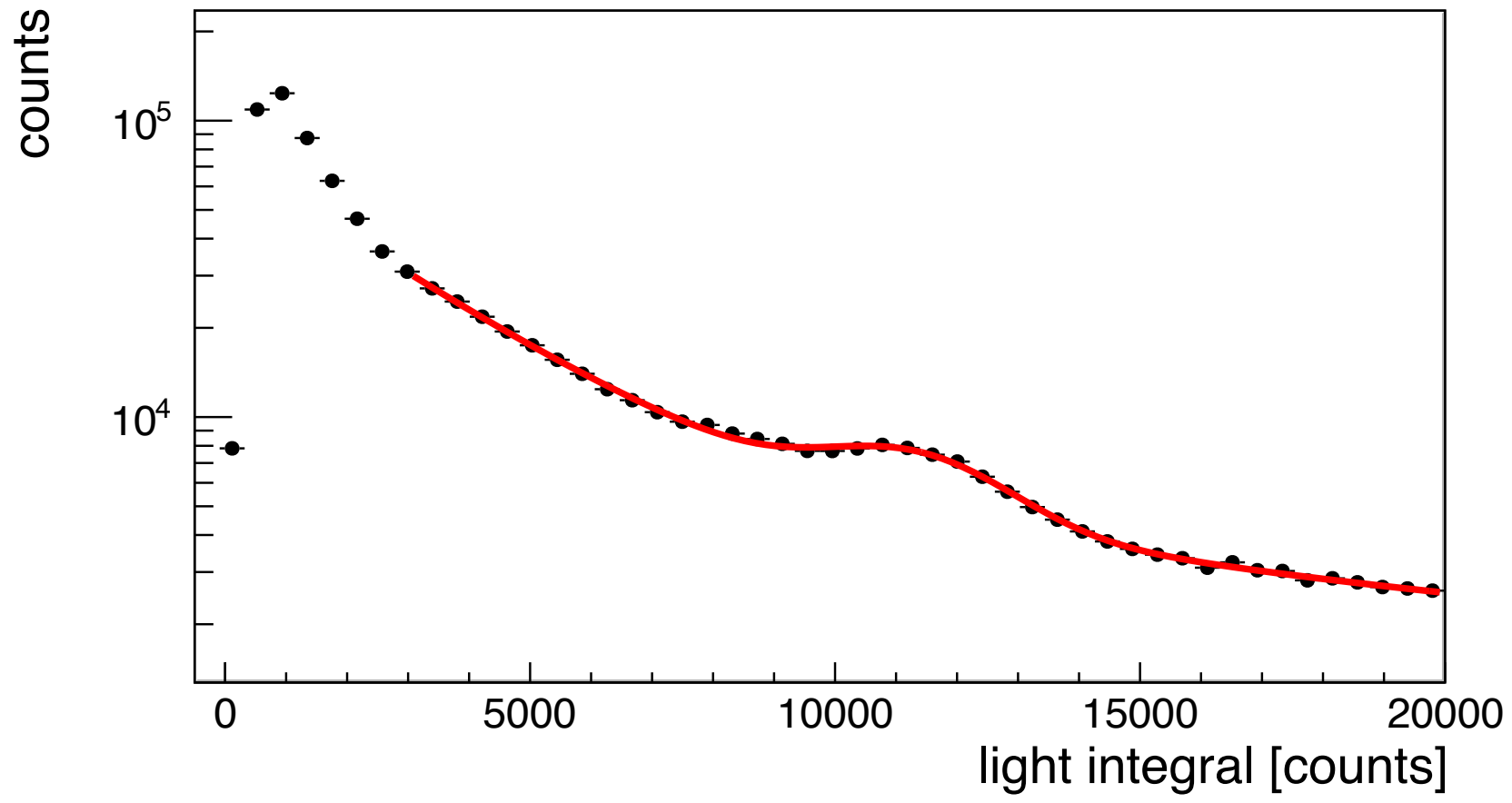

**Figure 6.15** The light integral distribution given by summing all the runs equalized where there is not the iron source.

## 6.3 Equalization of the Run3 dataset

After equalizing all the runs where no iron source was used, the light integral distribution is built, as shown in Fig. 6.15. A bump is clearly visible on a smoothly decreasing distribution. A fit combining an exponential and a Gaussian was performed, revealing a peak at $1.12 \times 10^4$ counts, with a 13 % width, which is consistent with what obtained with the toy Monte Carlo simulations. Knowing from the toy MC (Fig. 6.14) distribution that 10 k counts corresponds to the 5.9 KeV $^{55}$Fe peak, we compute that this peak corresponds to an energy of 7 keV, which can be attributed to the fluorence from the copper components of the setup. The copper events observed in the data are induced by particle interactions with the rings of the field cage and the shielding, both of which are made of copper. The copper-induced spots are uniformly distributed along the drift region.
The same copper contribution has been observed in data acquired with LIME during overground operation at LNF [101] and MC simulations (Section 4.6.1).

## 6.4 Application of the equalization method on the other Run

The same procedure used to evaluate the $^{55}Fe_{eq}$ has been applied to Run4, Run2 and Run1. For each run, the correction value has been evaluated to calculate the scaling factor.
By equalizing the light integral of each spot from the runs acquired with LIME underground at LNGS, where the iron source was positioned 25 cm from the GEMs, the resulting light

| Run | Width [%] |
|---|---|
| 1 | 12 |
| 2 | 12 |
| 3 | 13 |
| 4 | 13 |

**Table 6.2** The width of the iron peak after equalizing all the runs for each phase.

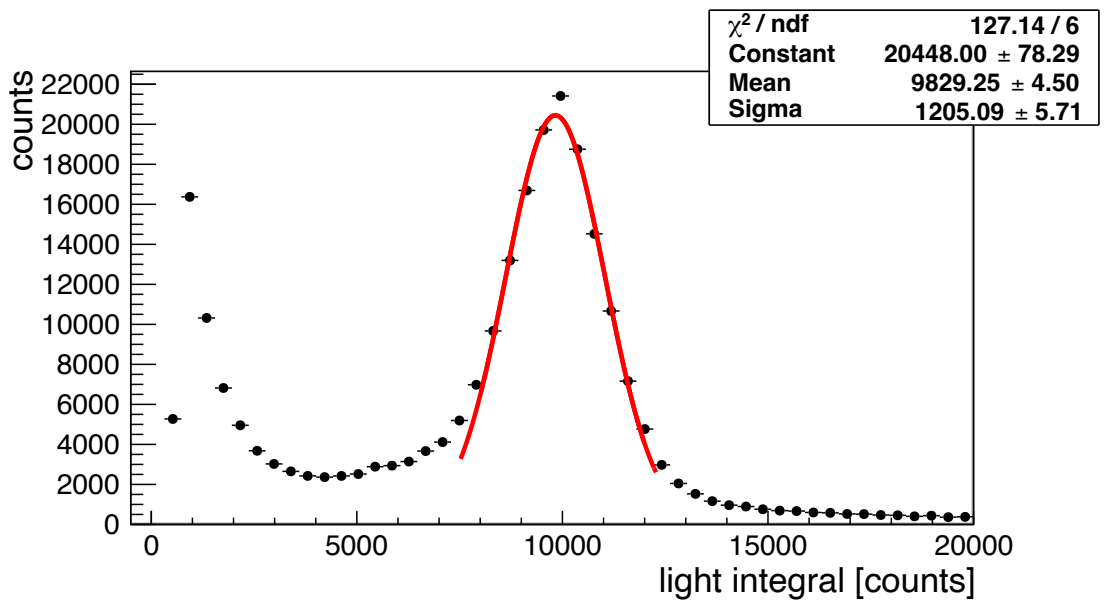

**(a)** Run1

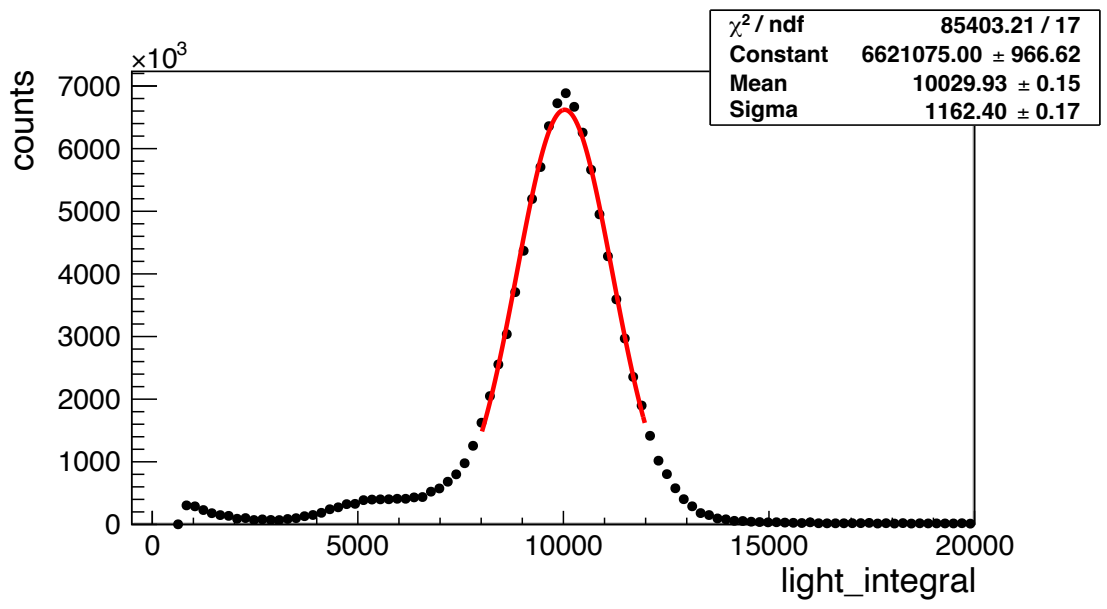

**(b)** Run2

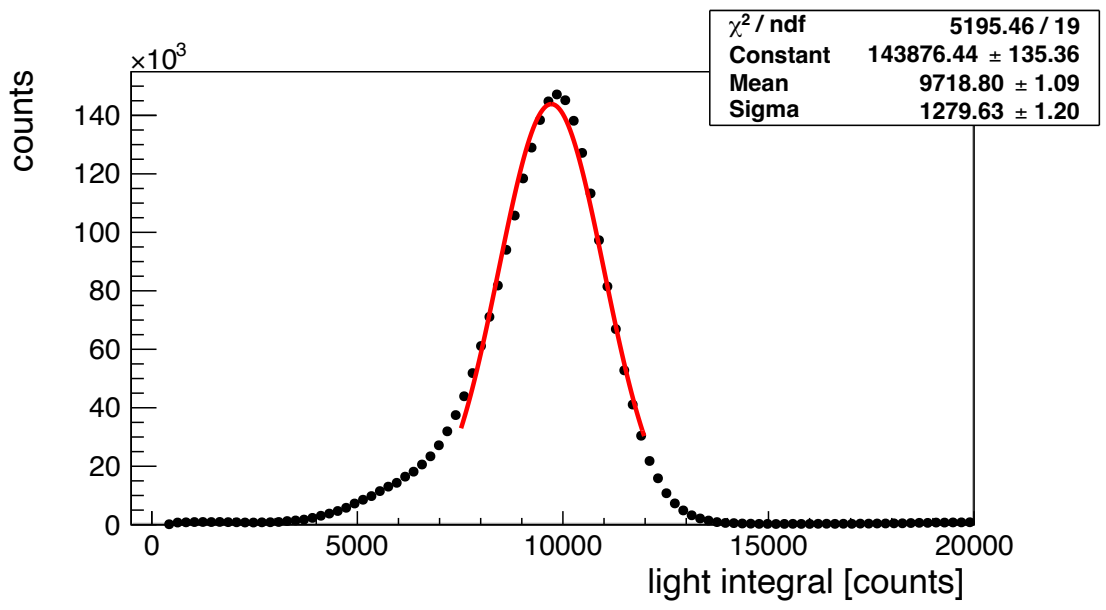

**(c)** Run4

**Figure 6.16** Equalized light integral distribution for the runs where the $^{55}Fe$ source is placed 25 cm from the GEMs, shown for Run1 (6.16a), Run2 (6.16b) and Run4 (6.16c).

integral distribution for Run4, Run2 and Run1 is build and it is shown in Fig. 6.16. A Gaussian fit is performed on each distribution, and the width of the iron peak for each phase is presented in Table 6.2.

Therefore the LY_30 variable allows for the equalization of all data, independently of environmental and operating conditions. After equalizing the data an energy resolution of approximately 12% has been demonstrated in each phase.

# 7 Preliminary studies in the search of a Dark Matter signal with LIME data

As already outlined, we were able to take data with the LIME prototype for a long period in stable conditions. With this data, it is interesting to start studying what we can achieve in the context of Dark Matter detection from all the available measurements. We know we cannot find any DM signal, being our exposure too small.
When no significant WIMPs induced nuclear recoils are observed above the expected background, it is only possible to set exclusion limits in the cross section versus WIMP mass parameter space. This chapter outlines the assumptions and methodologies used to evaluate these exclusion limits with the LIME prototype. The studies presented here are preliminary and represent the initial steps towards more in-depth analyses.
We know we cannot put more stringent exclusion limits with respect to the available ones presented in Section. 2.4.1.
It is the first attempt to put together all the different ingredients of the analysis to anticipate issues and potentiality.

## 7.1 Signal Model

In order to be able to compute an exclusion limit, we should be able to compute for a given WIMP mass value, the cross section resulting in a given number of events. This depends on several values and assumptions (the signal model) that will be described in this Section.
The energy region of interest [$E_{thr}$, $E_{max}$] defines the range in which Dark Matter interactions are expected to be detected. The upper energy limit, $E_{max}$, is set by the kinematics of WIMP-nucleus scattering for each element in the gas mixture. The limit is further constrained by the escape velocity of the Galaxy, $v_{esc}$, which sets the maximum recoil energy for a particular target element. The maximum energy, is given by:

$$E_{max} = \frac{1}{2} m_\chi r (v_{lab} \cos(\gamma) + v_{esc})^2 \tag{7.1}$$

where $r$ represent the efficiency of momentum transfer during recoil, which is maximized when the DM mass $m_\chi$ and the mass of the nucleus $m_A$ are equal, $v_{lab}$ is the velocity of the laboratory relative to the Galactic rest frame and $\gamma$ is the angle between the recoil direction and the laboratory's motion.
The energy threshold, E$_{thr}$, varies for each target material and it is influenced by the quenching effect, where part of the recoil energy is lost as non-visible ionization.
The expected number of DM-induced events for the Spin-Independent case, $N_{DM_{evt},i}$ is given by:

$$N_{DM_{evt},i} = tV\frac{P}{P_{atm}}\frac{T}{T_0}\rho_i\frac{N_0}{A_{mol,i}}\frac{2\rho_0\sigma_{n,SI}}{m_\chi^2 r_i}\frac{\mu_{A,i}^2}{\mu_n^2}A_i^2 I_i^{E\gamma}(m_\chi, E_{thr,i}) \tag{7.2}$$

where:

- $i$ represents the single monoatomic molecule;
- $t$ is the exposure time;
- $\rho_i$ is the gas density at atmospheric pressure and 0 degree Celsius;
- $V$ is the volume of the detector;
- $P$ and $P_{atm}$ are the working and atmospheric pressure of the gas;
- $T$ and $T_0$ are the working and reference temperature at (0°) expressed in Kelvin;
- $N_0$ is the Avogadro number;
- $A_{mol,i}$ is the molar mass of the gas;
- $\rho_0$ is the local DM density;
- $\mu_{A,i}$ is the reduced mass between the WIMP and the nucleus A;
- $\mu_n$ is the reduced mass between the WIMP and the nucleon;
- $A_i$ is the atomic mass of the nucleus;
- $I_i^{E\gamma}(m_\chi, E_{thr,i})$ is the integrated velocity distribution over energy and angle after the Radon transformation.

The integral form for the velocity distribution is:

$$I^{E\gamma}(m_\chi, E_{thr}) = \int_{E_{thr}}^{E_{max}} dE \int_{-1}^{1} d\cos(\gamma)\ S(E)\pi\frac{v_p^3}{v_{lab}}\alpha'\left(e^{-\frac{\left(\frac{\sqrt{2m_A E}}{2\mu_A} - v_{lab}\cos(\gamma)\right)^2}{v_p^2}} - e^{-\frac{v_{esc}^2}{v_p^2}}\right) \tag{7.3}$$

For a gas mixture composed of multiple molecular species, the total number of events is the sum of contributions from each element:

$$N_{DM_{evt}} = tV\frac{P}{P_{atm}}\frac{T_0}{T}\sum_{i}^{n_{mol}}\sum_{j}^{n_{el,i}}\rho_i k_i \frac{N_0}{A_{mol,i}}N_{at,i,j}\frac{2\rho_0\sigma_{n,SI}}{m_\chi^2 r_j}\frac{\mu_{A,j}^2}{\mu_n^2}A_j^2 I_j^{E\gamma}(m_\chi, E_{thr,j}) \quad (7.4)$$

where $k_i$ is the percentage of each molecule $i$ in the gas, $N_{at,i,j}$ is the number of atoms of $j$ in the molecule $i$, $n_{el,i}$ is the total number of elements in the gas $i$, and $n_{mol}$ the number of molecules composing the gas mixture.
The region of the DM velocity distribution, and thus the detectable DM masses, is limited at lower velocities by the detector's energy threshold E$_{thr}$ and at higher velocities by the Galactic escape velocity $v_{esc}$.
While, for the Spin-Dependent case, the expected number of DM-induced events is given by:

$$N_{DM_{evt},i} = tV\frac{P}{P_{atm}}\frac{T}{T_0}\rho_i\frac{N_0}{A_{mol,i}}\frac{2\rho_0\sigma_{p,SD}}{m_\chi^2 r_i}\frac{\mu_{A,i}^2}{\mu_n^2}\frac{4\langle S_p\rangle(J_i+1)}{3J_i}I_i^{E\gamma}(m_\chi, E_{thr,i}) \quad (7.5)$$

where $J$ is the nuclear spin, $\langle S_p\rangle$ is the expectation value of the proton spin inside the nucleus.

### 7.1.1 Estimation of the exposure time

As described in Section 4.3.2, the sCMOS sensor is open for a total of 484 ms, of which 184 ms are required to open the sensor, and 30 ms are needed to acquire the image. Consequently, the camera for the 40% of the time is not total efficient, implying an efficiency $\varepsilon$ of 0.60.
Even though the data from the PMTs are not utilized in the current analysis, the four PMTs operate in parallel, with each signal acquisition taking 0.01 s.
The effective exposure time $T$ for each run can be defined as:

$$T = (dt - 0.01n)\varepsilon \quad (7.6)$$

where $dt$ is the total duration of the run and $n$ is the number of PMT signals acquired.

### 7.1.2 Quenching Factor

The gas mixture, in the LIME prototype, consists of 60% helium and 40% carbon tetrafluoride at atmospheric pressure. Each element in this gas mixture contributes differently to the overall signal model due to their distinct kinematics and momentum transfer characteristics. When evaluating the signal model, it is critical to account for each element's individual contributions

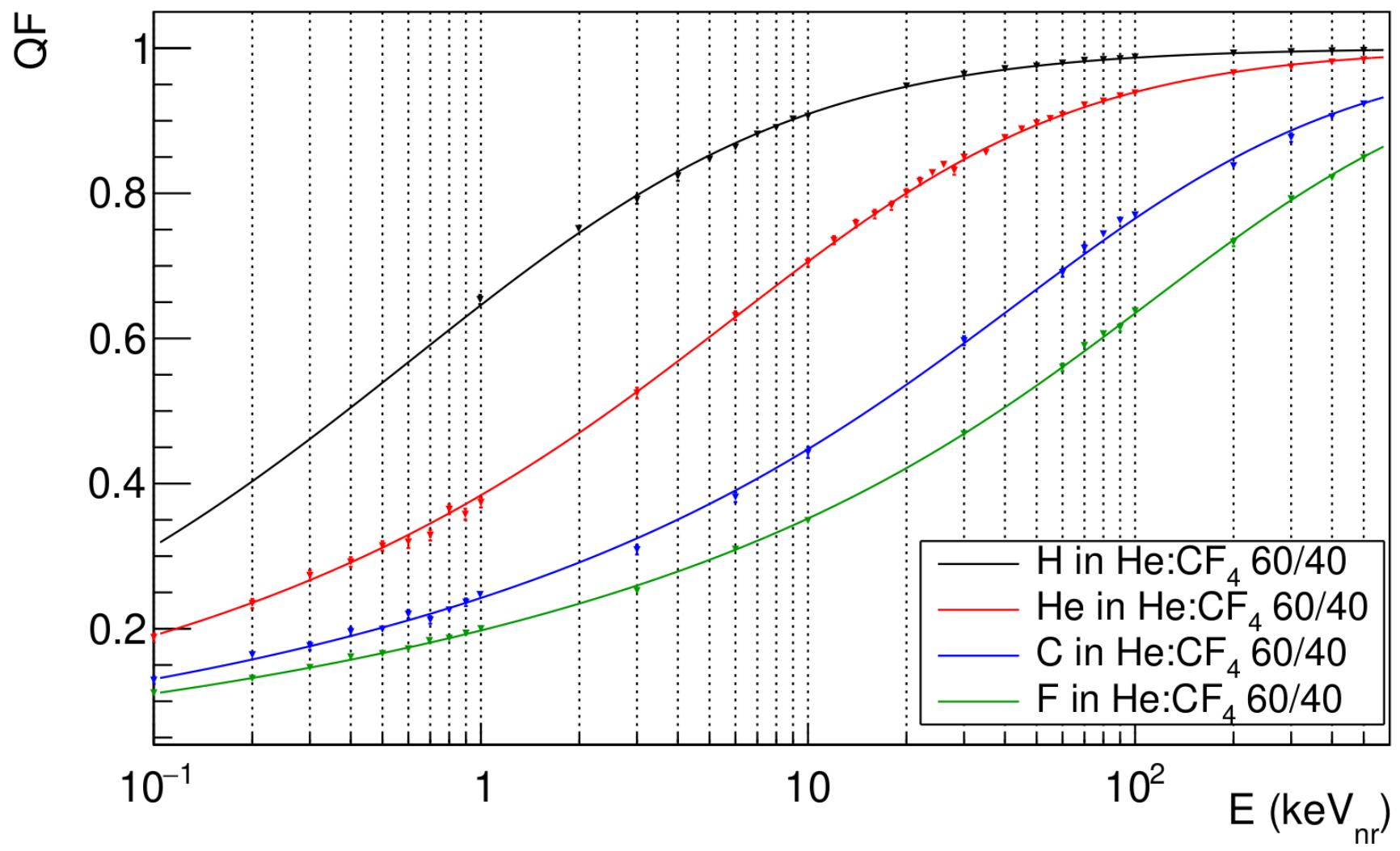

**Figure 7.1** Quenching factor as a function of nuclear recoil energy for the different target elements in the gas mixture, as simulated using SRIM software [114].

accurately.

Nuclear recoils deposit energy differently than electron recoils, especially in the low O(1) keV energy regime, relevant for direct DM searches. While electrons efficiently transfer their kinetic energy to gas ionization, low-energy nuclear recoils primarily lose energy through two mechanisms: inelastic Coulomb interactions with atomic electrons and elastic scattering in the screened field of nuclei.

The quenching factor (QF) is used to describe the fraction of kinetic energy released through ionization by a nuclear recoil. The relationship between the observed nuclear recoil energy $E_{ee}$ and the total kinetic nuclear recoil energy, $E_{nr}$, is given by:

$$E_{ee} = QF \times E_{nr} \tag{7.7}$$

The QF varies for each element in the gas mixture, affecting how much nuclear recoil energy can be detected via ionization. The collaboration estimated the QF for each element using SRIM simulations, with the results presented in Fig. 7.1. Simulations were performed also for hydrogen in view of the potential addition of H in the gas mixture for the CYGNO-30 setup. The QF as a function of nuclear recoil energy is modeled by the equation:

$$QF = \frac{a(E_{nr} + bE_{nr}^{c})}{1 + a(E_{nr} + bE_{nr}^{c})} \tag{7.8}$$

The QF differs for each nucleus, the effective threshold on the NR energy $E_{thr,nr}$, and the angular distribution will be affected differently for each target element. The minimum DM mass that can be detected by a particular element depends on its effective recoil energy threshold. This relationship allows for the evaluation of the minimum WIMP mass to which the detector is sensitive, assuming that $\cos(\gamma)$ = 1 and replacing $E_{max}$ with $E_{thr,nr}$. Table 7.1 summarizes the effective nuclear recoil energy threshold and the corresponding minimum detectable DM mass for each element in the gas mixture for two thresholds examined in this work. Lighter elements, such as hydrogen (H) and helium (He), are more sensitive to low-mass WIMPs due to better momentum transfer and higher QF, enabling the detector to probe lower DM mass ranges. Lowering the energy threshold enhances the experiment's sensitivity to lower-mass WIMPs. Lighter target elements not only allow better momentum transfer with low-mass WIMPs but also provide higher QF values, increasing the recoil energy and making it easier to detect smaller WIMP masses.

| | 1keV$_{ee}$ | | 1.5keV$_{ee}$ | |
|---|---|---|---|---|
| **Element** | **E$_{thr,nr}$ [keV$_{nr}$]** | **Min m$_{\chi}$ [GeV/c$^2$]** | **E$_{thr,nr}$ [keV$_{nr}$]** | **Min m$_{\chi}$ [GeV/c$^2$]** |
| He | 2.1 | 1.0 | 3.4 | 1.3 |
| C | 3.1 | 1.9 | 5.5 | 2.6 |
| F | 3.8 | 2.5 | 6.8 | 3.5 |

**Table 7.1** Summary for all the elements and the energy thresholds, of the effective threshold in nuclear recoil energy and the corresponding minimum DM mass detectable.

## 7.2 Estimation of the data samples

The experiment consists in counting the number of detected events. For this preliminary study, we will not exploit energy spectra nor direction of the observed events.
In this analysis, the data set used belongs to Run4, where 10 cm of copper and 40 cm of water shielding is installed. A total of 5394 runs, consisting of 2157600 images, were analyzed. For each image, as outlined in Section 5.3.1, the number of events was evaluated after applying the following selection criteria:

- fiducial cuts, only tracks falling within a predefined rectangular region are accepted, ensuring a more uniform detector response:
  - sc_xmin [pixel] $>$ 300;
  - sc_ymin [pixel] $>$ 300;
  - sc_xmax [pixel] $<$ 2000;
  - sc_ymax [pixel] $<$ 2000;
- quality cuts:

- sc_rms [counts] $> 6$ to avoid the sensor nosily pixel;
- $0.005 < \rho < 0.15$ where $\rho$ is defined as:

$$\rho = \frac{\text{sc_rms [counts]}}{\text{sc_nhits [counts]}} \tag{7.9}$$

This cut helps to filter out the MIP;

- sc_integral [counts] $> 1500$ or $> 2500$ if the energy threshold is 1 keV$_{ee}$ or 1.5 keV$_{ee}$, respectively.

In the hypothesis that DM interactions have not been observed in this set of data, the $20\%$ of the runs were randomly selected and used to estimate the background, while the remaining $80\%$ were used as the main data set for event analysis. This approach ensures that the background level is properly accounted for while maintaining the majority of the data for evaluating our exclusion limit.

## 7.3 Statistical approach

For the statistical analysis of the data, I decided to exploit a Bayesian approach.
In the Bayesian approach, boundary conditions or prior knowledge can be incorporated into the calculation by introducing prior probabilities, which reflect the information available before performing the experiment. The method is based on the Bayes theorem:

$$p(A|B) = \frac{p(B|A)p(A)}{p(B)} \tag{7.10}$$

This means that the probability of event $A$ occurring, given event $B$, is equal to the probability of event $B$ occurring given $A$, multiplied by the prior probability of $A$ and normalized by the total probability of $B$.
Let's assume a generic set of $n$ independent quantities $\vec{x}$ that described the state of a system or the outcome of a measurement. The probability density function $p(\vec{x})$ satisfies:

- $p(\vec{x})d\vec{x}$ gives the normalized probability of the system;
- $\int_D p(\vec{x})d\vec{x} = 1$, where D is the domain of $p$.

The Bayesian approach allows to compute the probability of a model given certain available information:

$$p(\vec{\mu},\vec{\theta}|\vec{x},H) = \frac{p(\vec{x}|\vec{\mu},\vec{\theta},H)\pi(\vec{\mu},\vec{\theta}|H)}{\int_\Omega \int_D p(\vec{x}|\vec{\mu},\vec{\theta},H)\pi(\vec{\mu},\vec{\theta}|H)d\vec{\mu}d\vec{\theta}}, \tag{7.11}$$

where:

- $\vec{\mu}$ represents the vector of free parameters of interest;
- $\vec{\theta}$ denotes the vector of nuisance parameters, describing experimental conditions or theoretical assumptions affecting the results but not directly of interest;
- $\vec{x}$ is the vector representing the data set;
- $H$ is the hypothesis under test;
- $\Omega$ is the nuisance parameter space;
- $D$ is the parameters space of $\vec{\mu}$;
- $p(\vec{\mu}|\vec{x})$ is the posterior probability function for the parameters $\vec{\mu}$ given $\vec{x}$;
- $p(\vec{x}|\vec{\mu},\vec{\theta},H)$ is the Likelihood function, describing how likely the data is to have been produced by a particular set of parameters;
- $\pi(\vec{\mu})$ represents the prior probability of the parameters, incorporating previous knowledge or constraints.

When the true value of a parameter is consistent with zero, our case for the expected number of DM interactions, the 90% credible interval (C.I.) is used to define the upper bound. This means that, with 90% probability, the true value of $\mu$ lies below the upper limit. The definition is:

$$\mu(90\%\text{C.I.}) = \int_0^{\mu(90\%\text{C.I.})} p(\mu|\vec{x},H)d\mu = 0.9 \tag{7.12}$$

where $p(\mu|\vec{x},H)$ is the posterior probability marginalized over the nuisance parameters, given by:

$$p(\mu|\vec{x},H) = \int_\Omega p(\mu,\vec{\theta}|\vec{x},H)d\vec{\theta} \tag{7.13}$$

### 7.3.1 Signal prior

The signal is the number of the interactions observed in the data. Signal prior is chosen to be a uniform distribution because we do not have any hint of the value of the DM cross-section.

### 7.3.2 Background prior

In Galactic coordinates, the angular distribution of background events can be reasonably assumed to be flat. Any localized background source that is not initially isotropic with respect to the detector will be diluted and smeared due to the Earth's rotation, ultimately resulting in an approximately uniform distribution. Unlike WIMP-induced events, which show a directional signature, background events that cause nuclear or indistinguishable electron recoils (ERs) in the detector do not alter their flat angular distribution. The probability of each target element recoiling due to background events is therefore assumed to be identical. The number of background events measured in the selected samples would follow a Poissonian distribution. However, as described in Section 7.2, it has been rescaled to match the signal exposure time. This implies that the background uncertainty should be correctly scaled by the ratio data to background exposure. Given that for a larger number of events a Poissonian distribution can be approximated with a Gaussian distribution, it is used the latter for the background prior to set the proper sigma values.

### 7.3.3 Likelihood

According to the Bayesian Networks (BN) method, the probability density function of a collection of random variables can be visually depicted as a network of nodes and arrows, where:

- The nodes represent the variables;
- A solid arrow between two node represents a probabilistic link between the two variables;
- A dashed arrow between two nodes represents a deterministic link between the two variables;
- A grey node indicates the corresponding variable has been observed.

Fig. 7.2 illustrates the BN for the measurement being examined. Here, $x$ denotes the observed data, $\mu$ represents the intensity of the total measured events, $\mu_b$ regulates the intensity of the background contribution and $\mu_S$ represents the signal strength parameter we want to determine. The objective is to infer $p(\mu_s|x)$, the *pdf* of $\mu_s$, conditioned on the observed data $x$. The Likelihood function for evaluating the confidence interval (C.I.) in this rare event analysis with a background contribution is defined as the Poisson distribution:

$$\mathcal{L}(\vec{x}|\mu_s,\mu_b,H) = \frac{(\mu_b+\mu_s)^{N_{\text{evt}}}}{N_{\text{evt}}!}e^{-(\mu_b+\mu_s)} \tag{7.14}$$

where, $N_{\text{evt}}$ is the observed total number of events, $\mu_s$ is the signal contribution from WIMP interactions, and $\mu_b$ is the expected background contribution.

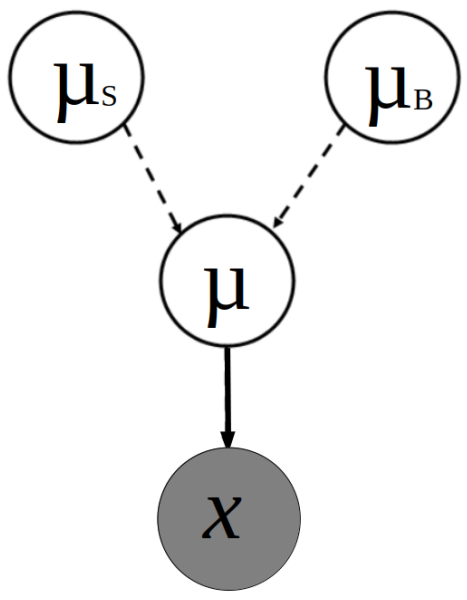

**Figure 7.2** Bayesian Network for a Poisson process with intensity $\mu$ and observed number of events $x$. The insentisy $\mu$ is the sum of the signal $\mu$ and background $\mu$ contribution.

## 7.4 Dark Matter exclusion limit estimation

By estimating the upper limit on WIMP-induced events $\mu_s$ from the 90% C.I. it becomes possible to calculate the upper limit on the WIMP-nucleon elastic cross-section using Eq. 7.2 for the SI case and Eq. 7.5 for the SD case, assuming $N_{DM_{evt}} = \mu_{s,90\%}$.
For this evaluation, the Bayesian Analysis Toolkit (BAT) [116; 117] is employed. BAT is a C++ software package designed to address statistical problems within the framework of Bayesian inference. Utilizing Bayes' Theorem, it implements Markov Chain Monte Carlo (MCMC) techniques, which allow access to the full posterior distribution, enabling comprehensive parameter estimation, limit setting, and uncertainty quantification.
A total of 17 days of exposure time is taken in exam. During this period, the pressure remained nearly constant at around 0.907 bar. The study for both SI and SD interactions was conducted using two energy thresholds ($E_{\text{thr}}$) of 1 keV$_{ee}$ and 1.5 keV$_{ee}$. As reported in Table. 7.1, higher thresholds increase the minimum detectable Dark Matter mass. Two scenarios for detector efficiency ($\varepsilon$) were examined: one assuming full efficiency of the LIME prototype and another assuming 0.8 efficiency. Reduced efficiency leads to less restrictive limits on the WIMP-nucleus elastic cross-section. The analysis also explored background evaluation uncertainty by considering also the case where the background sigma is 1.5 times larger.
Figure 7.3 shows the exclusion limits for the SD case. A comparison with the results from the DRIFT experiment [118], which employed similar technologies is reported.
Figure 7.4 presents the SI exclusion limits.
It is important to emphasize that this work shows preliminary results on the detector's sensitivity. Some assumptions needs to be reconsidered carefully to give a more solid result; for example, we are assuming that not DM signal is observed and that background can be computed from data. Also, the overall analysis can be refined to improve our performance. Still, the results we obtain are comparable with other encouraging findings.

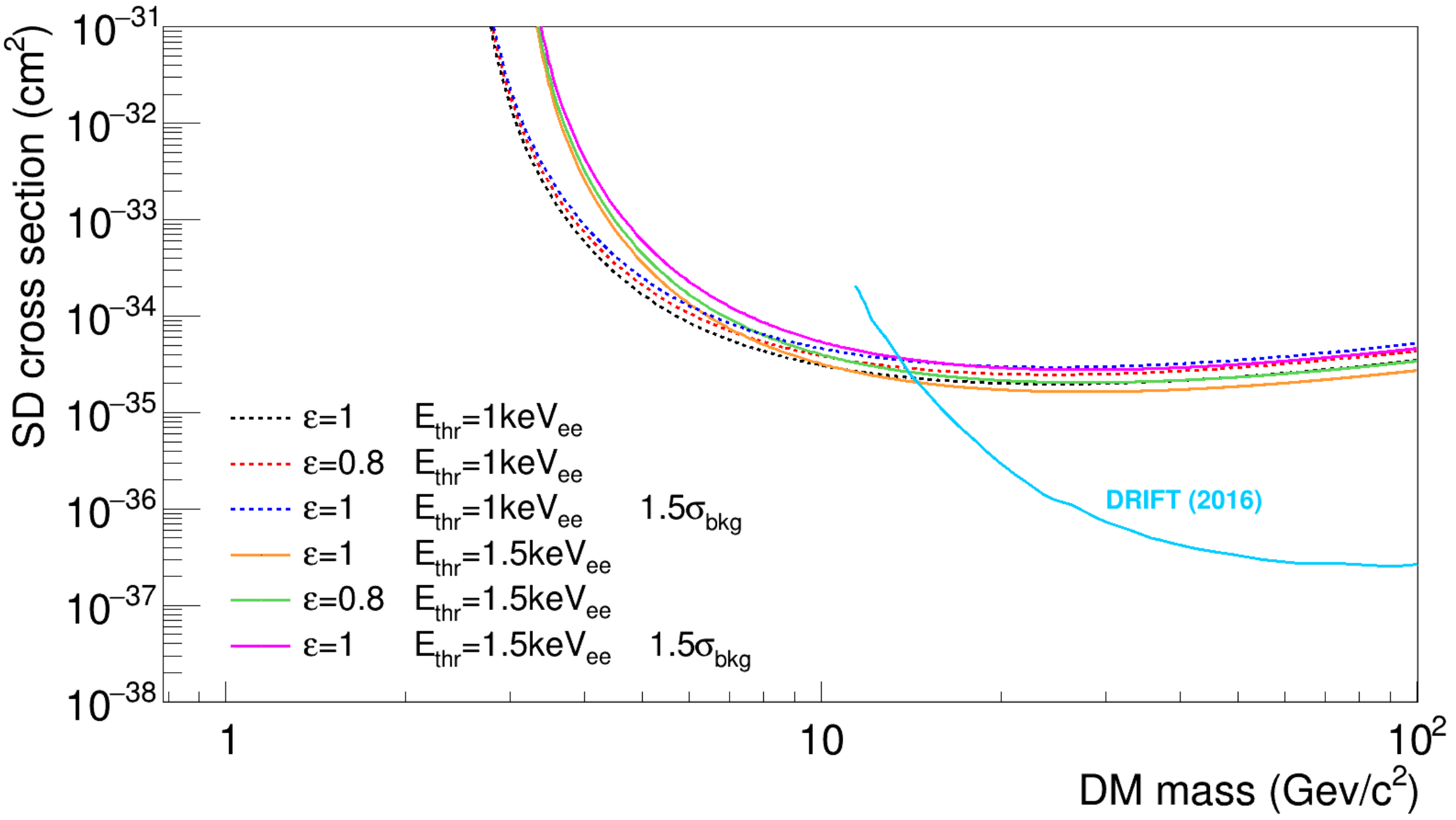

**Figure 7.3** Spin-Dependent 90% C.I. exclusion limits on the WIMP-proton cross-section for the LIME prototype, based on 17 days of exposure. Limits are shown for two energy thresholds, 1 $keV_{ee}$ and 1.5 $keV_{ee}$, and evaluated with two different detector efficiencies and background dispersions.

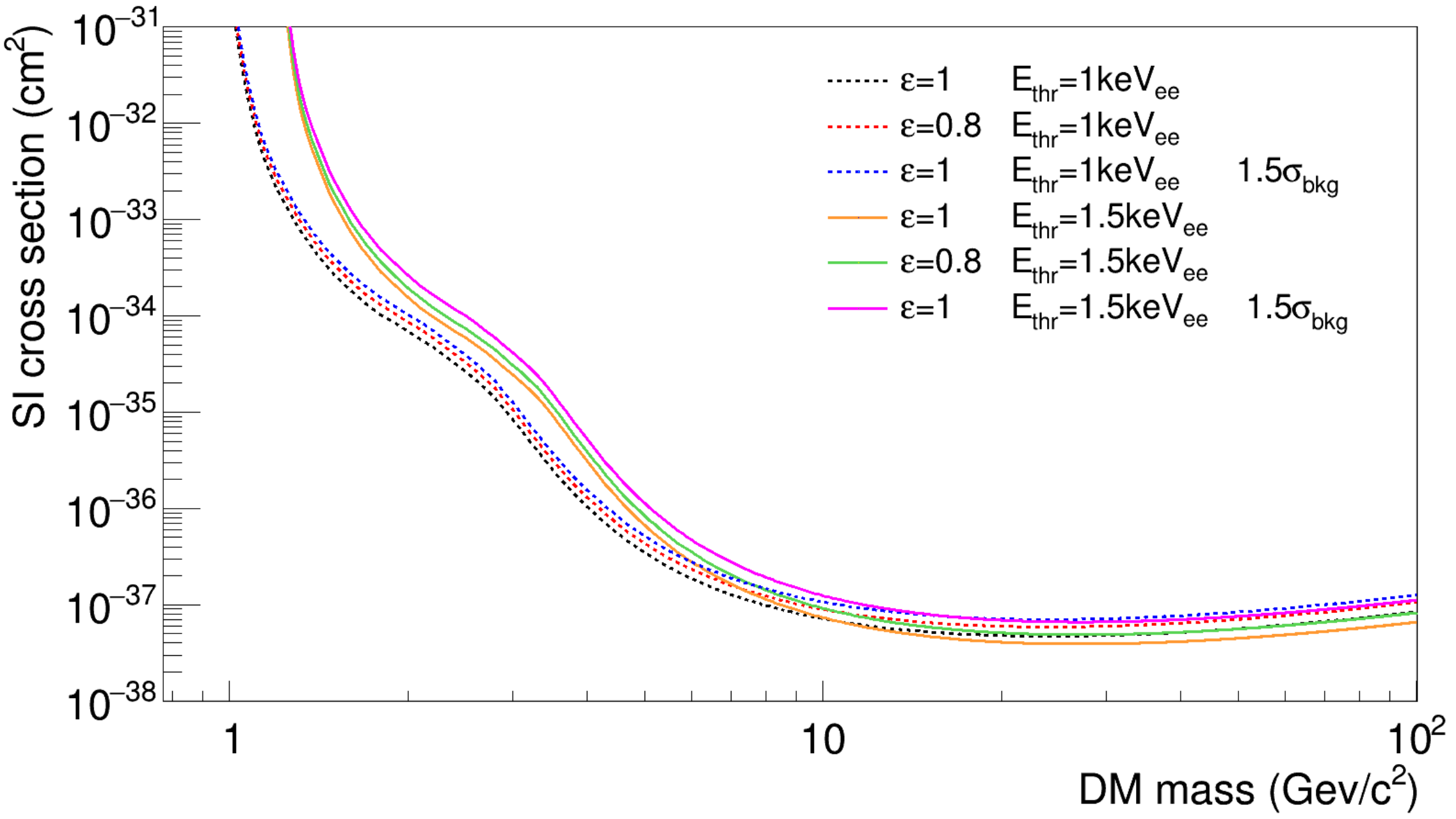

**Figure 7.4** Spin-Independent 90% C.I. exclusion limits on the WIMP-proton cross-section for the LIME prototype, based on 17 days of exposure. Limits are shown for two energy thresholds, 1 $keV_{ee}$ and 1.5 $keV_{ee}$, and evaluated with two different detector efficiencies and background dispersions.

# Conclusions

Over the past century, astrophysical and cosmological observations have indicated that our understanding of the Universe is incomplete. The Standard Model (SM) does not account for a substantial portion of unseen matter, referred to as Dark Matter (DM). One of the leading hypotheses suggests that DM is made up of particles distinct from those in the SM, with Weakly Interactive Massive Particles (WIMPs) being among the most promising candidates. Detecting Dark Matter particles represents one of the most difficult challenges in modern physics, leading to the development and pursuit of various experimental techniques. In the search for high-mass WIMPs, ton-scale detectors with the heavy-nuclei targets have become the predominant approach. On the other hand, the largely unexplored low-mass WIMP region requires a detector approach allowing sensitivity to extremely low energy deposits, for which the background rejection becomes increasingly difficult.
The use of gaseous TPC with directionality capability can not only break the limit of the neutrino fog, but also allows for a stronger identification of a DM signal due to the highly anisotropic nature of WIMP-induced recoils in the Galactic reference frame, that no background can mimic. The combined motion of the Sun and Earth produces a distinctive directional flux of WIMPs on Earth, creating a dipole-like angular distribution of nuclear recoils (NRs). By varying gas mixtures, these detectors can enhance sensitivity to both low- and high-mass WIMPs, while target nuclei with an odd number of protons or neutrons offer sensitivity to both Spin-Independent (SI) and Spin-Dependent (SD) WIMP-nuclei interactions.
The CYGNO project is developing gaseous TPCs with optical readout systems operating at atmospheric pressure, aiming to build a large-scale detector with a sensitive volume of O(30) $m^3$ for DM detection. Ionization signal amplification is achieved using a triple-GEM stack, enabling the detector to reach low energy thresholds O(1) keV. Secondary scintillation light emitted in electrons multiplication process is detected simultaneously by sCMOS cameras and PMTs, providing detailed measurements of energy deposition along the 3D path of each event.

An extensive R&D campaign led to development of the LIME prototype, making a pivotal step in the CYGNO project. LIME is a 50 L TPC with a 50 cm drift length and a $33 \times 33$ $cm^2$ triple-GEM amplification stage, read by an sCMOS camera and four PMTs operating with a He:$CF_4$ (60:40) gas mixture at atmospheric pressure. Initial testing and characterization of LIME took place over several months at LNF, where charge gain, light yield, and energy response were studied, including gain variation with pressure changes. Following this commissioning phase, LIME was installed underground at LNGS in February 2022, within a tunnel linking Halls A and B.
The main goal of LIME at LNGS includes validating Monte Carlo (MC) background simulations and performing spectral and directional measurements of the neutron flux within the LNGS environment. To achieve these goals, each requiring distinct configurations and shielding levels, the LIME underground program has been phased with progressively added shielding layers around the detector. This phase represents a milestone for CYGNO, marking its first underground prototype trial, essential for scientific performance studies and evaluations of costs and logistics. During this phase, critical systems, including DAQ, gas handling, high-voltage supply, and data quality monitoring, were developed and tested to support LIME's underground operations.
In this thesis, I analyzed the gain dependence on pressure variations for LNF data [101]. I further refined the analysis, extending it to assess gain variation in the underground environment. A 0.6% decrease in gain per mbar increase in pressure was observed when increasing the pressure, with this result remaining stable across different periods. This analysis was conducted for various gas flow rates; with low gas flow, specifically at 1 l/h, instabilities were observed, affecting the data because of impurities.
Starting with Run3, the gas recirculation system was activated and a humidity sensor was installed to monitor the relative humidity (RH) inside LIME. During this period, the pressure remained nearly constant while variations in relative humidity showed a 20% decrease in light yield for each unit increase in RH. This effect is attributed to impurities, such as $H_2O$, which are capable of electron attachment, thus reducing signal intensity.
To analyze data over extended periods, merging is essential, with careful consideration given to detector calibration consistency and response equalization over time. I developed an innovative data equalization method using signals from high-energy deposit events. This procedure equalizes all data independently of the environmental condition. An energy distribution constructed by equalizing data from runs without the iron source revealed a peak associated with copper, matching observations from both overground data and MC simulations. This peak is due to fluorescence from the copper components of the setup.

In Chapter 7, using the set of runs acquired with 10 cm of copper and 40 cm of water for external background shielding, I conducted a tentative Dark Matter search on 17 days of data taking. Given the small volume of LIME, the most significant achievable result is an exclusion limit on the cross-section. This work can be considered a promising preliminary result for future studies with data acquired using the CYGNO-04 demonstrator.